# Multi-Equalization in Conceptual Density Functional Theory: Beyond Electronegativity Equalization.

Jesús Sánchez-Márquez[1].

[1] Departamento de Química-Física, Facultad de Ciencias, Campus Universitario Río San Pedro, Universidad de Cádiz, 11510 Puerto Real, Cádiz, Spain.

*To whom correspondence should be addressed. E-mail: jesus.sanchezmarquez@uca.es

**Abstract**

The electronegativity equalization principle provides a simple framework to describe charge redistribution in molecules, yet its conventional formulation is inherently limited to a first-order description based on the equalization of the chemical potential. In this work, we introduce a generalized framework, termed *multi-equalization*, which extends this concept by incorporating higher-order responses of the electronic structure within the context of Conceptual Density Functional Theory.

The proposed approach is based on a flexible representation of molecules as sets of electron density partitions, allowing different electronic descriptions (e.g., atomic densities or localized orbitals) to be treated within a unified formalism. Within this framework, we demonstrate that correlations between energy derivatives and electron density derivatives with respect to the number of electrons naturally lead to the simultaneous equalization of multiple descriptors, including the chemical hardness and Fukui indices.

A constructive algorithm is introduced to determine the *ideal set of density partitions* that satisfies the multi-equalization conditions. This multi-equalization scheme provides a consistent description of both global charge transfer and local reactivity, overcoming the intrinsic limitations of traditional electronegativity equalization models. In particular, the inclusion of density response functions enables the emergence of local hardness equalization, introducing spatial resolution into the description of chemical reactivity. Under multi-equalization, defined as the selection of an *effective* set of density partitions, local reactivity descriptors become constrained functionals of the electron density defining the global electronic state of the system, reflecting the imposed equalization conditions.

The proposed framework establishes a deeper connection between charge equalization models and the formal structure of density functional theory, offering a theoretically grounded route toward improved predictions of molecular reactivity.

---

# 1. INTRODUCTION

## 1.1. Sanderson's principle: Electronegativity Equalization.

In 1951, R. T. Sanderson[1] introduced the first proposal of the equalization principle, in which he already assumes that electronegativity is equalised in the formation of the bond, In 1970, Hooydaak and Z. Eeckhaut proposed its use in chemical bond formation[2,3] and later, in 1988, Sanderson proposed some ideas[4] in relation with the principle of equalization, about what happens when different atoms combine to form a compound, namely that all the atoms adjust to the same intermediate electronegativity. He also describes a mechanism by which this equalization can occur. Sanderson points out that a widely used postulate of the principle of electronegativity equalization is that the intermediate electronegativity within the compound is the *geometric mean* of all the initial atomic electronegativities[5,6]. Finally, Parr[7,8] showed that electronegativity is the negative of the chemical potential, so the equalization principle corresponds to the equalization of the chemical potentials in a compound.

## 1.2. Global reactivity descriptors. Electrophilicity.

As is well known, the energy change ( $\Delta E$ )[9] due to the electron transfer ( $\Delta N$ ) satisfies the equation **Eq. (1)**.

$$\Delta E = \mu \, \Delta N + \frac{1}{2} \eta \, \Delta N^2 \qquad (1)$$

where $\mu$ and $\eta$ are the chemical potential (negative of the electronegativity) and chemical hardness, respectively defined by $\mu = \left( \frac{\partial E}{\partial N} \right)_\upsilon$ and $\eta = \left( \frac{\partial^2 E}{\partial N^2} \right)_\upsilon$, with $\upsilon(r)$ as the external potential. The Hohenberg-Kohn theorem can be extended to the fractional electron number[10] and the derivative discontinuity for conceptual DFT is discussed in references[11,12]. Using the finite difference method and according to Mulliken[13-17], working formulas for $\mu$ and $\eta$ may be given as: $\mu = -\frac{1}{2}(IP + EA)$ and $\eta = (IP - EA)$, where *EA* and *IP* are the electron affinity and the ionization energy. If the electrophile withdraws enough charge[18,19], it will be become saturated with electrons, $(dE/dN) = 0$, leading to the maximum amount of electron charge: $\Delta N_{\text{max}} = -\mu/\eta$, and a total energy reduction: $\Delta E_{\text{min}} = -\mu^2/2\eta$. The electrophilicity index ($\omega$) was introduced by Parr et al.[20-24] as the stabilization energy when atoms or molecules in their ground states acquire $\Delta N_{max}$ electronic charge from the environment. Finally, the new reactivity index, global electrophilicity index or electrophilicity index ($\omega$)[23] was proposed as:

$$\omega \equiv \frac{\mu^2}{2\eta} \qquad (2)$$

## 1.3. Previous models and correlations

In previous works, two independent models based on electronegativity equalization[26] (*cubic approximation of energy*) and an alternative operational formula for the Fukui function[27] (derivative of the density with respect to the total number of electrons and *parabolic approximation of density*) respectively, were developed to obtain local derivatives of the energy and net charge with respect to the total number of electrons for atoms in a molecule. Moreover, a model was developed to calculate reactivity descriptors for natural bond orbitals[28] based on Lagrange's multipliers[29]. Earlier work also studied strong correlations between the derivatives of the local energy and net charge with respect to the total number of electrons for the atoms in the molecule[30]. Subsequently, the ideas of these models for the atoms in the molecule have been transferred to a model based on natural bond orbitals[31], obtaining a much more complete model that enables the calculation of the aforementioned derivatives for the bond orbitals instead of the atoms. Significant correlations were also found between the derivatives of the bond orbitals, comparable to those obtained previously for the atoms of the molecule. The correlations studied in that work were **Eqs. (3-6)**.

$$\omega_i \approx a \cdot f_i + b \tag{3}$$

$$f_i \approx \frac{a}{\eta_i} + b \tag{4}$$

$$\Delta N_i^{\max}\left(parabolic\ \ approx.\ of\ density\right) \approx a \cdot \Delta N_i^{\max}\left(cubic\ \ approx.\ of\ energy\right) + b \tag{5}$$

where $\Delta N_i^{\max}\left(parabolic\ approx.\ of\ density\right)$ is based on the model of reference **[27]** and $\Delta N_i^{\max}\left(cubic\ approx.\ of\ energy\right)$ is based on the model of reference **[26]**. And finally, the correlation:

$$\Delta N_i^{\max} \approx \frac{a}{\eta_i} + b \tag{6}$$

## 2. COMPUTATIONAL DETAILS

All the structures included in this study were optimized at B3LYP/6-31G(d)[32,33] theory level using the Gaussian16 package[34]. The electronic densities used in the models were calculated at the same level of computation for the neutral molecule, the cation and anion through the Gaussian16 software. GaussView v5.0[35] was used to make the *.cub files of the Natural Hybrid Orbitals (NHOs) and Canonical Molecular Orbitals (CMOs) (see **Appendix I**) and the images of the molecular systems were generated with Chemcraft[34].

To perform the necessary calculations to obtain the parameters of the models, a computer code has been developed that has been integrated into an updated version of the UCA-FUKUI computer software (https://d127.uca.es/software/UCA-FUKUI_v2.exe)[37]. As an example, **Figures S1** and **S2** (in the supporting information) show the access to the corresponding calculation modules from the main program screen. **Figures S3** and **S4** show some calculation examples ($CH_2CHCl$) performed with the new code.

## 3. RESULTS AND DISCUSSION

The first part of this section will present the results of making partition groupings to obtain a set of partitions of smaller number of elements and how it affects the correlations of the final set of partitions; it will be discussed using the atoms in molecules of a set of representative molecules as an example. What these results try to show is the wide range of validity that the correlations studied have, which, as will be seen in a later section, gives an enormous field of application to the concept of multi-equalization that will be defined later. We have not been able to demonstrate that the correlations presented are general and valid for any ideal set of partitions that can be constructed, but we think that they probably are (perhaps with some exceptions). In any case, the range of validity in which they have been tested is wide enough to be very useful and we will continue working on this aspect (see **Future perspectives**).

The parameters of the models have been calculated for a representative sample of molecules. To obtain a set with different functional groups (aldehyde, halogen, amine, ester, carboxylic acid, amide, nitro, alcohol, thiol, -CN, and double bond) the following molecules were selected: $CH_2CHCN$, $CH_2CHOH$, $CH_3COOCH_3$, $CH_2CHNH_2$, $CH_2CHOCH_3$, $CH_2CHCHO$, $CH_2CHNO_2$, $CH_3CHSH$, $CH_2CHCl$, $CH_3COOH$ and $CH_3CONH_2$. This is the same set of test molecules that was used in reference **[30].**

### 3.1. Correlations between "partition sets". Partitions formed by sets of atoms in a molecule.

An interesting and potentially important point, as can be seen in the "Concept of multi-equalization" section, is the consequences of constructing new sets of partitions starting from a set that already satisfies the correlations we have seen in previous works for NBOs[31] or for atoms in molecules,[30]. Will the correlations of the original set still be valid? Logically the total number of partitions in the final set will be smaller than in the starting set and the dimensions of at least of some partitions will also change. Some of the parameters that fulfil the correlations are clearly additive, such as $f_i^{\pm}$, $\Delta N_i^{\max}$ and $\omega_i$, since they represent variations of charge or energy. In these cases, the variation corresponding to the sum of two partitions should be the sum of the variations of these partitions. This leads us to the conclusion that correlations of the type $\Delta N_i^{\max}$ **vs** $\Delta N_i^{\max}$**'** or $\omega_i$ **vs** $f_i$ should continue to hold even if a set is constructed where the final partitions are combinations (i.e. sums) of the starting partitions. For example, in the case of $f_i$ we could start from the definition of finite lateral differences:

$$f_i^{+} = N_i^{N+1} - N_i^{N} \quad (7)$$

$$f_i^{-} = N_i^{N} - N_i^{N-1} \quad (8)$$

and from the definition of the parameter $f_i = \left(f_i^{+} + f_i^{-}\right)/2$ for cases without degeneracy of the frontier orbitals, which we have used in the previous calculations. Combining **Eqs. (7)** and **(8)** we can obtain:

$$f_i = \frac{1}{2}\left(N_i^{N+1} - N_i^{N-1}\right) \quad (9)$$

If we consider partition $p_3$ as the sum of partitions $p_1$ and $p_2$, we have $N_3^{N+1} = N_1^{N+1} + N_2^{N+1}$ and $N_3^{N-1} = N_1^{N-1} + N_2^{N-1}$, where, $N_3^{N+1}$ and $N_3^{N-1}$ are the net charges of the corresponding partitions. Considering that $f_3 = \frac{1}{2}\left(N_3^{N+1} - N_3^{N-1}\right)$ and expanding this equality, we finally obtain:

$$f_3 = \frac{1}{2}\left[\left(N_1^{N+1} + N_2^{N+1}\right) - \left(N_1^{N-1} + N_2^{N-1}\right)\right] = \frac{1}{2}\left(N_1^{N+1} - N_1^{N-1}\right) + \frac{1}{2}\left(N_2^{N+1} - N_2^{N-1}\right) = f_1 + f_2 \quad (10)$$

The following graphs show the results of grouping atoms in the set of molecules of the sample. The grouping has been performed by types of atoms, that is, the atoms of the same type have been added, for example, all the hydrogens, all the carbons, all the oxygens, etc. and thus a new set of partitions has been formed for each molecule. For these sets, the new $f_i^+$ and $f_i^-$ values have been obtained, then **Eqs. (A7)-(A10)** and **(A12)-(A14)** of **Appendix I** have been applied by means of the calculation modules implemented in UCA-FUKUI. We have checked if the correlations found in ref. **[30]** are still fulfilled (using the parameters obtained from the model). **Figure 1** (left) shows the regression $\Delta N_i^{max}$ **vs** $\Delta N_i^{max}$'. The coefficient of determination is 0.957, which is not very different from the one obtained for the initial set ($R^2$=0.97). In any case, it shows a very high correlation between the parameters $\Delta N_i^{max}$ and $\Delta N_i^{max}$'.

The remaining linear regressions (**Figure 1 -** right and **Figures 2, 3**) also show very high correlations (coefficients of determination greater than 0.95), which leads us to believe that the partition sets we have constructed from the atoms in the molecule maintain the correlations of the starting set.

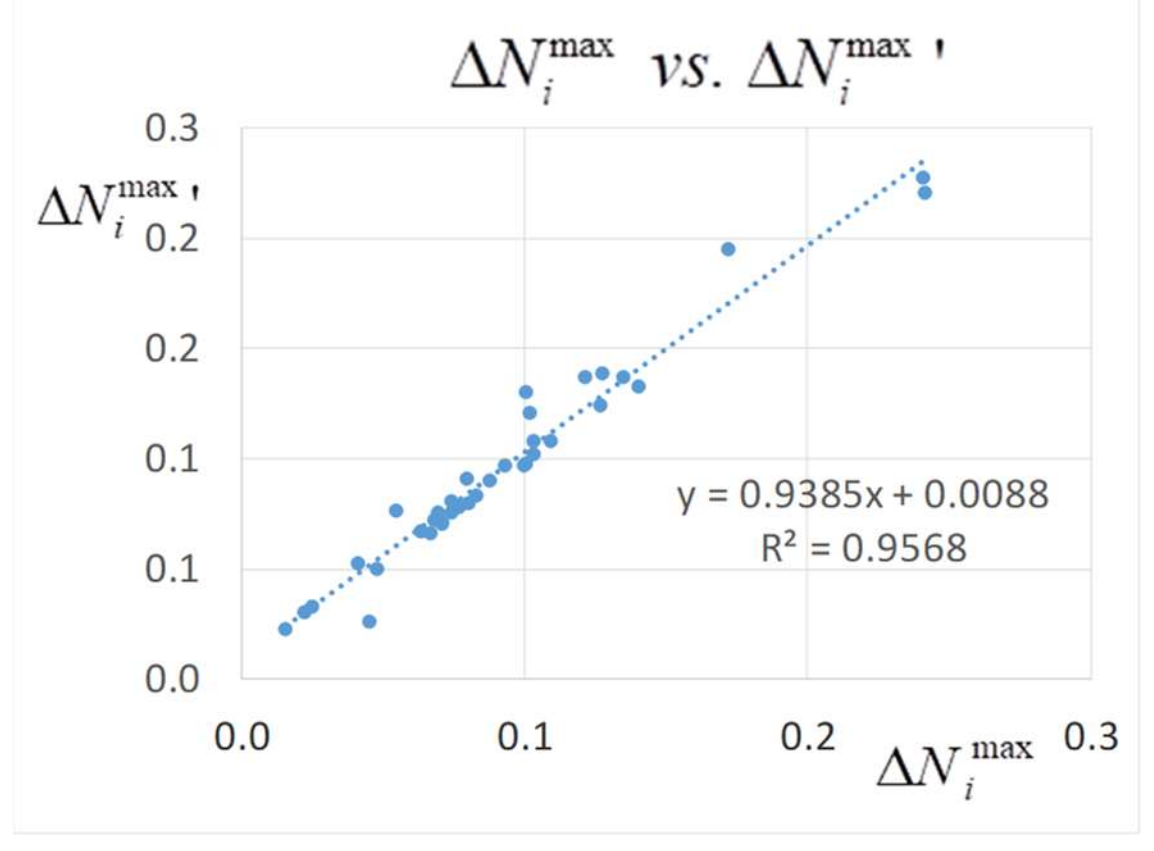

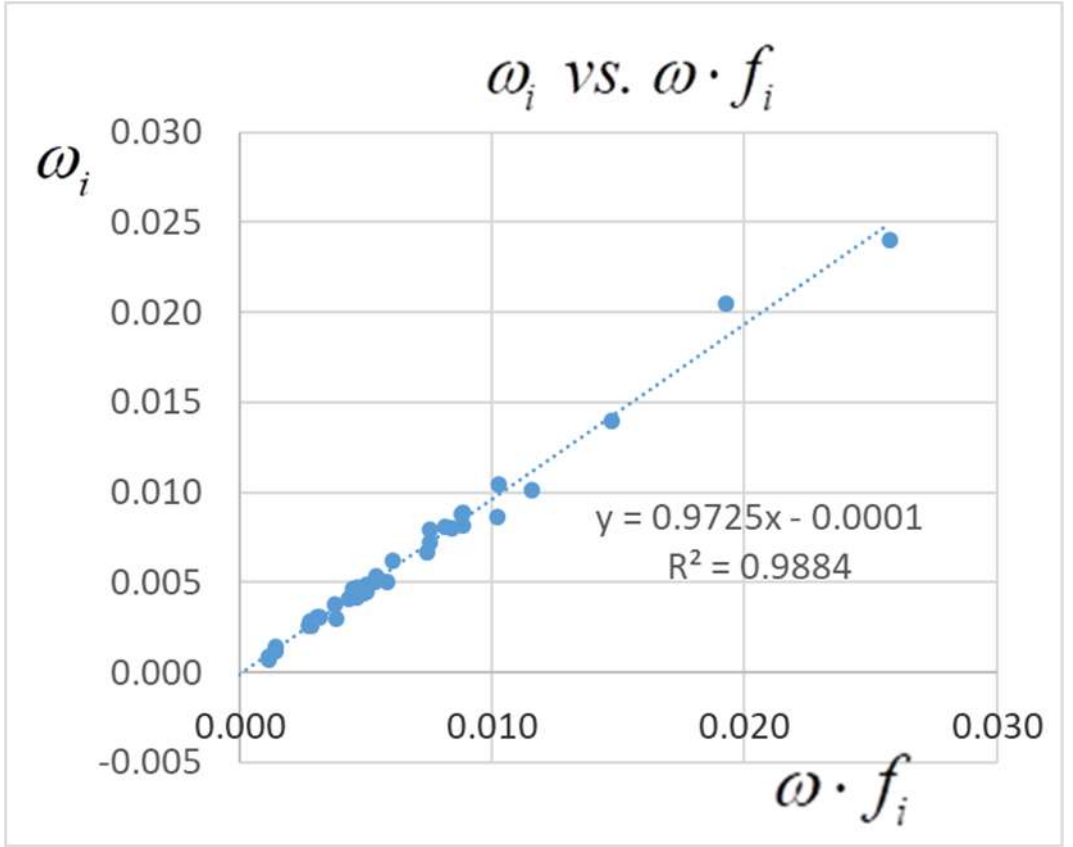

**Figure 1**. Left) Regression $\Delta N_i^{\max}$ *vs.* $\Delta N_i^{\max}{}'$ obtained for atom groups by using the sample set of molecules. Right) Regression $\omega_i$ *vs.* $\omega \cdot f_i$ obtained for atom groups by using the sample set of molecules.

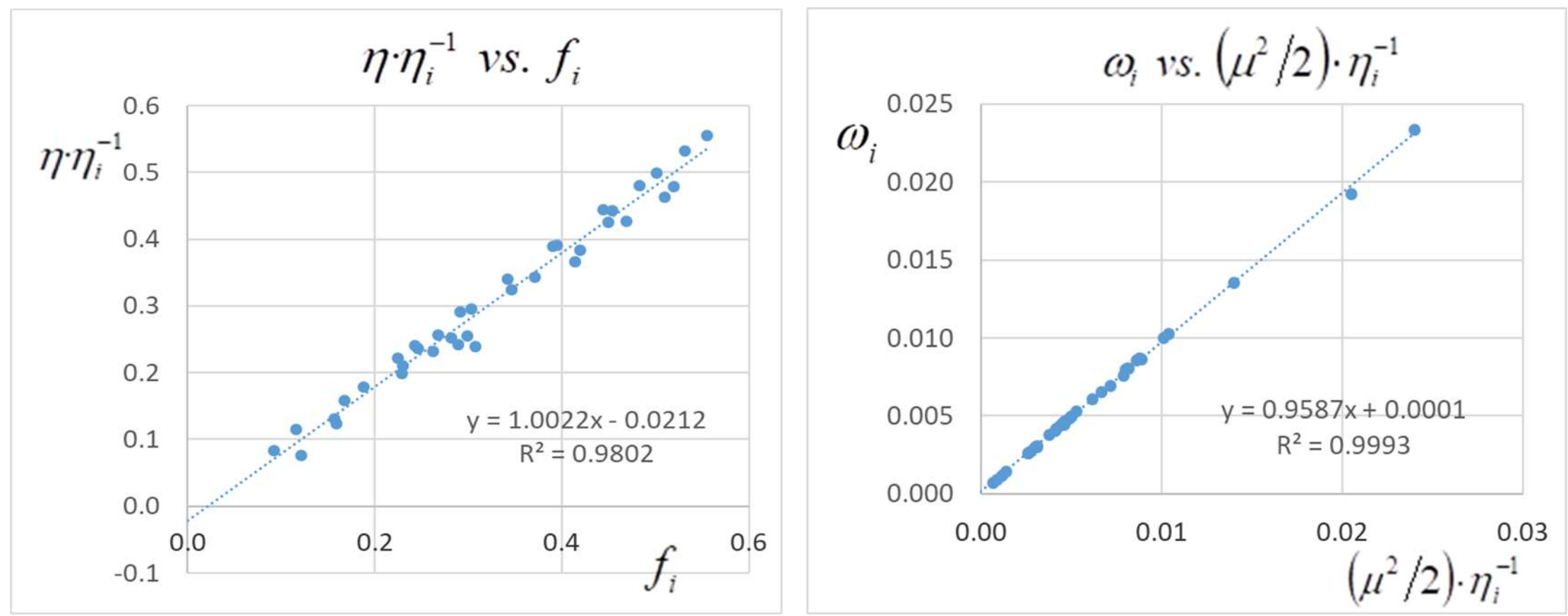

**Figure 2**. Left) Regression $\eta \eta_i^{-1}$ *vs.* $f_i$ obtained for atom groups by using the sample set of molecules. Right) Regression $\omega_i$ *vs.* $(\mu^2/2) \cdot \eta_i^{-1}$ obtained for atom groups by using the sample set of molecules.

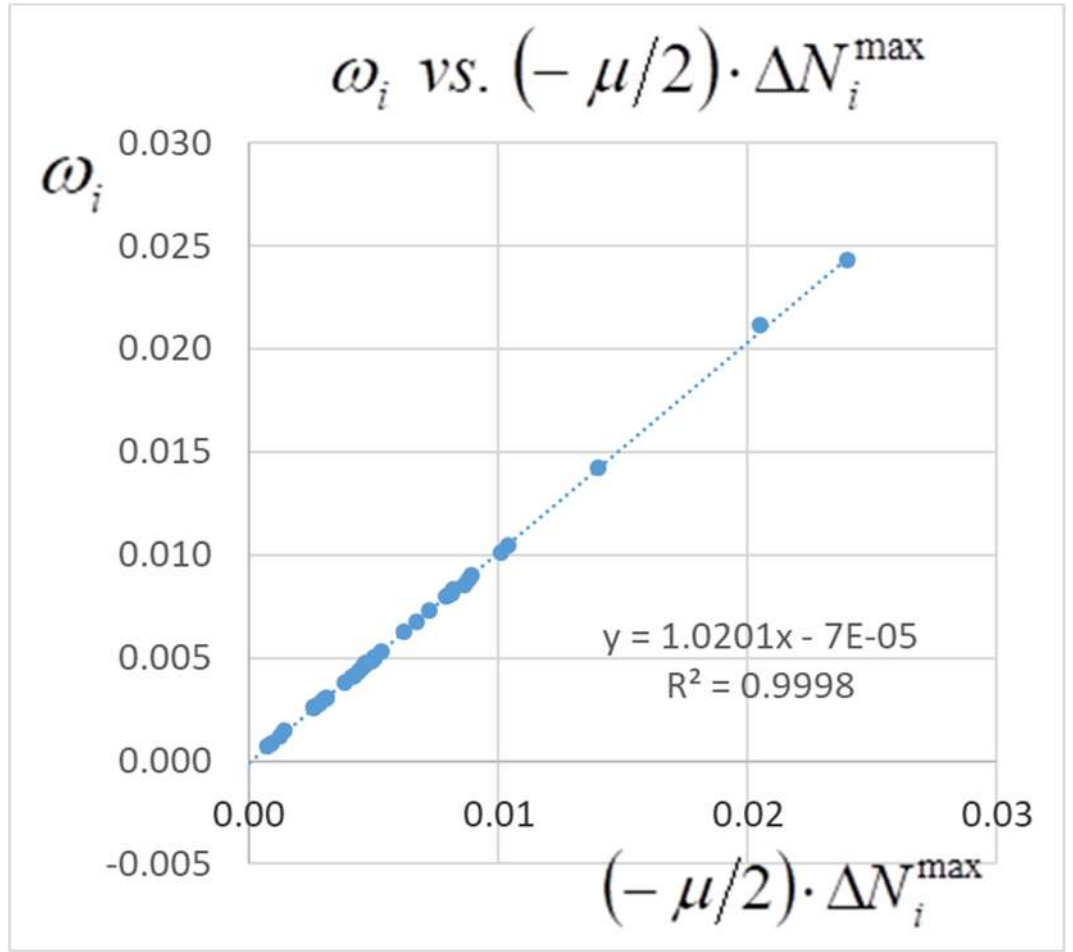

**Figure 3**. Regression $\omega_i$ *vs.* $(-\mu/2) \cdot \Delta N_i^{\max}$ obtained for atom groups by using the sample set of molecules.

### 3.2. Correlations between partition sets in larger molecules.

In this section, the reactivity index calculations have been studied as in **section 3.1**, but the sample set of molecules used is the series of increasing chain conjugated carbons shown in **Figure 4**. **Figure 5 - left** shows the linear regression of $\Delta N_i^{\max}$ *vs.* $\Delta N_i^{\max}{}'$ obtained for atom groups by using the sample set of molecules of **Figure 4. Figure 5 – right** shows the linear regression of $\omega_i$ *vs.* $\omega \cdot f_i$ obtained for atom groups by using the new sample set of molecules. In addition, **Figures 6** and **7** show the linear regressions $\eta\eta_i^{-1}$ *vs.* $f_i$, $\omega_i$ *vs.* $\left(\mu^2/2\right)\cdot\eta_i^{-1}$ and $\omega_i$ *vs.* $\left(-\mu/2\right)\cdot\Delta N_i^{\max}$. The results obtained for this new sample of molecules are equivalent to those obtained for the small-molecule sample of the previous section, the studied correlations are fulfilled in this set of larger molecules.

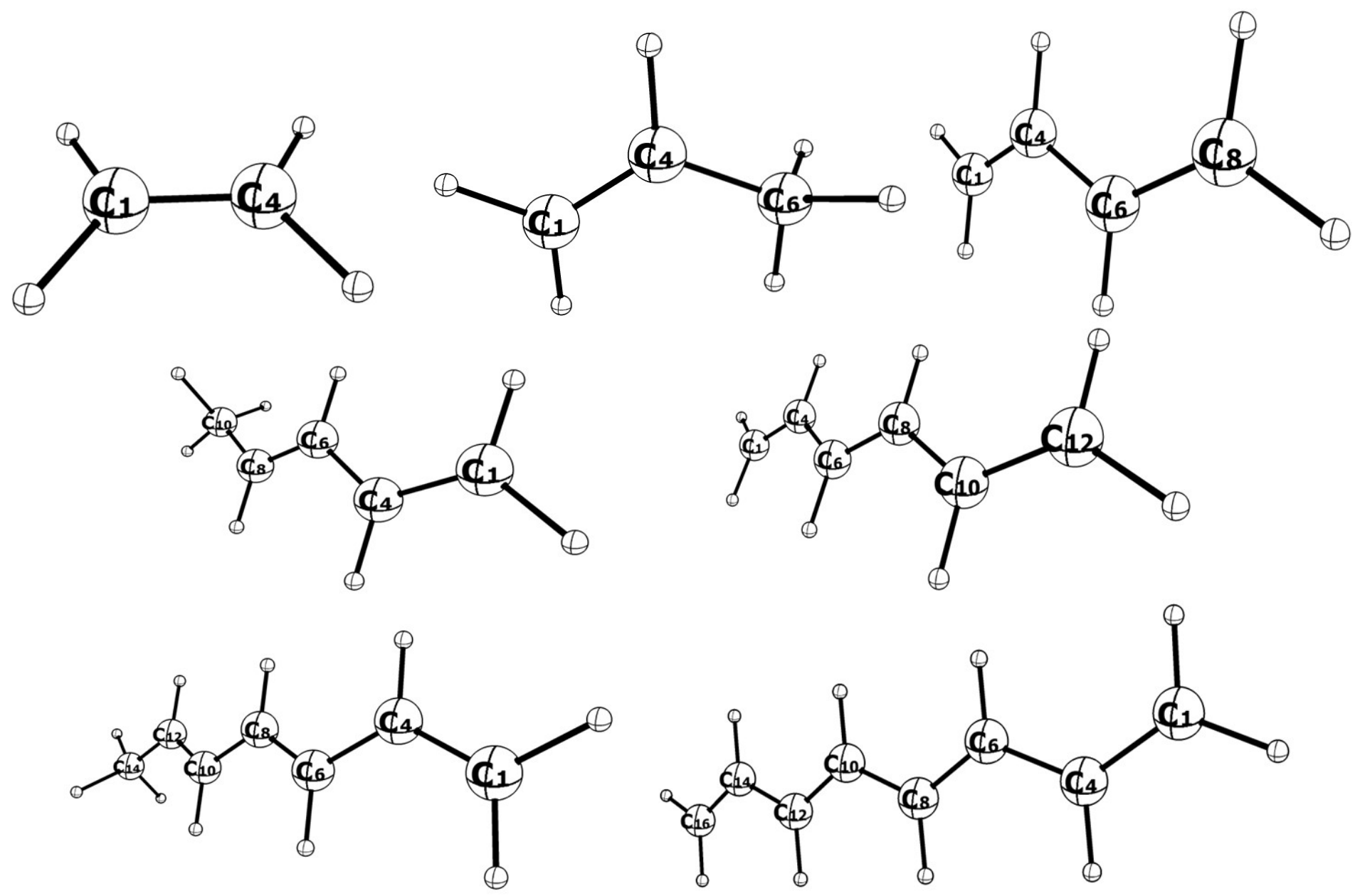

**Figure 4.** Sample set of test molecules: series of increasing chain of conjugated carbons.

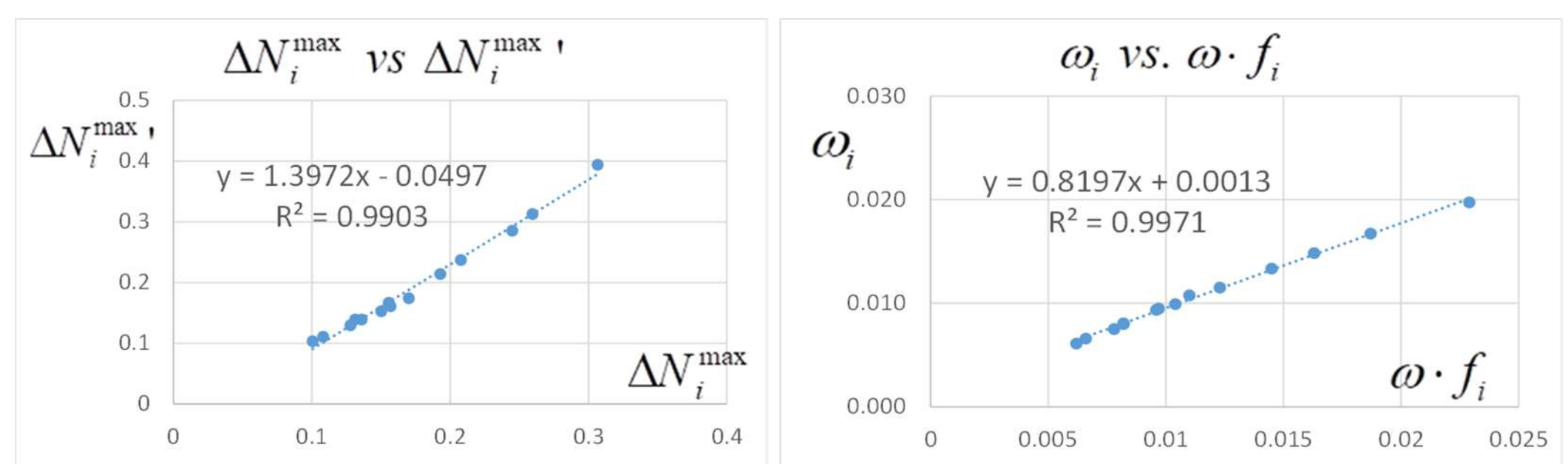

**Figure 5**. Left) Regression $\Delta N_i^{\max}$ *vs.* $\Delta N_i^{\max}$' obtained for atom groups by using the sample set of molecules of **Figure 4**. Right) Regression $\omega_i$ *vs.* $\omega\cdot f_i$ obtained for atom groups by using the sample set of **Figure 4**.

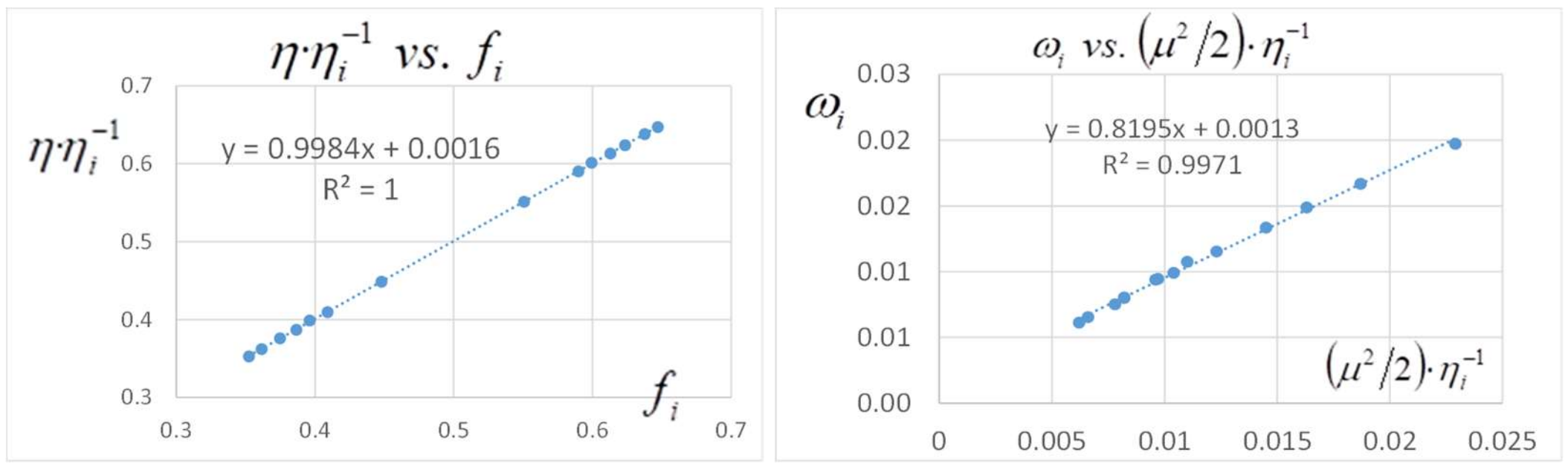

**Figure 6**. Left) Regression $\eta\eta_i^{-1}$ *vs.* $f_i$ obtained for atom groups by using the sample set of molecules of **Figure 4**. Right) Regression $\omega_i$ *vs.* $\left(\mu^2/2\right)\cdot\eta_i^{-1}$ obtained for atom groups by using the sample set of molecules of **Figure 4**.

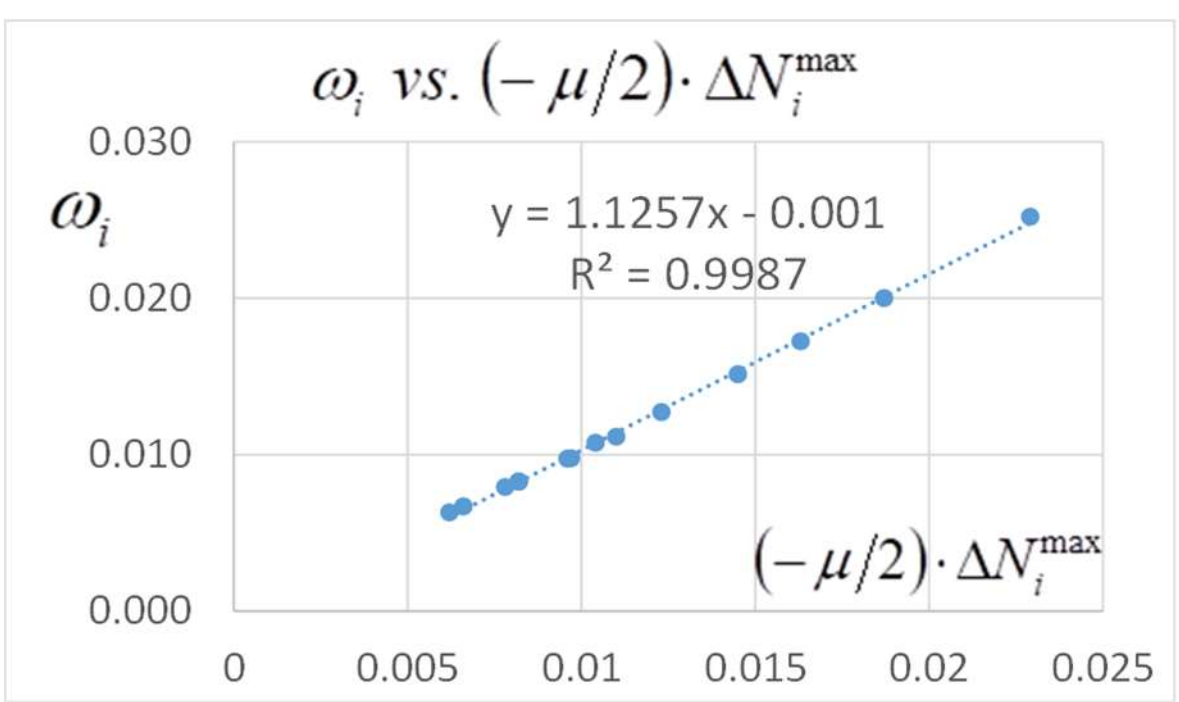

**Figure 7**. Regression $\omega_i$ *vs.* $(-\mu/2)\cdot\Delta N_i^{max}$ obtained for atom groups by using the sample set of molecules of **Figure 4**.

### 3.3. Density partitions formed by combinations of squared atomic or molecular orbitals.

**Appendix I** (**section A.1** in the supporting information) shows the method and the results of the calculation of reactivity descriptors applied to the case of natural hybrid orbitals (NHOs) and canonical molecular orbitals (CMOs), first describing the model used to calculate these descriptors and then presenting a study of the possible correlations between them. As can be seen in the **Appendix I** (**sections A.2** and **A.3**), there are strong indications that the correlations found in previous work for atoms in the molecule and NBOs are also observed in NHOs and CMOs (and quite possibly not the only partitions). With this in mind, it would be interesting to describe the partitions that can be formed from these orbitals. To construct new density partition sets based on combinations of squared atomic or molecular orbitals, we start from the following approximation:

$$\rho(r)\approx\sum_i \alpha_i\left|\phi_i^X(r)\right|^2 \tag{11}$$

where $\phi_i^X(r)=\phi_i^{NHO}(r),\phi_i^{NBO}(r),\phi_i^{CMO}(r),\ldots$, and taking into account that

$$\Delta\rho^{\pm}(r)\approx\sum_i X_i^{\pm}\cdot\left|\phi_i^X(r)\right|^2 \tag{12}$$

where $\Delta\rho^{+}(r)=\rho^{N+1}(r)-\rho^{N}(r)$, $\Delta\rho^{-}(r)=\rho^{N}(r)-\rho^{N-1}(r)$, $X$= NHO, NBO or CMO, and $X_i^{\pm}$ are the corresponding coefficients. Below, for $X_i^{\pm}$ coefficients, we will take the symbology $C_i^{\pm}$ when CMO type orbitals are used, $B_i^{\pm}$ for NBOs and $H_i^{\pm}$ for NHOs. Also, we will admit the approximations:

$$\phi_j^{CMO}(r) \approx \sum_i BC_{ij} \cdot \phi_i^{NBO}(r)$$
$$\phi_j^{NBO}(r) \approx \sum_i HB_{ij} \cdot \phi_i^{NHO}(r) \qquad (13)$$
$$\phi_j^{CMO}(r) \approx \sum_i HC_{ij} \cdot \phi_i^{NHO}(r)$$

where $BC_{ij}$, $HB_{ij}$ and $HC_{ij}$ are the coefficients of the corresponding developments. Considering for example the Canonical orbitals and taking into account the approximations of **Eqs. (12)** and **(13)**, we can obtain:

$$\Delta\rho^{\pm}(r) \approx \sum_j C_j^{\pm} \cdot \left[\sum_i BC_{ij} \cdot \phi_i^{NBO}(r)\right]^2 = \qquad (14)$$
$$= \sum_j C_j^{\pm} \cdot \left(\left[\sum_i BC_{ij}^2 \cdot \left|\phi_i^{NBO}(r)\right|^2\right] + \left[\sum_i \sum_{k \neq i} BC_{ij} \cdot BC_{kj} \cdot \phi_i^{NBO}(r) \cdot \phi_k^{NBO}(r)\right]\right)$$

If we integrate this equality

$$\int \Delta\rho^{\pm}(r) d\tau \approx \sum_j C_j^{\pm} \cdot \left(\left[\sum_i BC_{ij}^2 \cdot \int \left|\phi_i^{NBO}(r)\right|^2 d\tau\right] + \left[\sum_i \sum_{k \neq i} BC_{ij} \cdot BC_{kj} \cdot \int \phi_i^{NBO}(r) \cdot \phi_k^{NBO}(r) d\tau\right]\right) \qquad (15)$$

Considering now that the $\phi_i^{NBO}(r)$ orbitals are orthogonal, **Eq. (15)** becomes:

$$\int \Delta\rho^{\pm}(r) d\tau \approx \sum_j C_j^{\pm} \cdot \sum_i BC_{ij}^2 \cdot \int \left|\phi_i^{NBO}(r)\right|^2 d\tau \qquad (16)$$

By integrating the approximation **Eq. (12)** where $X_i^{\pm} = B_i^{\pm}$, that is $\Delta\rho^{\pm}(r) \approx \sum_i B_i^{\pm} \cdot \left|\phi_i^{NBO}(r)\right|^2$, we obtain:

$$\int \Delta\rho^{\pm}(r) d\tau \approx \sum_i B_i^{\pm} \cdot \int \left|\phi_i^{NBO}(r)\right|^2 d\tau \qquad (17)$$

If we compare **Eq. (16)** with **(17)** we can see that:

$$B_i^{\pm} \approx \sum_j C_j^{\pm} \cdot BC_{ij}^2 \qquad (18)$$

and similarly, the following can be obtained:

$$H_i^{\pm} \approx \sum_j B_j^{\pm} \cdot HB_{ij}^2 \qquad (19)$$

Note that as the parameters $B_i^{\pm}$, $C_i^{\pm}$ and $H_i^{\pm}$ have been defined, they correspond to the $f_i^{\pm}$ parameters that we have calculated with the UCA-FUKUI software in this and previous works to study correlations between local derivatives. Using a nomenclature closer to the one we have used previously, we would have:

$$f_i^{\pm NBO} \approx \sum_j f_j^{\pm CMO} \cdot BC_{ij}^2 \qquad (20)$$

and:

$$f_i^{\pm NHO} \approx \sum_j f_j^{\pm NBO} \cdot HB_{ij}^2 \tag{21}$$

From the equalities **Eqs. (20)** and **(21)**, it can be noted that the same linear combination of $BC_{ij}^2$ or $HB_{ij}^2$ values is the same for $f_i^+$ and $f_i^-$, which we think is important for the correlations, between derivatives that we have studied for the different types of orbitals, to be fulfilled. For a set of partitions where each partition is formed by a linear combination of a starting set (although, as will be seen in the next section, we will focus on a particular type of combination consisting of the sum of some partitions of the starting set), we would have a relation like this:

$$f_i^{\pm E} \approx \sum_j f_j^{\pm X} \cdot EX_{ij}^2 \tag{22}$$

where "*E*" represents the final partition set and "*X*" represents the original or starting partition set. Note that "*X*" can represent, for example, NHO, NBO or CMO and that we have focused on sets of partitions "*X*" that we have studied previously and that satisfy the correlations we have seen. The $f_j^{\pm X}$ values can be calculated for the starting set. If any criterion was known to obtain the $f_i^{\pm E}$ values, as we will see in the next section, all that remains would be to find the $EX_{ij}^2$ values that satisfy the equalities **Eq. (22)** to determine which combinations of the starting set that the partitions of the final set are obtained with.

#### 3.3.1. Strategy to design a new partition set with better correlations between reactivity descriptors. Examples of modified partition sets formed by canonical molecular orbitals or natural hybrid orbitals.

In this section, we will outline a strategy for forming a new set of partitions that consists of linear combinations of partitions from a starting set. The starting partitions are typically atomic or molecular orbitals. The new set of partitions is designed to yield improved linear regressions, with higher correlation than the initial set of partitions. To select the partitions to be combined, we will use some specific criteria, which we will describe in detail later in this section. We select combinations of partitions such that the dual-descriptor $f_i^{(2)}$ parameter approaches zero. The dual-descriptor parameter is the final term in the expansion of one of the models used in calculating descriptors (see **Eq. A10** of **Appendix I** in the **supporting information**). By minimizing this parameter, we can improve the convergence of the expansion and significantly reduce the model error, which in turn leads to improved correlations. We illustrate this approach using a detailed example, and provide further justification in **Section 3.4.1**.

In exploring alternatives to the dual descriptor parameter, we considered using the hypersoftness parameter $\gamma_i$, which is also the final term in the expansion of one of the models used in calculating reactivity descriptors (see **Eq. A11** of **Appendix I** in the **supporting information**). However, upon examining **Figures S5-S13** and **S41-S49** in the **supporting information**, where we compared graphs **B** and **C**, we found that the $\left|\Delta N_i^{\max} - \Delta N'^{\max}_i\right|$ errors followed the trend of $f_i^{(2)}$ more closely than that of $\gamma_i$. We believe that this is because the expansion given in **Eq. (A10)** in **Appendix I** is shorter than the expansion in **Eq. (A11)**, and therefore has lower error

To illustrate the discussion of the previous section, we will use the occupied canonical molecular orbitals (CMOs) obtained for the molecule $CH_3CONH_2$. In **Figure 8**, we present the $\left|\Delta N_i^{\max} - \Delta N'^{\max}_i\right|$ values calculated using the models described in **Appendix I**, **section A.1**, as well as the absolute values of the dual-descriptor parameters ($f_i^{(2)}$), which are defined in **Eqs. (A9)** and **(A10)** of **Appendix I**, and the hypersoftness parameters ($\gamma_i$), which are defined in **Eqs. (A11)** and **(A12)** of **Appendix I**. We chose this example because it showed the worst correlation in the starting set (see **Figures 9-A**, **C**, and **E**). In **Figure 8-A**, we observed that partitions 11, 14, 16, and 17 had the most significant values, with corresponding dual-descriptor parameter values of 0.1966, -0.3845, -0.5355, and 0.5196, respectively. We then constructed two new partitions, 11+14 and 16+17, to replace the previous four, resulting in a total of 24 partitions (down from 26). **Figure 9-B**, **D**, and **F** show the improved correlations after the partition modifications. As can be seen, the improvement is significant, with the coefficient of determination increasing from approximately 0.2 to 0.95-0.96 in all cases. It is worth noting that further improvements could be achieved by continuing to combine partitions in a way that reduces the absolute values of the dual-descriptor parameter.

A) B)

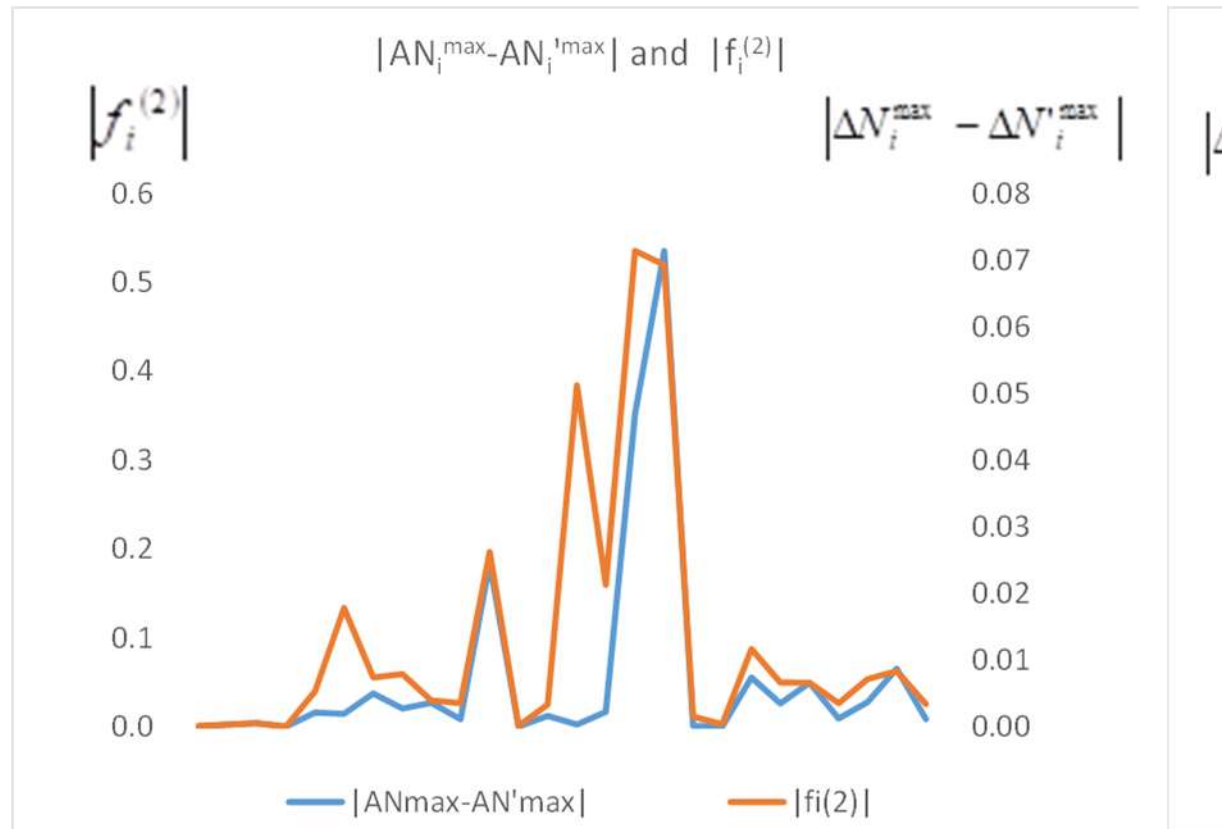

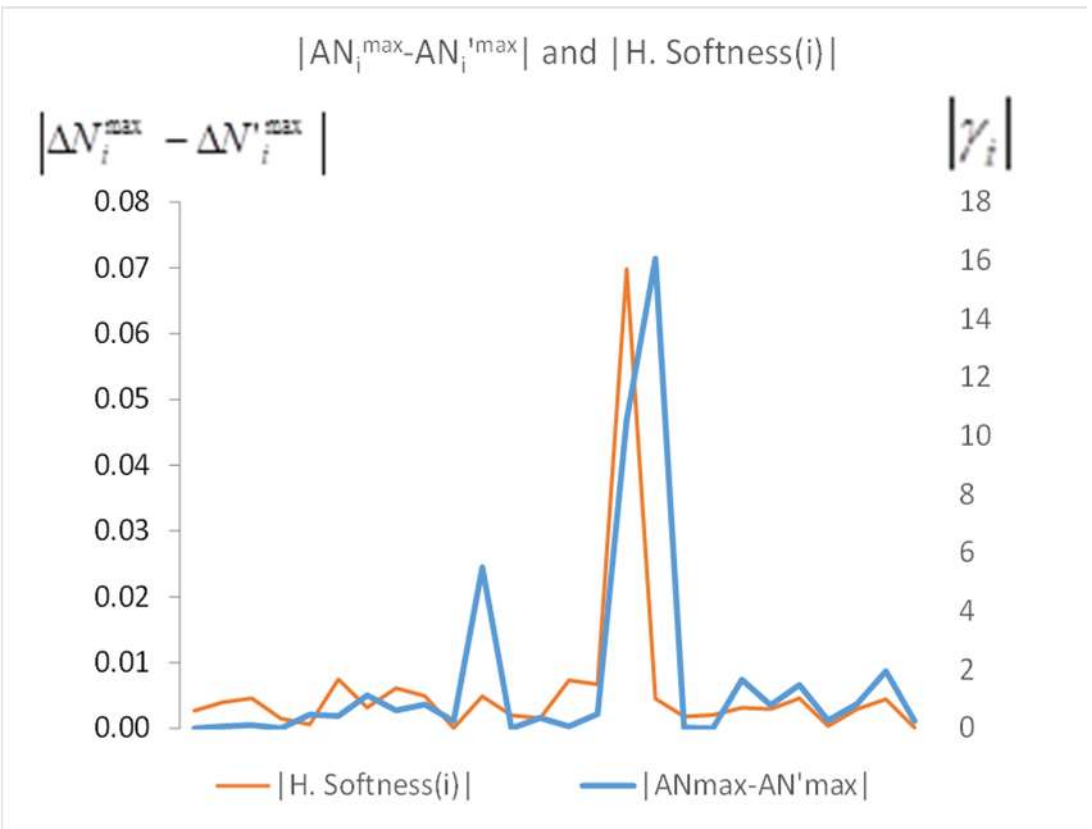

**Figure 8**. A) Comparison of $\left|\Delta N_i^{\max} - \Delta N'^{\max}_i\right|$ and $\left|f_i^{(2)}\right|$ values. B) Comparison of $\left|\Delta N_i^{\max} - \Delta N'^{\max}_i\right|$ and $\left|\gamma_i\right|$ values. In both cases for the occupied CMOs of the $CH_3CONH_2$ molecule.

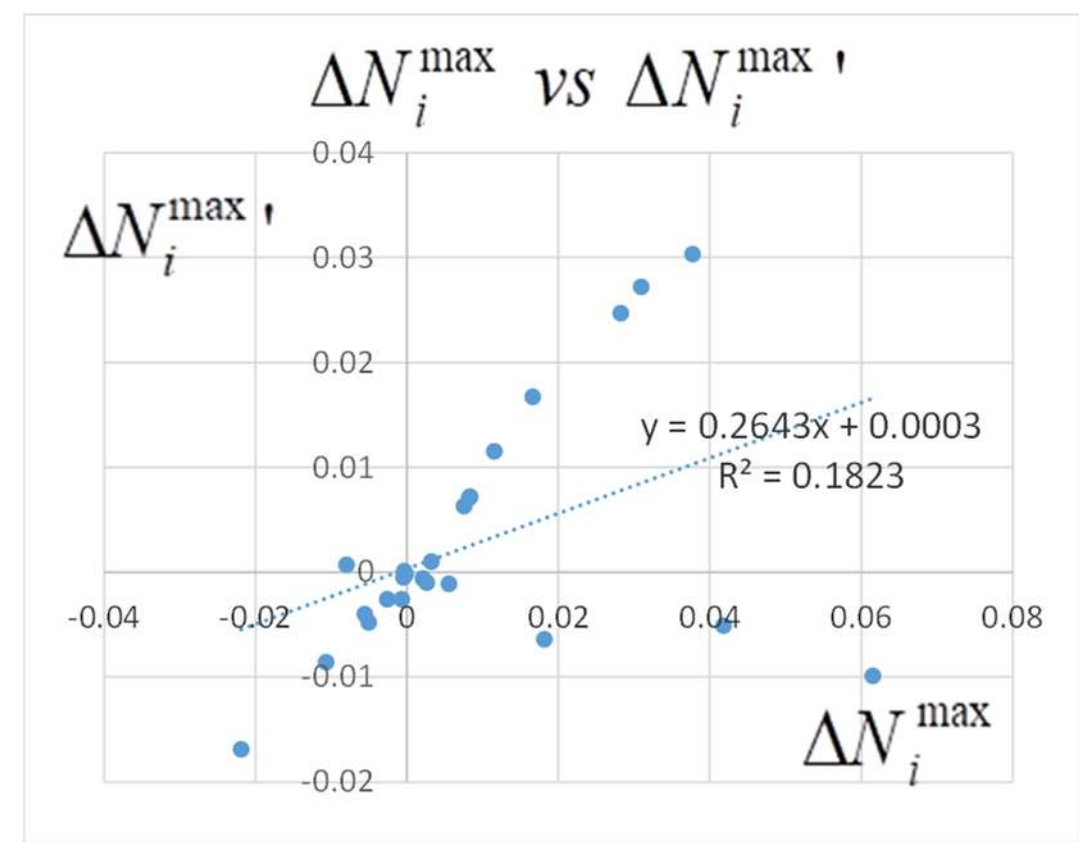

**A)** *Original* set of partitions.

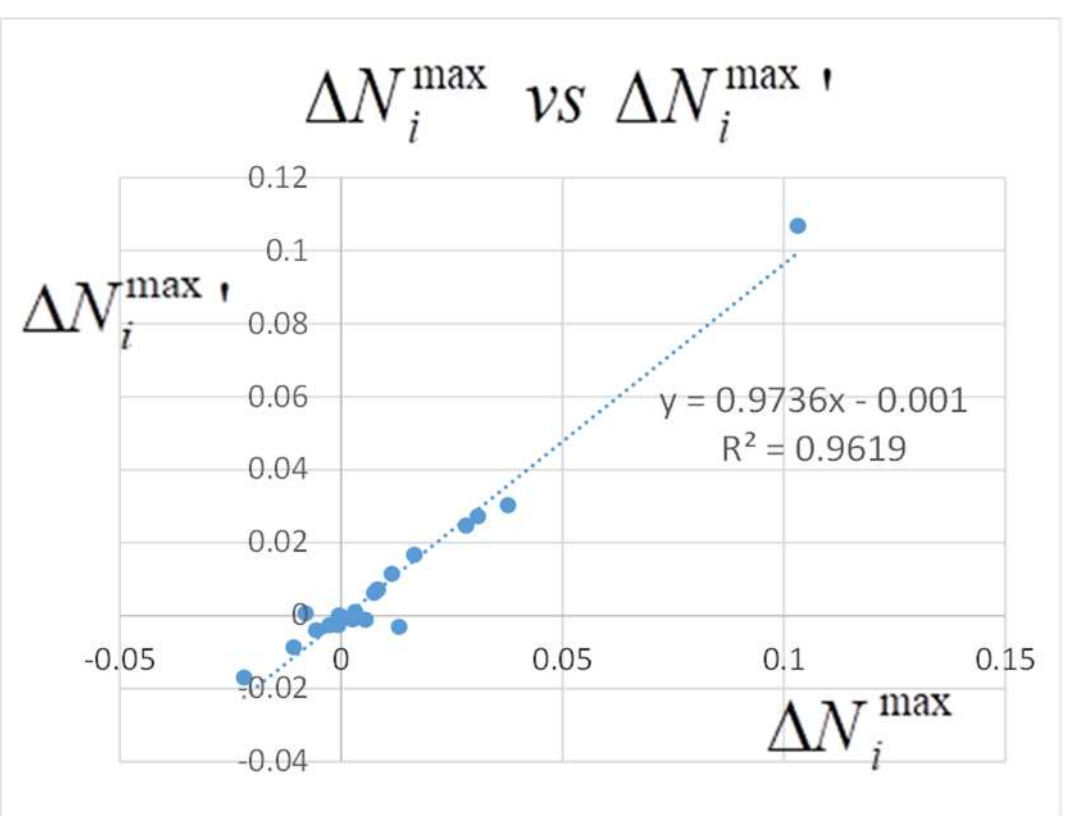

**B)** *Improved* set of partitions.

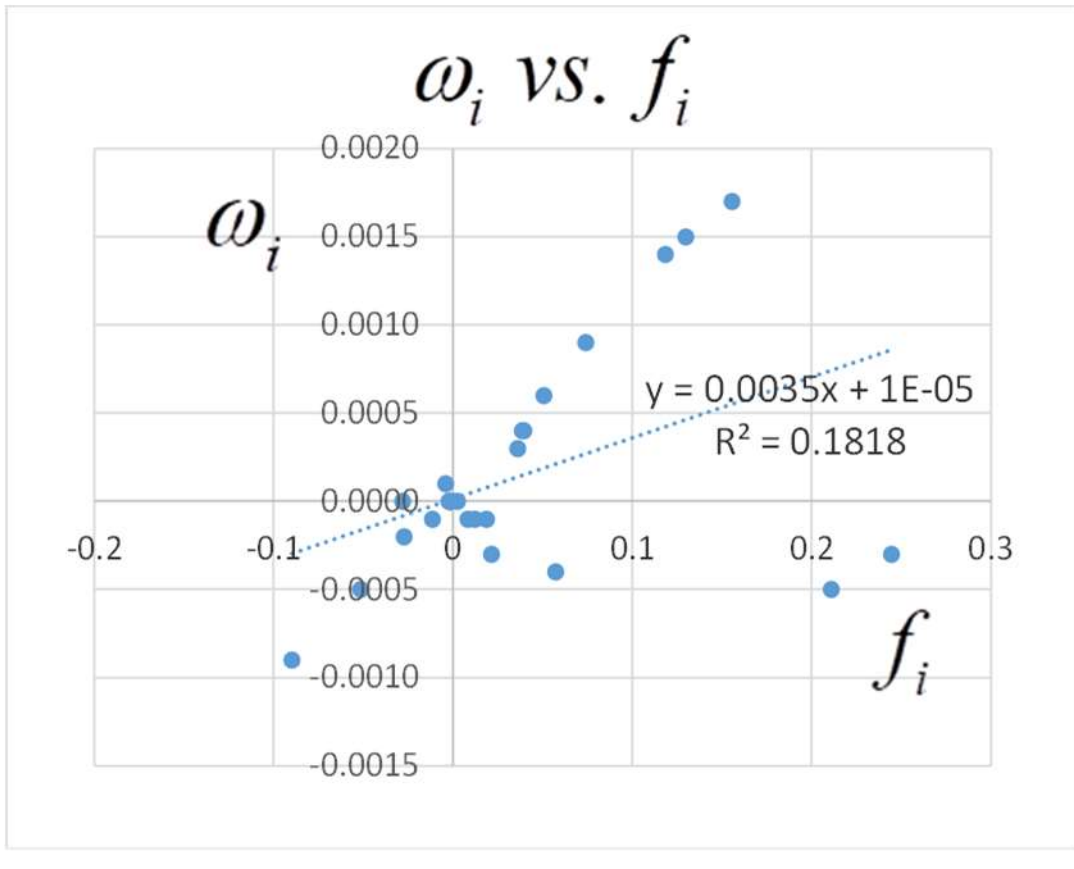

**C)** *Original* set of partitions.

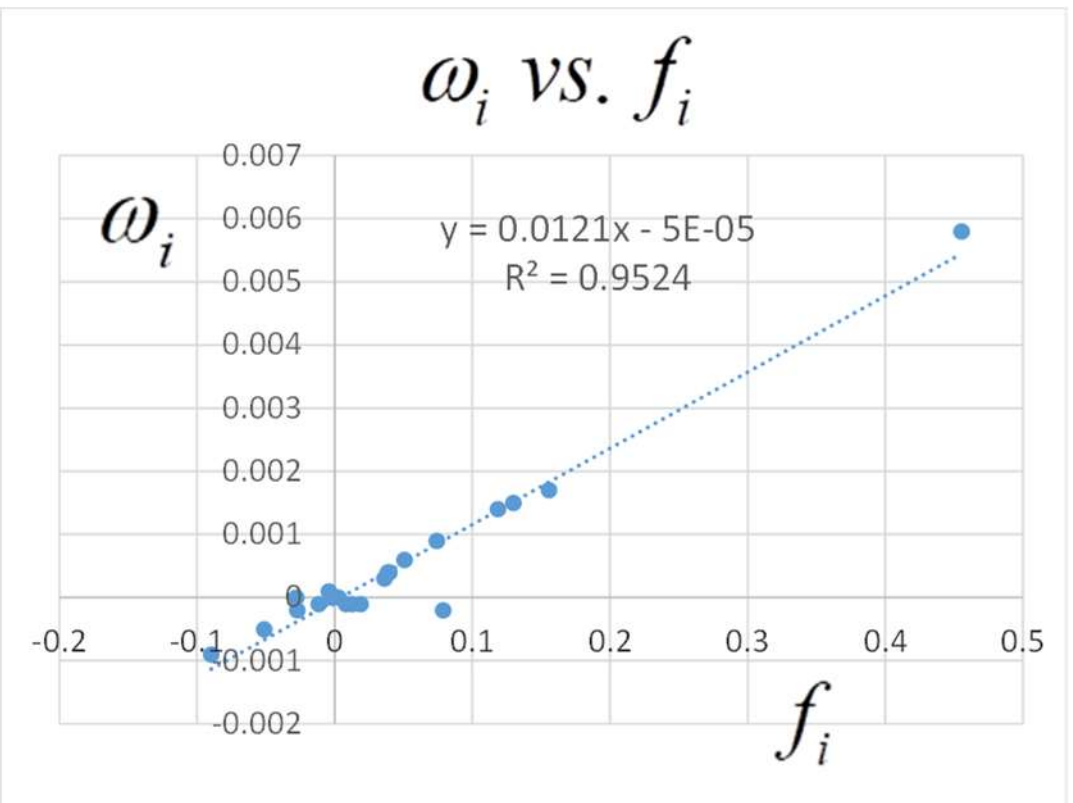

**D)** *Improved* set of partitions.

**E)** *Original* set of partitions.

**F)** *Improved* set of partitions.

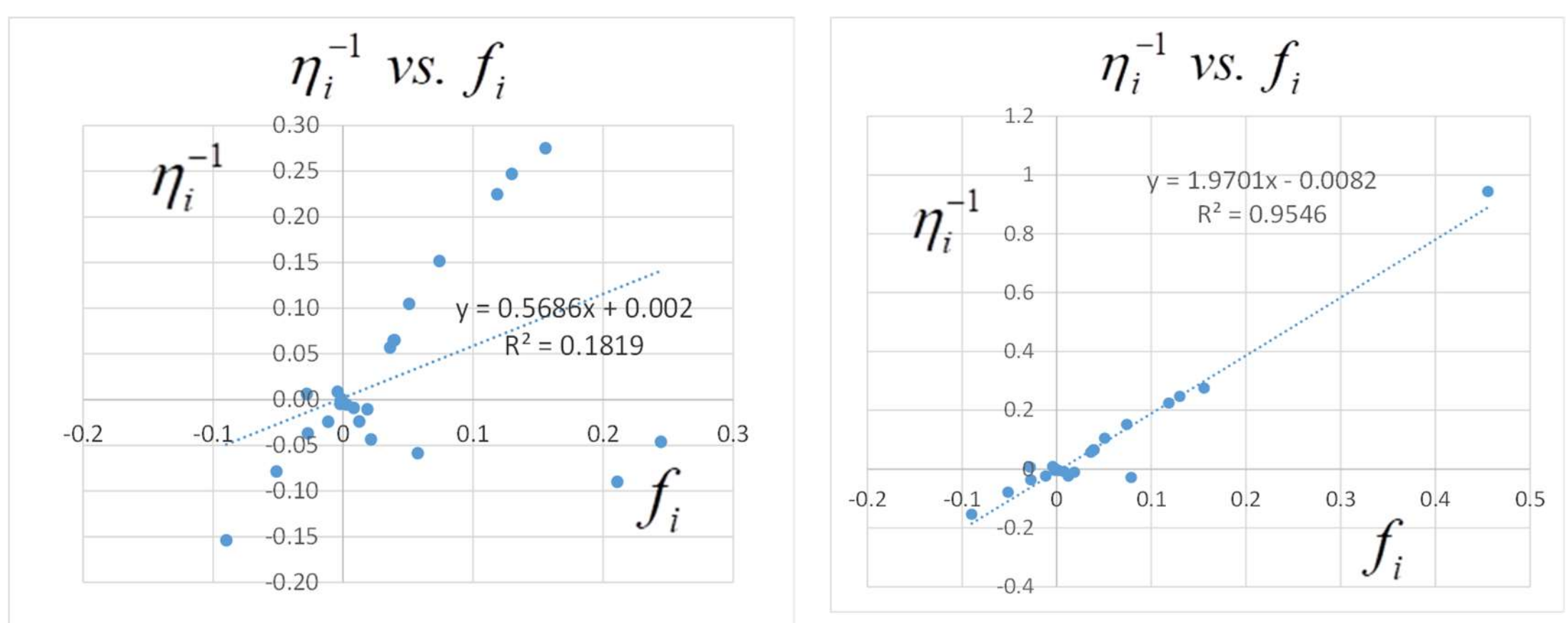

**Figure 9**. The following figures depict the regression results for the occupied CMOs of the $CH_3CONH_2$ molecule: **A**) Regression $\Delta N_i^{\max}$ *vs.* $\Delta N_i^{\max}{}'$, **B**) Regression $\Delta N_i^{\max}$ *vs.* $\Delta N_i^{\max}{}'$ using the new set of partitions, **C**) Regression $\omega_i$ *vs.* $f_i$, **D**) Regression $\omega_i$ *vs.* $f_i$ with the new partition, **E**) Regression $\eta_i^{-1}$ *vs.* $f_i$, and **F**) Regression $\eta_i^{-1}$ *vs.* $f_i$ with the new partition.

**Figure 10 - A** and **B** display graphs similar to those in **Figure 8 - A** and **B**, except the set of partitions has been modified as described above. The most notable difference in **Figure 10** is the scale. For example, the maximum value of the left axis in **Figure 10 - A** is 0.2 compared to 0.6 in **Figure 8 - A,** and for the right axis, it goes from 0.018 in **Figure 10 - A** to 0.08 in **Figure 8 - A**. A similar trend is observed in **Figure 10 - B**. The changes in the set of partitions have led to lower values of local dual-descriptor and local hypersoftness in absolute terms. These findings align with the improved correlations seen in **Figure 9 - B**, **D**, and **F**.

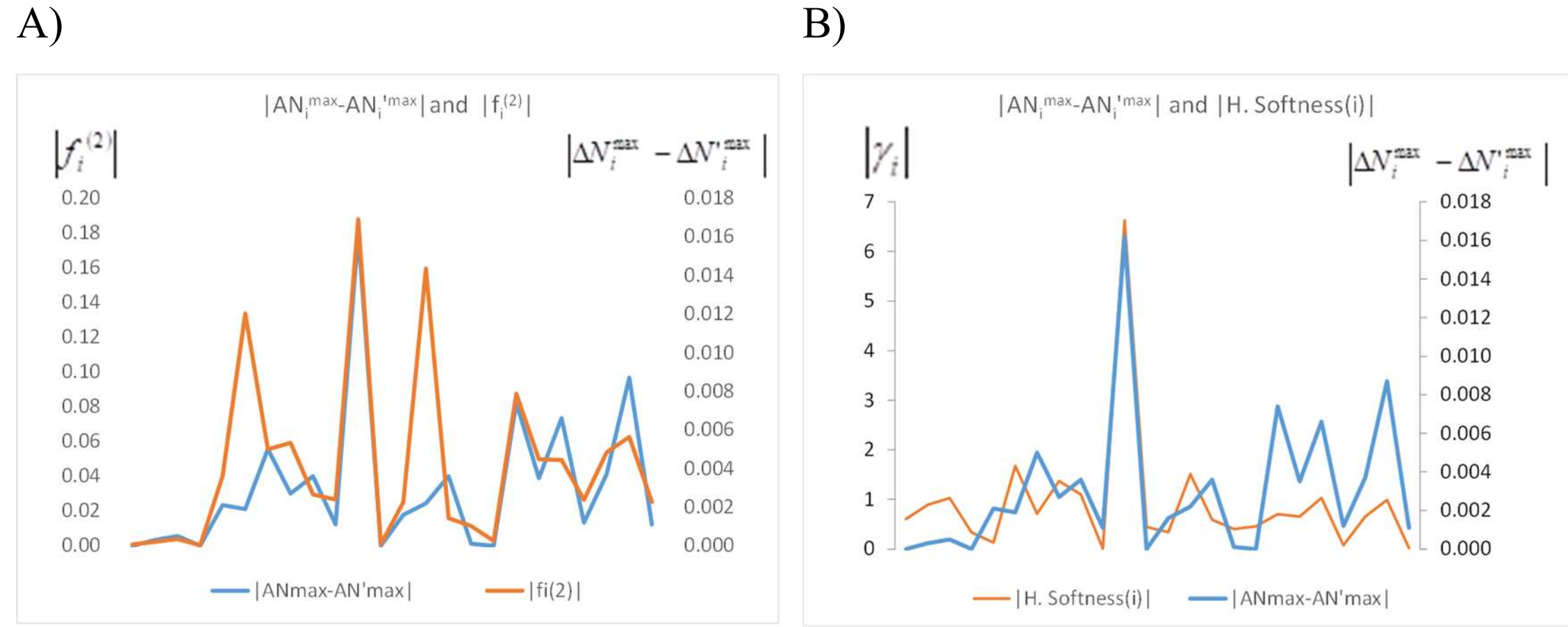

**Figure 10**. Comparison of partitions for $CH_3CONH_2$ molecule. **A**) Same parameters as in **Figure 8-A** but for the new partition. **B**) Same parameters as in **Figure 8-B** but for the new partition. The occupied CMOs of the molecule were used as partitions in all cases.

To further illustrate this point, we will present two additional examples. As with the previous case, we have chosen molecules with poor correlations. In the first example, we examine the molecule $CH_3COOH$ and use its natural hybrid orbitals (NHO) as partitions. Similar to the previous example, we select partitions 3, 4, and 21 with the highest absolute dual-descriptor values and combine them to form a new partition. This new partition results in a substantial decrease in the absolute value of $f_i^{(2)}$. The results show a significant improvement in correlations. For example, in the first case, the coefficient of determination increased from 0.3842 to 0.8969 for the $\Delta N_i^{\max}$ $vs.$ $\Delta N_i^{\max}{}'$ correlation, from 0.2903 to 0.9375 for the $\omega_i$ $vs.$ $f_i$ correlation, and from 0.3040 to 0.9997 for the $\eta_i^{-1}$ $vs.$ $f_i$ correlation. In the second example, we used the canonical molecular orbitals (CMO) of the molecule $CH_3COOCH_3$ as partitions. We identified three new partitions by adding partitions 12+14, 20+21, and 26+28, respectively, based on the largest absolute $f_i^{(2)}$ values. The resulting correlations were greatly improved, with the coefficient of determination increasing from 0.130 to 0.956 for the $\Delta N_i^{\max}$ $vs.$ $\Delta N_i^{\max}{}'$ correlation, from 0.1746 to 0.9641 for the $\omega_i$ $vs.$ $f_i$ correlation, and from 0.1761 to 0.9662 for the $\eta_i^{-1}$ $vs.$ $f_i$ correlation. It's worth noting that the correlations could be enhanced by continuing to combine partitions in a manner that minimizes the absolute values of the dual-descriptor parameter.

**3.4 *Multi-equalization* concept.**

In this work and in the previous ones, we have checked correlations **Eqs. (3)-(5)** in several sets of very different partitions (atoms in molecules, NBOs, NHOs and CMOs) and it has been possible to justify that if we group these partitions, we obtain a new set of partitions where these correlations are most likely still fulfilled. **Section 3.4.1** will first explore a general partition set and justify the correlations observed above. **Section 3.4.1** will define the concept of *multi-equalization* and specify the relations a multi-equalised partition set must fulfill. The effect of multi-equalization on correlations will then be studied using approximate models. In **section 3.4.2**, the study will show that correlations *become exact relations* in an ideal multi-equalised partition set, without the use of approximations.

### 3.4.1. Justification of correlations using approximate models in NON-multi-equalized partitions.

To justify the correlation between $\omega_i$ and $f_i$ we will start from **Eq. (23)**, which was used in reference **[27]**

$$\Delta E_i \approx \mu^0 \cdot \Delta N_i + \frac{1}{2}\eta_i \Delta N_i^2 + \frac{1}{6} \cdot \gamma_i \cdot \Delta N_i^3 \quad (23)$$

But, for simplicity, we will truncate the cubic expansion to the first two terms (**Eq. 24**), this approach is clearly very crude; it will only be used to justify the correlations studied in this work. The hypersoftness $\gamma_i$ is the last term of this expansion, the value of this term is closely related to the convergence of the expansion and the model error.

$$\Delta E_i \approx \mu^0 \cdot \Delta N_i + \frac{1}{2}\eta_i \Delta N_i^2 \quad (24)$$

and therefore:

$$\omega_i \approx -\left(\mu^0 \cdot \Delta N_i^{\max} + \frac{1}{2}\eta_i \left(\Delta N_i^{\max}\right)^2\right) \quad (25)$$

In reference **[27]** we obtained an estimation of $\Delta N_i^{\max}$ like:

$$\Delta N_i^{\max} \approx f_i \cdot \Delta N_{\max} + \frac{1}{2} f_i^{(2)} \cdot \Delta N_{\max}^2 \quad (26)$$

where, $\Delta N_{\max}$ is the global parameter corresponding to the maximum amount of stabilizing charge and $\Delta N_i^{\max}$ the fraction of $\Delta N_{max}$ corresponding to each atom.

We will also truncate the expansion **Eq. (26)** (second model) to its first term (**Eq. 27**), as in the previous case, this approach is also *very rude* and will only be used to justify the correlations studied in this work. In this case, the local parameter $f_i^{(2)}$ is the last term of this expansion, the value of this term is closely related to the convergence of the power series and the model error. It is similar to what has been said about the hypersoftness after **Eq. (23)**, but as we have seen in **section 3.3.1** the quality of the correlations depends more on parameter $f_i^{(2)}$ than on parameter $\gamma_i$, probably because the expansion **Eq. (26)** is shorter and for that reason the error committed in this expansion is greater.

$$\Delta N_i^{\max} \approx f_i \cdot \Delta N_{\max} \quad (27)$$

Deriving **Eq. (24)** we get:

$$\mu \approx \mu^0 + \eta_i \Delta N_i \quad (28)$$

At the minimum of the energy curve, you get:

$$0 \approx \mu^0 + \eta_i \Delta N_i^{\max} \quad (29)$$

and:

$$\Delta N_i^{\max} \approx -\frac{\mu^0}{\eta_i} \quad (30)$$

Substituting **Eqs. (27)** and **(30)** on **Eq. (25)** we get:

$$\omega_i \approx -\left(\mu^0 \cdot f_i \cdot \Delta N_{\max} + \frac{1}{2}\eta_i \left(f_i \cdot \Delta N_{\max}\right) \cdot \left(-\frac{\mu^0}{\eta_i}\right)\right) \quad (31)$$

Simplifying:

$$-\omega_i \approx \frac{1}{2}\mu^0 \cdot f_i \cdot \Delta N_{\max} \quad (32)$$

As is well known $\Delta N_{\max} = -\mu^0 / \eta$, and substituting this expression in **Eq. (32)** we finally reach:

$$\omega_i \approx \frac{\left(\mu^0\right)^2}{2\eta} \cdot f_i = \omega \cdot f_i \quad (33)$$

Now, if we start from the correlation **Eq. (5)** and replace **Eq. (30)** and **Eq. (27)** we get:

$$-\frac{\mu^0}{\eta_i} \approx f_i \cdot \Delta N_{\max} \quad (34)$$

It is well known that $\Delta N_{\max} = -\mu^0 / \eta$, and substituting this expression in **Eq. (34)** we get:

$$f_i \approx \frac{\eta}{\eta_i} \quad (35)$$

when replacing **Eq. (30)** in **Eq. (25)** you get:

$$\omega_i \approx -\left(\mu^0 \cdot \left(-\frac{\mu^0}{\eta_i}\right) + \frac{1}{2}\eta_i \left(-\frac{\mu^0}{\eta_i}\right)^2\right) \quad (36)$$

operating you get:

$$-\omega_i \approx -\frac{\left(\mu^0\right)^2}{\eta_i} + \frac{1}{2}\frac{\left(\mu^0\right)^2}{\eta_i} \quad (37)$$

and simplifying:

$$\omega_i \approx \frac{\left(\mu^0\right)^2}{2 \cdot \eta_i} \quad (38)$$

Finally, clearing $\eta_i$ from **Eq. (30)** we get:

$$\eta_i \approx -\frac{\mu^0}{\Delta N_i^{\max}} \tag{39}$$

now, replacing $\eta_i$ in **Eq. (25)** we get:

$$-\omega_i \approx \mu^0 \cdot \Delta N_i^{\max} - \frac{1}{2}\mu^0 \cdot \Delta N_i^{\max} \tag{40}$$

and simplifying:

$$\omega_i \approx -\frac{\mu}{2} \cdot \Delta N_i^{\max} \tag{41}$$

Note that for these justifications to be valid, the $f_i^{(2)}$ and $\gamma_i$ values must be negligible This means that the correlations will be better fulfilled the smaller these values are, as discussed in **section 3.3.1**.

**3.4.2. Justification of correlations using the approximate models in *an ideal set of multi-equalised partitions*.**

At this point, it seems interesting to propose the following idea: let us imagine that by grouping partitions in a set we obtain another set of partitions where some of the parameters that we calculate with the models are equalized, i.e., equal in all the partitions of the set. For example, we could be talking about equalizing $f_i$ or $\Delta N_i^{\max}$ or $\omega_i$ or $\eta_i$ ; in any case, the electronic chemical potential will always be equalized by Sanderson's principle (electronegativity equalization principle). In the event that the previous correlations are fulfilled we have an interesting effect. When equalizing a parameter ( $f_i$ ), in addition to the chemical potential, we will find that by effect of the correlations *some other parameters are equalized*.

Let's consider the following case; let's start from the parameters $f_i^+$ and $f_i^-$ of a set of partitions (they can be atoms in a molecule or orbitals, as we have already seen). Let's imagine that by grouping partitions we get another set (evidently, with less partitions) where the following is simultaneously fulfilled:

$$\begin{aligned} f_1^+ &= f_2^+ = \ldots = f_n^+ \\ f_1^- &= f_2^- = \ldots = f_n^- \end{aligned} \tag{42}$$

where *n* is the number of partitions of the new set. Note that condition **Eq. (42)** leads to:

$$f_1^{(2)} = f_2^{(2)} = \ldots = f_n^{(2)} = 0 \tag{43}$$

and this makes the truncation of **Eq. (27)** exact.

Considering that $f_i^+$ and $f_i^-$ are normalized parameter sets and that the sum of the $f_i^+$ and the $f_i^-$ values are equal to one, it can be concluded that $f_i^+ = f_i^- = \frac{1}{n}$. Substituting these values in **Eq. (A7)** (of **Appendix I**) we obtain $f_i = \frac{1}{n}$. Equivalently, we substitute in **Eq. (A9)** (of **Appendix I**) and finally we obtain $f_i^{(2)} = 0$.

Substituting now the values obtained for $f_i$ and $f_i^{(2)}$ in **Eq. (A10)** (of **Appendix I**), we obtain $\Delta N_i = \Delta N / n$, and therefore $\Delta N_i^{\max} = \Delta N^{\max} / n$. In addition, substituting $f_i$ and $f_i^{(2)}$ in **Eq. (A12)** (of **Appendix I**) we obtain:

$$\begin{aligned} \eta_i &= n \cdot \frac{\left(\mu^+ - \mu^-\right)}{2} \\ \gamma_i &= n^2 \cdot \left(\mu^+ + \mu^- - 2\mu\right) \end{aligned} \tag{44}$$

In **Eq. (44)** it can be noted that the local hardness is inversely proportional to the Fukui index (which in this case is 1/n), that the hypersoftness is inversely proportional to the squared Fukui index and that it is closely related to **Eq. (35)**. Substituting the values obtained for $\eta_i$ and $\gamma_i$ of **Eq. (44)** in **Eq. (A13)** (of **Appendix I**) we have:

$$\Delta N_i^{\max} \approx \frac{1}{n} \cdot \frac{\left(\frac{\left(\mu^- - \mu^+\right)}{2}\right) + \sqrt{\frac{1}{4}\left(\mu^+ - \mu^-\right)^2 - 2\mu \cdot \left(\mu^+ + \mu^- - 2\mu\right)}}{\left(\mu^+ + \mu^- - 2\mu\right)} \tag{45}$$

In this last equation it can be noted that the local maximum charge variation is proportional to the Fukui index (which in this case is 1/n) and is closely related to **Eq. (27)**. Now if we substitute the value $\Delta N_i^{\max}$ obtained (**Eq. 45**) in **Eq. (A14)** (of **Appendix I**) we obtain:

$$\begin{aligned} \omega_i \approx -\frac{1}{n} \Bigg( & \mu \cdot \frac{\left(\frac{\left(\mu^- - \mu^+\right)}{2}\right) + \sqrt{\frac{1}{4}\left(\mu^+ - \mu^-\right)^2 - 2\mu \cdot \left(\mu^+ + \mu^- - 2\mu\right)}}{\left(\mu^+ + \mu^- - 2\mu\right)} + \\ & + \frac{1}{2}\frac{\left(\mu^+ - \mu^-\right)}{2} \left( \frac{\left(\frac{\left(\mu^- - \mu^+\right)}{2}\right) + \sqrt{\frac{1}{4}\left(\mu^+ - \mu^-\right)^2 - 2\mu \cdot \left(\mu^+ + \mu^- - 2\mu\right)}}{\left(\mu^+ + \mu^- - 2\mu\right)} \right)^2 + \\ & + \frac{1}{6} \cdot \left(\mu^+ + \mu^- - 2\mu\right) \cdot \left( \frac{\left(\frac{\left(\mu^- - \mu^+\right)}{2}\right) + \sqrt{\frac{1}{4}\left(\mu^+ - \mu^-\right)^2 - 2\mu \cdot \left(\mu^+ + \mu^- - 2\mu\right)}}{\left(\mu^+ + \mu^- - 2\mu\right)} \right)^3 \Bigg) \end{aligned} \tag{46}$$

Note that the latter equation is closely related to **Eq. (33)**. It can also be noted that the local philicity is also proportional to the Fukui index (which in this case is equal to 1/n).

As a consequence, local philicity and maximum charge variation are proportional, and both are inversely proportional to hardness.

Finally, if we substitute the values shown in **Eq. (24)** in **Eq. (A11)** (of **Appendix I**) we can obtain:

$$\Delta E_i \approx \frac{1}{n}\left[\mu \cdot \Delta N + \frac{1}{2}\left(\frac{\mu^+ - \mu^-}{2}\right)\Delta N^2 + \frac{1}{6}\cdot\left(\mu^+ + \mu^- - 2\mu\right)\cdot \Delta N^3\right] \tag{47}$$

**Eqs. (45)**, **(46)** and **(47)** can be decomposed into a term $\frac{1}{n}$ that multiplies another term that only depends on global parameters and that is independent of the number of partitions ($n$). In view of these results, it is easy to conclude that if **Eq. (42)** is satisfied all the descriptors that we have calculated have the same value in any of the partitions. This leads us to the conclusion that the correlations that we have studied are satisfied regardless of the type of partitions used.

#### 3.4.3. Demonstration that the above correlations become exact relationships in a set of ideal multi-equalised partitions.

Until now, we have used approximate models to justify the correlations found in this and previous works, but now we will see that, in the case of multi-equalised partitions, the correlations are exact, for this we will introduce four statements with their corresponding demonstrations.

**Statement 1**. Equalisation of the electronic hardness in a set of "*i*" partitions, where the Fukui indices and the dual descriptor indices are equalised.

Given a system of "*i*" partitions partially equalised such that the Fukui indices are equalised:

$$\left(\frac{\partial N_1}{\partial N}\right)_\upsilon = \left(\frac{\partial N_2}{\partial N}\right)_\upsilon = \ldots = \left(\frac{\partial N_i}{\partial N}\right)_\upsilon \tag{48}$$

where $N_1$, $N_2$, …, $N_i$ are the charges corresponding to partitions 1, 2, ..., $i$, and $N$ is the total charge of the molecule being partitioned. The *dual descriptor* index is also equalised as follows:

$$\left(\frac{\partial^2 N_1}{\partial N^2}\right)_\upsilon = \left(\frac{\partial^2 N_2}{\partial N^2}\right)_\upsilon = \ldots = \left(\frac{\partial^2 N_i}{\partial N^2}\right)_\upsilon = 0 \tag{49}$$

and the rest of the higher order derivatives are also equalised in the same way as the dual descriptor index:

$$\left(\frac{\partial^j N_1}{\partial N^j}\right)_\upsilon = \left(\frac{\partial^j N_2}{\partial N^j}\right)_\upsilon = \ldots = \left(\frac{\partial^j N_i}{\partial N^j}\right)_\upsilon = 0;\ j \geq 2 \tag{50}$$

then the electronic hardnesses of the partitions ($\eta_i$) are equalised in the form:

$$\left(\frac{\partial^2 E_1}{\partial N_1^2}\right)_\upsilon = \left(\frac{\partial^2 E_2}{\partial N_2^2}\right)_\upsilon = \ldots = \left(\frac{\partial^2 E_i}{\partial N_i^2}\right)_\upsilon = \eta_i \tag{51}$$

where $E_i$ is the energy corresponding to partition "*i*".

**Demonstration 1**: We start from an *infinite* expansion of the charge variation of the partition "*i*" of the form:

$$\Delta N_i = \left(\frac{\partial N_i}{\partial N}\right)\cdot \Delta N + \frac{1}{2}\left(\frac{\partial^2 N_i}{\partial N^2}\right)\cdot \Delta N^2 + \ldots + \frac{1}{j!}\left(\frac{\partial^j N_i}{\partial N^j}\right)\cdot \Delta N^j ;\forall i \quad (52)$$

taking into account the premises of the theorem, the infinite expansion is of the form:

$$\Delta N_i = \left(\frac{\partial N_i}{\partial N}\right)\cdot \Delta N;\ \forall i \quad (53)$$

taking into account **Eq. (49)** it can be obtained that $\Delta N_1 = \Delta N_2 = \ldots = \Delta N_i$. If we now perform an *infinite* expansion for the energy variation $\Delta E_i$ of partition "*i*" of the form:

$$\Delta E_i = \left(\frac{\partial E_i}{\partial N_i}\right)_{N=0}\cdot \Delta N_i + \frac{1}{2}\left(\frac{\partial^2 E_i}{\partial N_i^2}\right)_{N=0}\cdot \Delta N_i^2 + \ldots + \frac{1}{j!}\left(\frac{\partial^j E_i}{\partial N_i^j}\right)_{N=0}\cdot \Delta N_i^j ;\forall i$$

(54)

deriving this expansion, we obtain:

$$\left(\frac{\partial E_i}{\partial N_i}\right) = \mu_i = \left(\frac{\partial E_i}{\partial N_i}\right)_{N=0} + \left(\frac{\partial^2 E_i}{\partial N_i^2}\right)_{N=0}\cdot \Delta N_i + \ldots + \frac{1}{(j-1)!}\left(\frac{\partial^{j-1} E_i}{\partial N_i^{j-1}}\right)_{N=0}\cdot \Delta N_i^{j-1} =$$

$$= \mu_i^0 + \left(\frac{\partial \mu_i}{\partial N_i}\right)_{N=0}\cdot \Delta N_i + \frac{1}{2}\left(\frac{\partial^2 \mu_i}{\partial N_i^2}\right)_{N=0}\cdot \Delta N_i^2 + \ldots + \frac{1}{(j-1)!}\left(\frac{\partial^{j-1} \mu_i}{\partial N_i^{j-1}}\right)_{N=0}\cdot \Delta N_i^{j-1};\forall i$$

(55)

where $\mu_i^0$ is the chemical potential of the "*i*" partition of the neutral molecule. According to Sanderson's principle, once equilibrium is reached it is satisfied that $\mu_1 = \mu_2 = \ldots = \mu_i$, independently of the value $\Delta N_i$, furthermore, as $\Delta N_1 = \Delta N_2 = \ldots = \Delta N_i$ we obtain that the functions are equal in all partitions: $\mu_1(\Delta N_1) = \mu_2(\Delta N_2) = \ldots = \mu_i(\Delta N_i)$. This leads to the fact that the derivatives of **Eq. (55)** must be equal in all partitions, since the equality of chemical potentials holds for arbitrary values of Δ*N*, the corresponding functional forms must be identical, which implies the equality of their derivatives, then:

$$\left(\frac{\partial^2 E_1}{\partial N_1^2}\right)_{N=0} = \left(\frac{\partial^2 E_2}{\partial N_2^2}\right)_{N=0} = \ldots = \left(\frac{\partial^2 E_i}{\partial N_i^2}\right)_{N=0} \quad (56)$$

in other words, $\eta_1 = \eta_2 = \ldots = \eta_i$, the electronic hardnesses of the neutral molecule are equalised (whenever the requirements of the statement are met).

**Statement 2**. Correlation between $\omega_i$ and $f_i$.

If we start from a multi-equalised system where:

$$\begin{aligned} f_1 &= f_2 = \ldots = f_i \\ f_1^{(2)} &= f_2^{(2)} = \ldots = f_i^{(2)} = 0 \end{aligned} \quad (57)$$

where the parameters $f_i$ are the local Fukui indices and the $f_i^{(2)}$ are the dual descriptor indices. Then it has to be fulfilled that $\omega_i = \omega \cdot f_i$.

**Demonstration 2**: Taking into account the starting assumptions and that the Fukui indices are normalised of the form $\sum_i f_i = 1$ one can arrive at:

$$f_1 = f_2 = \ldots = f_i = \frac{1}{n} \tag{58}$$

where "$n$" is the number of partitions.

On the other hand, from the previous statement it can be extracted that $\omega_1 = \omega_2 = \ldots = \omega_i$, since $\omega_i$ is the depth of the curve $\Delta E_i(\Delta N_i)$ in the partition "$i$", and as already seen, the derivatives of **Eq. (55)** must be equal in all partitions, furthermore Sanderson's principle guarantees that the first derivatives ($\mu_i^0$) are also equalised.

Finally, from the principle of conservation of energy it can be extracted that $\omega = \sum_i \omega_i$ and as we have already seen that $\omega_1 = \omega_2 = \ldots = \omega_i$, it can be obtained that $\omega_i = \frac{\omega}{n}$. We have already seen that $f_i = \frac{1}{n}$, then it must be fulfilled:

$$\omega_i = \omega \cdot \frac{1}{n} = \omega \cdot f_i \tag{59}$$

Note that the latter equation is closely related to **Eqs. (33)** and **(46)**.

**Statement 3**. Correlation between $\Delta N_i^{\max}$ and $f_i$.

If we start from a multi-equalised system where **Eqs. (57)** are fulfilled, then it must be fulfilled that $\Delta N_i^{\max} = \Delta N^{\max} \cdot f_i$

**Demonstration 3**: from statement 1 it can be extracted that $\Delta N_1^{\max} = \Delta N_2^{\max} = \ldots = \Delta N_i^{\max}$, since $\Delta N_i^{\max}$ it is $\Delta N_i$ in the curve $\Delta E_i(\Delta N_i)$ at the minimum of the curve and as already seen, the derivatives of **Eq. (55)** must be equal in all partitions, furthermore Sanderson's principle guarantees that the first derivatives ($\mu_i^0$) are also equalised.

On the other hand, from the conservation of charge it can be extracted that $\Delta N^{\max} = \sum_i \Delta N_i^{\max}$ and as we have already seen that $\Delta N_1^{\max} = \Delta N_2^{\max} = \ldots = \Delta N_i^{\max}$, then it can be obtained that $\Delta N_i^{\max} = \frac{\Delta N^{\max}}{n}$. We have already seen that $f_i = \frac{1}{n}$, then it must be fulfilled:

$$\Delta N_i^{\max} = \Delta N^{\max} \cdot \frac{1}{n} = \Delta N^{\max} \cdot f_i \tag{60}$$

Note that the latter equation is closely related to **Eqs. (27)** and **(45)**.

**Statement 4**. Correlation between $\eta_i$ and $f_i$.

If we start from a multi-equalised system where **Eqs. (57)** are satisfied, then it must be satisfied that $\eta_i = \eta / f_i$

**Demonstration 4**: if we start from the infinite expansion **Eq. (54)** and carry out the sum $\Delta E = \sum_i \Delta E_i$ we obtain:

$$\Delta E = \sum_i \Delta E_i = \sum_i \left( \left( \frac{\partial E_i}{\partial N_i} \right)_{N=0} \cdot \Delta N_i + \frac{1}{2} \left( \frac{\partial^2 E_i}{\partial N_i^2} \right)_{N=0} \cdot \Delta N_i^2 + \ldots + \frac{1}{j!} \left( \frac{\partial^j E_i}{\partial N_i^j} \right)_{N=0} \cdot \Delta N_i^j \right) =$$

$$= \sum_i \mu \cdot \Delta N_i + \frac{1}{2} \sum_i \eta_i \cdot \Delta N_i^2 + \ldots + \frac{1}{j!} \sum_i \left( \frac{\partial^j E_i}{\partial N_i^j} \right)_{N=0} \cdot \Delta N_i^j =$$

$$= \mu \cdot \Delta N + \frac{1}{2} \eta \Delta N^2 + \ldots + \frac{1}{j!} \left( \frac{\partial^j E}{\partial N^j} \right)_{N=0} \cdot \Delta N^j \tag{61}$$

grouping terms we obtain that $\eta \Delta N^2 = \sum_i \eta_i \cdot \Delta N_i^2$, in the first statement it was concluded that $\eta_1 = \eta_2 = \ldots = \eta_i$; then $\eta \Delta N^2 = \eta_i \sum_i \Delta N_i^2$. On the other hand we have already seen that $\Delta N_1 = \Delta N_2 = \ldots = \Delta N_i$; then $\sum_i \Delta N_i^2 = n \Delta N_i^2$, and furthermore by the conservation of charges it must be fulfilled that $\Delta N = n \Delta N_i$ and therefore $\Delta N^2 = n^2 \Delta N_i^2$. This leads us to the conclusion that

$$\eta = \eta_i \frac{\sum_i \Delta N_i^2}{\Delta N^2} = \eta_i \frac{n \Delta N_i^2}{n^2 \Delta N_i^2} = \frac{\eta_i}{n} \tag{62}$$

as we have already seen that $f_i = \frac{1}{n}$ it is easy to reach:

$$\eta_i = \eta / f_i \tag{63}$$

Note that the latter equation is closely related to **Eqs. (35)** and (**44)**. With the equalities **Eqs. (56)**, **(59)**, **(60)** and **(63)** one can easily arrive at the correlations **Eqs. (3)-(6)**.

As can be seen, a set of multi-equalized partitions presents a very large degree of simplification in relation to the number of variables necessary to describe the reactivity of the molecule. The values of all the local variables are the same in each of the partitions, and they have strong correlations between them. Basically only a few global parameters and the number of partitions ($n$) are necessary to describe the reactivity of the system since all the local parameters are perfectly determined. At the local level, only one parameter remains that has different values in different partitions and that is the partition size (i.e. the partition volumes if they remain independent in the multi-equalized set). It is easy to connect the partition size with the partition reactivity by the concept of electronic localization, all other parameters being equal. Smaller partitions will localize more charge in the case of an attacker transferring electronic charge to the reactant in a chemical reaction. Covalent bonding is known to be closely related to electronic localization, so it is easy to think of a relationship between partition size and the tendency to form covalent bonds.

### 3.4.4. Matrix form of the multi-equalization concept.

The concept of multi-equalization allows us to achieve a partition set where the main local descriptors of reactivity (mainly the derivatives of the energy and the local charge with regard to the number of electrons) coincide in all partitions. This is possibly an ideal concept, and it is probably necessary to start from a set with infinite partitions to achieve an accurately multi-equalized set of partitions, but we believe that qualitative results can be obtained by an approximation to this set of partitions, as will be seen in one of the examples presented in the following sections.

Finally, it has been seen that the combinations of multi-equalized partitions must fulfil **Eq. (42)**, also taking into account **Eq. (22)** in its matrix form; we can summarize the equations that the multi-equalized combinations must fulfil in:

$$\begin{pmatrix} EX_{11}^{2} & \dots & EX_{1n}^{2} \\ \dots & \dots & \dots \\ EX_{1n}^{2} & \dots & EX_{nm}^{2} \end{pmatrix} \begin{pmatrix} f_1^{+X} \\ \dots \\ f_m^{+X} \end{pmatrix} \approx \begin{pmatrix} f_1^{+E} \\ \dots \\ f_n^{+E} \end{pmatrix} = \begin{pmatrix} \frac{1}{n} \\ \dots \\ \frac{1}{n} \end{pmatrix}_{nx1} \quad (64)$$

and

$$\begin{pmatrix} EX_{11}^{2} & \dots & EX_{1n}^{2} \\ \dots & \dots & \dots \\ EX_{1n}^{2} & \dots & EX_{nm}^{2} \end{pmatrix} \begin{pmatrix} f_1^{-X} \\ \dots \\ f_m^{-X} \end{pmatrix} \approx \begin{pmatrix} f_1^{-E} \\ \dots \\ f_n^{-E} \end{pmatrix} = \begin{pmatrix} \frac{1}{n} \\ \dots \\ \frac{1}{n} \end{pmatrix}_{nx1} \quad (65)$$

where $X$ corresponds with the set of starting partitions, $E$ to the final multi-equalized set, and the coefficients $EX_{ij}^{2}$ determine the combinations of the starting partitions. For example, in the case where the final partitions are obtained by grouping partitions from the starting set, the coefficients $EX_{ij}^{2}$ will have values "1" or "0", which can greatly simplify the search process.

### 3.5. First examples of multi-equalization.

The first example to be discussed will be the benzene molecule (**Figure 11 - left**). We will consider a multi-equalized partition set consisting of six partitions corresponding to a carbon and its bonded hydrogen. It is evident that the six partitions considered are symmetrically equivalent and the $f_i^{+}$ and $f_i^{-}$ values are all equal to $\frac{1}{6}$. The overall values obtained were (in a.u.) $\mu$ : -0.1434, $\eta$: 0.3938, $\Delta N^{\max}$ : 0.3642 and $\omega$: 0.0261. The $f_i^{+}$ and $f_i^{-}$ indices were entered in the UCA-FUKUI software and the values shown in **Table 1** were obtained. As can be seen, the equalities that we obtained previously $f_i = \frac{1}{n}$, $f_i^{(2)}$ =0, $\Delta N_i^{\max} = \Delta N^{\max}/n$ ; $\forall i$, are verified, and in addition, taking into account the correlations that we have verified previously, it can be seen that $\omega_i = \omega \cdot f_i$ and $f_i \cdot \eta_i = \eta$.

| | $f_i$ | $f_i^{(2)}$ | $\Delta N_i^{\max}$ | $\eta_i$ | $\gamma_i$ | $\Delta N_i^{\max\prime}$ | $\omega_i$ |
|---|---|---|---|---|---|---|---|
| **{C+H}** | 0.1667 | 0.0000 | 0.0607 | 2.3628 | -0.4051 | 0.0610 | 0.0044 |

**Table 1**. Reactivity descriptors calculated for $C_6H_6$, using a set of multi-equalized 6-elements partitions. The global values obtained were $\mu$ : -0.1434, $\eta$ : 0.3938, $\Delta N^{\max}$ : 0.3642 and $\omega$: 0.0261 (in a.u.).

The second example we have studied is the $CH_2CHCH_3$ molecule (**Figure 11 - right**). Four partitions have been formed starting from the 9 atoms of the molecule and we have grouped them as follows {$C_1$, $C_2$+$H_8$, $H_3$+$H_7$+$H_9$, $H_4$+$H_5$+$C_6$}. This new set has the values $f_i^{-}$ and $f_i^{+}$ as close as could be obtained to the atomic values. This new set was introduced in the UCA-FUKUI software and the values shown in **Table 2** were obtained. The $f_i^{-}$ and $f_i^{+}$ values are not equal to each other, but if they are compared with the values of **Table S1** of the Supporting Information, it is clear that the new set shows much more equalized values than the original one. It is interesting to note that the remaining parameters follow the same trend, as the descriptors in **Table 2** are much more equalized than those in **Table S1**.

| | **$C_1$** | **$C_2$+ $H_8$** | **$H_3$+ $H_7$+ $H_9$** | **$H_4$+ $H_5$+ $C_6$** |
|---|---|---|---|---|
| $f_i^{-}$ | 0.281 | 0.264 | 0.236 | 0.219 |
| $f_i^{+}$ | 0.263 | 0.261 | 0.239 | 0.237 |
| $f_i$ | 0.272 | 0.263 | 0.238 | 0.228 |
| $f_i^{(2)}$ | -0.018 | -0.003 | 0.003 | 0.018 |
| $\Delta N_i^{\max}$ | 0.065 | 0.063 | 0.057 | 0.055 |
| $\eta_i$ | 1.753 | 1.811 | 2.000 | 2.096 |
| $\gamma_i$ | -0.481 | -0.505 | -0.510 | -0.538 |
| $\Delta N_i^{\max\prime}$ | 0.066 | 0.064 | 0.057 | 0.055 |
| $\omega_i$ | 0.0037 | 0.0036 | 0.0033 | 0.0031 |
| Volume: | 81.22 | 107.54 | 108.33 | 136.76 |

**Table 2.** Reactivity descriptors calculated for $CH_2CHCH_3$ using a 4-elements partition set, all the values are in a.u.. The global values obtained were (in a.u.) $\mu$ : -0.1139, $\eta$: 0.4756, $\Delta N^{\max}$ : 0.2396, and $\omega$ : 0.0136.

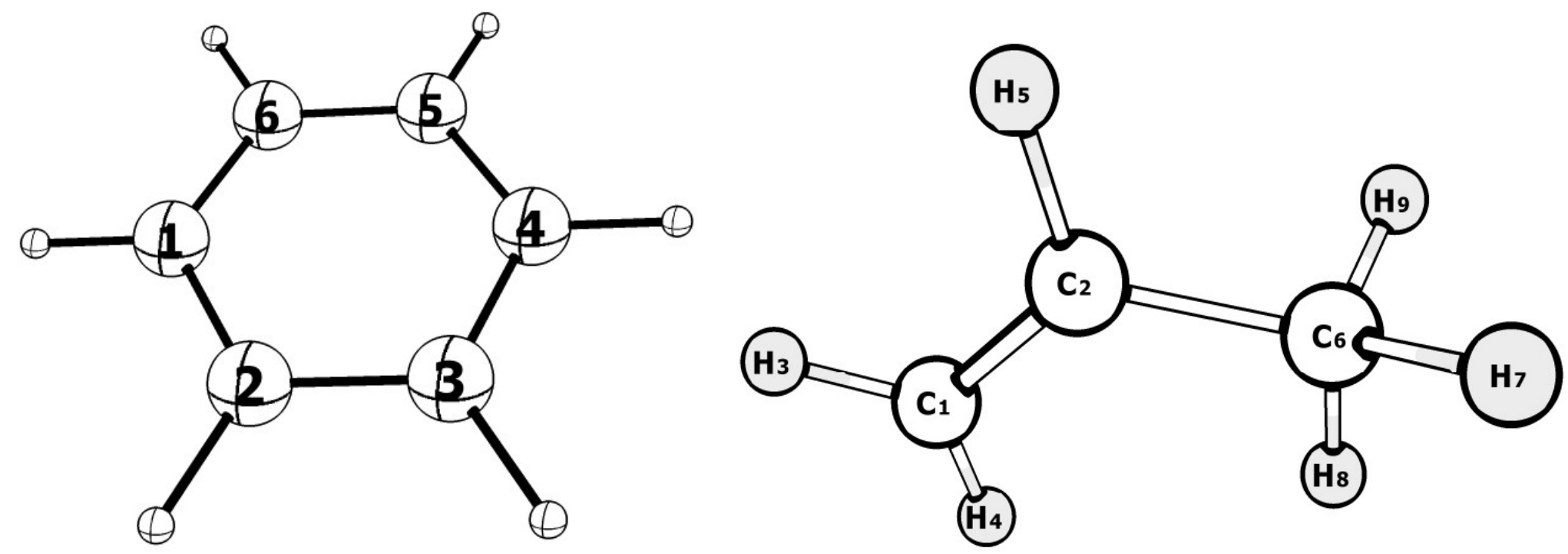

**Figure 11**. Left) Numbering corresponding to the benzene molecule in the example. Right) Numbering corresponding to the example molecule $CH_2CHCH_3$

### 3.6. Partial multi-equalization.

If we restrict **Eq. (42)** to only a subset of partitions (the first *a* elements) of the starting set (of *m* partitions), we obtain the equalities:

$$\begin{pmatrix} EX_{11}^2 & \dots & EX_{11}^2 & 0 & \dots & 0 \\ \dots & \dots & \dots & \dots & \dots & \dots \\ EX_{11}^2 & \dots & EX_{ab}^2 & 0 & \dots & 0 \\ 0 & \dots & 0 & 1 & \dots & 0 \\ \dots & \dots & \dots & \dots & \dots & \dots \\ 0 & \dots & 0 & 0 & \dots & 1 \end{pmatrix} \begin{pmatrix} f_1^{-X} \\ \dots \\ f_b^{-X} \\ f_{b+1}^{-X} \\ \dots \\ f_m^{-X} \end{pmatrix} \approx \begin{pmatrix} f_1^{-E} \\ \dots \\ f_a^{-E} \\ f_{a+1}^{-X} \\ \dots \\ f_n^{-X} \end{pmatrix} = \begin{pmatrix} \varphi \\ \dots \\ \varphi \\ f_{a+1}^{-X} \\ \dots \\ f_n^{-X} \end{pmatrix} \tag{66}$$

and

$$\begin{pmatrix} EX_{11}^2 & \dots & EX_{11}^2 & 0 & \dots & 0 \\ \dots & \dots & \dots & \dots & \dots & \dots \\ EX_{11}^2 & \dots & EX_{ab}^2 & 0 & \dots & 0 \\ 0 & \dots & 0 & 1 & \dots & 0 \\ \dots & \dots & \dots & \dots & \dots & \dots \\ 0 & \dots & 0 & 0 & \dots & 1 \end{pmatrix} \begin{pmatrix} f_1^{+X} \\ \dots \\ f_b^{+X} \\ f_{b+1}^{+X} \\ \dots \\ f_m^{+X} \end{pmatrix} \approx \begin{pmatrix} f_1^{+E} \\ \dots \\ f_a^{+E} \\ f_{a+1}^{+X} \\ \dots \\ f_n^{+X} \end{pmatrix} = \begin{pmatrix} \varphi \\ \dots \\ \varphi \\ f_{a+1}^{+X} \\ \dots \\ f_n^{+X} \end{pmatrix} \tag{67}$$

where *X* corresponds to the set of starting partitions and *E* to the final multi-equalised set and $\varphi$ is the value corresponding to the parameters $f_i^{-E}$ and $f_i^{+E}$ of the multi-equalized partition subset. It can be verified that the correlations are still fulfilled in this subset since substituting the calculated values $f_i^{-E}$ and $f_i^{+E}$ in **Eqs. (A12)**, **(A13)**, **(A14)**, **(A11)** and **(A10)** of **Appendix I** we obtain results equivalent to **Eqs. (44)**, **(45)**, **(46)** and **(47)**.

#### 3.6.1. The simplest version of partial multi-equalization: multi-equalization of two partitions into a set of three or more partitions.

This is the simplest example of partial multi-equalization that can be studied. It consists of constructing two multi-equalized partitions within a set of three or more partitions. Let us consider that the multi-equalized partitions are those corresponding to indexes $i$=1 and $i$=2 (partitions $p_1$ and $p_2$ respectively), then $\forall i \; / \; 2 < i : f_1^+ = f_2^+ \neq f_i^+$ and $\forall i \; / \; 2 < i : f_1^- = f_2^- \neq f_i^-$. Note that in this case $f_i^\alpha \left(\alpha = + or -\right)$ need not be equal to $1/n$. For the two partitions $p_1$ and $p_2$ to be multi-equalized, the following conditions must be fulfilled:

$$\begin{aligned} f_1^+ &= f_1^{'-} = \delta \\ f_2^+ &= f_2^{'-} = \delta \end{aligned} \tag{68}$$

where $\delta$ is an arbitrary value. Let's consider that there is no generation in the frontier orbitals, in that case it is fulfilled:

$$\begin{aligned} f_1 &= \frac{1}{2}\left(f_1^+ + f_1^-\right) = \delta \\ f_1^{(2)} &= f_1^+ - f_1^- = 0 \end{aligned} \tag{69}$$

On the other hand, taking into account **Eq. (A10)** (of **Appendix I**) we can obtain:

$$\begin{aligned} \Delta N_1 &= f_1 \Delta N + \frac{1}{2} f_1^{(2)} \Delta N^2 = \delta \, \Delta N \\ \Delta N_2 &= f_2 \Delta N + \frac{1}{2} f_2^{(2)} \Delta N^2 = \delta \, \Delta N = \Delta N_1 \end{aligned} \tag{70}$$

Since we have considered that there is no generation, we can reach:

$$\begin{aligned} f_2 &= \frac{1}{2}\left(f_2^+ + f_2^-\right) = \delta = f_1 \\ f_2^{(2)} &= f_2^+ - f_2^- = 0 = f_1^{(2)} \end{aligned} \tag{71}$$

and it can be written as follows:

$$\begin{aligned} f_1 &= f_2 = f \\ f_1^{(2)} &= f_2^{(2)} = f^{(2)} \end{aligned} \tag{72}$$

then **Eq. (70)** can be written as follows:

$$\Delta N_1 \approx \Delta N_2 \approx f \Delta N + \frac{1}{2} f^{(2)} \Delta N^2 \tag{73}$$

Substituting $f_1^+$ and $f_1^{'-}$ in **Eq. (A12)** (of **Appendix I**) we obtain:

$$\eta_1 = \frac{\left(\mu^+ - \mu\right)\cdot\left(-f_1^-\right)^2 - \left(\mu^- - \mu\right)\cdot\left(f_1^+\right)^2}{f_1^- \cdot f_1^+ \cdot \left(f_1^- + f_1^+\right)} \tag{74}$$

where $\mu$, $\mu^+$ and $\mu^-$ are global parameters (and equal for $i$ = 1 and $i$ = 2). Substituting the parameters with their value we obtain:

$$\eta_1 = \frac{(\mu^+ - \mu)\cdot(-\delta)^2 - (\mu^- - \mu)\cdot(\delta)^2}{\delta\cdot\delta\cdot(\delta+\delta)} = \frac{\mu^+ - \mu^-}{2\delta} = \eta_2 \quad (75)$$

and equivalently, we can finally obtain that $\gamma_1 = \gamma_2$. Finally, substituting in **Eqs. (A13)** and **(A14)** of **Appendix I**, we have also been able to verify that:

$$\Delta N_1^{max} = \frac{-\eta_1 + \sqrt{\eta_1^{\,2} - 2\gamma_1\ \mu}}{\gamma_1} = \frac{-\eta_2 + \sqrt{\eta_2^{\,2} - 2\gamma_2\ \mu}}{\gamma_2} = \Delta N_2^{max} \quad (76)$$

and that

$$\begin{aligned}\omega_1 &\approx -\left(\mu\cdot\Delta N_1^{max} + \frac{1}{2}\eta_1\left(\Delta N_1^{max}\right)^2 + \frac{1}{6}\cdot\gamma_1\cdot\left(\Delta N_1^{max}\right)^3\right) = \\ &= -\left(\mu\cdot\Delta N_2^{max} + \frac{1}{2}\eta_2\left(\Delta N_2^{max}\right)^2 + \frac{1}{6}\cdot\gamma_2\cdot\left(\Delta N_2^{max}\right)^3\right) \approx \omega_2\end{aligned} \quad (77)$$

### 3.6.2 An example of partial multi-equalization: the molecule $CH_2CHCN$.

**Scheme 1** shows the reaction of acronitrile with a protonated nucleophile, cyanoethylation, is a process for the attachment of $CH_2CH_2CN$ group to another organic substrate. The method is used in the synthesis of organic compounds where protic nucleophiles are alcohols, thiols, and amines[38-41]. A general mechanism for this type of reaction can be seen in **Scheme 1**. **Table 3** shows the reactivity descriptors calculated for $CH_2CHCN$ using a 3-element partition set. The partitions have been formed in this way so that the $f_i$ , $f_i^+$, $f_i^-$ values are as equalised as possible in two of the three sets. Note that the index analysis only describes the first step of the reaction. The partitions that obtain similar values are $C_1$+$C_2$ and $C_6$+$N_7$, similar reactivity would be expected from these two sets in the first stage, as can be seen these sets can be related to the simultaneous rupture of the C-C double bond and the C-N triple bond in the first stage of the reaction mechanism.

HY : + CH₂═CH—C≡N : ⟶ $^{\oplus}$Y(H)—CH₂—CH═C═N$^{\ominus}$ : ⟶ CH₂Y—CH₂—C≡N :

**Scheme 1**. Reaction of acronitrile with a protonated nucleophile HY:.

| | $C_1$+ $C_2$ | $H_3$+ $H_4$+ $H_5$ | $C_6$+ $N_7$ |
|---|---|---|---|
| $f_i^-$ | 0.4011 | 0.2064 | 0.3926 |
| $f_i^+$ | 0.4132 | 0.2434 | 0.3434 |
| $f_i$ | 0.4072 | 0.2249 | 0.3680 |
| $f_i^{(2)}$ | 0.0121 | 0.0370 | -0.0492 |

| $\Delta N_i^{max}$ | 0.1695 | 0.0962 | 0.1481 |
|---|---|---|---|
| $\eta_i$ | 1.0400 | 1.9076 | 1.1605 |
| $\gamma_i$ | -0.5192 | -0.5093 | -0.4287 |
| $\Delta N_i^{max}$' | 0.1761 | 0.0930 | 0.1554 |
| $\omega_i$ | 0.0152 | 0.0081 | 0.0135 |
| Volume: | 213.64 | 188.75 | 286.55 |

**Table 3**. Reactivity descriptors calculated for $CH_2CHCN$ using a 3-elements partition set, all the values are in a.u.. The global values obtained were (in a.u.) $\mu$ : -0.1751, $\eta$: 0.4233, $\Delta N^{max}$ : 0.4138, and $\omega$ : 0.0362. The isovalue used to calculate the volume is 0.0004.

## CONCLUSIONS

The concept of multi-equalization, which is based on the studied correlations and the principle of equalization, has been introduced and justified, and the first examples have been presented and analysed. It has also been demonstrated that in a set of multi-equalised partitions, the correlations we have studied in this and previous works are fulfilled, independently of the type of partitions studied. The conclusion that can be reached is that the model used in this work to calculate reactivity descriptors provides approximate correlations in the case of non-multi-equalised systems; however, it has been demonstrated that for multi-equalised systems the correlations obtained are exact. What has not yet been clarified is whether the correlations in non-multi-equalised systems are not accurate because of the approximations of the model used or because the correlations are not exact. We are inclined to think that the correlations have been masked by the model approximations but that they are perfectly valid and the drastic simplification in multi-equalised systems makes the model provide accurate values despite its approximations.

The case of making groupings of atoms in molecules has also been studied. Groupings have been made by atom type, adding atoms of the same type and thus forming a new set of partitions for each molecule. The new sets also show very high correlations, of the same order as the original partitions, which leads us to believe that the sets constructed by grouping atoms in the molecule maintain the correlations of the starting set.

Finally, it has been shown that the concept of multi-equalization allows the number of variables in the system under study to be drastically simplified. Basically, the behaviour of the partitions is determined by a single variable, the size of the partition. Many studies, in this context of work, focus on one or a few parameters to describe the reactivity of chemical systems, ignoring the remaining variables to a great extent. Using the concept of multi-equalization (an ideal set of partitions), the chemical behaviour of the system can be described using a single variable but with the great advantage that the remaining variables do not need to be considered because they *are equalised*.

## FUTURE PERSPECTIVES

We are currently developing a computer code, in the form of a calculation module, which will be implemented in the UCA-FUKUI software. This software, starting from a set of partitions (such as a set of orbitals) and the corresponding parameters, will allow us to obtain a new set of partitions (logically with a smaller number of partitions) that is as close as possible to the multi-equalised (an ideal set of partitions). As soon as the program is running, we intend to study the set of example molecules we have discussed in this work, looking for multi-equalised partition sets.

Furthermore we are also thinking of making new sets of partitions using the basin analysis[42, 43] and using as starting functions the electron density, the spin density (in radicals or ions), the Fukui or dual descriptor functions, the HOMO, LUMO orbitals, their squares or linear combinations of these (such as those corresponding to the Fukui and dual-descriptor function under the FMO approximation), the ALIE (average local ionization energy), the ELF, or even the PES, using the Multiwfn software[44]. The study with the new partition sets is focused on checking that the correlations found for the case of atoms in molecules and the types of orbitals we have studied are fulfilled.

In reference **[45]**, the models used in this work were used to calculate reactivity descriptors *along the IRC*. Currently, we are studying how the Multi-equalised concept can be applied to partitions along the IRC of a chemical reaction, but with the inclusion of terms representing the variation of the external field. Most interestingly, a set of partitions remains Multi-equalised even if the total charge changes, indicating that the effect of varying the external field can be observed in the deformation of the Multi-equalised set of partitions. There is a clear distinction here, as the load variation will affect the global parameter values without deforming the Multi-equalised partition set, while the external field variation will be the only factor affecting the deformation of the Multi-equalised partition set.

**ACKNOWLEDGEMENTS**

The calculations were performed using the computational facilities from the supercomputational area of University of Cadiz and CICA (Centro Informático Científico de Andalucía).

**REFERENCES**

[1] R. T. Sanderson, An Interpretation of Bond Lengths and a Classification of Bonds. SCIENCE **1951** vol 114 670-672.

[2] G. Van Hooydonk and Z. Eeckhaut, On the Principle of Electronegativity Equalization and its Usein the Theory of the Chemical Bond. Berichte der Bunsengesellschaft für physikalische Chemie Vol 74, 4, **1970**, 323-326 https://doi.org/10.1002/bbpc.19700740405

[3] Van Hooydonk, G. Calculation of bond energies in diatomic molecules. Theoret. Chim. Acta 22, 157–166 (1971). https://doi.org/10.1007/BF00537624

[4] R. T. Sanderson, Principles of Electronegativity, Journal of Chemical Education, **1988**, Volume 65, Number 2, 112-118.

[5] R. T. Sanderson. J. Chem. Educ. **1952**, 29,539544,1954.31,2-7,238-245.

[6] R. G. Parr, L. J. Bartalotti, J. Am. Chem. Soc. **1982**, 104, 3801-3803.

[7] R. G. Parr, R. A. Donnelly, M. Levy, W. E. Palke, J. Chem. Phys. **1978**, 68, 38013807.

[8] R. G. Parr, In Electron Distributions and the Chemical Bond; P. Cappens,: Hall. M. 6. Mullay, J. J. Am. Chem. Soc. **1984**,106, 5842-5847. B.. Eds.; Pienun: NewYork, 1982; pp95-1W.

[9] Parr R, Yang W (1989) Density-Functional Theory of Atoms and Molecules. Oxford University Press

[10] Perdew JP, Parr RG, Levy M, Balduz JL, Jr. (1982) Density-functional theory for fractional particle number: derivative discontinuities of the energy. Phys Rev Lett 49:1691-1694

[11] Yang WT, Zhang YK, Ayers PW (2000) Degenerate ground states and fractional number of electrons in density and reduced density matrix functional theory. Phys Rev Lett 84:5172-5175

[12] Ayers PW (2008) The continuity of the energy and other molecular properties with respect to the number of electrons. J Math Chem 43:285-303

[13] Mulliken RS (1934) A New Electroaffinity Scale; Together with Data on Valence States and on Valence Ionization Potentials and Electron Affinities J. Chem. Phys. 2 :782

[14] Iczkowski RP, Margrave JL (1961) Electronegativity J. Am. Chem. Soc. 83:3547-3551

[15] Sen KD, Jørgensen CK (1987) Electronegativity, Structure and Bonding (Springer, Berlin,

[16] Pearson RG (1997) Chemical Hardness: Applications from Molecules to Solids. Wiley-VCH, Weinheim,

[17] Geerlings P, De Proft F, Langenaeker W (2003) Conceptual Density Functional Theory. Chem. Rev. 103:1793-1874

[18] Perdew JP, Parr RG, Levy M, Balduz JL, Jr. (1982) Density-functional theory for fractional particle number: derivative discontinuities of the energy. Phys Rev Lett 49:1691-1694

[19] Yang WT, Zhang YK, Ayers PW (2000) Degenerate ground states and fractional number of electrons in density and reduced density matrix functional theory. Phys Rev Lett 84:5172-5175

[20] Parr RG, Szentpaly LV, Liu SB (1999) Electrophilicity Index. J. Am. Chem. Soc. 121:1922-1924

[21] Liu SB (2009) Electrophilicity. In: Chattaraj PK (ed) Chemical reactivity theory: A density functional view. Taylor and Francis, Boca Raton, p 179

[22] Chattaraj PK, Giri S (2009) Electrophilicity index within a conceptual DFT framework. Annual Reports of Progress in Chemistry C 105:13-39

[23] Chattaraj PK, Sarkar U, Roy DR (2006) Electrophilicity index. Chem Rev 106:2065-2091

[24] Chattaraj PK, Maiti B, Sarkar U (2003) Philicity: A unified treatment of chemical reactivity and selectivity. J Phys Chem A 107:4973-4975

[25] Shubin Liu (2009) Chemical Reactivity Theory a Density Functional View, CRC Press, Chapter 13

[26] J. Sánchez-Márquez, V. García, D. Zorrilla, M. Fernández. New insights in conceptual DFT: New model for the calculation of local reactivity indices based on the Sanderson's principle. Int J Quantum Chem. **2018**; e25844.

[27] Sánchez-Márquez, J. New advances in conceptual-DFT: an alternative way to calculate the fukui function. *J. Mol. Mod.* **2019**, *25*, 123.

[28] J. Sánchez-Márquez. Introducing new reactivity descriptors: "Bond reactivity indices." Comparison of the new definitions and atomic reactivity indices. J. Chem. Phys. 145, 194105 (2016).

[29] R. Draper Norman, H. Smith, Applied Regression Analysis (third edition) Wiley series in Probability and Stadistics, John Wiley & Sons; ISBN: 0-471-17082-8

[30] J. Sánchez-Márquez. Correlations between Fukui Indices and Reactivity Descriptors. Based on Sanderson's Principle. J. Phys. Chem. A 2019, 123, 8571−8582.

[31] Jesús Sánchez-Márquez, David Zorrilla Cuenca, Manuel Fernández Núñez, Víctor García Hernández. Introducing a new model based on electronegativity equalization principle for the analysis of the natural bond orbital reactivity in the c-DFT background. Int J Quantum Chem. 2022; e26993. https://doi.org/10.1002/qua.26993

[32] Becke, A.D. Density-functional thermochemistry. III The role of exact exchange. *J. Chem. Phys.* **1993**, *98*, 5648-52.

[33] Frisch, M. J., Pople, J. A. and Binkley, J. S. Self-consistent molecular orbital methods. 25. Supplementary functions for gaussian basis sets. *J. Chem. Phys.* **1984**, *80*, 3265-3269.

[34] Gaussian 16, Revision B.01, M. J. Frisch, G. W. Trucks, H. B. Schlegel, G. E. Scuseria, M. A. Robb, J. R. Cheeseman, G. Scalmani, V. Barone, G. A. Petersson, H. Nakatsuji, X. Li, M. Caricato, A. V. Marenich, J. Bloino, B. G. Janesko, R. Gomperts, B. Mennucci, H. P. Hratchian, J. V. Ortiz, A. F. Izmaylov, J. L. Sonnenberg, D. Williams-Young, F. Ding, F. Lipparini, F. Egidi, J. Goings, B. Peng, A. Petrone, T. Henderson, D. Ranasinghe, V. G. Zakrzewski, J. Gao, N. Rega, G. Zheng, W. Liang, M. Hada, M. Ehara, K. Toyota, R. Fukuda, J. Hasegawa, M. Ishida, T. Nakajima, Y. Honda, O. Kitao, H. Nakai, T. Vreven, K. Throssell, J. A. Montgomery, Jr., J. E. Peralta, F. Ogliaro, M. J. Bearpark, J. J. Heyd, E. N. Brothers, K. N. Kudin, V. N. Staroverov, T. A. Keith, R. Kobayashi, J. Normand, K. Raghavachari, A. P. Rendell, J. C. Burant, S. S. Iyengar, J. Tomasi, M. Cossi, J. M. Millam, M. Klene, C. Adamo, R. Cammi, J. W. Ochterski, R. L. Martin, K. Morokuma, O. Farkas, J. B. Foresman, and D. J. Fox, Gaussian, Inc., Wallingford CT, 2016.

[35] Gauss View 5.0. https://wiki.crc.nd.edu/w/images/d/d7/Gaussview-5-ref.pdf

[36] Zhurko G. A. Chemcraft - graphical program for visualization of quantum chemistry computations. Ivanovo, Russia, 2005. https://chemcraftprog.com

[37] Sánchez-Márquez, J., Zorrilla, D., Sánchez-Coronilla, A.M., de los Santos, D., Navas J., Fernández-Lorenzo, C., Alcántara, R. and Martín-Calleja, J. Introducing UCA-FUKUI software: reactivity-index calculations. *J. Mol. Model.* **2014**, *20*, 2492.

[38] J. Cymerman-Craig, M. Moyle. *Organic Syntheses Collective* **4**, 205 (1963)

[39] S. A. Heininger. *Organic Syntheses Collective* **4**, 146 (1963)

[40] T. E. Snider, D. L. Morris, K. C. Srivastava, K. D. Berlin. *Organic Syntheses Collective* **6**, 932 (1988)

[41] E. C. Horning, A. F. Finelli. *Organic Syntheses Collective* **4**, 776 (1963)

[42] Biegler-König, F. (2000), Calculation of atomic integration data. J. Comput. Chem., 21: 1040-1048.

https://doi.org/10.1002/1096-987X(200009)21:12<1040::AID-JCC2>3.0.CO;2-8

[43] Paul L. A. Popelier Fully Analytical Integration Over the 3D Volume Bounded by the β Sphere in Topological Atoms J. Phys. Chem. A, 115, 45, 13169–13179 (2011)

[44] Tian Lu, Feiwu Chen, Multiwfn: A Multifunctional Wavefunction Analyzer, J. Comput. Chem. 33, 580-592 (2012).

[45] Jesús Sánchez-Márquez, Himangshu Mondal, Shanti Gopal Patra, Alejandro Morales-Bayuelo, Pratim Kumar Chattaraj. Local reactivity descriptors of the important atoms in chelotropic reactions provide insight into their global variants along the reaction path. Int J Quantum Chem. e27129 (2023). https://doi.org/10.1002/qua.27129

# Multi-Equalization in Conceptual Density Functional Theory: Beyond Electronegativity Equalization.

Jesús Sánchez-Márquez[1].

[1] Departamento de Química-Física, Facultad de Ciencias, Campus Universitario Río San Pedro, Universidad de Cádiz, 11510 Puerto Real, Cádiz, Spain.

*To whom correspondence should be addressed. E-mail: jesus.sanchezmarquez@uca.es

## Supporting Information

---

**Table of contents:**

**1.-UCA-FUKUI software:**

**2.- Maximum local charge variations (obtained with a cubic expansion of the local energy) versus the maximum local charge variations (achieved with a quadratic expansion of the local charge variation) for NHOs:**

**3.- Local electrophilicities versus local Fukui indices for NHOs:**

**4.- Local values $\eta_i^{-1}$ versus the local Fukui indices for NHOs:**

**5.- Local values $\eta_i^{-1}$ versus the maximum local charge variations for NHOs:**

**6.- Maximum local charge variations (obtained with a cubic expansion of the local energy) versus the maximum local charge variations (achieved with a quadratic expansion of the local charge variation) for CMOs:**

**7.- Local electrophilicities versus local Fukui indices for CMOs:**

**8.- Local values $\eta_i^{-1}$ versus the local Fukui indices for CMOs:**

**9.- Local values $\eta_i^{-1}$ versus the maximum local charge variations for CMOs:**

**10.- Tables:**

## Appendix I. Partitions formed by atomic or molecular orbitals: natural hybrid orbitals and canonical molecular orbitals.

In this appendix we will present the method to calculate reactivity descriptors for atomic and molecular orbitals, the results of the derivative calculation applied to the case of natural hybrid orbitals (NHO) and canonical molecular orbitals (CMO), and the correlations found between the derivatives of the newly proposed models. As will be seen in this appendix, there are *strong indications* that the correlations found in previous work between reactivity descriptors for atoms in the molecule and NBOs can also be observed in NHOs and CMOs. We first begin by describing the model used to calculate these derivatives.

### A.1. Calculation of reactivity descriptors for atomic and molecular orbitals $f_i^{\alpha}$ $(\alpha = + or -)$.

In ref. **[26]**, a methodology was defined to obtain bond reactivity indices for NBOs ( $f_i^{-(NBO)}$ and $f_i^{+(NBO)}$) based on the regression that supplies reactivity indices for bond orbitals instead of the atoms in the molecule. Previous indices satisfy: $\sum_{i=1}^{all\ orbitals} f_i^{-(NBO)} = 1$ and $\sum_{i=1}^{all\ orbitals} f_i^{+(NBO)} = 1$. Lagrange's multipliers[27] were used to normalize the indices of Eq. (A1) and we obtained:

$$\int\left(f^{-}(\vec{r}) - \sum_{i=1}^{all\ orbitals} f_i^{-(NBO)}\left|\phi_i^{(NBO)}(\vec{r})\right|^2\right)^2 d\vec{r} + \lambda\left[\sum_{i=1}^{all\ orbitals} f_i^{-(NBO)} - 1\right] = MINIMUM$$
$$\int\left(f^{+}(\vec{r}) - \sum_{i=1}^{all\ orbitals} f_i^{+(NBO)}\left|\phi_i^{(NBO)}(\vec{r})\right|^2\right)^2 d\vec{r} + \lambda\left[\sum_{i=1}^{all\ orbitals} f_i^{+(NBO)} - 1\right] = MINIMUM$$

(A1)

The more general version that we have obtained based on **Eq. (A1)** and that has been used to calculate indices corresponding to natural hybrid orbitals (NHO) and canonical molecular orbitals (CMO), as can be seen in the Results and Discussion section, was:

$$\int\left(f^{-}(\vec{r}) - \sum_{i=1}^{all\ orbitals} f_i^{-(NHO/CMO)}\left|\phi_i^{(NHO/CMO)}(\vec{r})\right|^2\right)^2 d\vec{r} + \lambda\left[\sum_{i=1}^{all\ orbitals} f_i^{-(NHO/CMO)} - 1\right] = MINIMUM \qquad (A2)$$

$$\int\left(f^{+}(\vec{r}) - \sum_{i=1}^{all\ orbitals} f_i^{+(NHO/CMO)}\left|\phi_i^{(NHO/CMO)}(\vec{r})\right|^2\right)^2 d\vec{r} + \lambda\left[\sum_{i=1}^{all\ orbitals} f_i^{+(NHO/CMO)} - 1\right] = MINIMUM \qquad (A3)$$

In order to perform calculations of $f_i^{\alpha(NHO/CMO)}$ $(\alpha = + or -)$, a computer code has been created in the form of a calculation module that has been implemented in the UCA-FUKUI software (see section *Computational details*).

#### A.1.1. Local charge derivatives with regards to the total number of electrons (*N*).

If we start from a second-order approximation for the electron density as in **Eq. (A4)**,

$$\Delta\rho_N(\mathbf{r}) \approx \left(\frac{\partial\rho_N(\mathbf{r})}{\partial N}\right)_\upsilon \Delta N + \frac{1}{2}\left(\frac{\partial^2\rho_N(\mathbf{r})}{\partial N^2}\right)_\upsilon \Delta N^2 \quad \text{(A4)}$$

and substitute the values $\Delta N = 1$ ($\Delta\rho_N^-(\mathbf{r})$) and $\Delta N = -1$ ($\Delta\rho_N^+(\mathbf{r})$) in **Eq. (A4)** we can obtain:

$$\begin{aligned}\Delta\rho_N^-(\mathbf{r}) &= \rho_{N-1}(\mathbf{r}) - \rho_N(\mathbf{r}) \approx -\left(\frac{\partial\rho_N(\mathbf{r})}{\partial N}\right)_\upsilon + \frac{1}{2}\left(\frac{\partial^2\rho_N(\mathbf{r})}{\partial N^2}\right)_\upsilon \\ \Delta\rho_N^+(\mathbf{r}) &= \rho_{N+1}(\mathbf{r}) - \rho_N(\mathbf{r}) \approx \left(\frac{\partial\rho_N(\mathbf{r})}{\partial N}\right)_\upsilon + \frac{1}{2}\left(\frac{\partial^2\rho_N(\mathbf{r})}{\partial N^2}\right)_\upsilon\end{aligned} \quad \text{(A5)}$$

and by solving the system **Eq. (A5)** we obtain **Eq. (A6)**, where, $f(\mathbf{r})$ is the Fukui function and $f^{(2)}(\mathbf{r})$ is the dual-descriptor function.

$$\begin{aligned}\left(\frac{\partial\rho_N(\mathbf{r})}{\partial N}\right)_\upsilon &= f(\mathbf{r}) \approx \frac{1}{2}\cdot(\rho_{N+1}(\mathbf{r}) - \rho_{N-1}(\mathbf{r})) \\ \left(\frac{\partial^2\rho_N(\mathbf{r})}{\partial N^2}\right)_\upsilon &= f^{(2)}(\mathbf{r}) \approx \rho_{N+1}(\mathbf{r}) - 2\cdot\rho_N(\mathbf{r}) + \rho_{N-1}(\mathbf{r})\end{aligned} \quad \text{(A6)}$$

Taking into account the $f_i^{\alpha(NHO/CMO)}$ $(\alpha = +or-)$ parameters of **Eqs. (A2)** and **(A3)**, and operating in the same way as described in ref. **[29]**, the following can be obtained:

$$\begin{aligned}f(\mathbf{r}) = \left(\frac{\partial\rho_N(\mathbf{r})}{\partial N}\right)_\upsilon &\approx \frac{1}{2}\cdot\left(\sum_{i=1}^{all\ orbitals}\left(f_i^{+(NHO/CMO)} + f_i^{-(NHO/CMO)}\right)\left|\phi_i^{(NHO/CMO)}(\vec{r})\right|\right) \\ f^{(2)}(\mathbf{r}) = \left(\frac{\partial^2\rho_N(\mathbf{r})}{\partial N^2}\right)_\upsilon &\approx \sum_{i=1}^{all\ orbitals}\left(f_i^{+(NHO/CMO)} - f_i^{-(NHO/CMO)}\right)\left|\phi_i^{(NHO/CMO)}(\vec{r})\right|\end{aligned} \quad \text{(A7)}$$

Considering that $f_i$ and $f_i^{(2)}$ are the contributions to the functions $f(r)$ and $f^{(2)}(r)$ respectively, corresponding to the "$i$" orbital, we can obtain the following:

$$f_i = \left(\frac{\partial N_i}{\partial N}\right)_\upsilon \approx \frac{1}{2}\left(f_i^{+(NHO/CMO)} + f_i^{-(NHO/CMO)}\right) \quad \text{(A8)}$$

$$f_i^{(2)} = \left(\frac{\partial^2 N_i}{\partial N^2}\right)_\upsilon \approx f_i^{+(NHO/CMO)} - f_i^{-(NHO/CMO)} \quad \text{(A9)}$$

since the orbitals $\phi_i^{NHO/CMO}(\vec{r})$ are normalized and so their squared integral is one. Thus, we can introduce the next approach:

$$\Delta N_i \approx f_i\Delta N + \frac{1}{2} f_i^{(2)}\, \Delta N^2 \quad \text{(A10)}$$

where $\Delta N_i$ is the charge variation corresponding to orbital "$i$".

### A.1.2. Local energy derivatives with regards to the total number of electrons.

To calculate the energy derivatives $\eta_i$ and $\gamma_i$ of the "$i$" orbital, as in ref. **[29]** the starting point was the expansion:

$$\Delta E_i \approx \mu \cdot \Delta N_i + \frac{1}{2}\eta_i \Delta N_i^2 + \frac{1}{6} \cdot \gamma_i \cdot \Delta N_i^3 \tag{A11}$$

and following a process similar to that used with atoms in molecules[24], we can obtain:

$$\begin{aligned} \eta_i &= \frac{(\mu^+ - \mu)\cdot\left(-f_i^{-(NHO/CMO)}\right)^2 - (\mu^- - \mu)\cdot\left(f_i^{+(NHO/CMO)}\right)^2}{f_i^{-(NHO/CMO)} \cdot f_i^{+(NHO/CMO)} \cdot \left(f_i^{-(NHO/CMO)} + f_i^{+(NHO/CMO)}\right)} \\ \gamma_i &= 2\frac{(\mu^+ - \mu)\cdot\left(f_i^{-(NHO/CMO)}\right) + (\mu^- - \mu)\cdot\left(f_i^{+(NHO/CMO)}\right)}{f_i^{-(NHO/CMO)} \cdot f_i^{+(NHO/CMO)} \cdot \left(f_i^{-(NHO/CMO)} + f_i^{+(NHO/CMO)}\right)} \end{aligned} \tag{A12}$$

and the local parameters $\Delta N_i^{\max}$ and $\omega_i$:

$$\Delta N_i^{\max} = \frac{-\eta_i + \sqrt{\eta_i^2 - 2\gamma_i\ \mu}}{\gamma_i} \tag{A13}$$

$$\omega_i \approx -\left(\mu \cdot \Delta N_i^{\max} + \frac{1}{2}\eta_i \left(\Delta N_i^{\max}\right)^2 + \frac{1}{6} \cdot \gamma_i \cdot \left(\Delta N_i^{\max}\right)^3\right) \tag{A14}$$

**A.2. Correlations found for the derivatives calculated for natural hybrid orbitals.**

**Table A1** shows the coefficients of determination ($R^2$) obtained in regressions **Eqs. (3)-(5)** but using the values $f_i^+$ and $f_i^-$, obtained for NHOs instead of NBOs. As can be seen, the sample of molecules described above has been used. The Supporting Information (**Figures S5-S40**) presents the graphs of the correlations corresponding to the coefficients of determination. The values obtained indicate that the correlations of **Eqs. (3)-(5)** are fulfilled.

| | $\omega_i$ **vs** $f_i$ | $\eta_i$ **-1 vs** $f_i$ | $\Delta N_i^{\max}$ **vs** $\Delta N_i^{\max}$ **'** | $\eta_i$ **-1 vs** $\Delta N_i^{\max}$ |
|---|---|---|---|---|
| $CH_2CHCN$ | 0.997 | 0.997 | 0.993 | 0.999 |
| $CH_2CHOH$ | 0.975 | 0.979 | 0.978 | 1.000 |
| $CH_2CHCl$ | 0.942 | 0.945 | 0.963 | 1.000 |
| $CH_2CHNH_2$ | 0.950 | 0.941 | 0.940 | 1.000 |
| $CH_3COOCH_3$(***) | 0.906 | 0.901 | 0.893 | 1.000 |
| $CH_2CHOCH_3$ (*) | 0.982 | 0.984 | 0.971 | 1.000 |
| $CH_2CHCHO$ (**) | 0.953 | 0.955 | 0.929 | 1.000 |
| $CH_2CHNO_2$ (**) | 0.864 | 0.864 | 0.806 | 1.000 |
| $CH_3CHSH$ (**) | 0.830 | 0.857 | 0.769 | 1.000 |
| $CH_3COOH$ (***) | 0.869 | 0.885 | 0.866 | 1.000 |
| $CH_3CONH_2$ (**) | 0.796 | 0.797 | 0.773 | 0.994 |

(*) One outlier has been eliminated from the regression.
(**) two outliers have been eliminated from the regression.
(***) three outliers have been eliminated from the regression.

**Table A1**. Determination coefficients ($R^2$) obtained for the regressions of NHO parameters by using of the sample set of molecules.

### A.3. Correlations found for the derivatives calculated for canonical molecular orbitals.

**Table A2** shows the coefficients of determination ($R^2$) obtained in regressions **Eqs. (3)-(5)** but using the values obtained for CMOs instead of NBOs or NHOs. The graphs corresponding to each of the coefficients of determination can be seen in the Supporting Information (**Figures S41-S76**). The values obtained indicate that the correlations of **Eqs. (3)-(5)** are also fulfilled for this type of orbitals.

| | $\omega_i$ vs $f_i$ | $\eta_i^{-1}$ vs $f_i$ | $\Delta N_i^{max}$ vs $\Delta N_i^{max\,\prime}$ | $\eta_i^{-1}$ vs $\Delta N_i^{max}$ |
|---|---|---|---|---|
| $CH_2CHCN$(*) | 0.966 | 0.968 | 0.930 | 1.000 |
| $CH_2CHOH$(*) | 0.955 | 0.956 | 0.947 | 1.000 |
| $CH_2CHCl$(*) | 0.985 | 0.985 | 0.987 | 1.000 |
| $CH_2CHNH_2$(*) | 0.960 | 0.965 | 0.962 | 1.000 |
| $CH_3COOCH_3$(***) | 0.793 | 0.800 | 0.822 | 1.000 |
| $CH_2CHOCH_3$(*) | 0.950 | 0.947 | 0.933 | 1.000 |
| $CH_2CHCHO$(*) | 0.946 | 0.945 | 0.971 | 1.000 |
| $CH_2CHNO_2$(*) | 0.890 | 0.892 | 0.871 | 1.000 |
| $CH_3CHSH$(*) | 0.950 | 0.948 | 0.962 | 1.000 |
| $CH_3COOH$(*) | 0.961 | 0.959 | 0.937 | 1.000 |
| $CH_3CONH_2$(**) | 0.928 | 0.933 | 0.955 | 1.000 |

(*) two outliers have been eliminated from the regression.
(**) three outliers have been eliminated from the regression.
(***) four outliers have been eliminated from the regression.

**Table A2**. Determination coefficients ($R^2$) obtained for the regressions of CMO parameters by using of the sample set of molecules.

### 1.-UCA-FUKUI software:

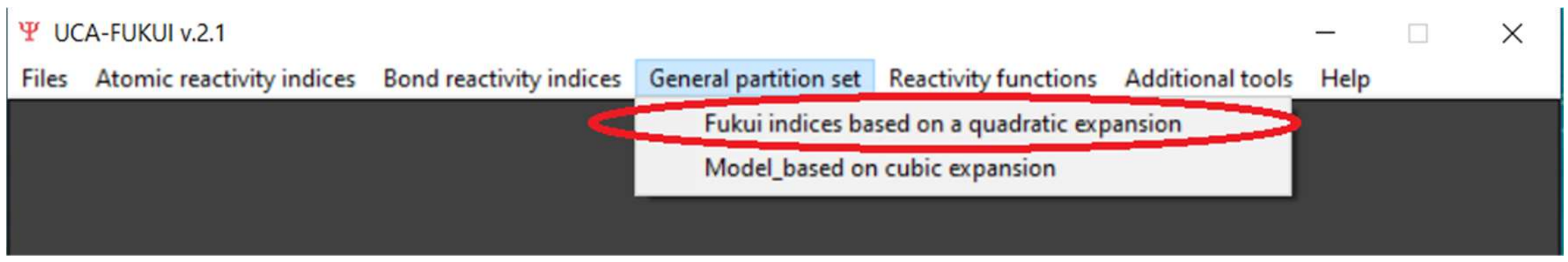

**Figure S1**. Main menu of the UCA-FUKUI software showing the calculation of local Fukui indices and other parameters for NHOs/CMOs.

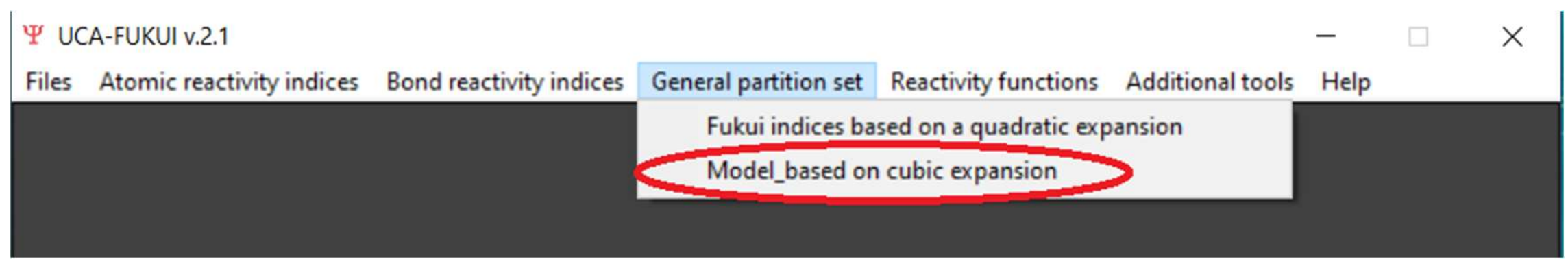

**Figure S2**. Main menu of the UCA-FUKUI software showing the calculation of local hardnesses and other parameters for NHOs/CMOs.

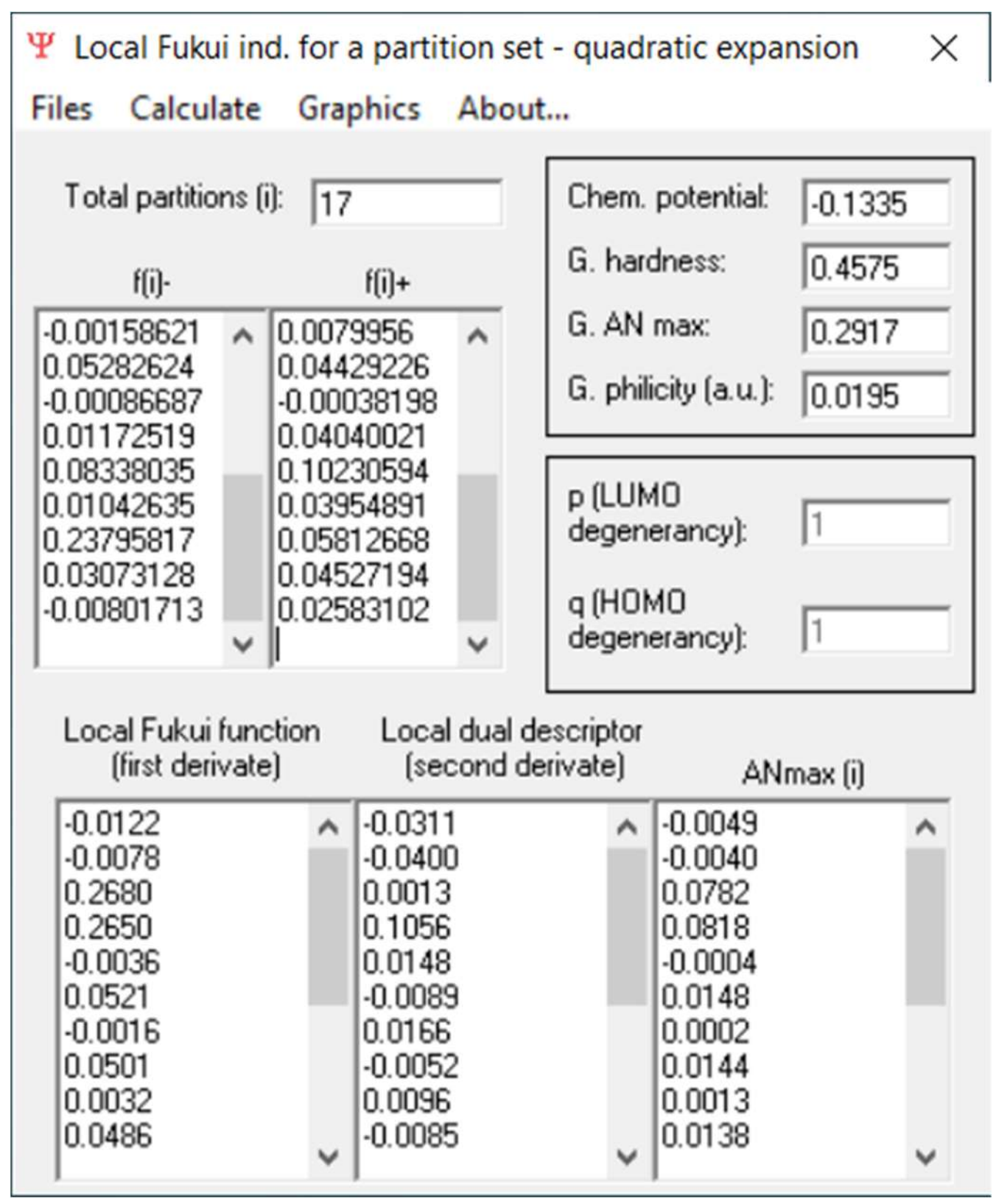

**Figure S3**. Example of a screen of the UCA-FUKUI software showing the calculation module for obtaining local Fukui indices and other parameters for NHOs of $CH_2CHCl$.

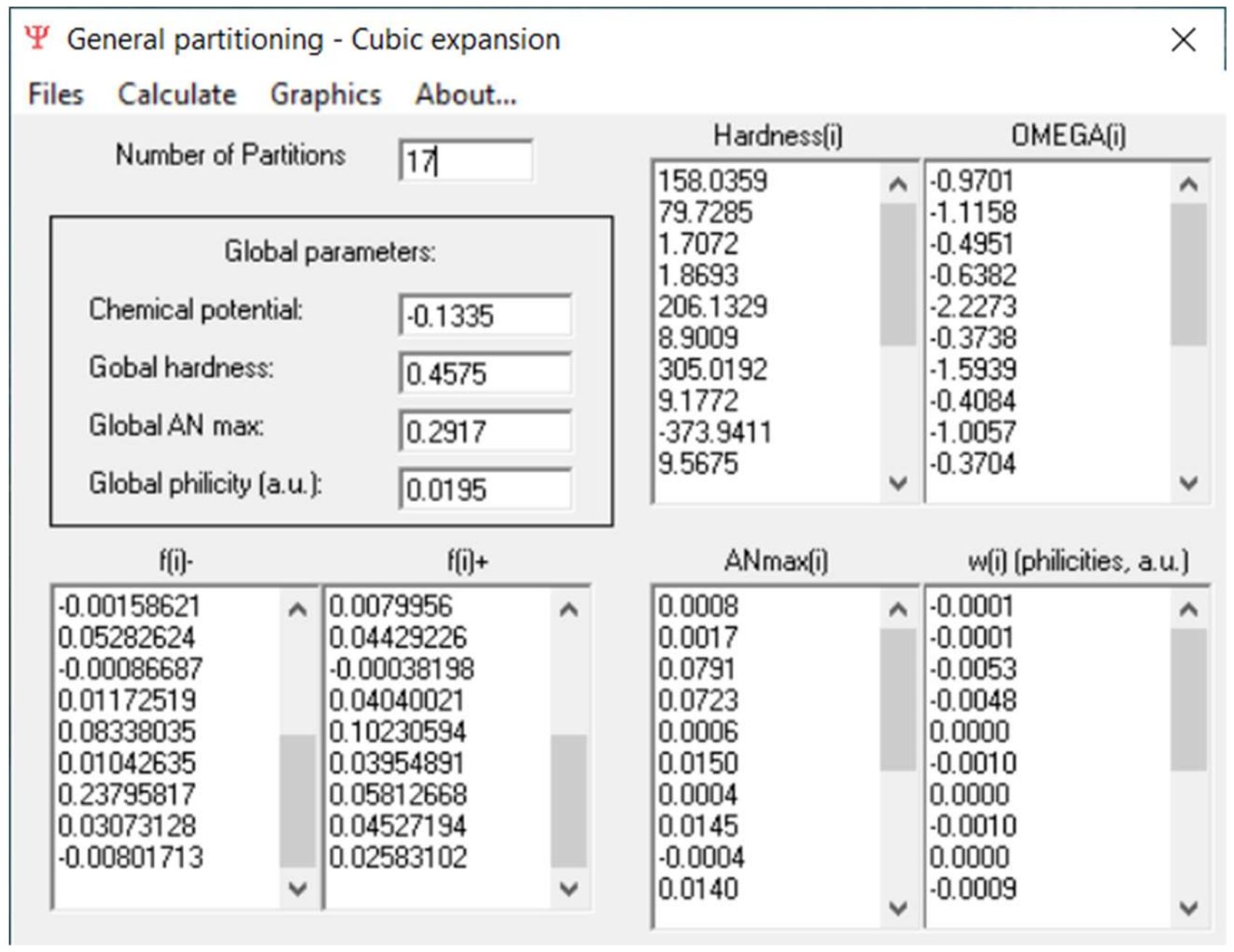

**Figure S4**. Window of the UCA-FUKUI software showing the calculation module for obtaining local hardnesses and other parameters for NHOs of $CH_2CHCl$.

**2. Maximum local charge variations (obtained with a cubic expansion of the local energy) versus the maximum local charge variations (achieved with a quadratic expansion of the local charge variation):**

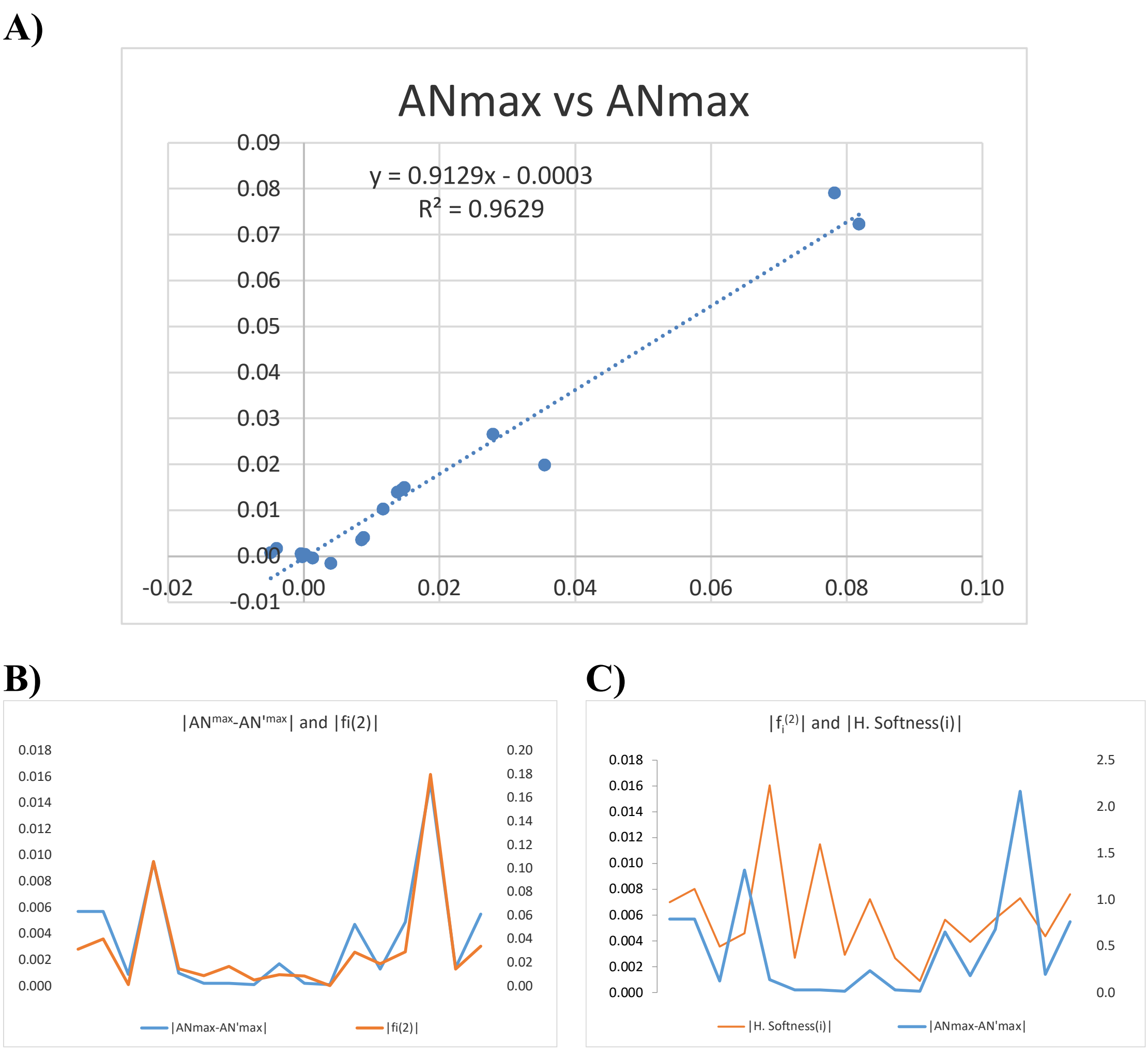

Figure S5. A) Linear regression of the maximum local charge variations (obtained with a cubic expansion of the local energy) versus the maximum local charge variations (achieved with a quadratic expansion of the local charge variation). B) Comparison of $\left|\Delta N_i^{\max} - \Delta N'^{\max}_i\right|$ and $\left|f_i^{(2)}\right|$ values. C) Comparison of $\left|\Delta N_i^{\max} - \Delta N'^{\max}_i\right|$ and $\left|\gamma_i\right|$ values. In all cases for the NHOs of the $CH_2CHCl$ molecule.

A)

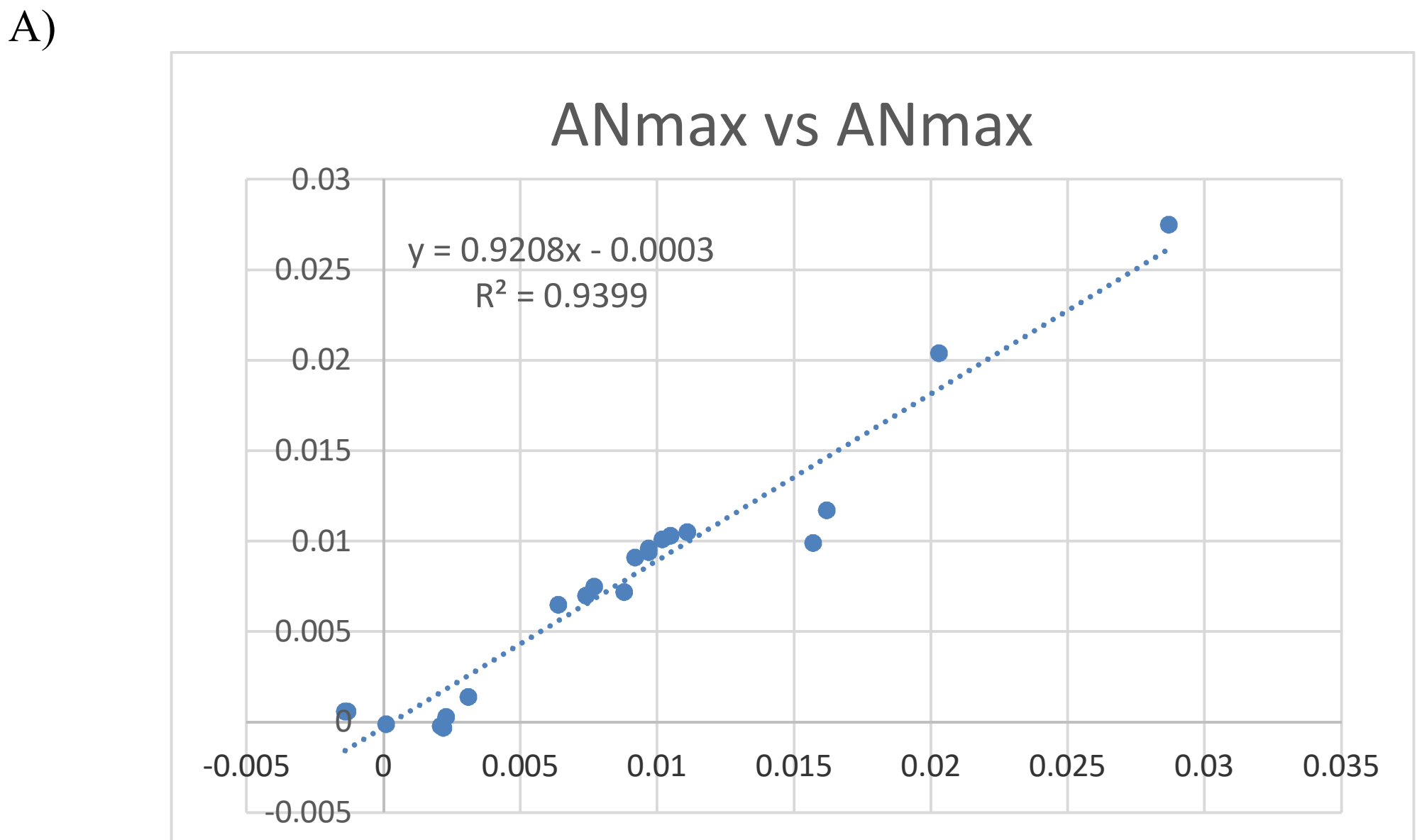

B) C)

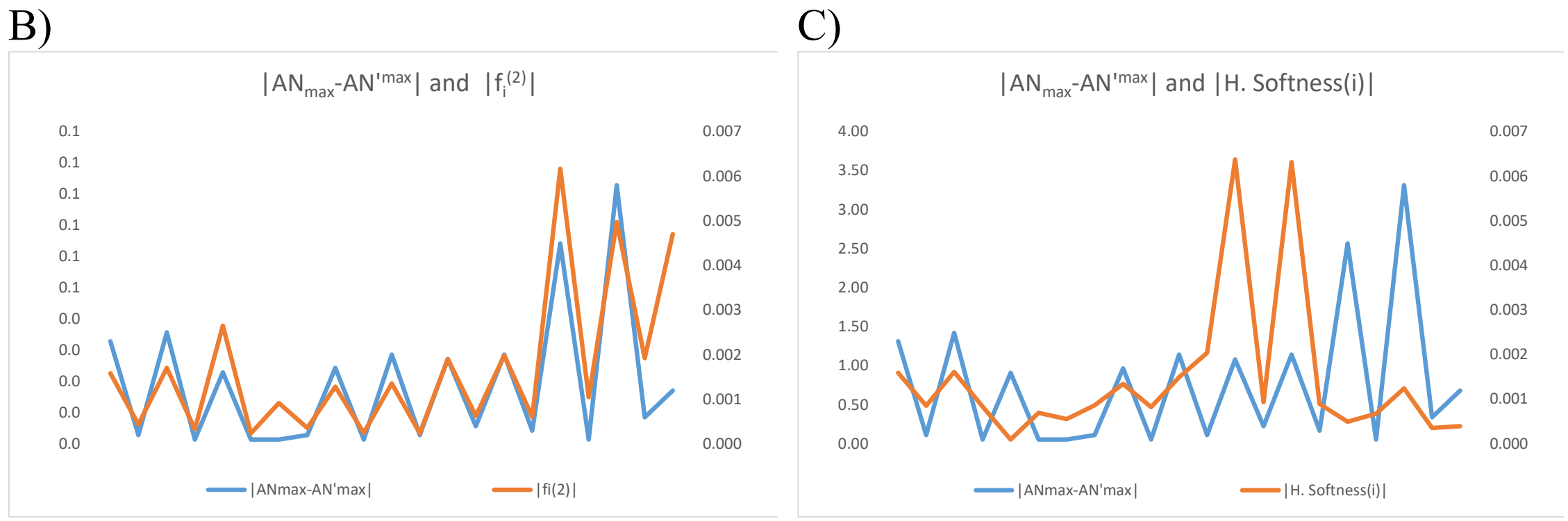

Figure S6. Linear regression of the maximum local charge variations (obtained with a cubic expansion of the local energy) versus the maximum local charge variations (achieved with a quadratic expansion of the local charge variation). B) Comparison of $\left|\Delta N_i^{\max} - \Delta N'^{\max}_i\right|$ and $\left|f_i^{(2)}\right|$ values. C) Comparison of $\left|\Delta N_i^{\max} - \Delta N'^{\max}_i\right|$ and $\left|\gamma_i\right|$ values. In all cases for the NHOs of the $CH_2CHNH_2$ molecule.

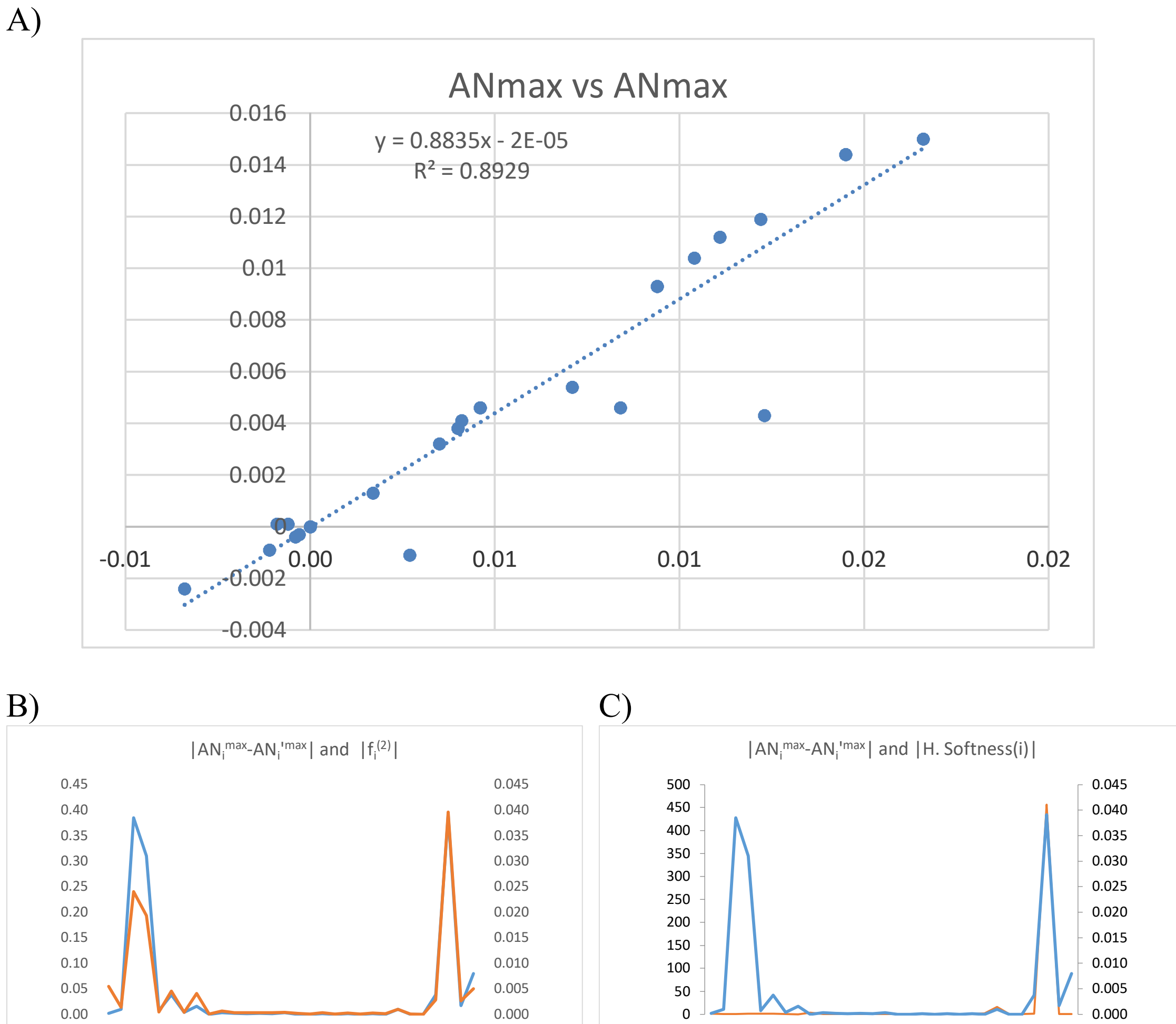

Figure S7. Linear regression of the maximum local charge variations (obtained with a cubic expansion of the local energy) versus the maximum local charge variations (achieved with a quadratic expansion of the local charge variation). Three outliers corresponding to NHOs 3,4 and 28 have been removed. B) Comparison of $\left|\Delta N_i^{\max} - \Delta N'^{\max}_i\right|$ and $\left|f_i^{(2)}\right|$ values. C) Comparison of $\left|\Delta N_i^{\max} - \Delta N'^{\max}_i\right|$ and $\left|\gamma_i\right|$ values. In all cases for the NHOs of the $CH_3COOCH_3$ molecule.

A)

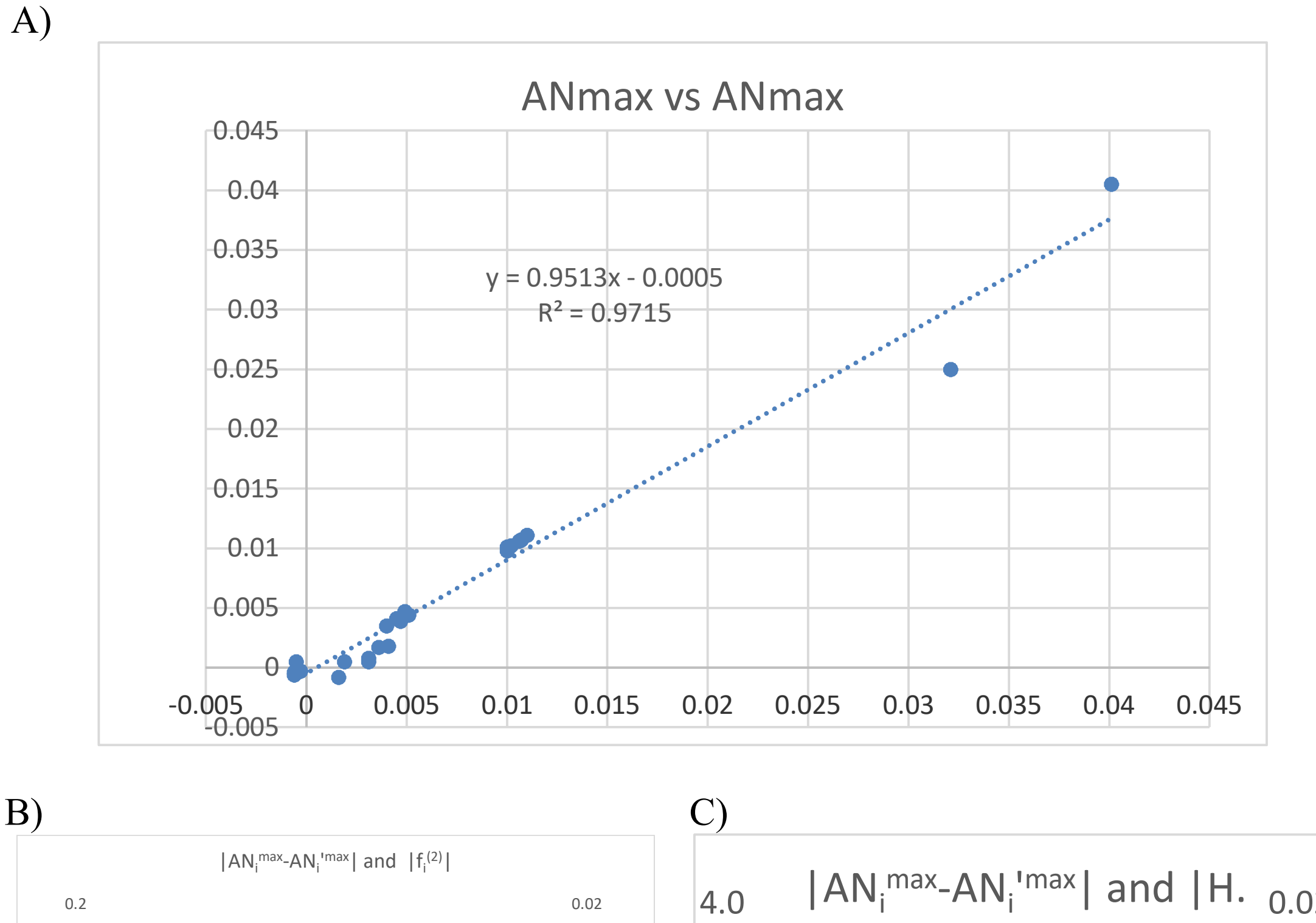

B) C)

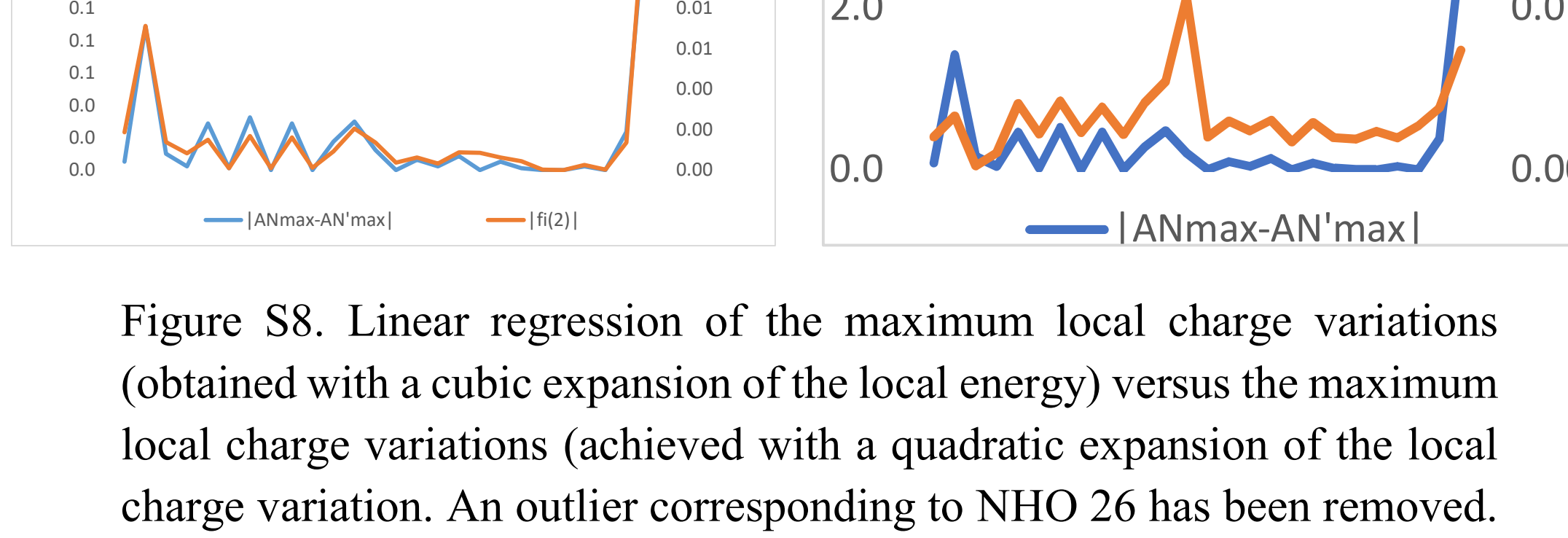

Figure S8. Linear regression of the maximum local charge variations (obtained with a cubic expansion of the local energy) versus the maximum local charge variations (achieved with a quadratic expansion of the local charge variation. An outlier corresponding to NHO 26 has been removed. B) Comparison of $\left|\Delta N_i^{\max} - \Delta N'^{\max}_i\right|$ and $\left|f_i^{(2)}\right|$ values. C) Comparison of $\left|\Delta N_i^{\max} - \Delta N'^{\max}_i\right|$ and $\left|\gamma_i\right|$ values. In all cases for the NHOs of the $CH_2CHOCH_3$ molecule.

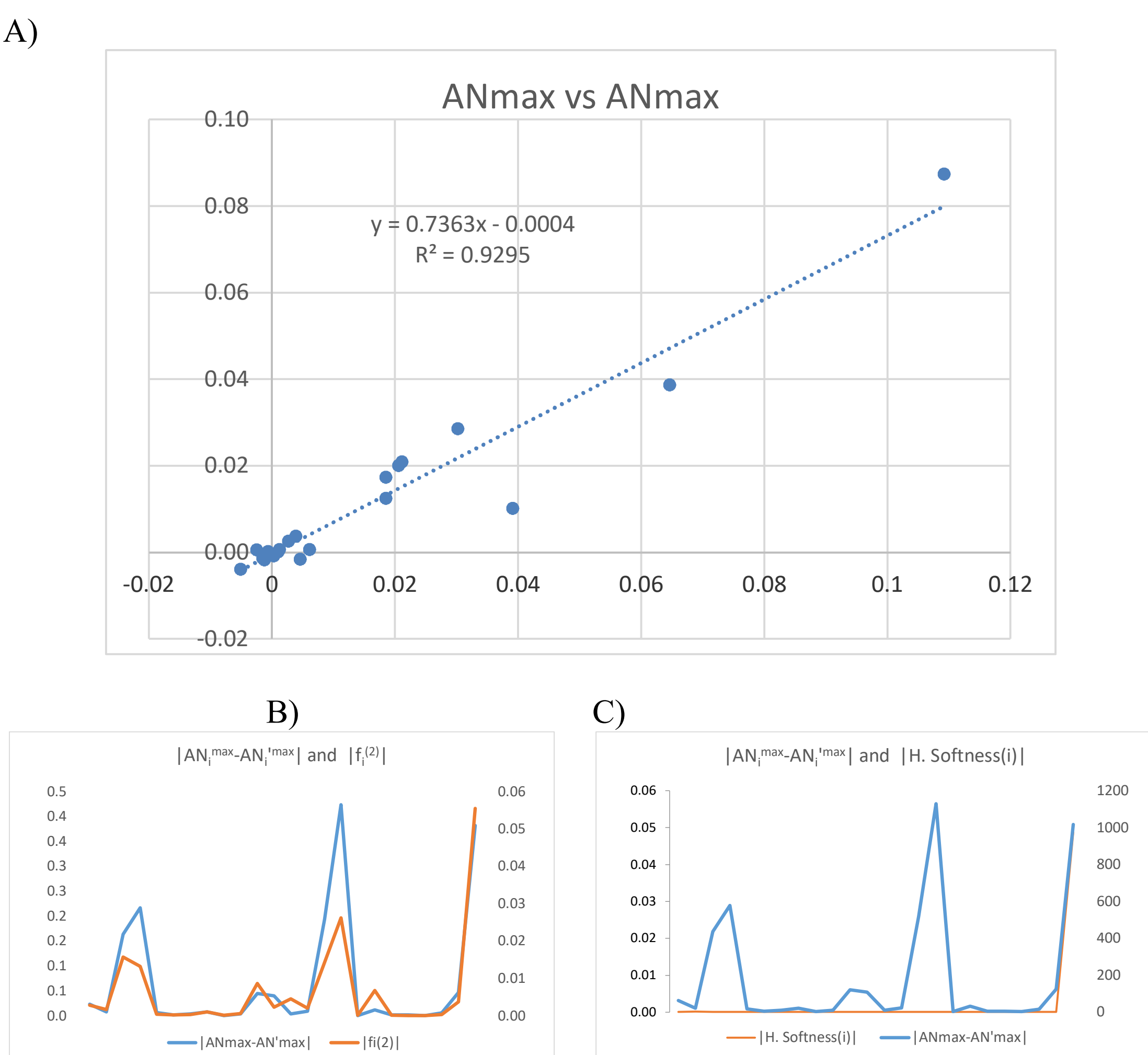

Figure S9. Linear regression of the maximum local charge variations (obtained with a cubic expansion of the local energy) versus the maximum local charge variations (achieved with a quadratic expansion of the local charge variation). Two outliers corresponding to NHOs 16 and 24 have been removed. B) Comparison of $\left|\Delta N_i^{\max} - \Delta N'^{\max}_i\right|$ and $\left|f_i^{(2)}\right|$ values. C) Comparison of $\left|\Delta N_i^{\max} - \Delta N'^{\max}_i\right|$ and $\left|\gamma_i\right|$ values. In all cases for the NHOs of the $CH_2CHCHO$ molecule.

A)

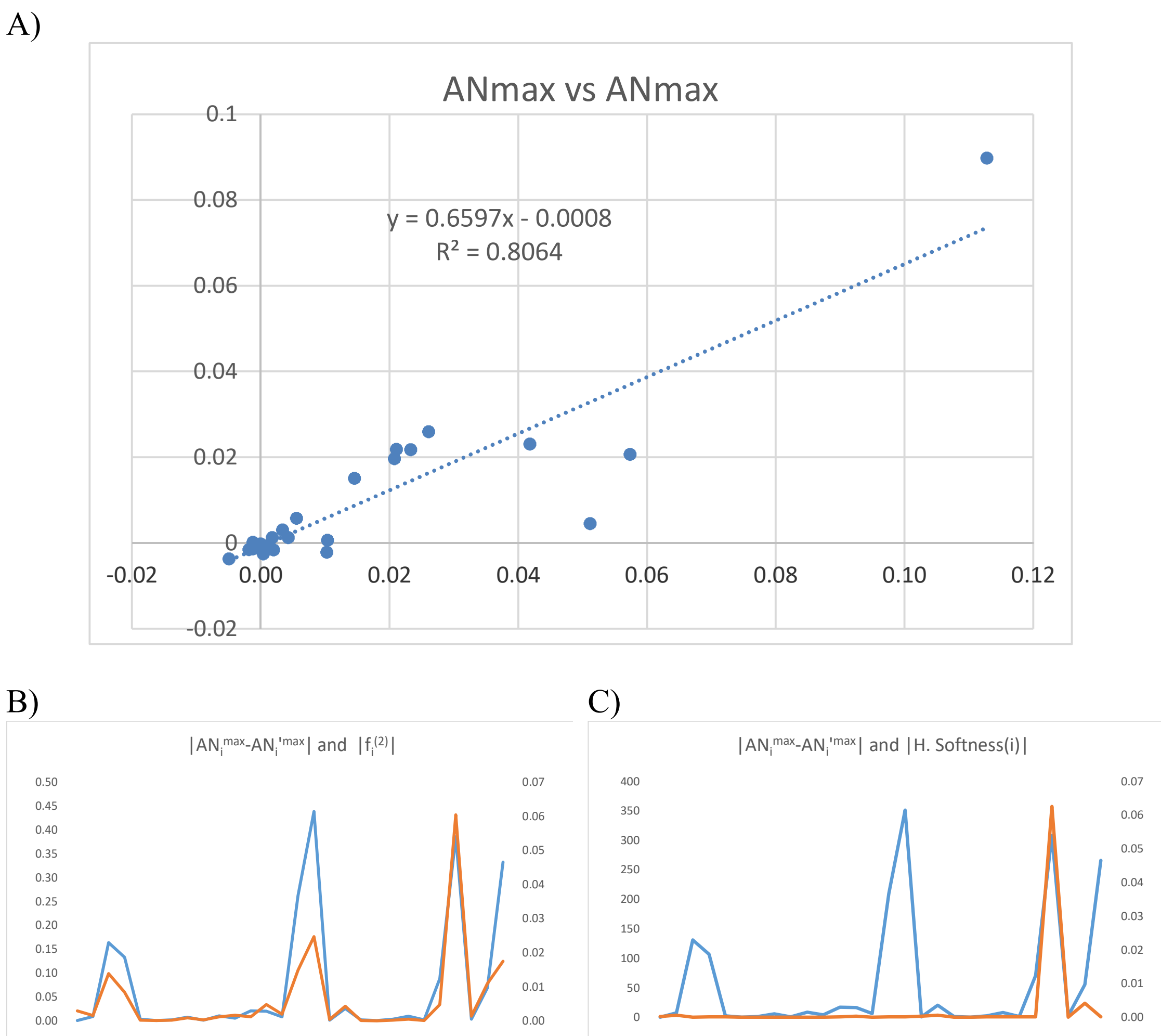

Figure S10. Linear regression of the maximum local charge variations (obtained with a cubic expansion of the local energy) versus the maximum local charge variations (achieved with a quadratic expansion of the local charge variation). Two outliers corresponding to NHOs 16 and 25 have been removed. B) Comparison of $\left|\Delta N_i^{\max} - \Delta N'^{\max}_i\right|$ and $\left|f_i^{(2)}\right|$ values. C) Comparison of $\left|\Delta N_i^{\max} - \Delta N'^{\max}_i\right|$ and $\left|\gamma_i\right|$ values. In all cases for the NHOs of the $CH_2CHNO_2$ molecule.

A)

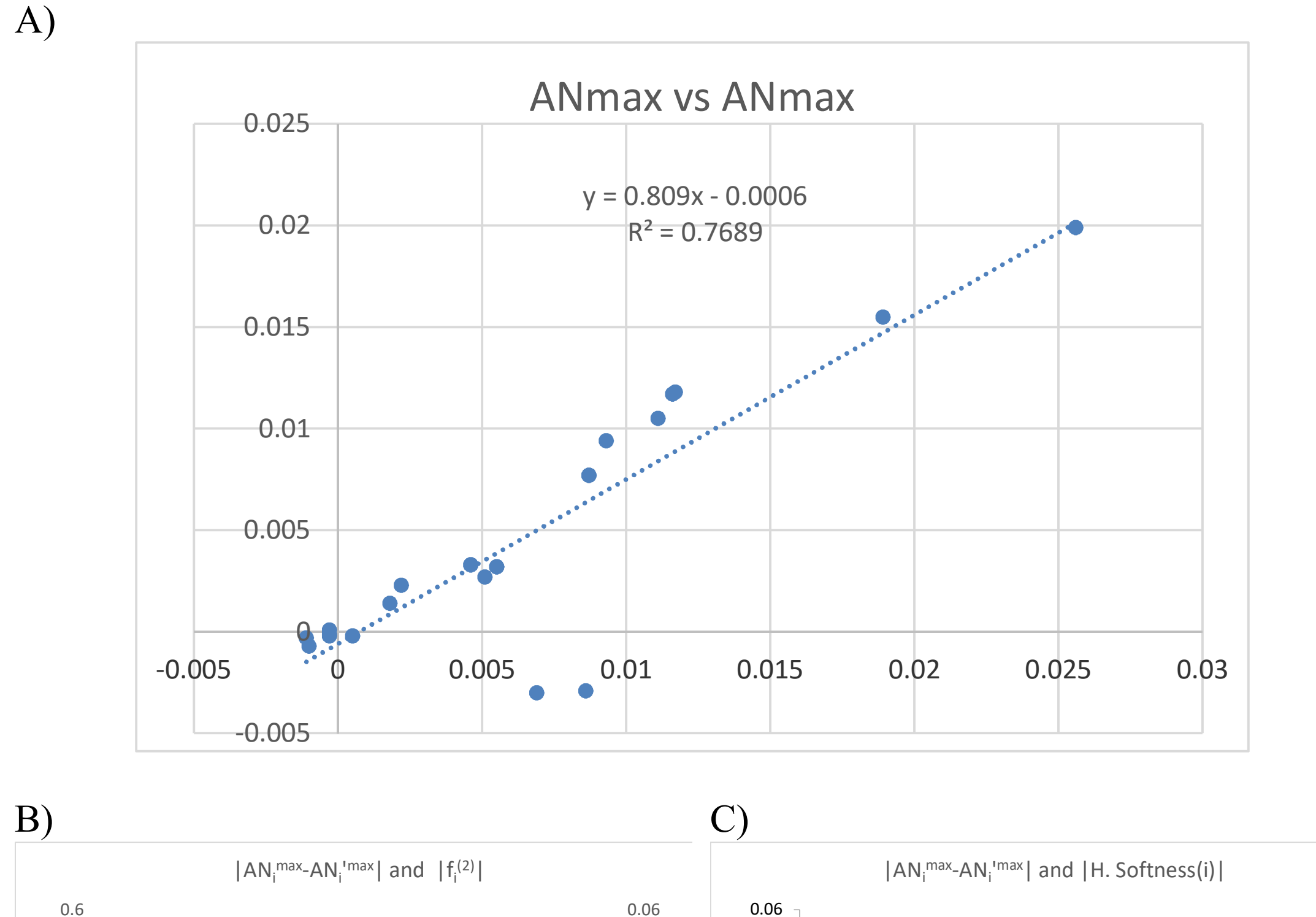

B) C)

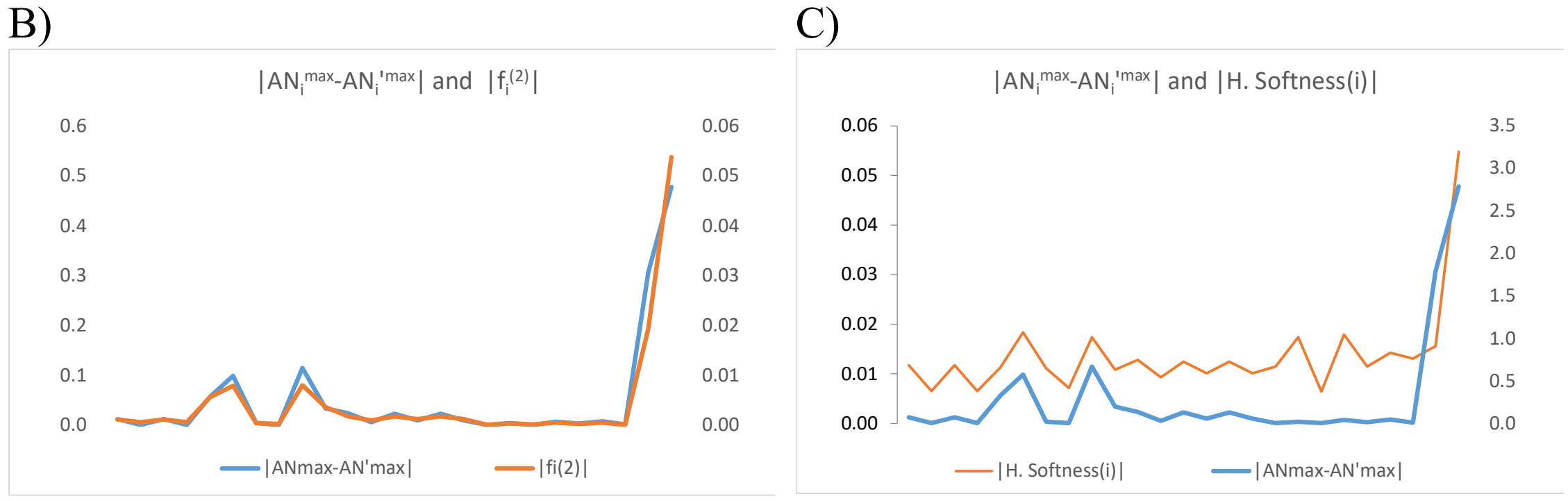

Figure S11. Linear regression of the maximum local charge variations (obtained with a cubic expansion of the local energy) versus the maximum local charge variations (achieved with a quadratic expansion of the local charge variation). Two outliers corresponding to NHOs 24 and 25 have been removed. B) Comparison of $\left|\Delta N_i^{\max} - \Delta N'^{\max}_i\right|$ and $\left|f_i^{(2)}\right|$ values. C) Comparison of $\left|\Delta N_i^{\max} - \Delta N'^{\max}_i\right|$ and $\left|\gamma_i\right|$ values. In all cases for the NHOs of the $CH_3CH_2SH$ molecule.

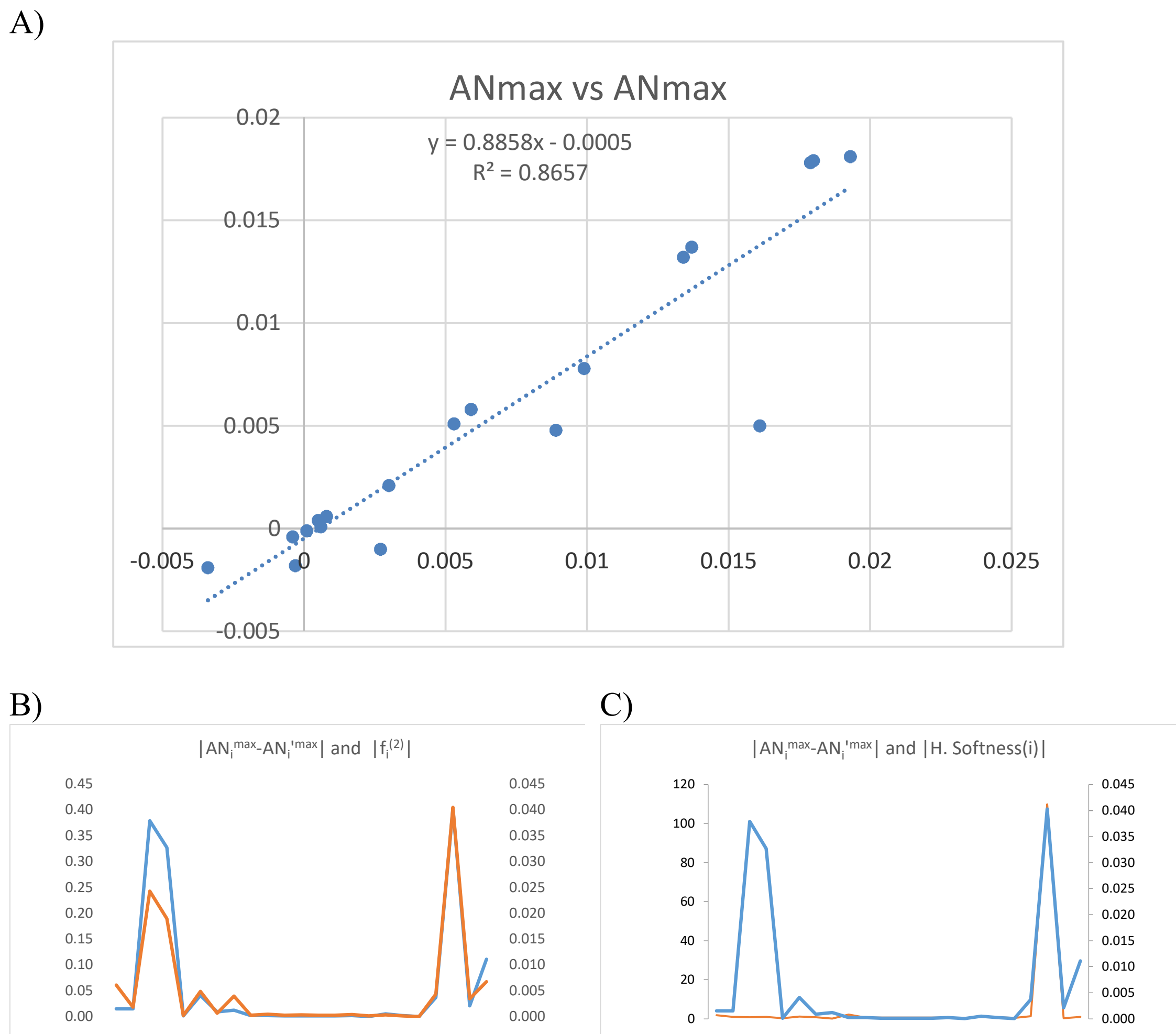

Figure S12. Linear regression of the maximum local charge variations (obtained with a cubic expansion of the local energy) versus the maximum local charge variations (achieved with a quadratic expansion of the local charge variation). Two outliers corresponding to NHOs 24 and 25 have been removed. B) Comparison of $\left|\Delta N_i^{\max} - \Delta N'^{\max}_i\right|$ and $\left|f_i^{(2)}\right|$ values. C) Comparison of $\left|\Delta N_i^{\max} - \Delta N'^{\max}_i\right|$ and $\left|\gamma_i\right|$ values. In all cases for the NHOs of the $CH_3COOH$ molecule.

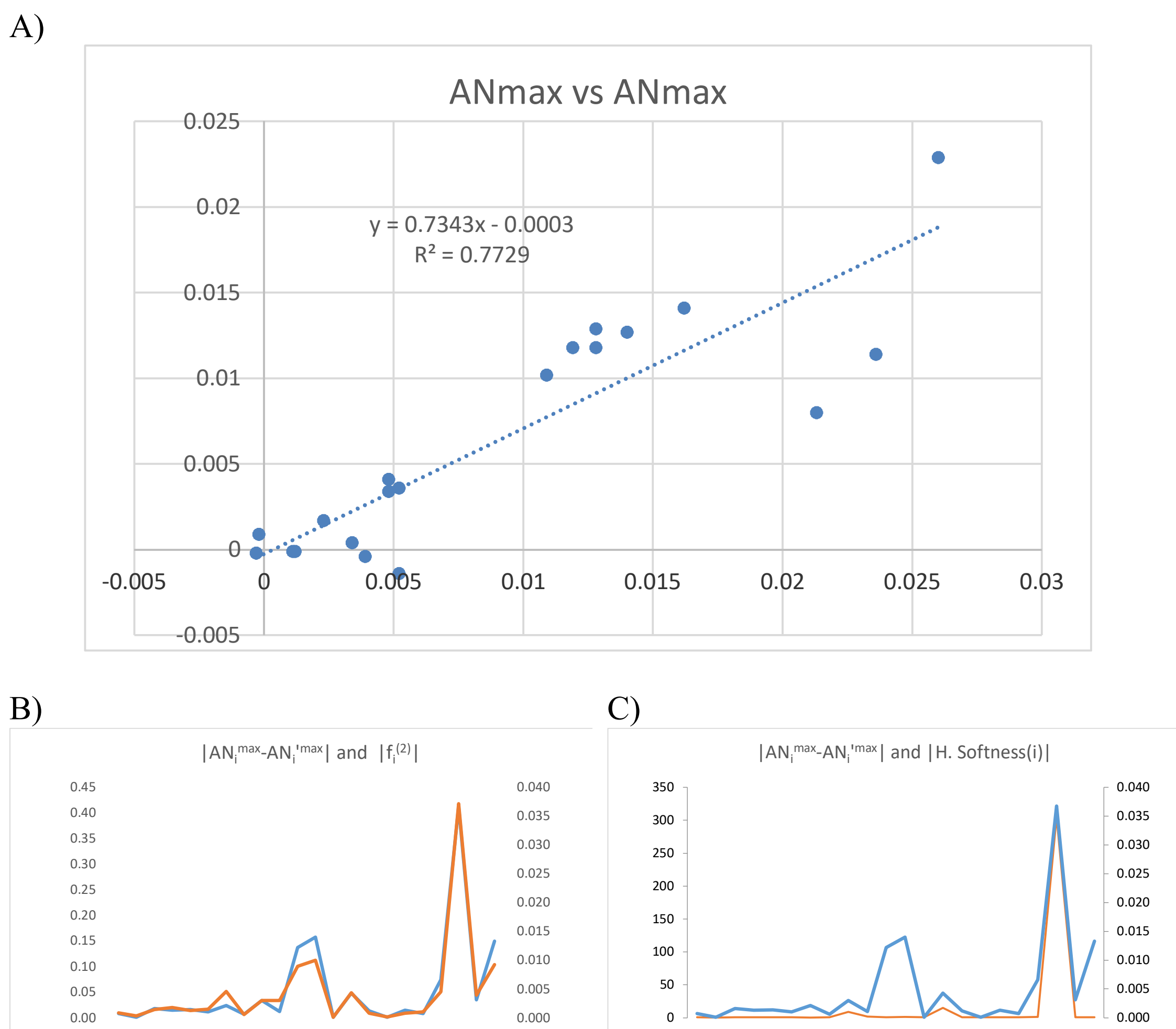

Figure S13. Linear regression of the maximum local charge variations (obtained with a cubic expansion of the local energy) versus the maximum local charge variations (achieved with a quadratic expansion of the local charge variation). Two outliers corresponding to NHOs 12 and 20 have been removed. B) Comparison of $\left|\Delta N_i^{\max} - \Delta N'^{\max}_i\right|$ and $\left|f_i^{(2)}\right|$ values. C) Comparison of $\left|\Delta N_i^{\max} - \Delta N'^{\max}_i\right|$ and $\left|\gamma_i\right|$ values. In all cases for the NHOs of the $CH_3CONH_2$ molecule

## 3. Local electrophilicities against local Fukui indices for NHOs:

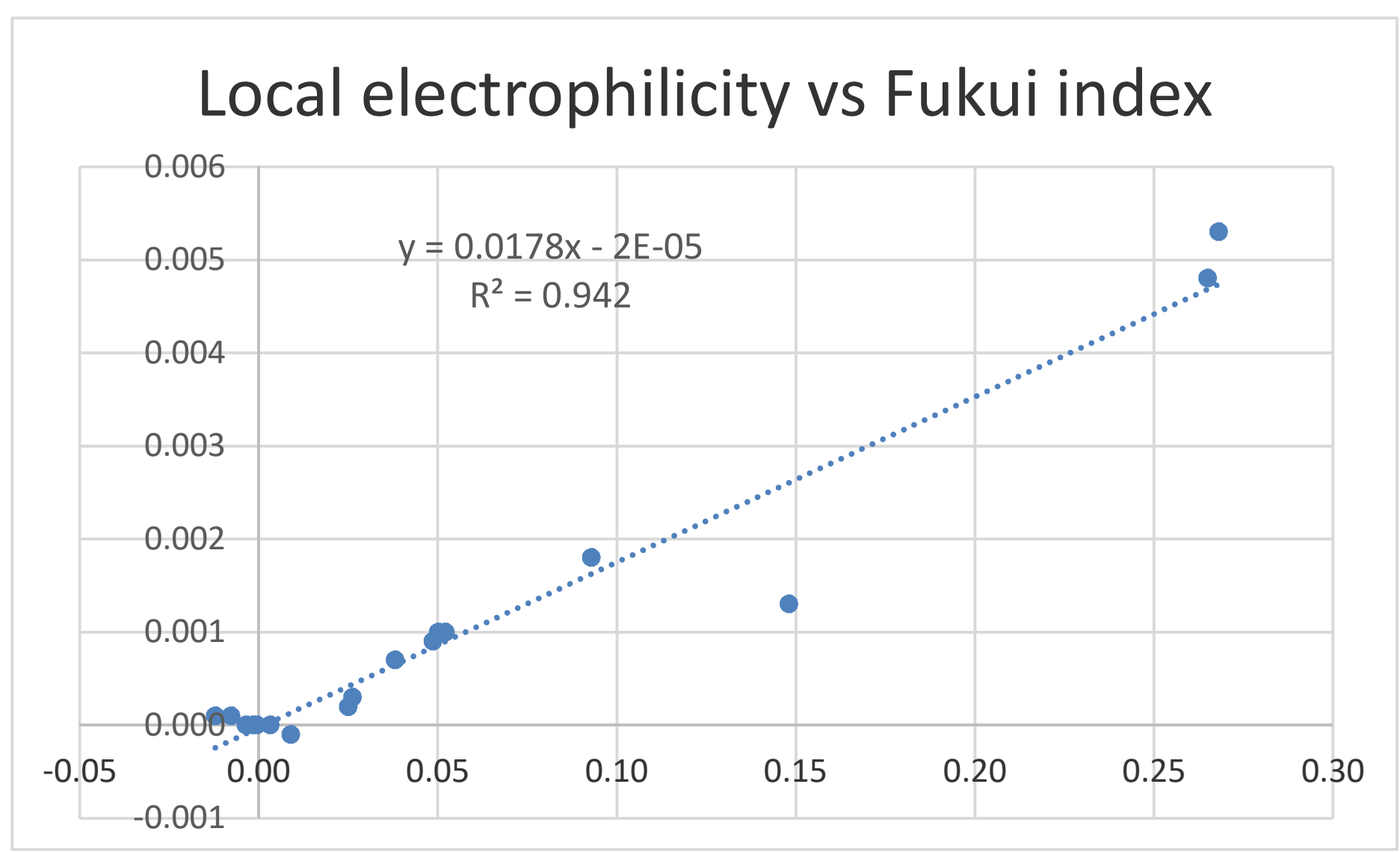

Figure S14. Linear regression of local electrophilicities against local Fukui indices for the NHOs of the $CH_2CHCl$ molecule.

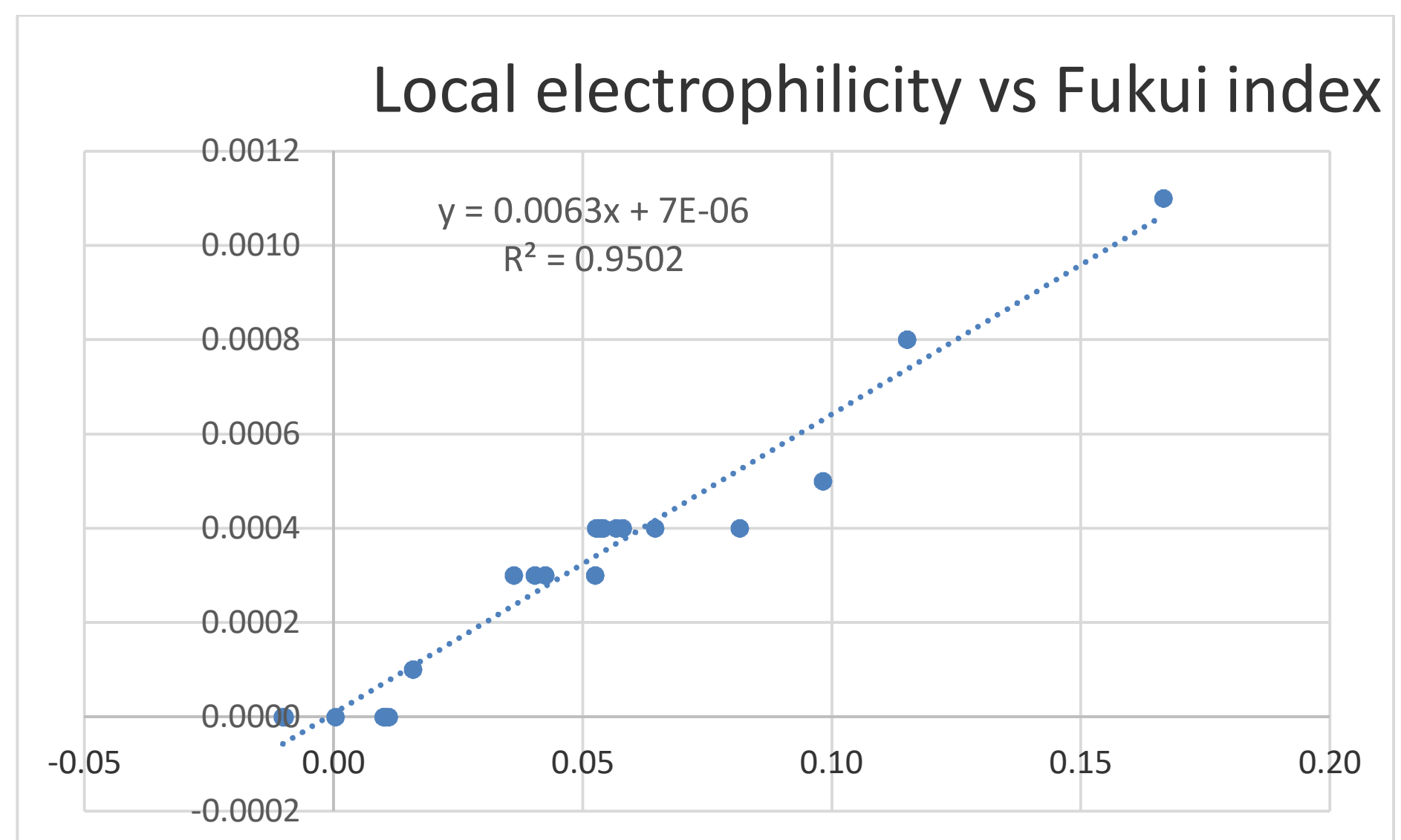

Figure S15. Linear regression of local electrophilicities against local Fukui indices for the NHOs of the $CH_2CHNH_2$ molecule.

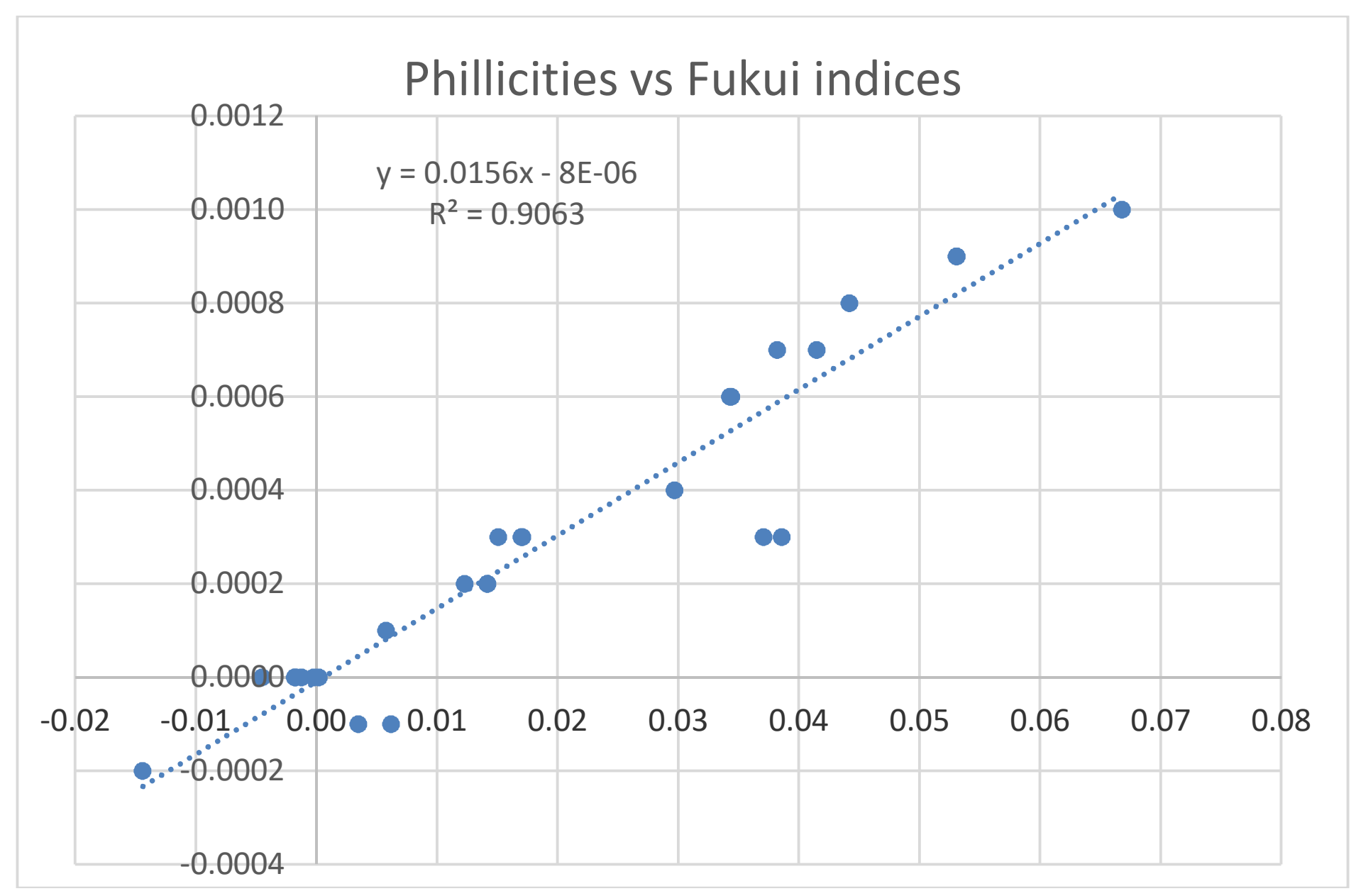

Figure S16. Linear regression of local electrophilicities against local Fukui indices for the NHOs of the $CH_3COOCH_3$ molecule. Three outliers corresponding to NHOs 3,4 and 28 have been removed.

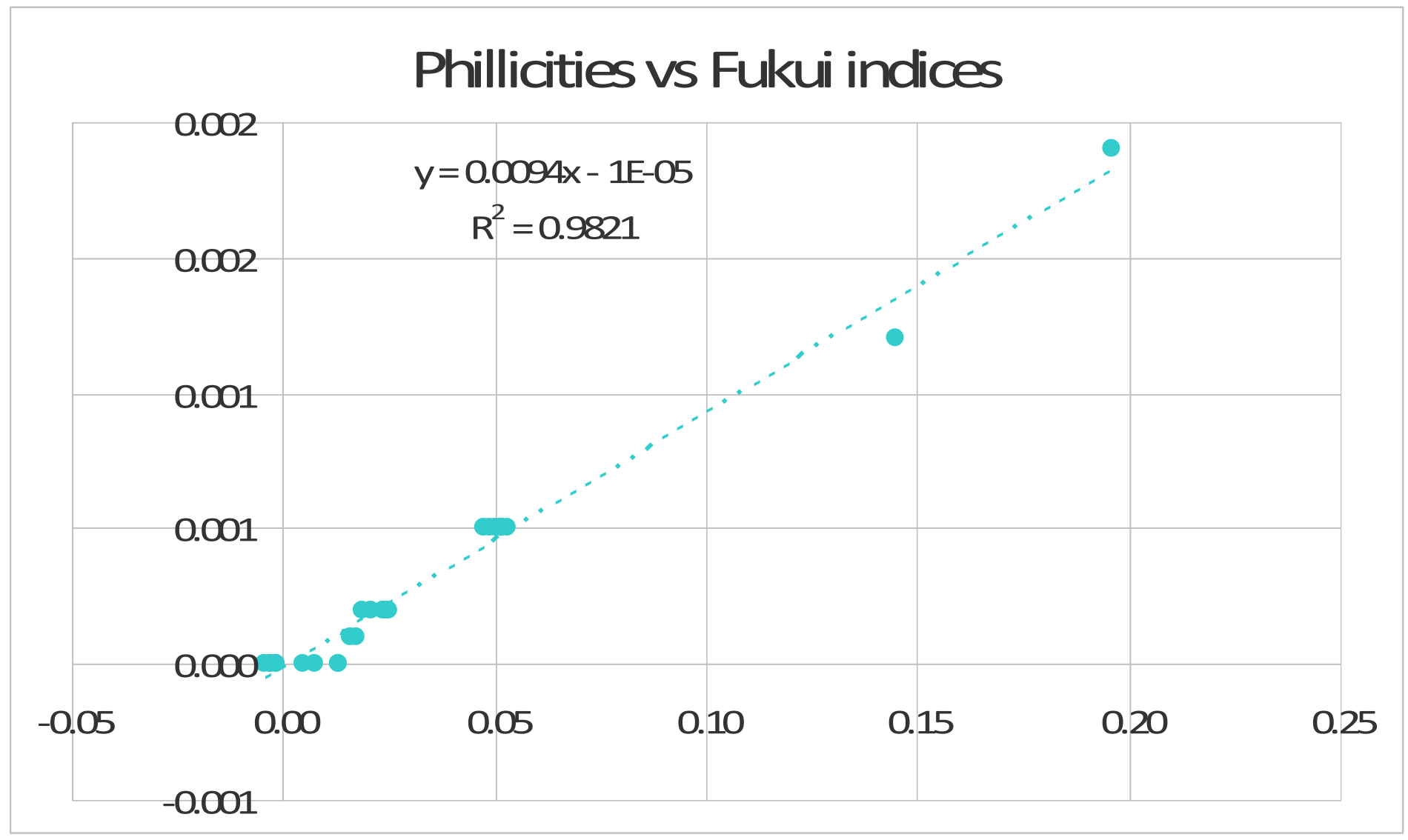

Figure S17. Linear regression of local electrophilicities against local Fukui indices for the NHOs of the $CH_2CHOCH_3$ molecule. An outlier corresponding to NHO 26 has been removed.

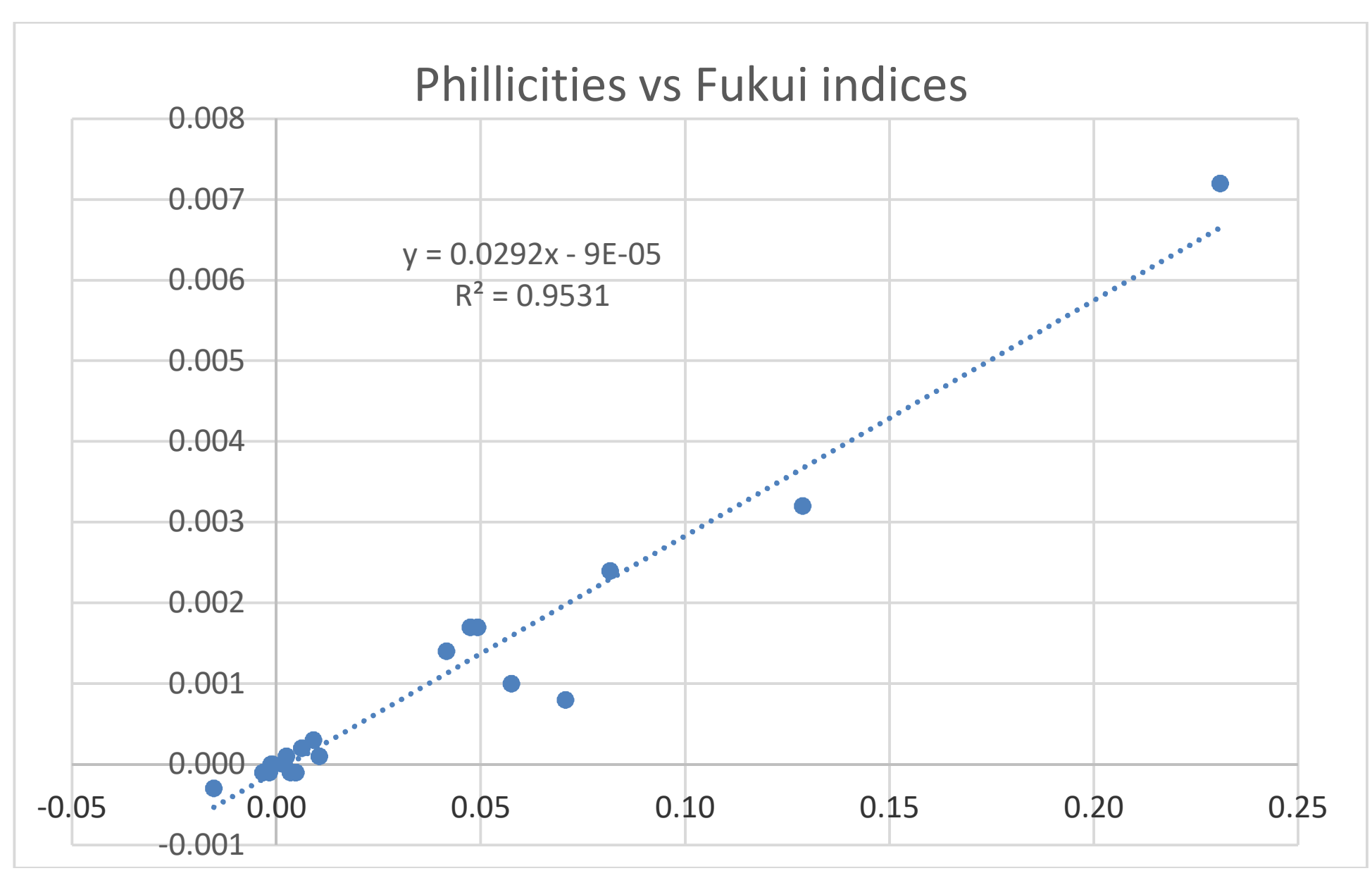

Figure S18. Linear regression of local electrophilicities against local Fukui indices for the NHOs of the $CH_2CHCHO$ molecule. Two outliers corresponding to NHOs 16 and 24 have been removed.

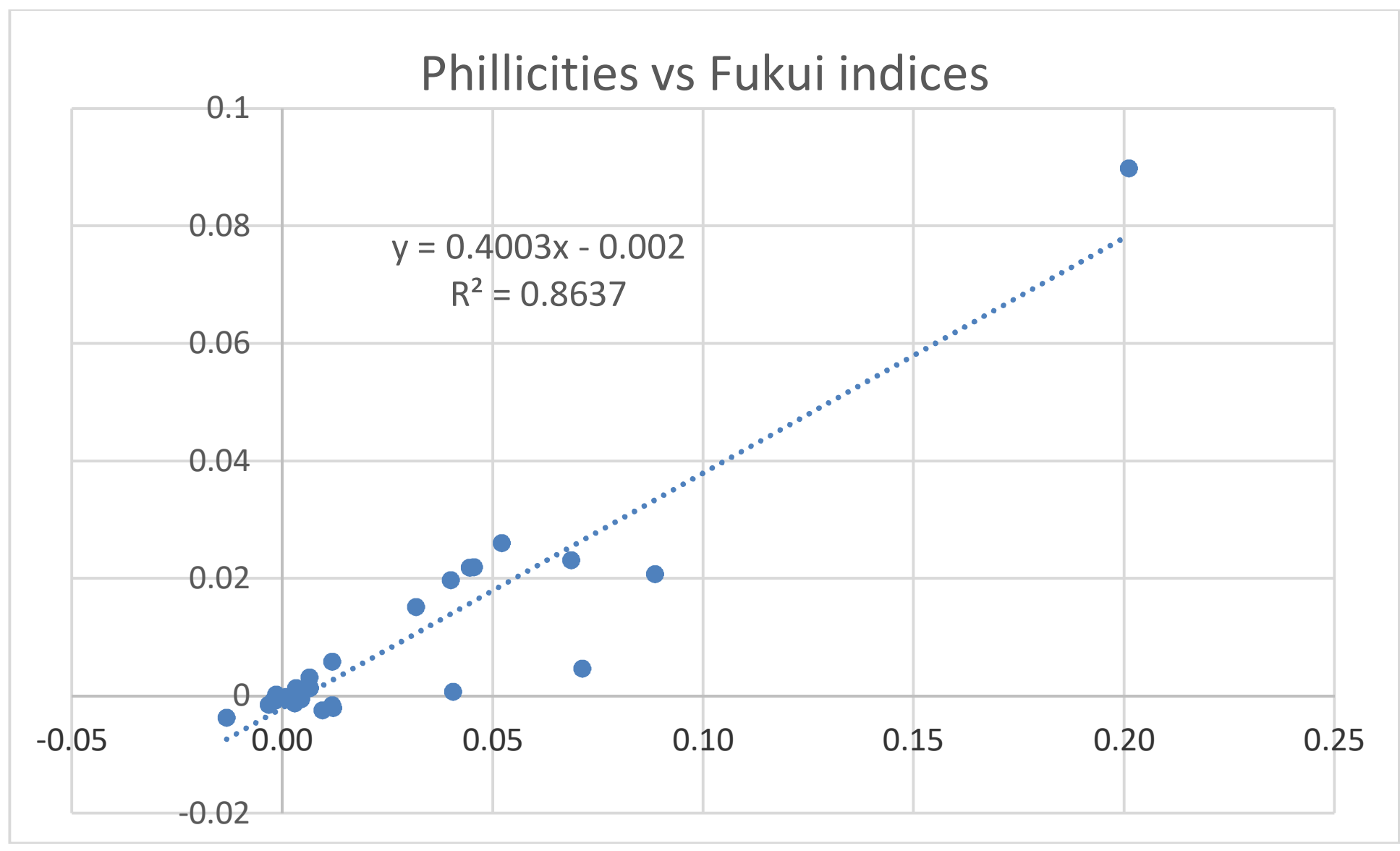

Figure S19. Linear regression of local electrophilicities against local Fukui indices for the NHOs of the $CH_2CHNO_2$ molecule. Two outliers corresponding to NHOs 16 and 25 have been removed.

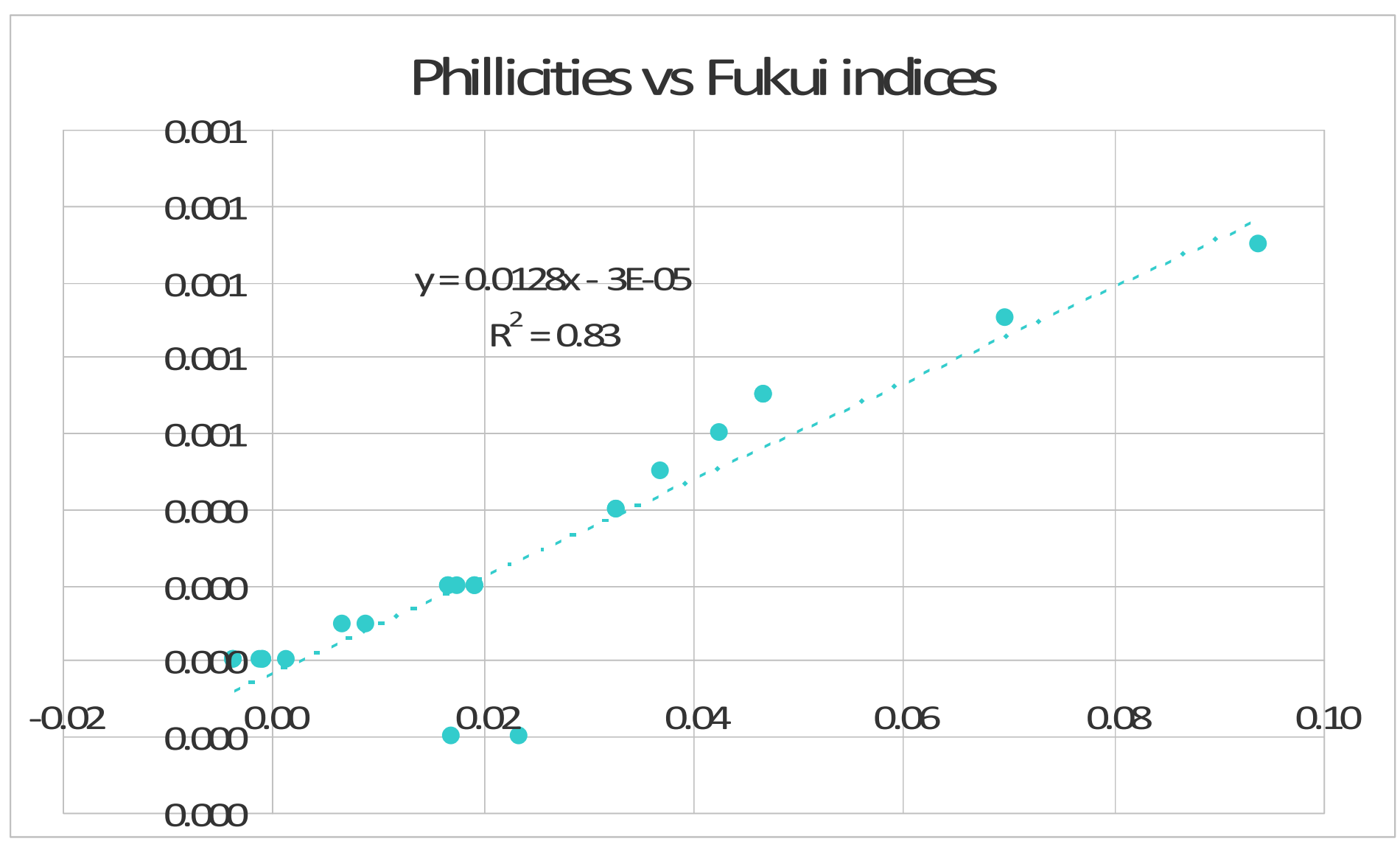

Figure S20. Linear regression of local electrophilicities against local Fukui indices for the NHOs of the $CH_3CHSH$ molecule. Two outliers corresponding to NHOs 24 and 25 have been removed.

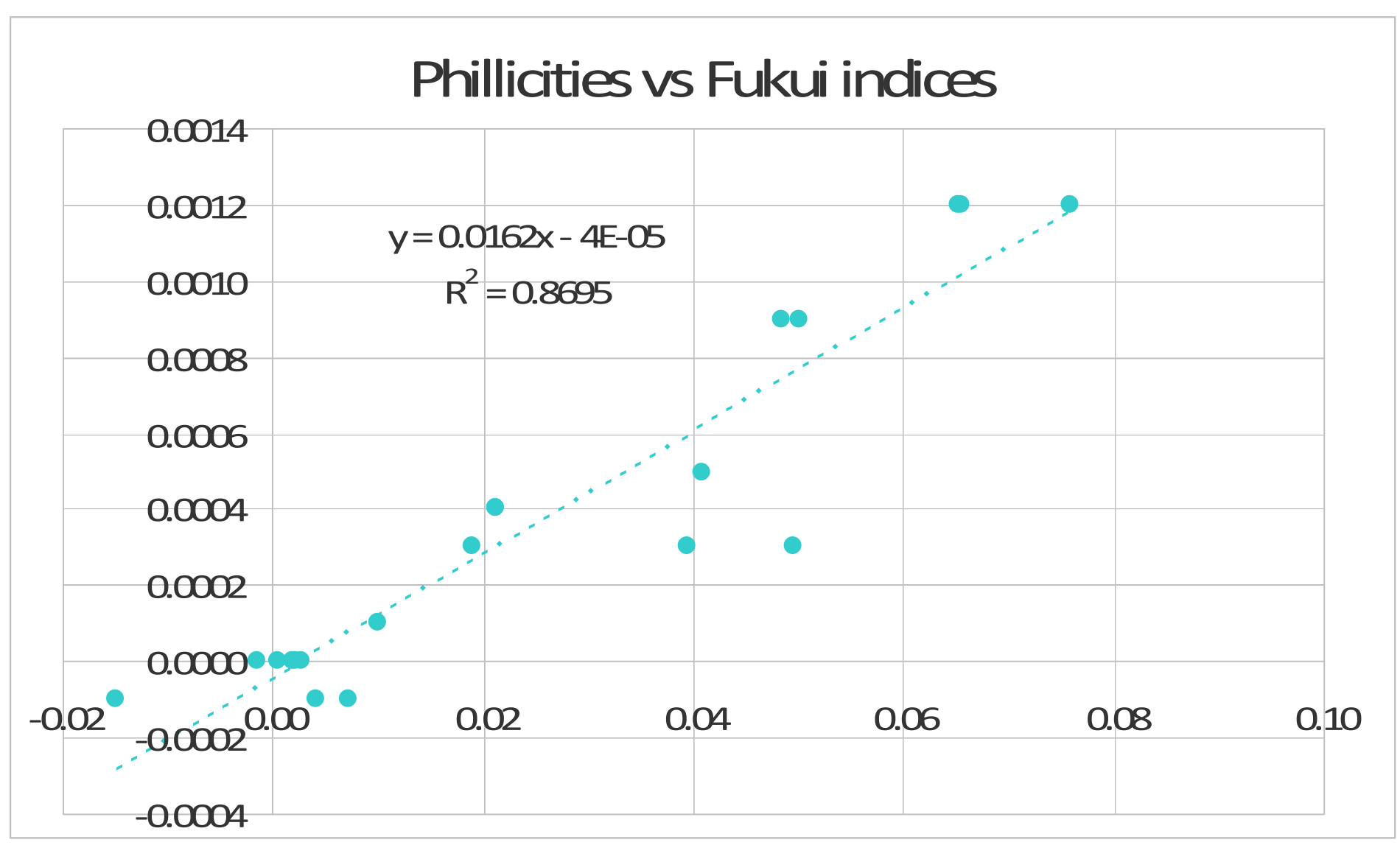

Figure S21. Linear regression of local electrophilicities against local Fukui indices for the NHOs of the $CH_3COOH$ molecule. Three outliers corresponding to NHOs 3, 4 and 21 have been removed.

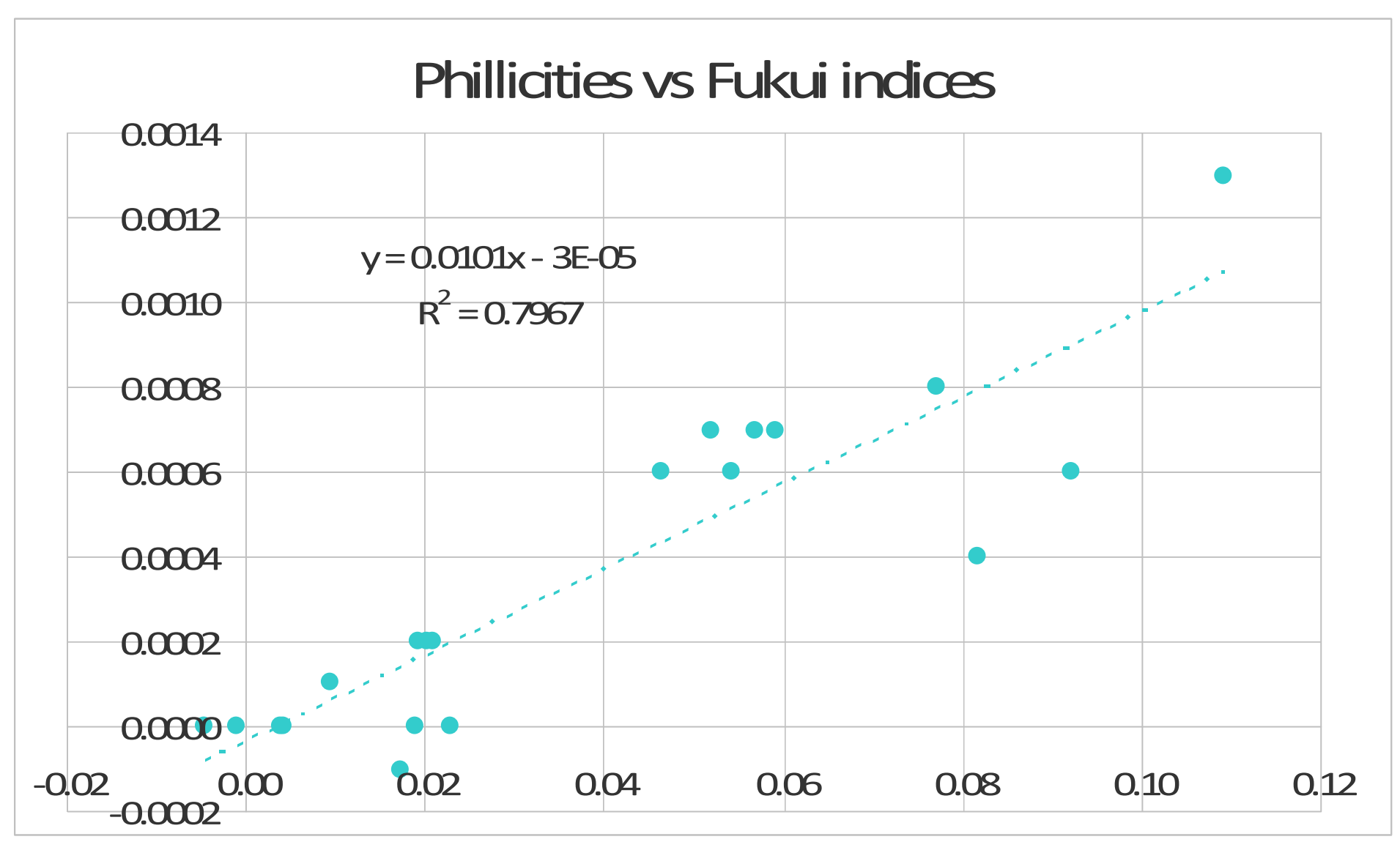

Figure S22. Linear regression of local electrophilicities against local Fukui indices for the NHOs of the $CH_3CONH_2$ molecule. Two outliers corresponding to NHOs 12 and 20 have been removed.

## 4. Local $\eta_i^{-1}$ values versus the local Fukui indices:

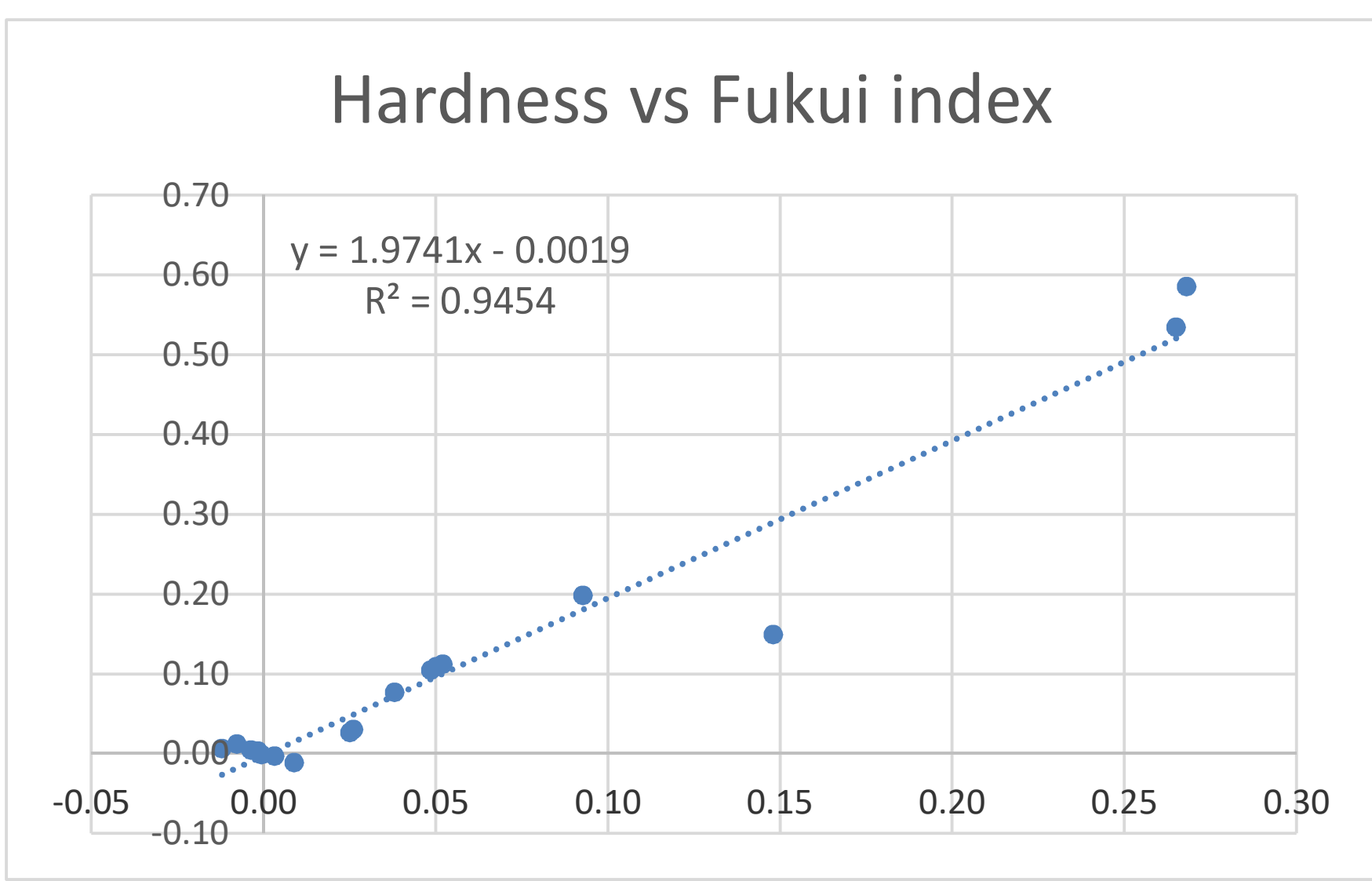

Figure S23. Lineal regression for local $\eta_i^{-1}$ values versus the local Fukui indices for the NHOs of the $CH_2CHCl$ molecule.

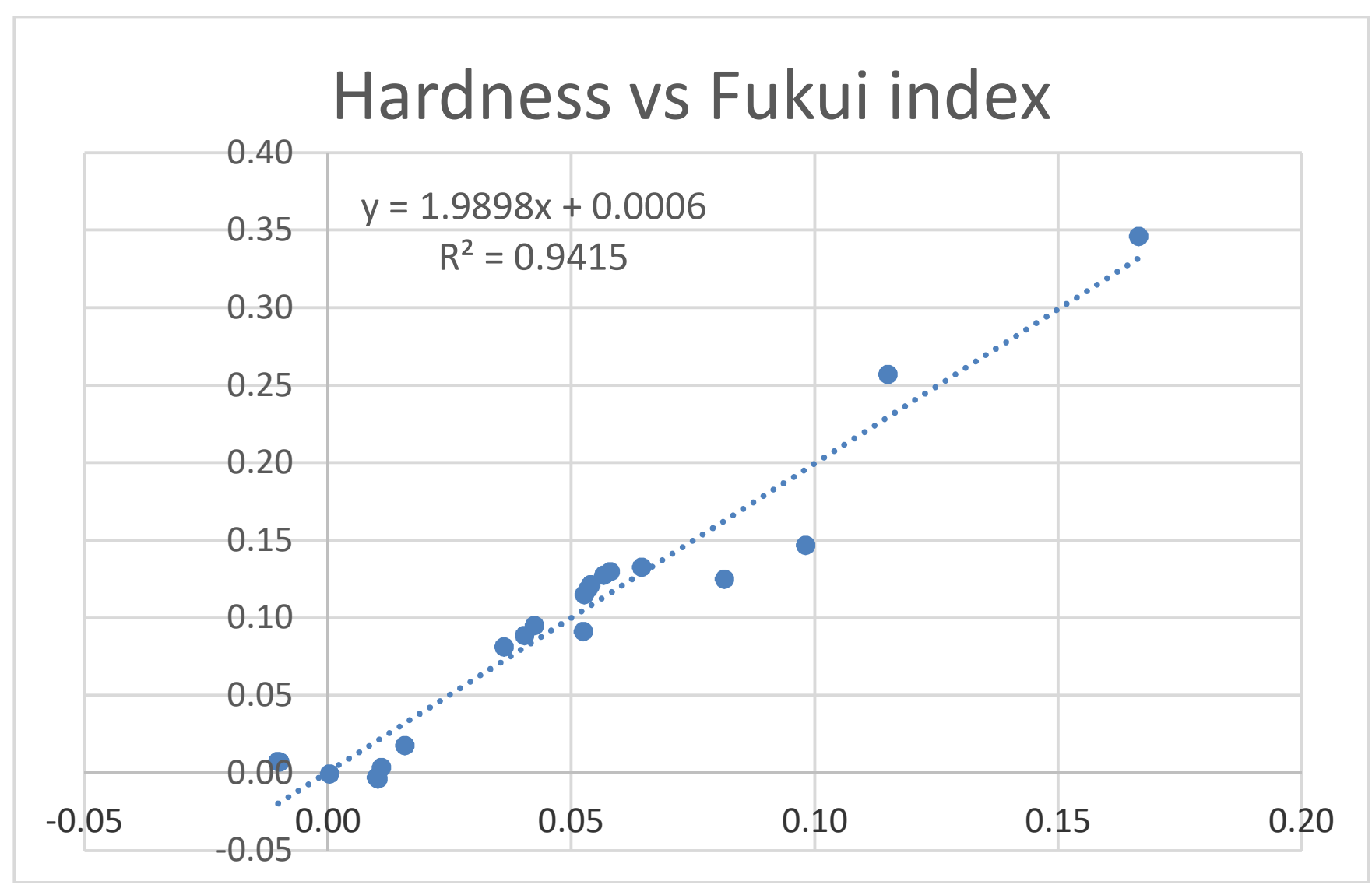

Figure S24. Lineal regression for local $\eta_i^{-1}$ values versus the local Fukui indices for the NHOs of the $CH_2CHNH_2$ molecule.

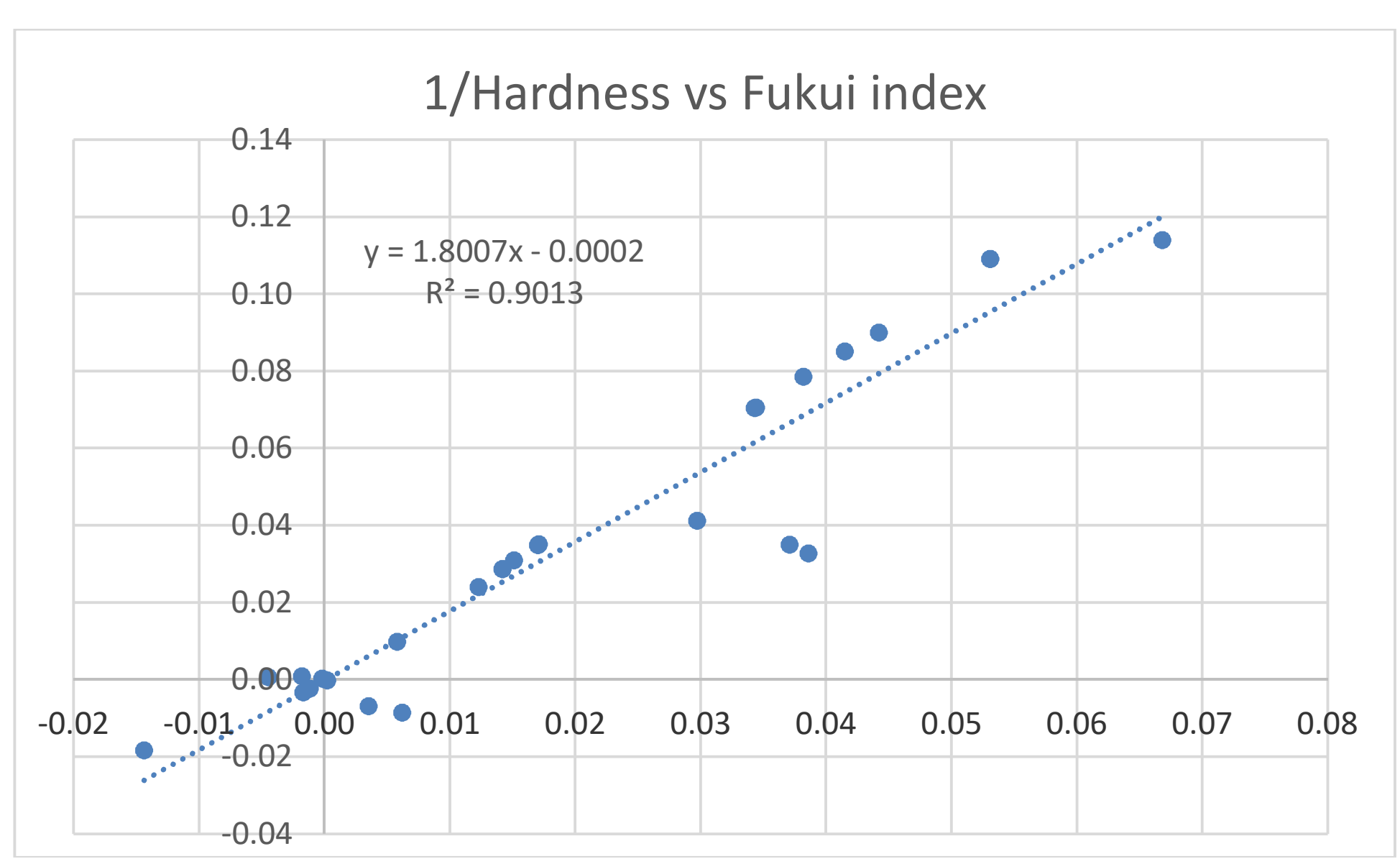

Figure S25. Lineal regression for local $\eta_i^{-1}$ values versus the local Fukui indices for the NHOs of the $CH_3COOCH_3$ molecule. Three outliers corresponding to NHOs 3,4 and 28 have been removed.

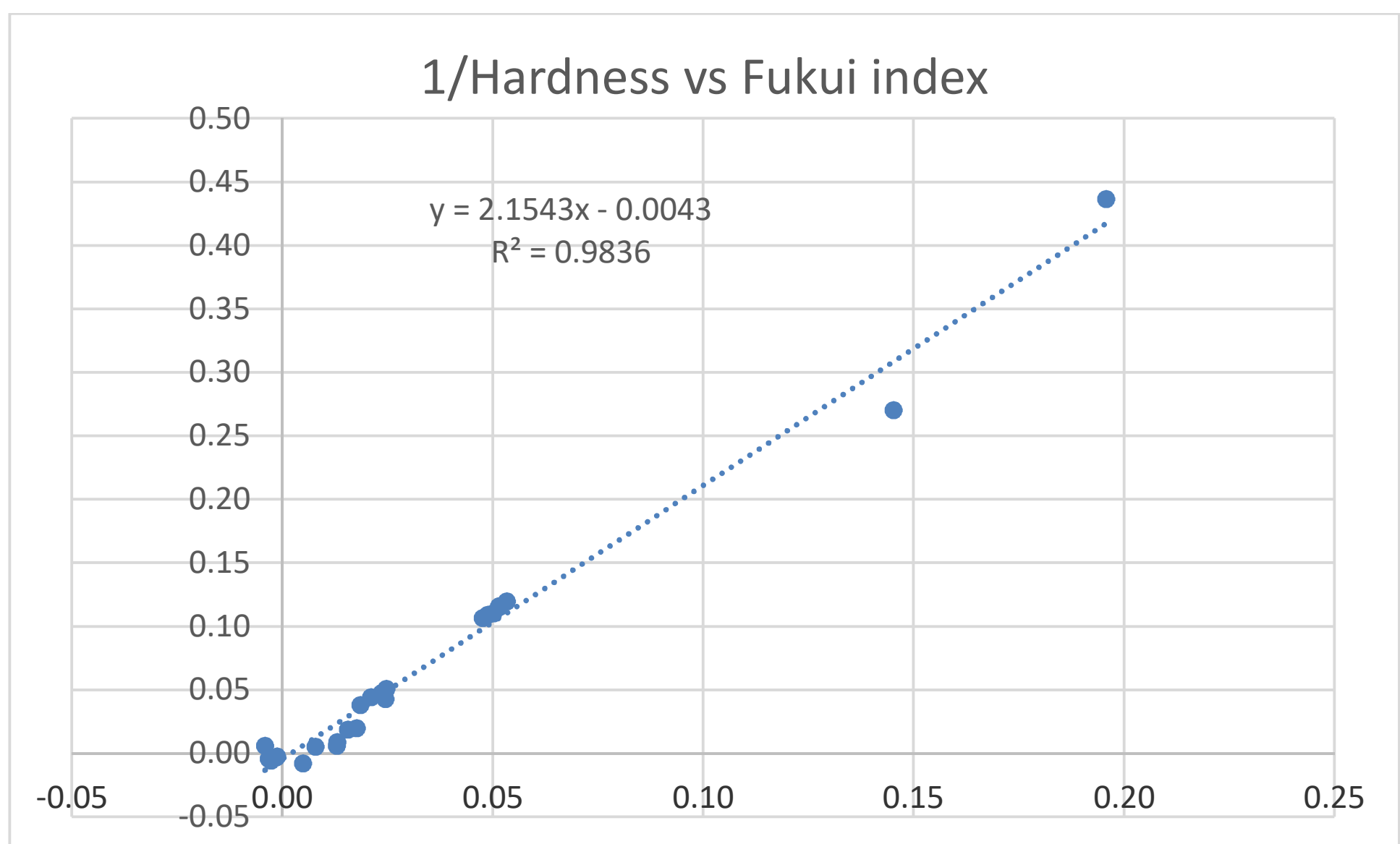

Figure S26. Lineal regression for local $\eta_i^{-1}$ values versus the local Fukui indices for the NHOs of the $CH_2CHOCH_3$ molecule. An outlier corresponding to NHO 26 has been removed.

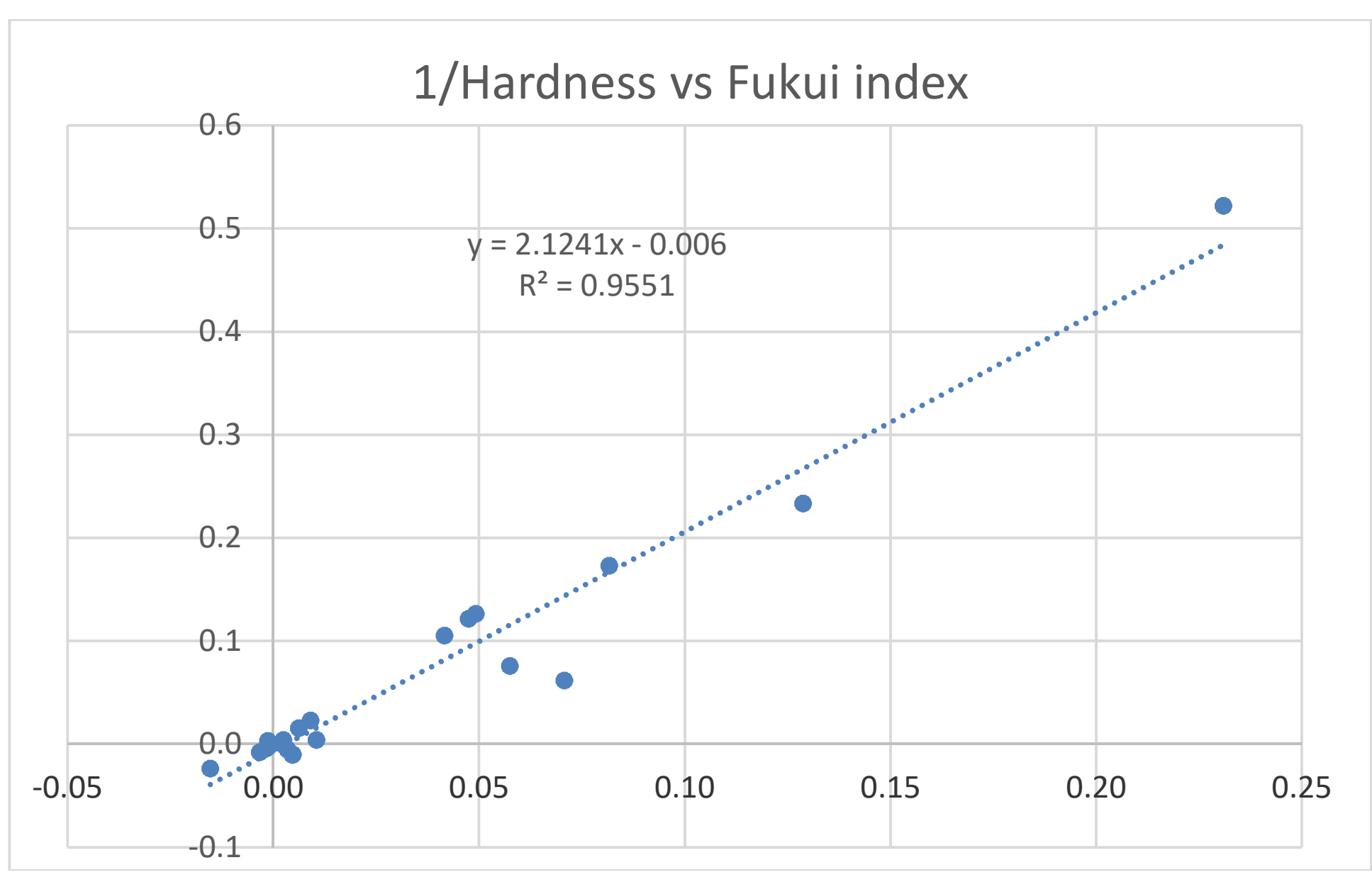

Figure S27. Lineal regression for local $\eta_i^{-1}$ values versus the local Fukui indices for the NHOs of the $CH_2CHCHO$ molecule. Two outliers corresponding to NHOs 16 and 24 have been removed.

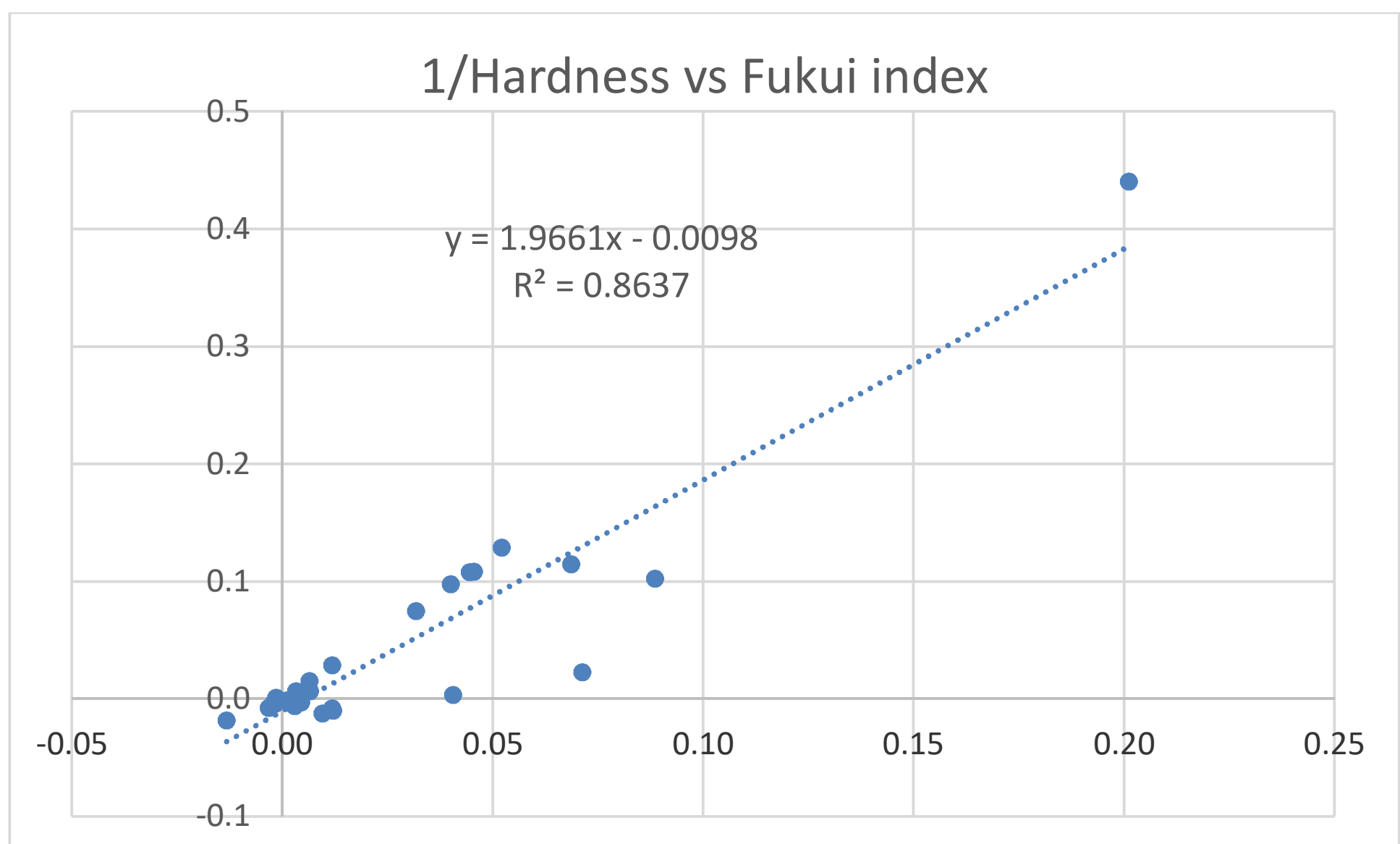

Figure S28. Lineal regression for local $\eta_i^{-1}$ values versus the local Fukui indices for the NHOs of the $CH_2CHNO_2$ molecule. Two outliers corresponding to NHOs 16 and 25 have been removed.

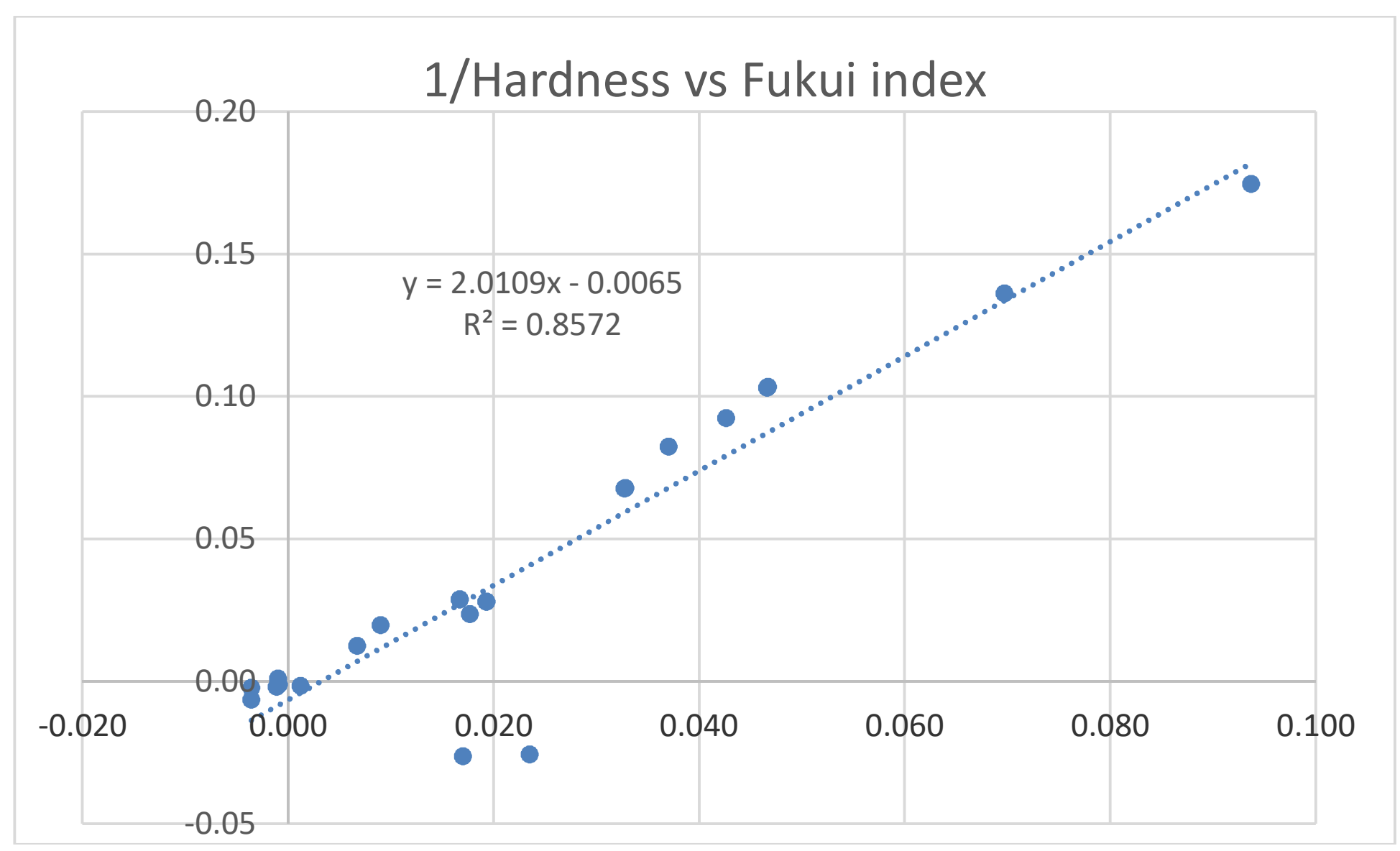

Figure S29. Lineal regression for local $\eta_i^{-1}$ values versus the local Fukui indices for the NHOs of the $CH_3CHSH$ molecule. Two outliers corresponding to NHOs 24 and 25 have been removed.

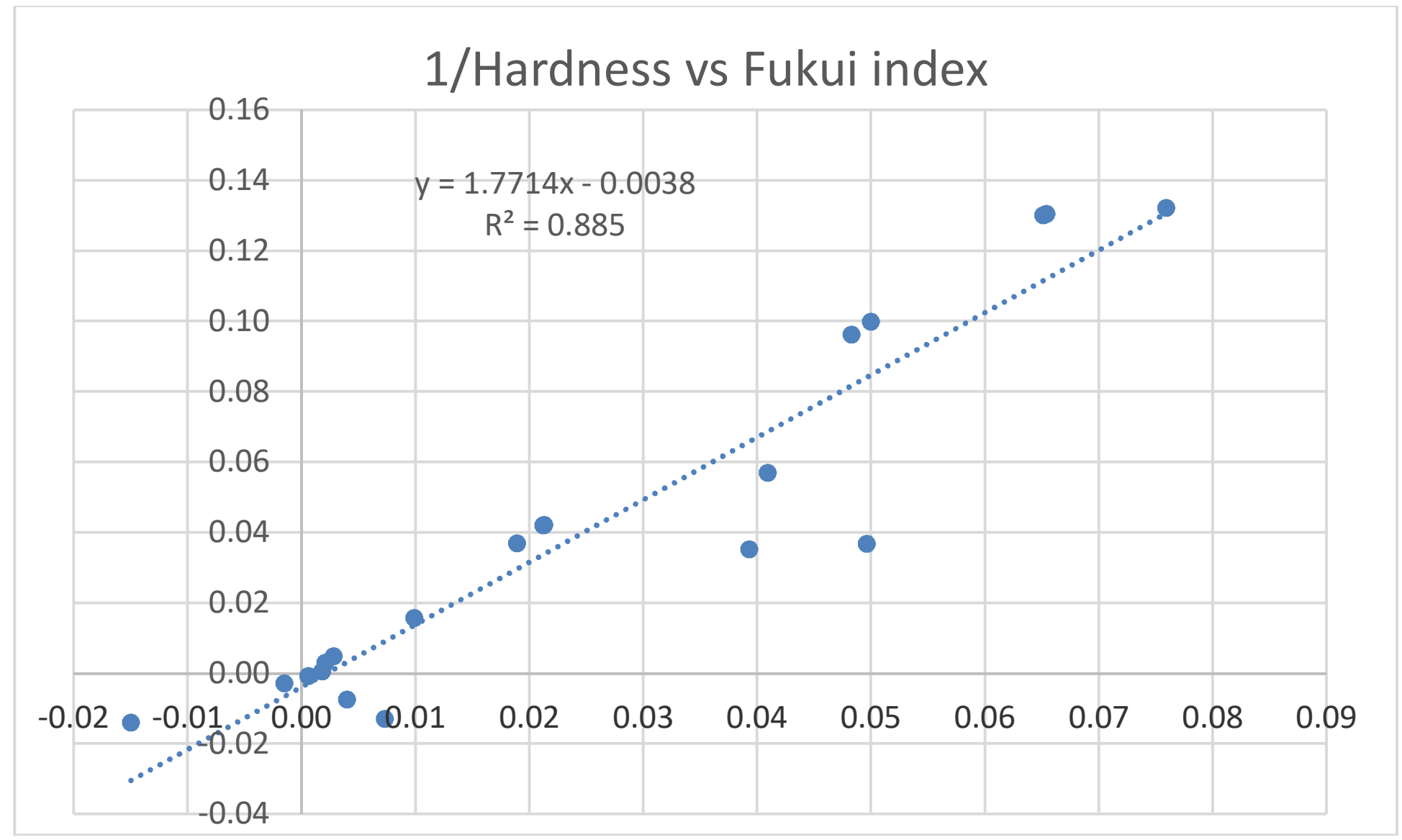

Figure S30. Lineal regression for local $\eta_i^{-1}$ values versus the local Fukui indices for the NHOs of the $CH_3COOH$ molecule. Three outliers corresponding to NHOs 3, 4 and 21 have been removed.

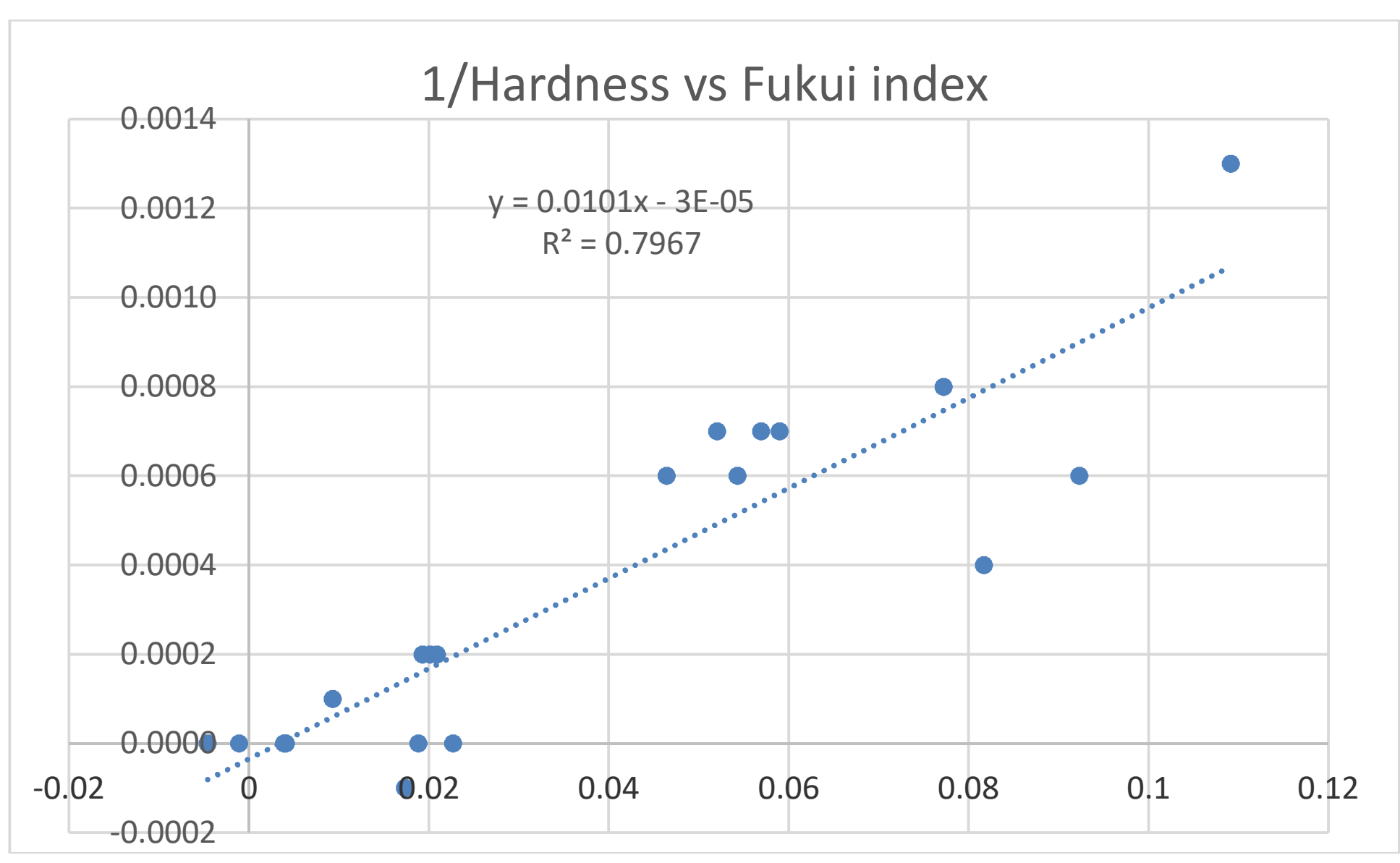

Figure S31. Lineal regression for local $\eta_i^{-1}$ values versus the local Fukui indices for the NHOs of the $CH_3CONH_2$ molecule. Two outliers corresponding to NHOs 12 and 20 have been removed.

## 5. Local $\eta_i^{-1}$ values versus the maximum local charge variations:

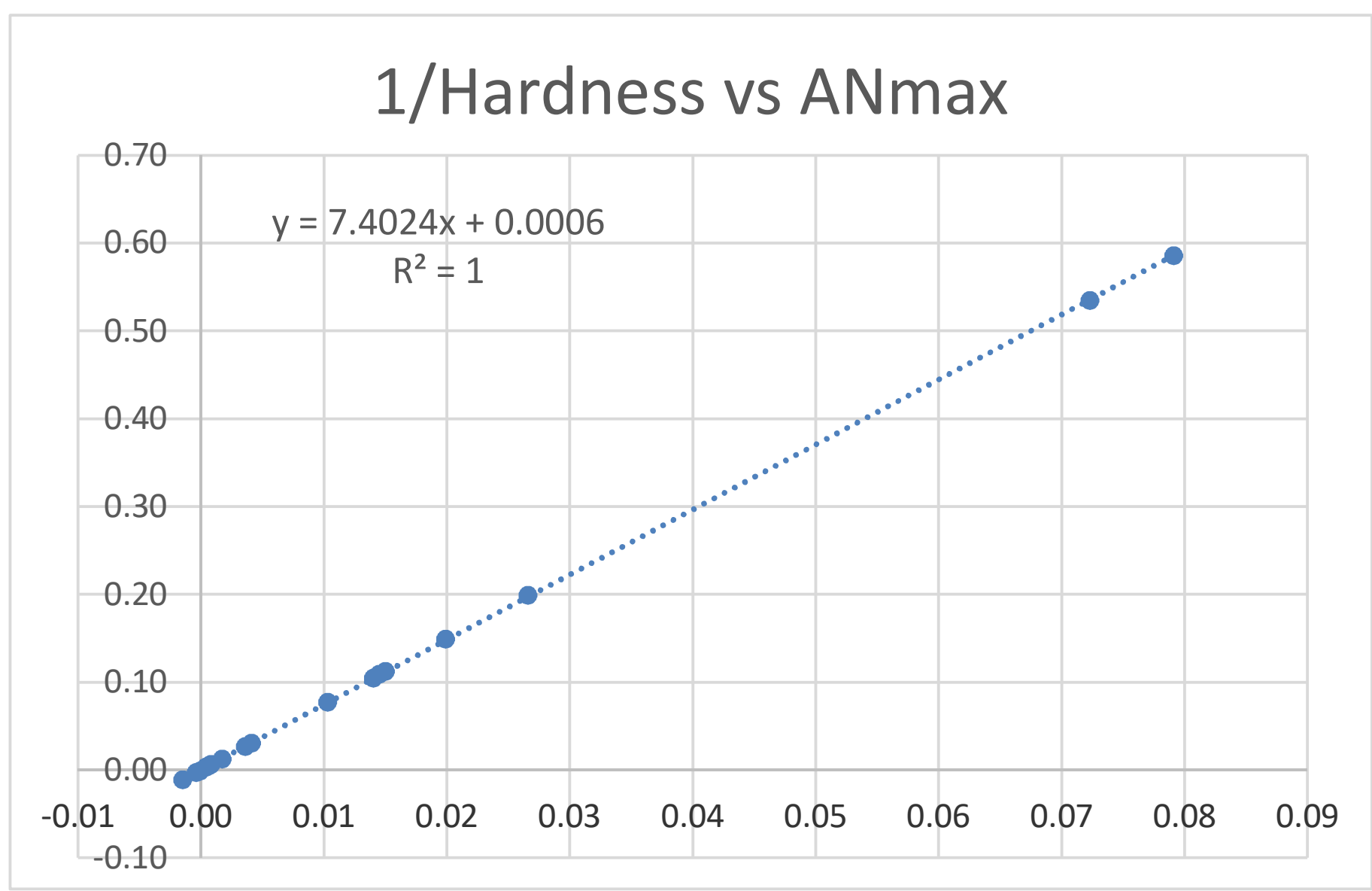

Figure S32. Lineal regression for local values $\eta_i^{-1}$ versus the maximum local charge variations for the NHOs of the $CH_2CHCl$ molecule.

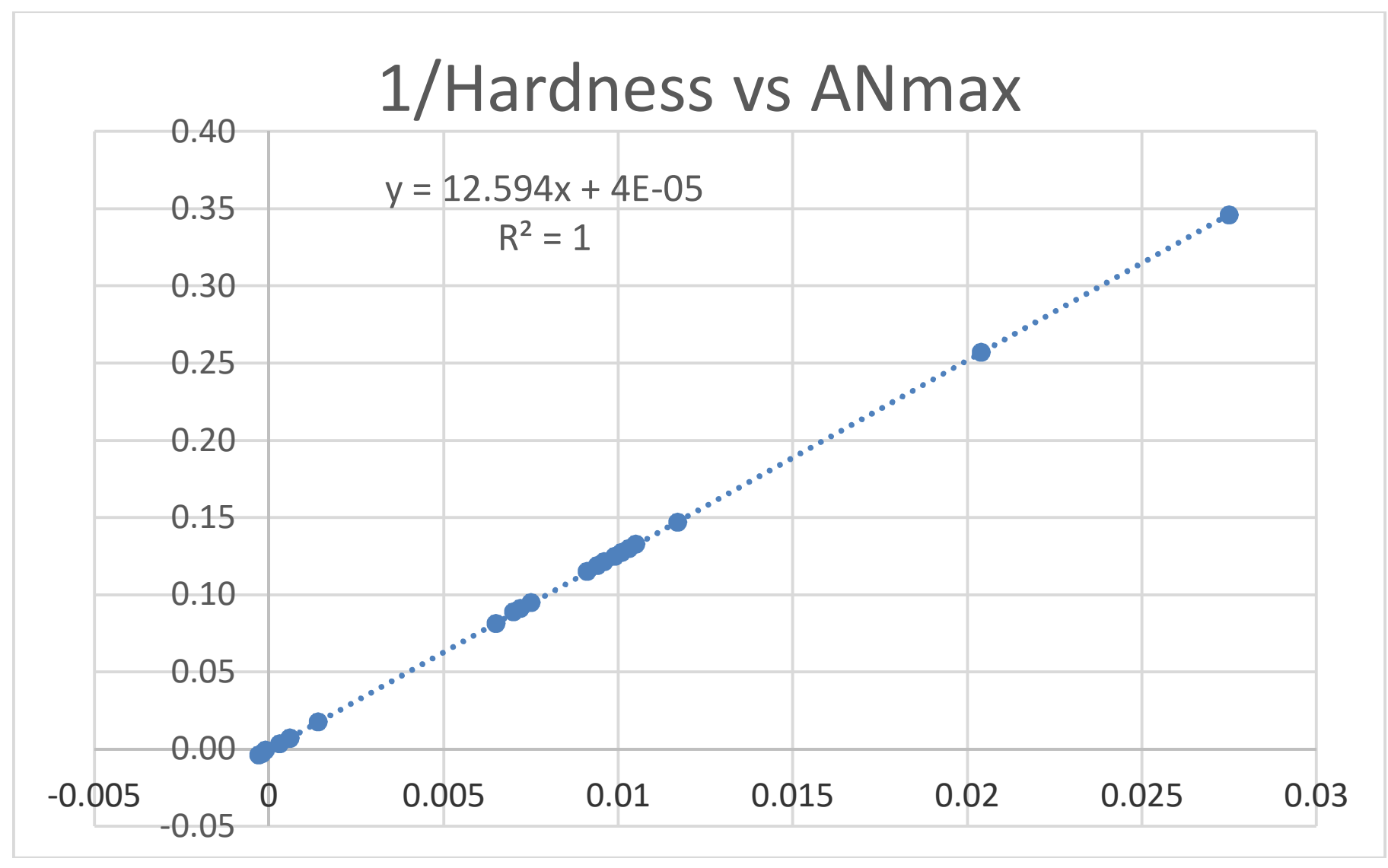

Figure S33. Lineal regression for local values $\eta_i^{-1}$ versus the maximum local charge variations for the NHOs of the $CH_2CHNH_2$ molecule.

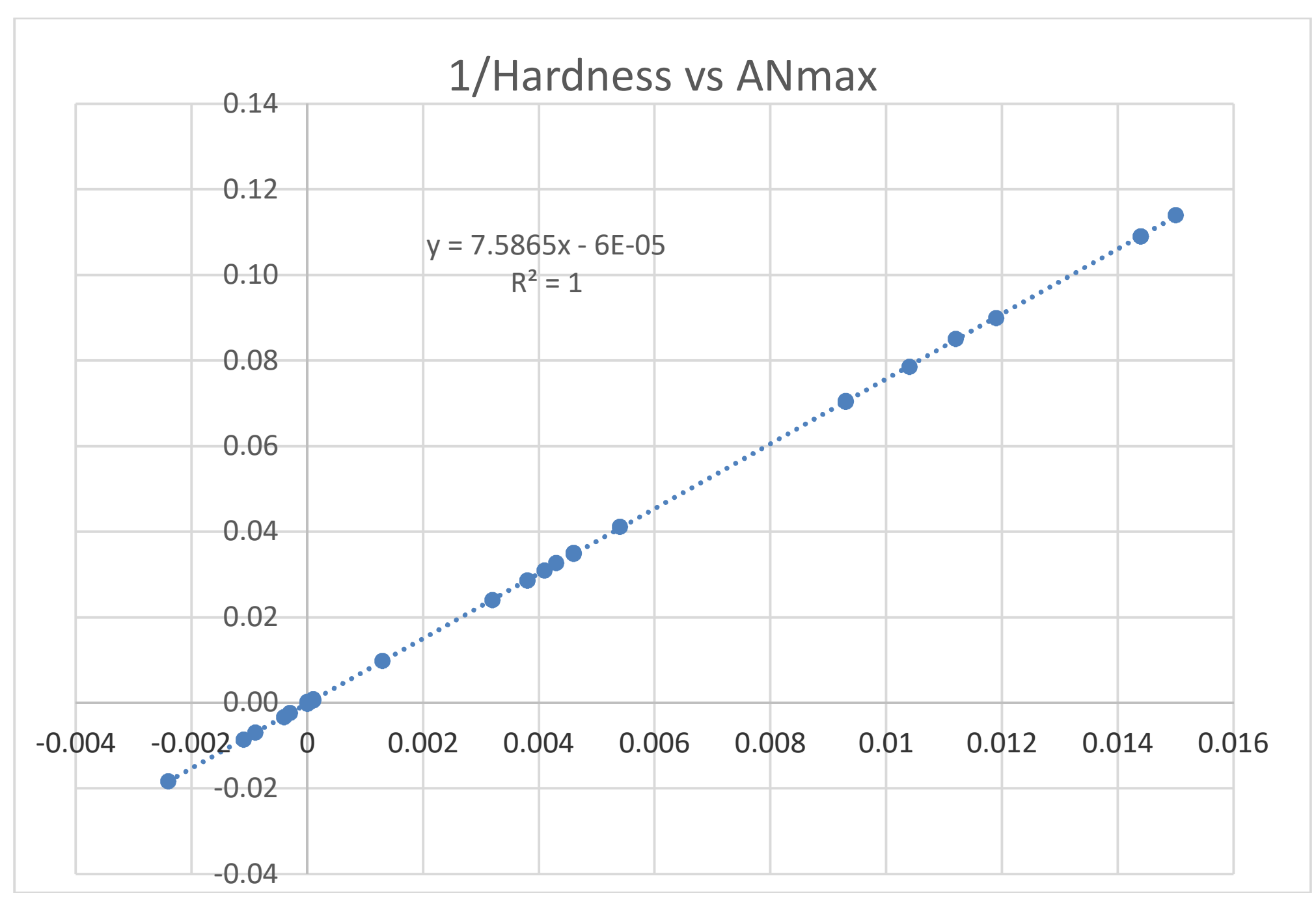

Figure S34. Lineal regression for local values $\eta_i^{-1}$ versus the maximum local charge variations for the NBOs of the $CH_3COOCH_3$ molecule. Three outliers corresponding to NHOs 3,4 and 28 have been removed.

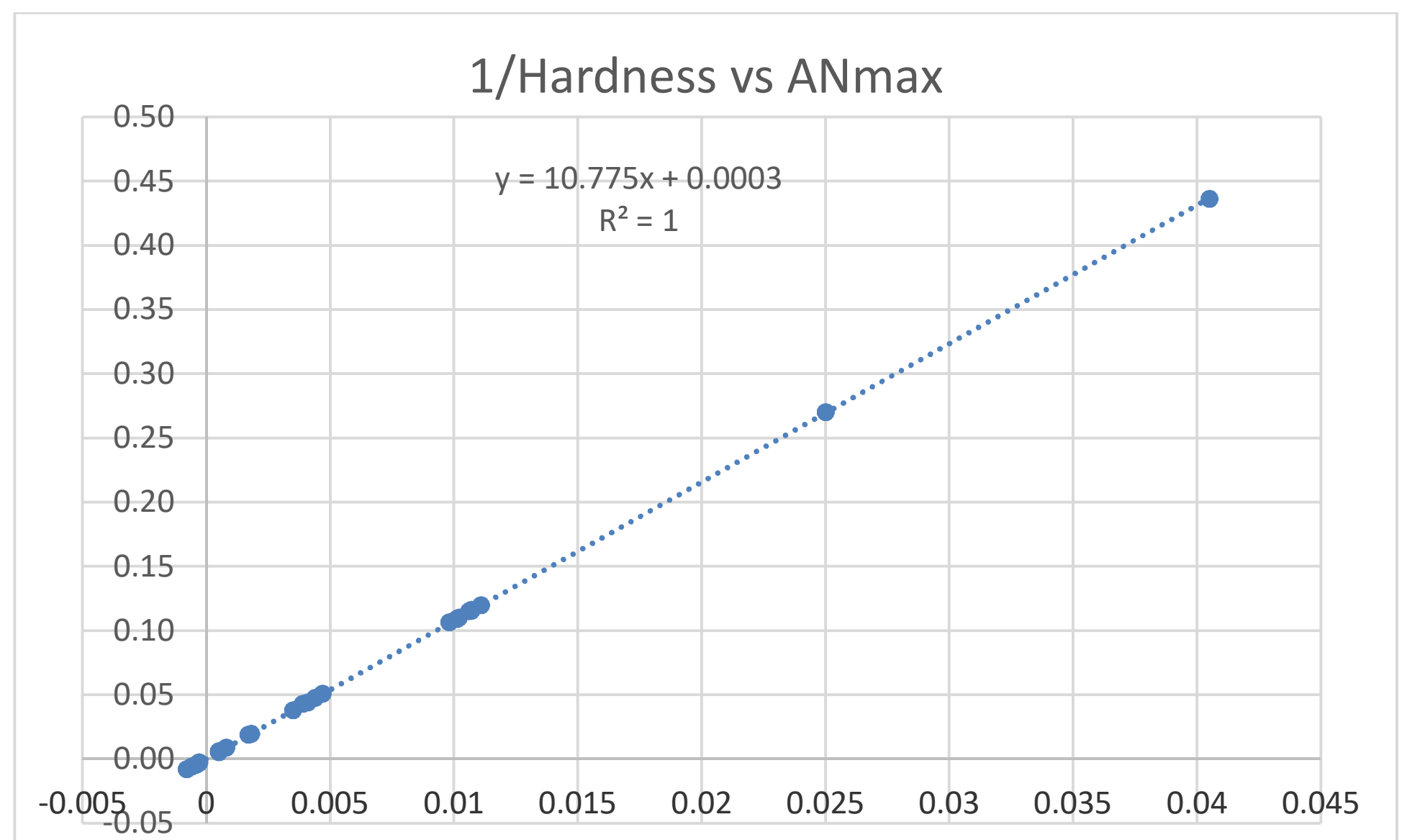

Figure S35. Lineal regression for local values $\eta_i^{-1}$ versus the maximum local charge variations for the NHOs of the $CH_2CHOCH_3$ molecule. An outlier corresponding to NHO 26 has been removed.

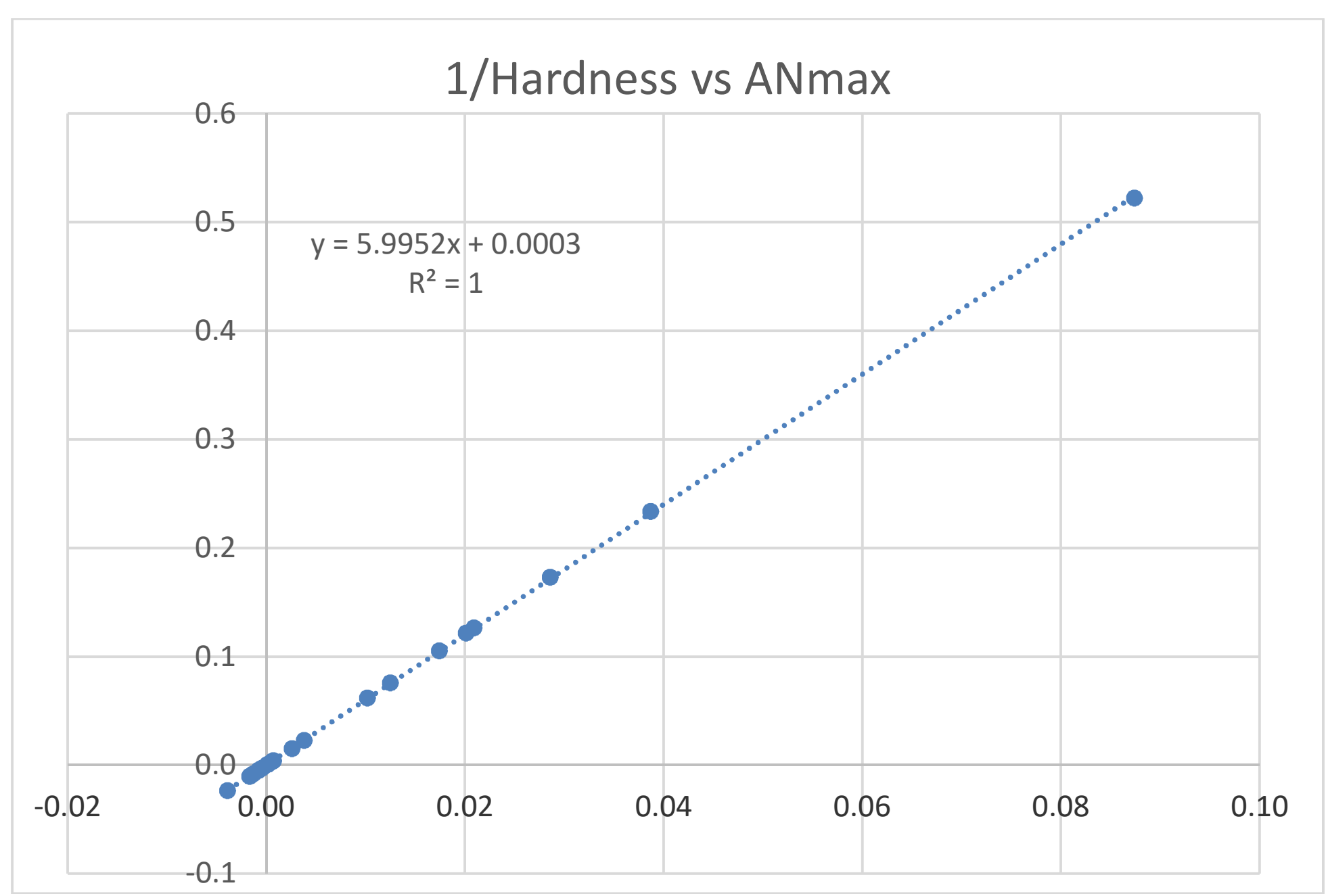

Figure S36. Lineal regression for local values $\eta_i^{-1}$ versus the maximum local charge variations for the NHOs of the $CH_2CHCHO$ molecule. Two outliers corresponding to NHOs 16 and 24 have been removed.

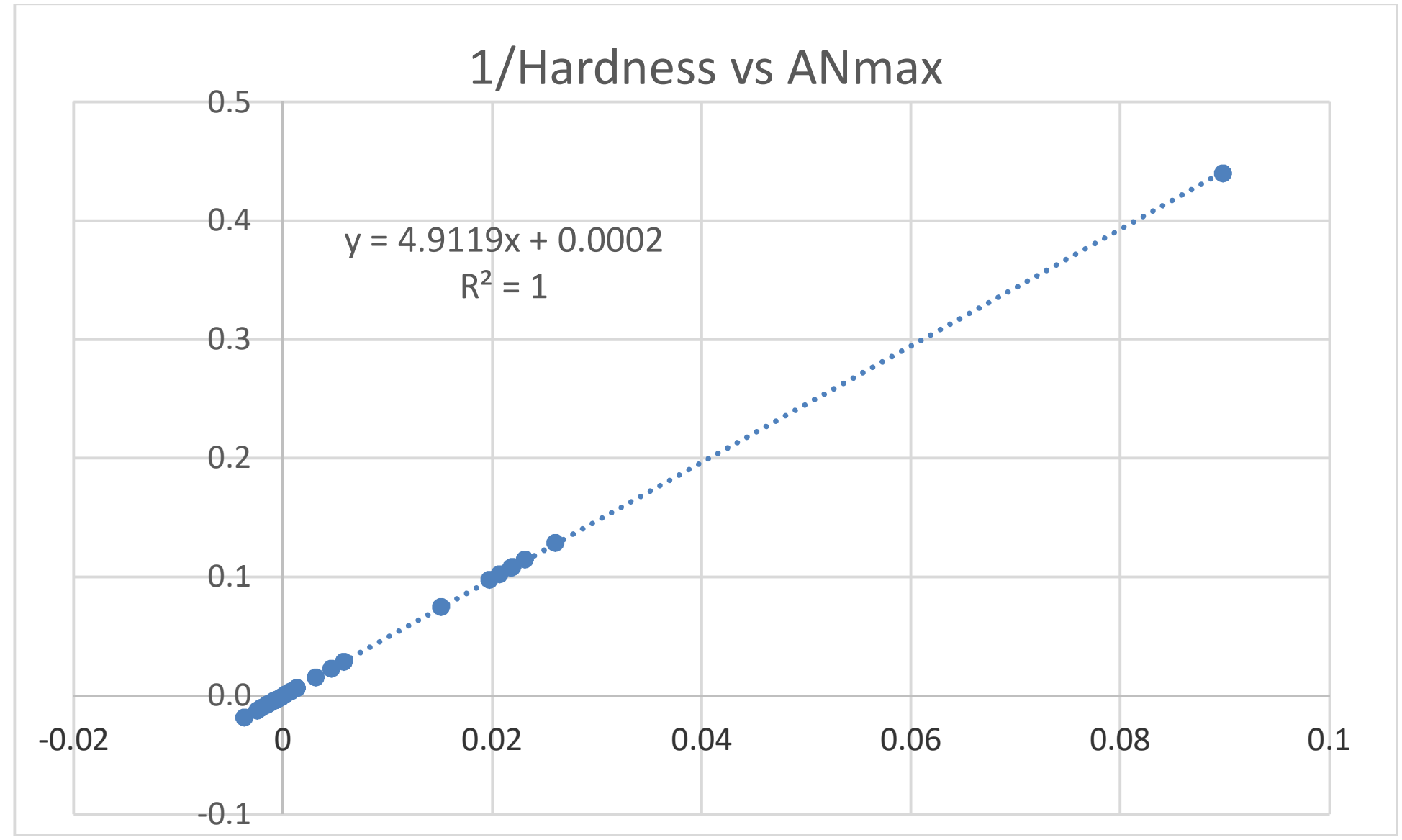

Figure S37. Lineal regression for local values $\eta_i^{-1}$ versus the maximum local charge variations for the NHOs of the $CH_2CHNO_2$ molecule. Two outliers corresponding to NHOs 16 and 25 have been removed.

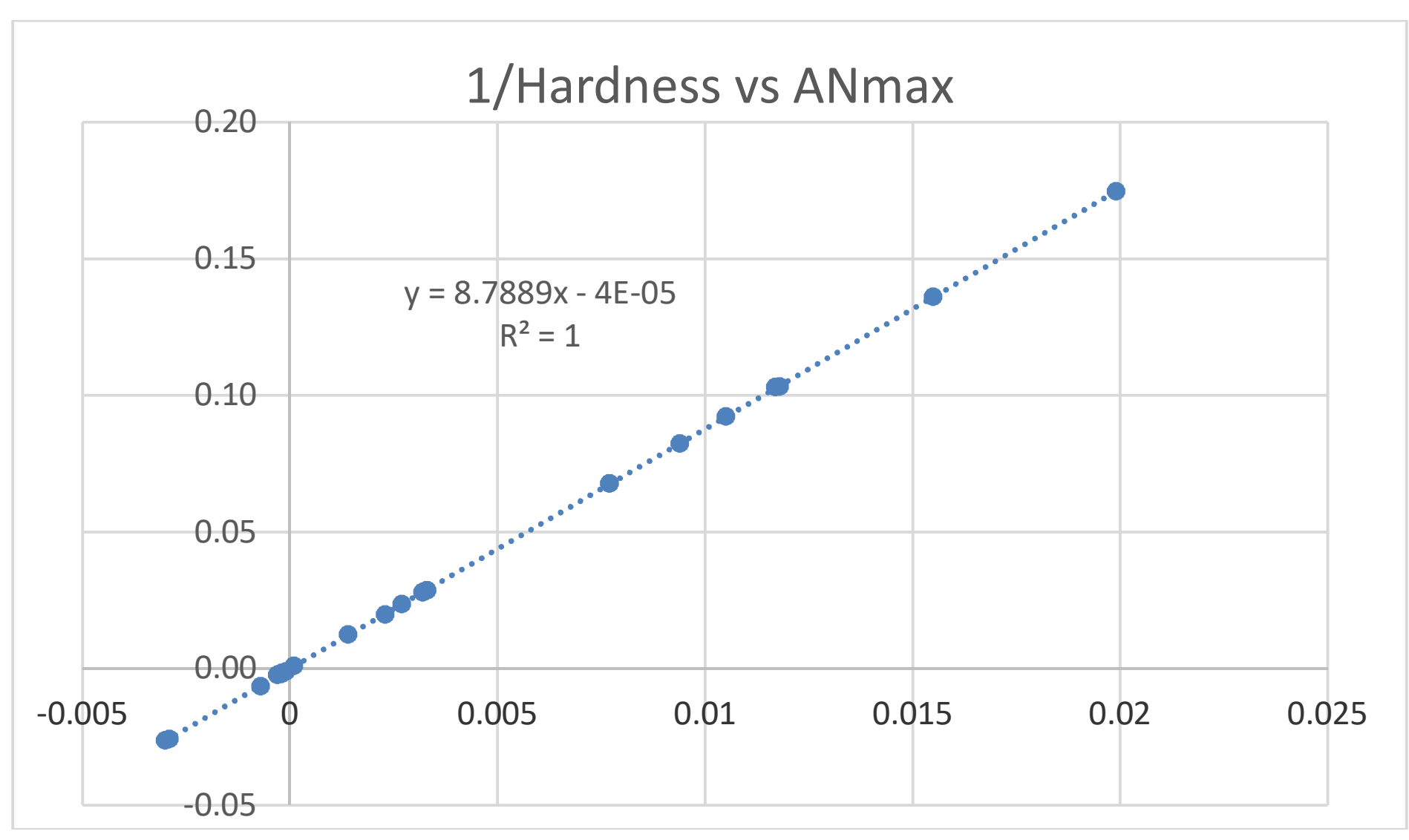

Figure S38. Lineal regression for local values $\eta_i^{-1}$ versus the maximum local charge variations for the NHOs of the $CH_3CHSH$ molecule. Two outliers corresponding to NHOs 24 and 25 have been removed.

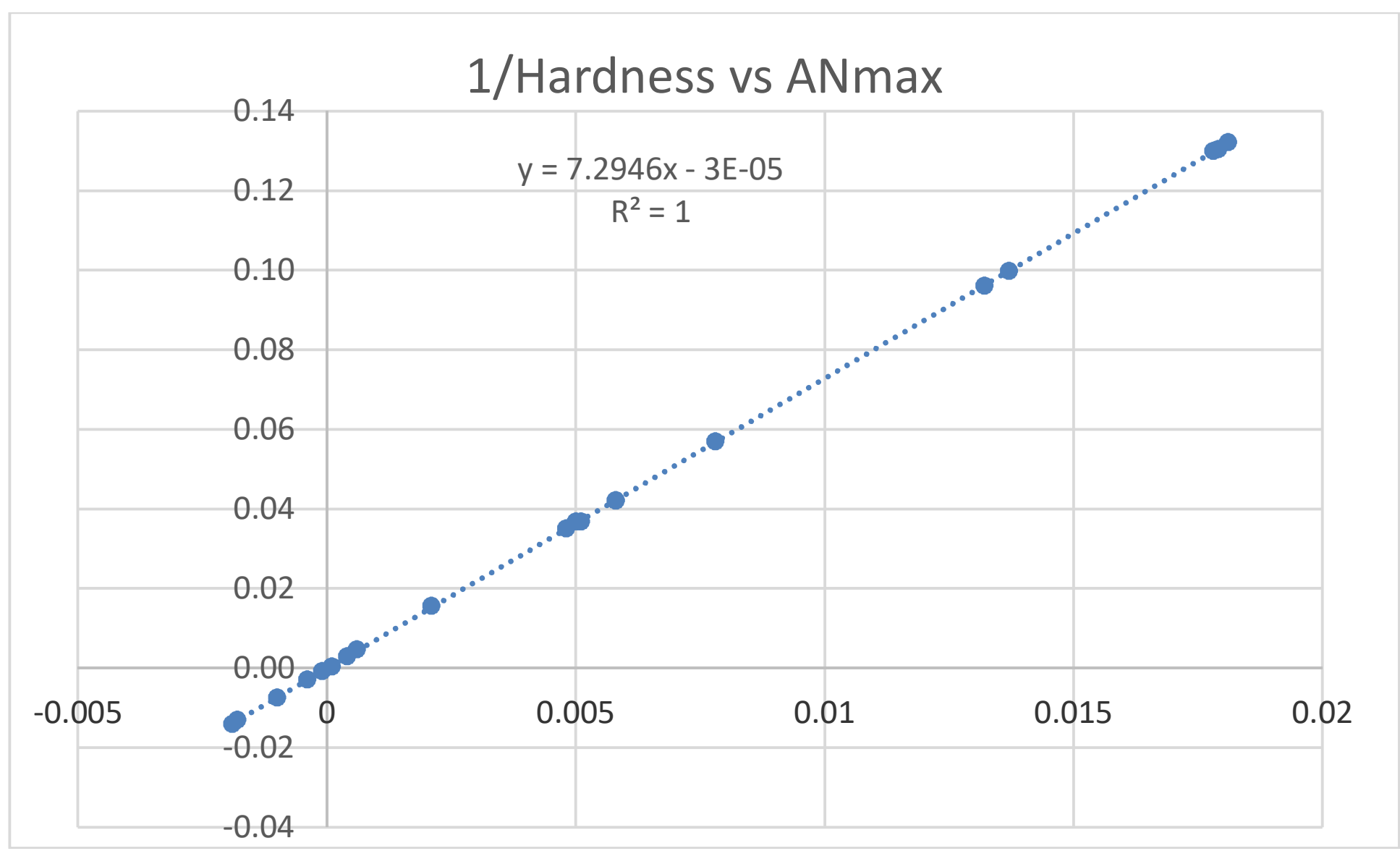

Figure S39. Lineal regression for local values $\eta_i^{-1}$ versus the maximum local charge variations for the NHOs of the $CH_3COOH$ molecule. Three outliers corresponding to NHOs 3, 4 and 21 have been removed.

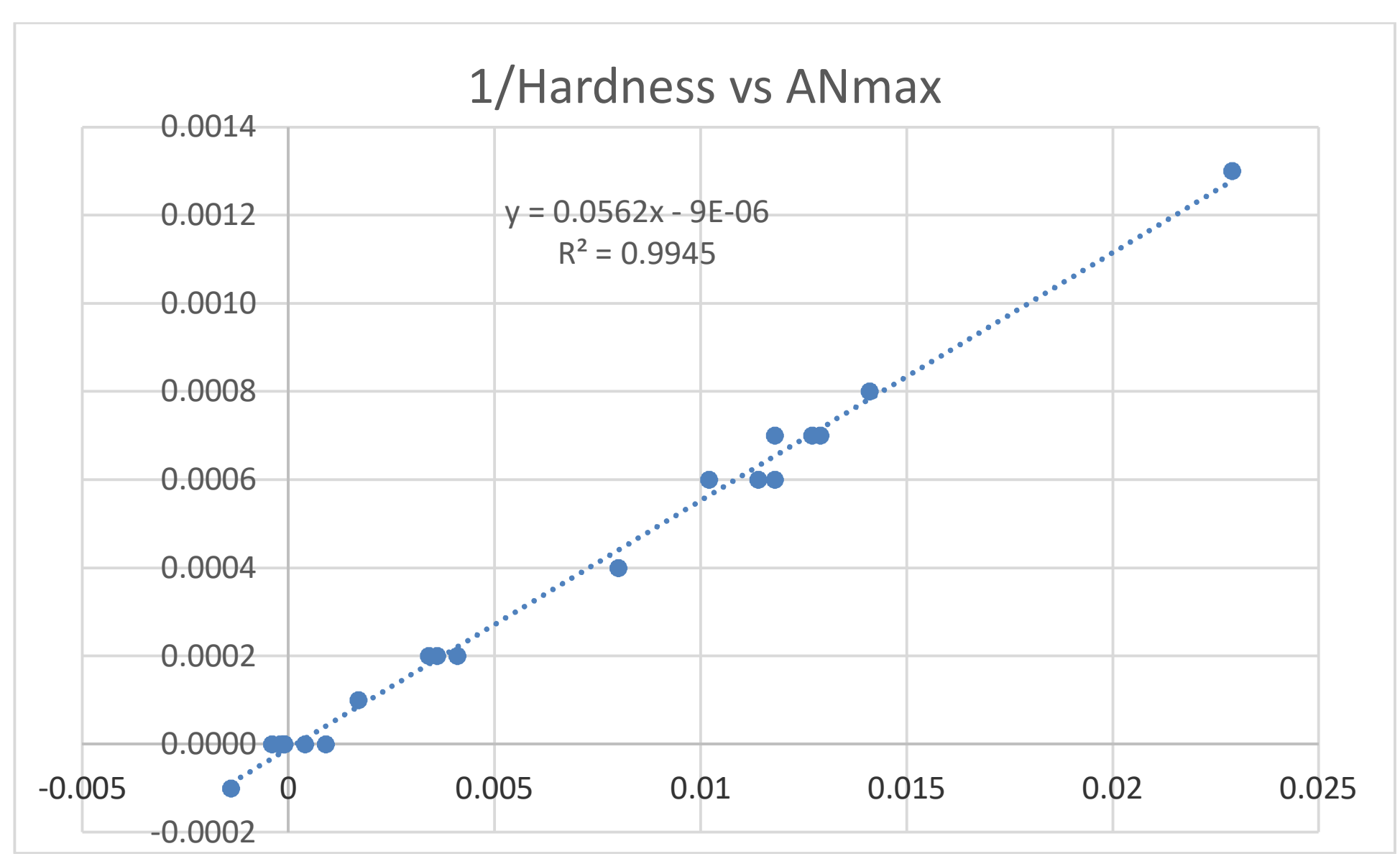

Figure S40. Lineal regression for local values $\eta_i^{-1}$ versus the maximum local charge variations for the NHOs of the $CH_3CONH_2$ molecule. Two outliers corresponding to NHOs 12 and 20 have been removed.

## 6. Maximum local charge variations (obtained with a cubic expansion of the local energy) versus the maximum local charge variations (achieved with a quadratic expansion of the local charge variation) for CMOs:

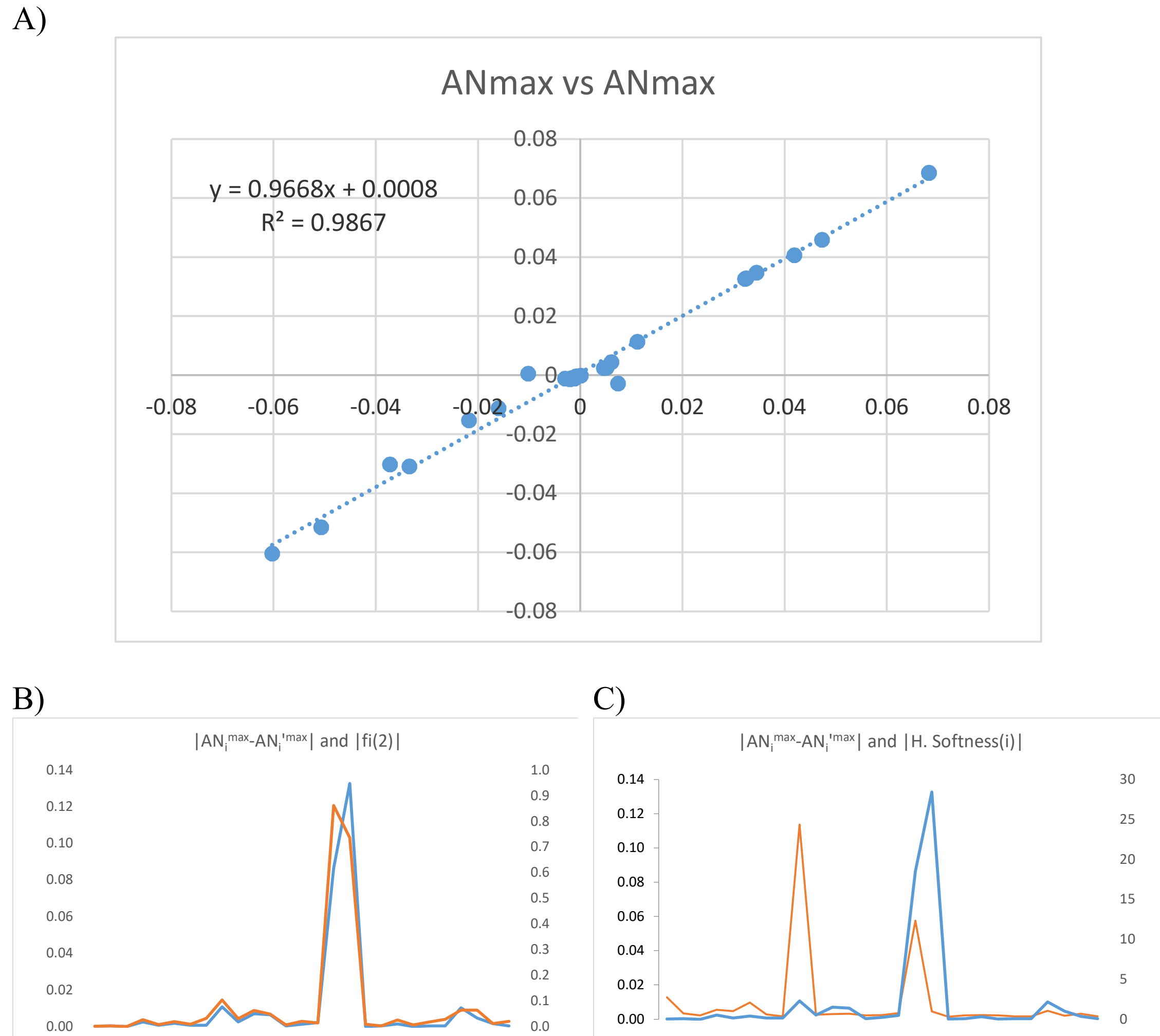

Figure S41. Linear regression of the maximum local charge variations (obtained with a cubic expansion of the local energy) versus the maximum local charge variations (achieved with a quadratic expansion of the local charge variation). Two outliers corresponding to CMOs 16 and 17 have been removed. B) Comparison of $\left|\Delta N_i^{\max} - \Delta N'^{\max}_i\right|$ and $\left|f_i^{(2)}\right|$ values. C) Comparison of $\left|\Delta N_i^{\max} - \Delta N'^{\max}_i\right|$ and $\left|\gamma_i\right|$ values. In all cases for the occupied CMOs of the $CH_2CHCl$ molecule.

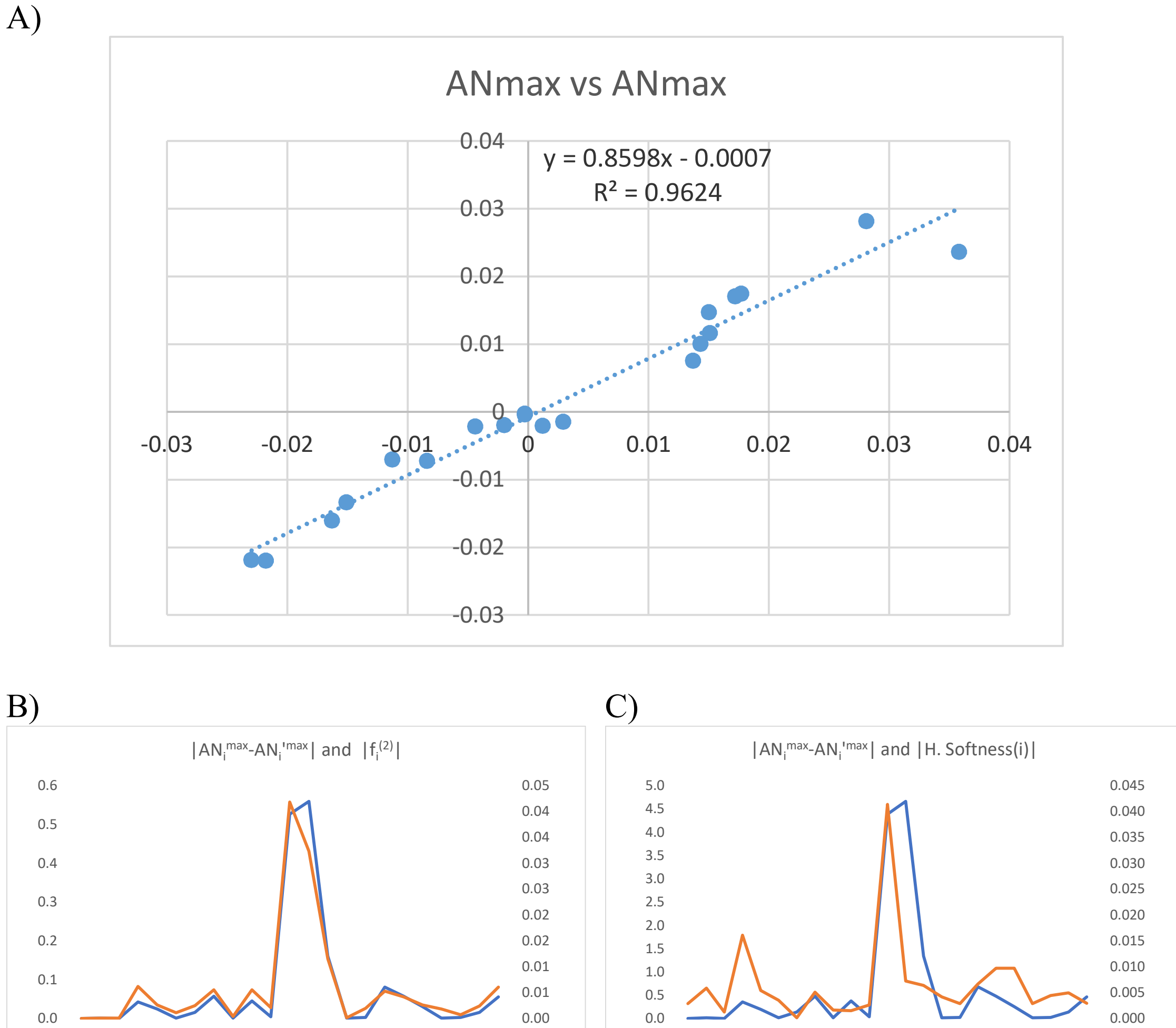

Figure S42. Linear regression of the maximum local charge variations (obtained with a cubic expansion of the local energy) versus the maximum local charge variations (achieved with a quadratic expansion of the local charge variation). Two outliers corresponding to CMOs 12 and 13 have been removed. B) Comparison of $\left|\Delta N_i^{\max} - \Delta N'^{\max}_i\right|$ and $\left|f_i^{(2)}\right|$ values. C) Comparison of $\left|\Delta N_i^{\max} - \Delta N'^{\max}_i\right|$ and $\left|\gamma_i\right|$ values. In all cases for the occupied CMOs of the $CH_2CHNH_2$ molecule.

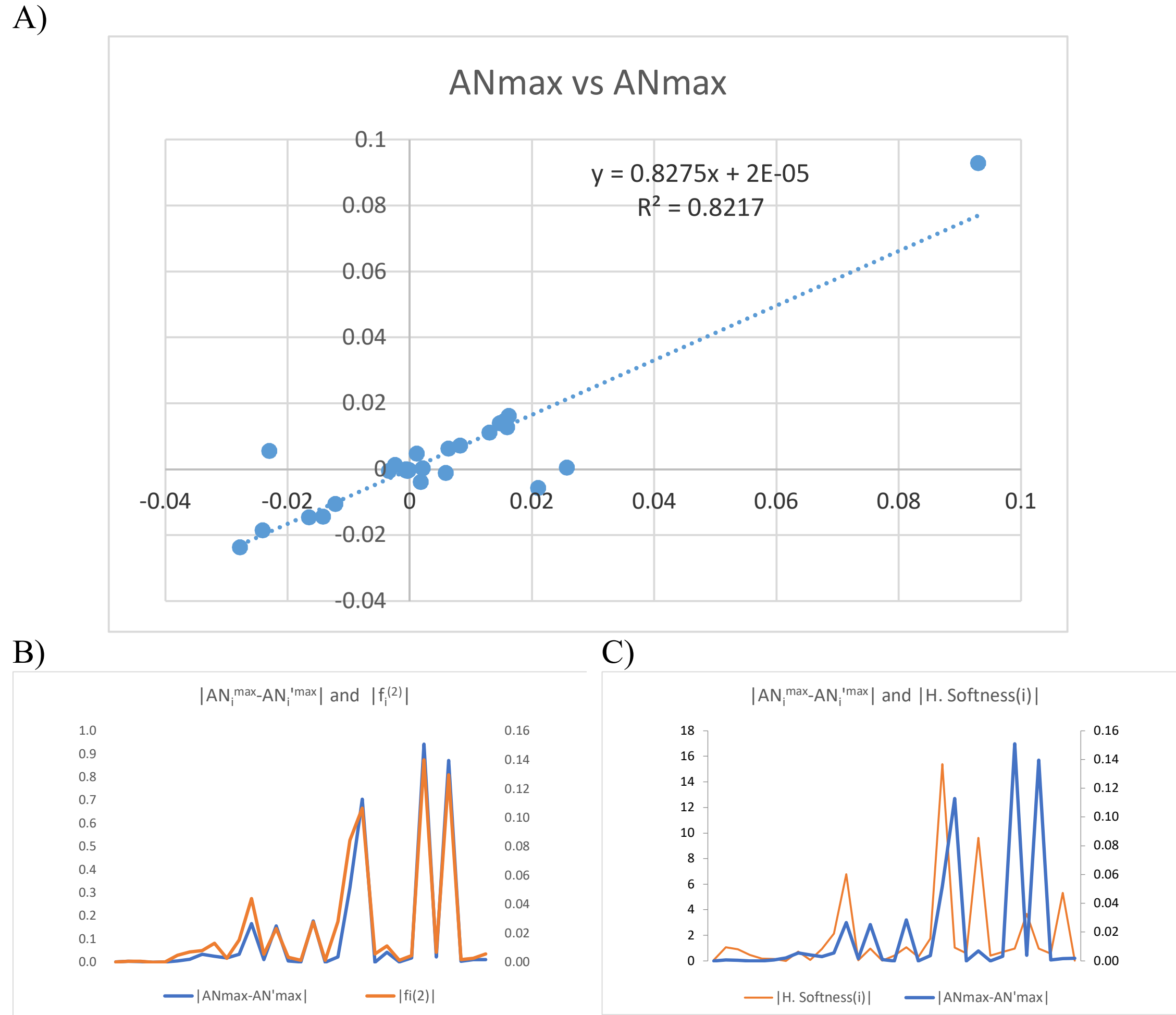

Figure S43. Linear regression of the maximum local charge variations (obtained with a cubic expansion of the local energy) versus the maximum local charge variations (achieved with a quadratic expansion of the local charge variation). Four outliers corresponding to CMOs 20, 21, 26 and 28 have been removed. B) Comparison of $\left|\Delta N_i^{\max} - \Delta N'^{\max}_i\right|$ and $\left|f_i^{(2)}\right|$ values. C) Comparison of $\left|\Delta N_i^{\max} - \Delta N'^{\max}_i\right|$ and $\left|\gamma_i\right|$ values. In all cases for the occupied CMOs of the $CH_3COOCH_3$ molecule.

A)

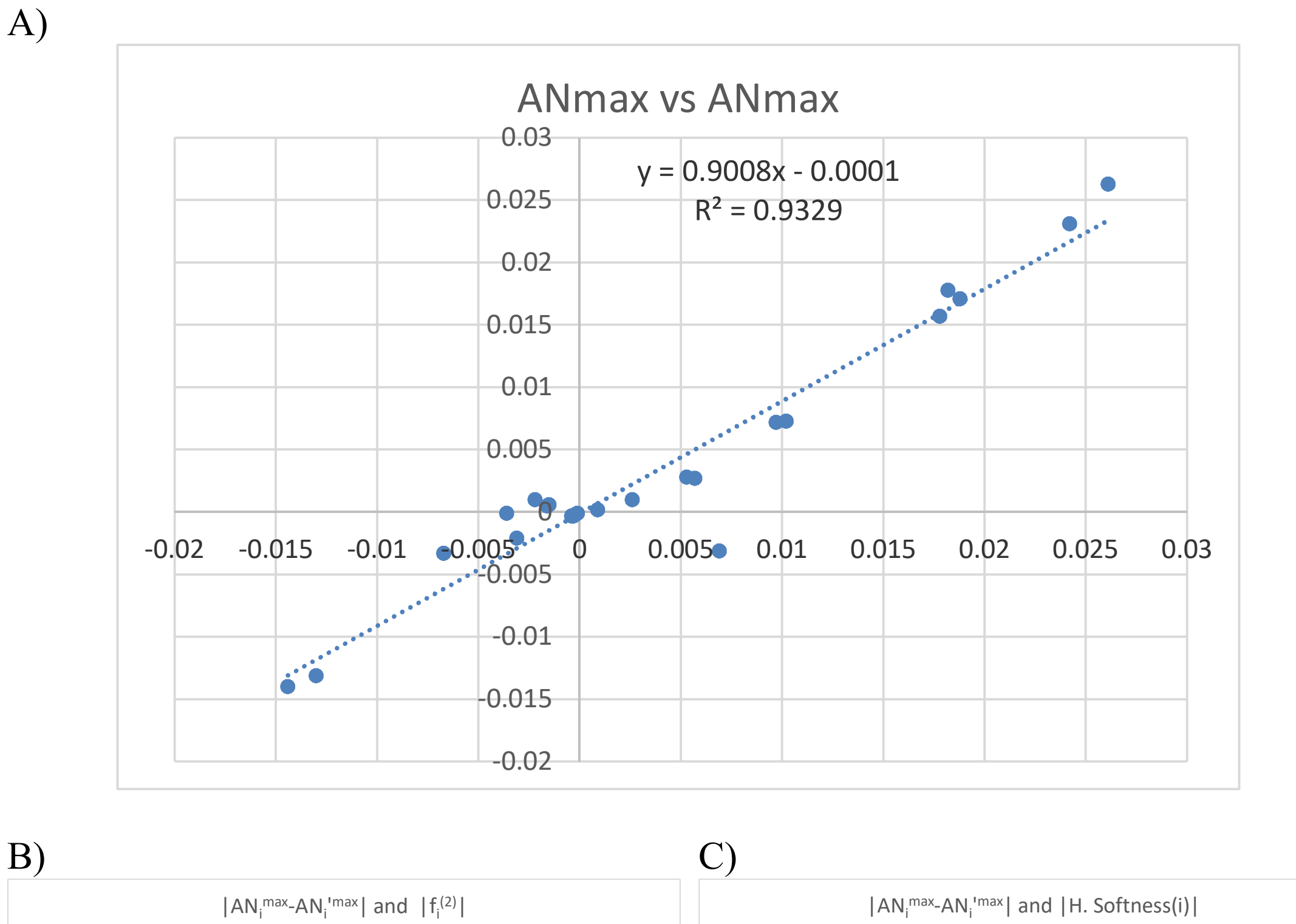

B) C)

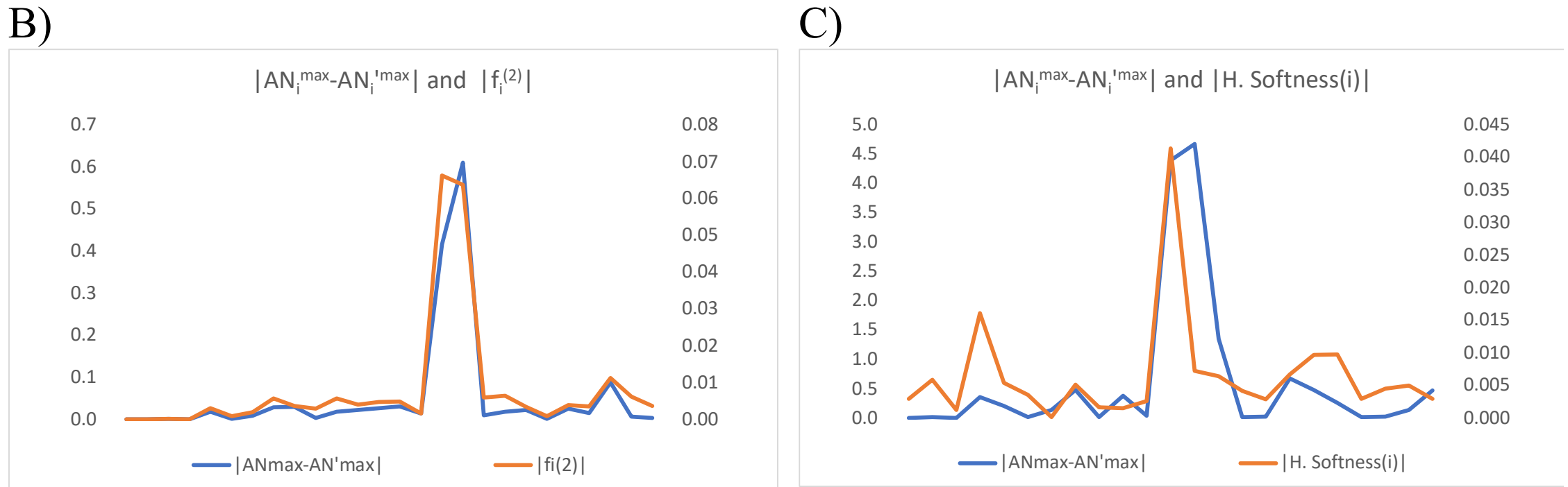

Figure S44. Linear regression of the maximum local charge variations (obtained with a cubic expansion of the local energy) versus the maximum local charge variations (achieved with a quadratic expansion of the local charge variation). Two outliers corresponding to CMOs 16 and 17 have been removed. B) Comparison of $\left|\Delta N_i^{max} - \Delta N'^{max}_i\right|$ and $\left|f_i^{(2)}\right|$ values. C) Comparison of $\left|\Delta N_i^{max} - \Delta N'^{max}_i\right|$ and $\left|\gamma_i\right|$ values. In all cases for the occupied CMOs of the $CH_2CHOCH_3$ molecule.

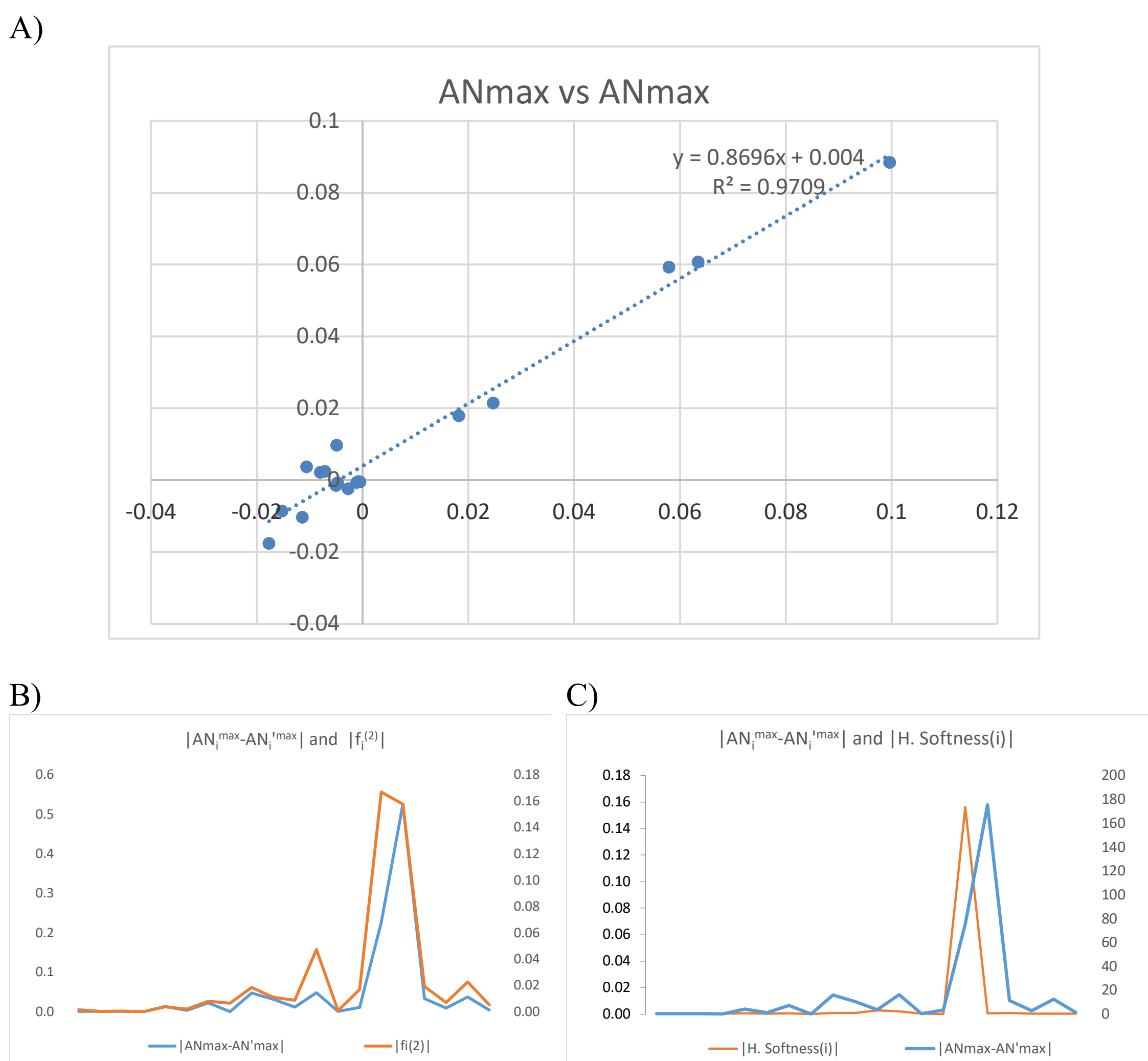

Figure S45. Linear regression of the maximum local charge variations (obtained with a cubic expansion of the local energy) versus the maximum local charge variations (achieved with a quadratic expansion of the local charge variation). Two outliers corresponding to CMOs 15 and 16 have been removed. B) Comparison of $\left|\Delta N_i^{\max} - \Delta N'^{\max}_i\right|$ and $\left|f_i^{(2)}\right|$ values. C) Comparison of $\left|\Delta N_i^{\max} - \Delta N'^{\max}_i\right|$ and $\left|\gamma_i\right|$ values. In all cases for the occupied CMOs of the $CH_2CHCHO$ molecule.

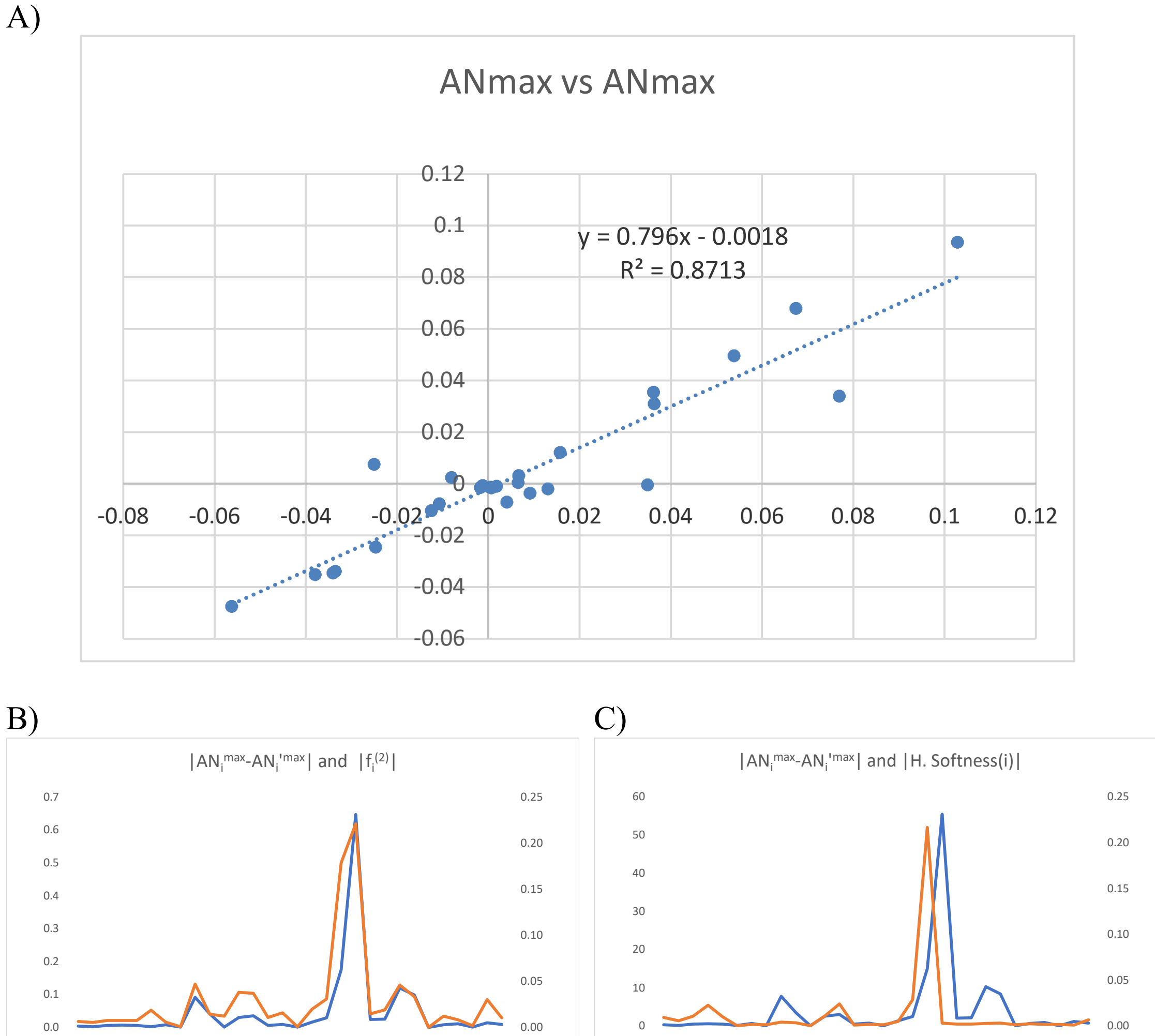

Figure S46. Linear regression of the maximum local charge variations (obtained with a cubic expansion of the local energy) versus the maximum local charge variations (achieved with a quadratic expansion of the local charge variation). Two outliers corresponding to CMOs 19 and 20 have been removed. B) Comparison of $\left|\Delta N_i^{\max} - \Delta N'^{\max}_i\right|$ and $\left|f_i^{(2)}\right|$ values. C) Comparison of $\left|\Delta N_i^{\max} - \Delta N'^{\max}_i\right|$ and $\left|\gamma_i\right|$ values. In all cases for the occupied CMOs of the $CH_2CHNO_2$ molecule.

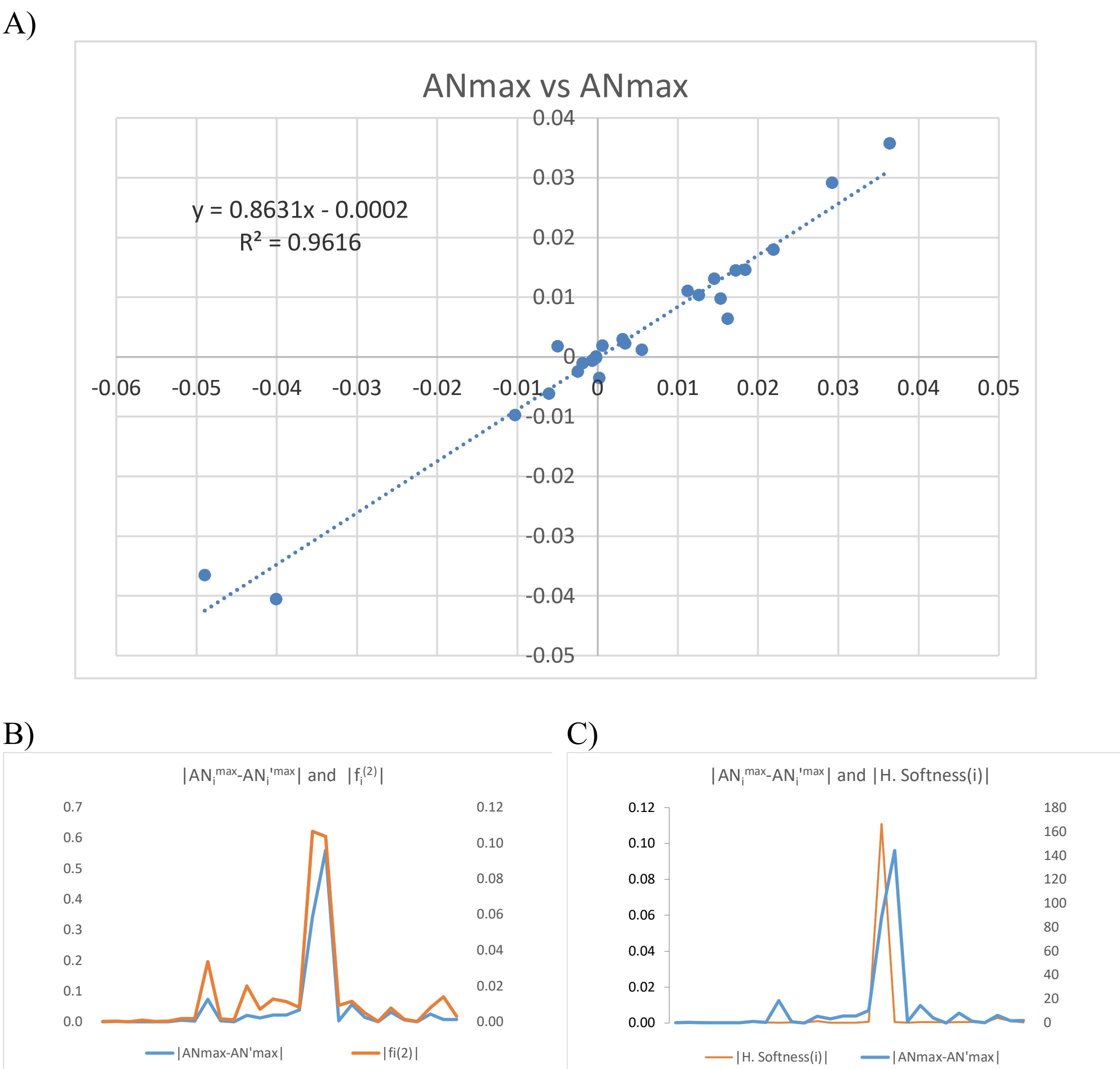

Figure S47. Linear regression of the maximum local charge variations (obtained with a cubic expansion of the local energy) versus the maximum local charge variations (achieved with a quadratic expansion of the local charge variation). Two outliers corresponding to CMOs 17 and 18 have been removed. B) Comparison of $\left|\Delta N_i^{\max} - \Delta N'^{\max}_i\right|$ and $\left|f_i^{(2)}\right|$ values. C) Comparison of $\left|\Delta N_i^{\max} - \Delta N'^{\max}_i\right|$ and $\left|\gamma_i\right|$ values. In all cases for the occupied CMOs of the $CH_3CH_2SH$ molecule.

A)

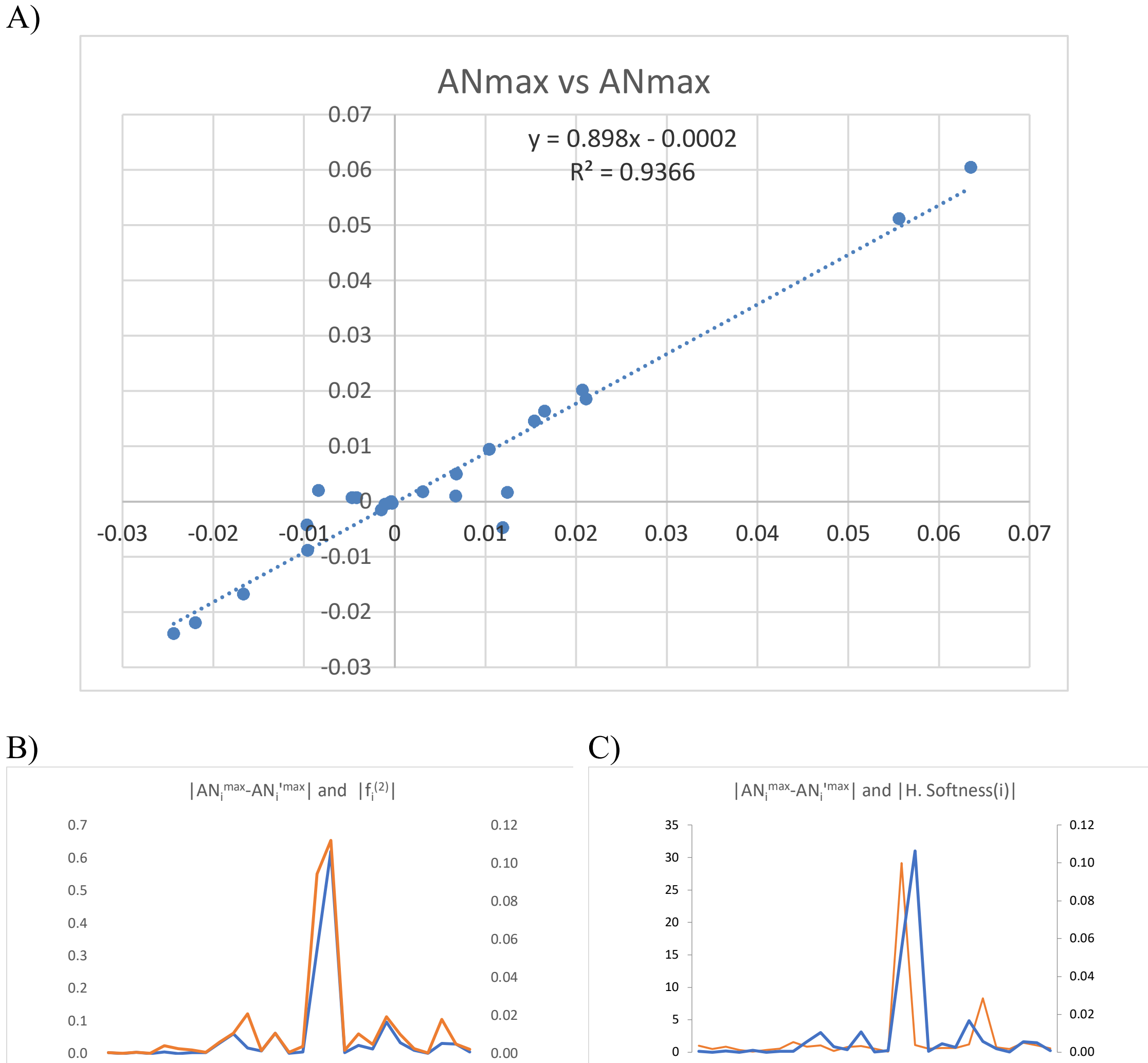

Figure S48. Linear regression of the maximum local charge variations (obtained with a cubic expansion of the local energy) versus the maximum local charge variations (achieved with a quadratic expansion of the local charge variation). Two outliers corresponding to CMOs 16 and 17 have been removed. B) Comparison of $\left|\Delta N_i^{\max} - \Delta N'^{\max}_i\right|$ and $\left|f_i^{(2)}\right|$ values. C) Comparison of $\left|\Delta N_i^{\max} - \Delta N'^{\max}_i\right|$ and $\left|\gamma_i\right|$ values. In all cases for the occupied CMOs of the $CH_3COOH$ molecule.

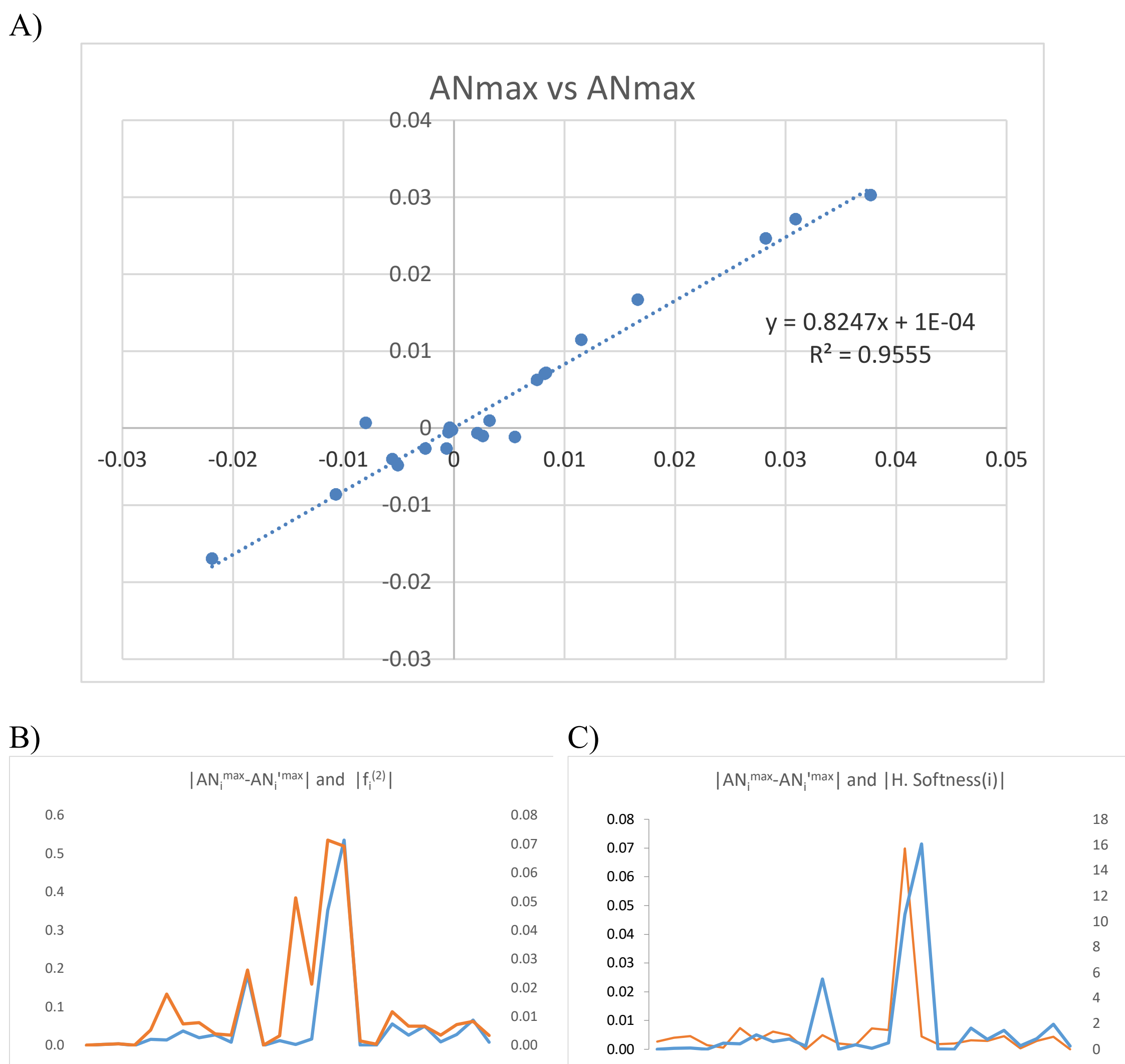

Figure S49. Linear regression of the maximum local charge variations (obtained with a cubic expansion of the local energy) versus the maximum local charge variations (achieved with a quadratic expansion of the local charge variation). Three outliers corresponding to CMOs 11, 16 and 17 have been removed. B) Comparison of $\left|\Delta N_i^{\max} - \Delta N'^{\max}_i\right|$ and $\left|f_i^{(2)}\right|$ values. C) Comparison of $\left|\Delta N_i^{\max} - \Delta N'^{\max}_i\right|$ and $\left|\gamma_i\right|$ values. In all cases for the occupied CMOs of the $CH_3CONH_2$ molecule.

## 7. Local electrophilicities against local Fukui indices:

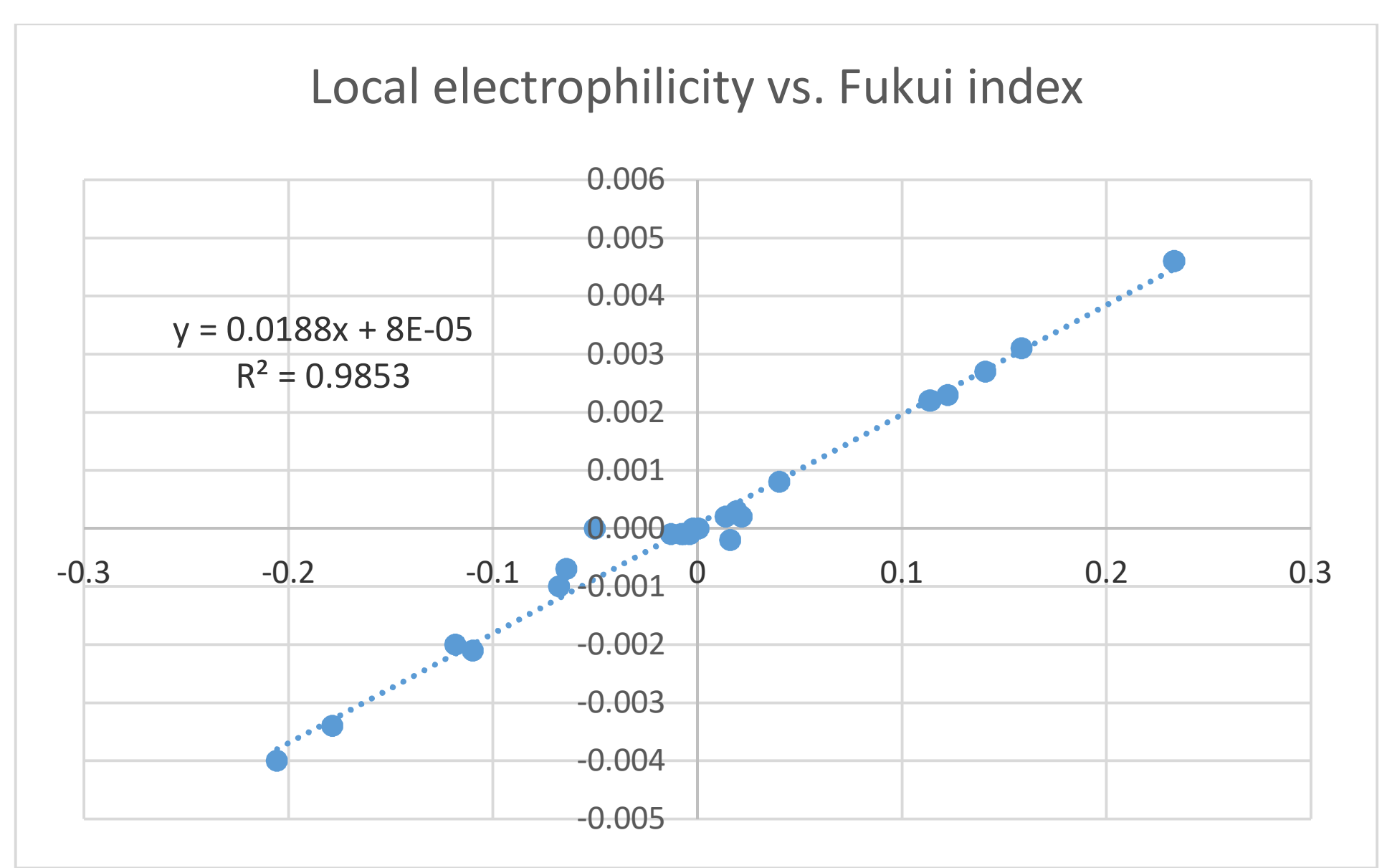

Figure S50. Linear regression of local electrophilicities against local Fukui indices for the CMOs of the $CH_2CHCl$ molecule. Two outliers corresponding to CMOs 16 and 17 have been removed.

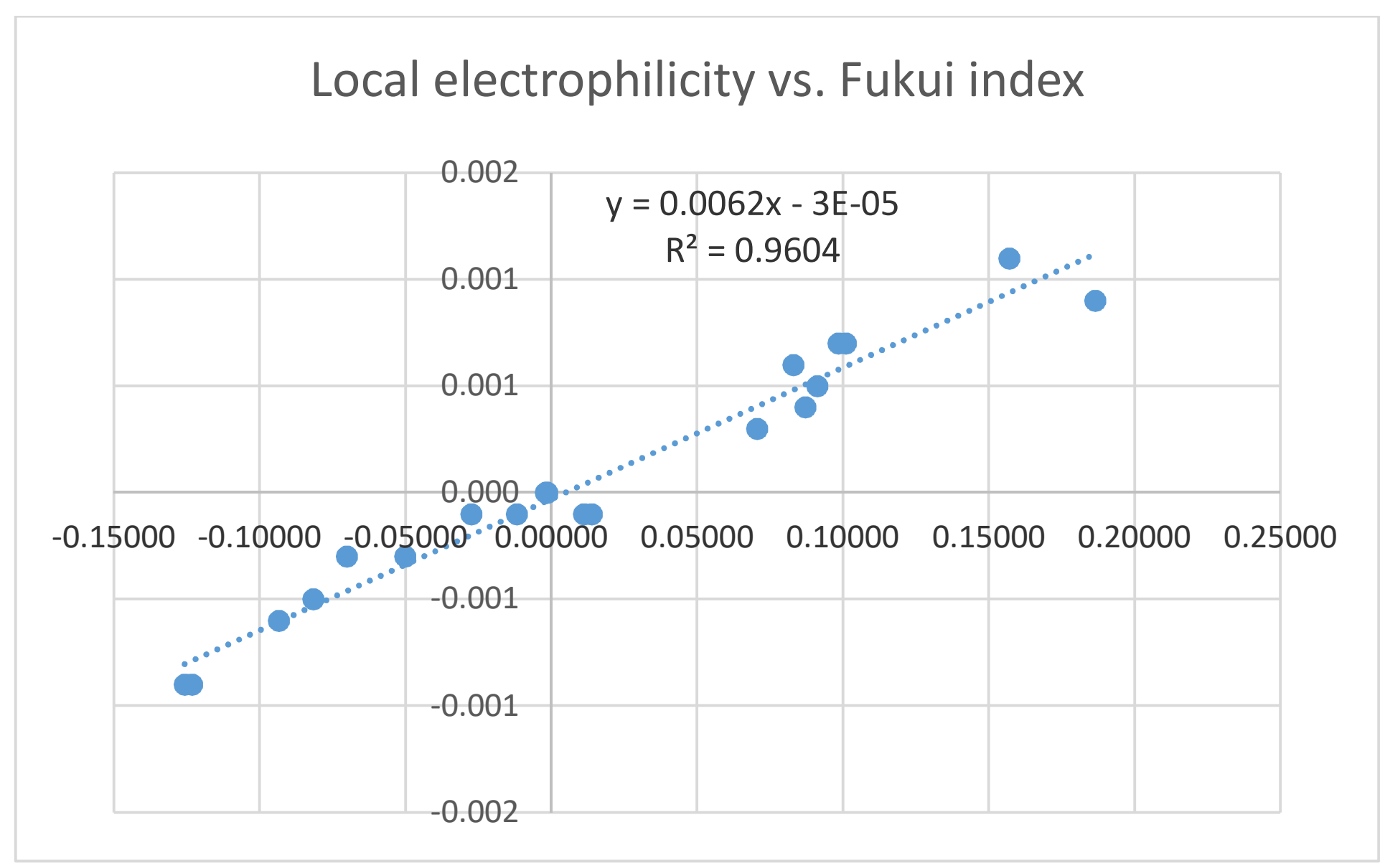

Figure S51. Linear regression of local electrophilicities against local Fukui indices for the CMOs of the $CH_2CHNH_2$ molecule. Two outliers corresponding to CMOs 12 and 13 have been removed.

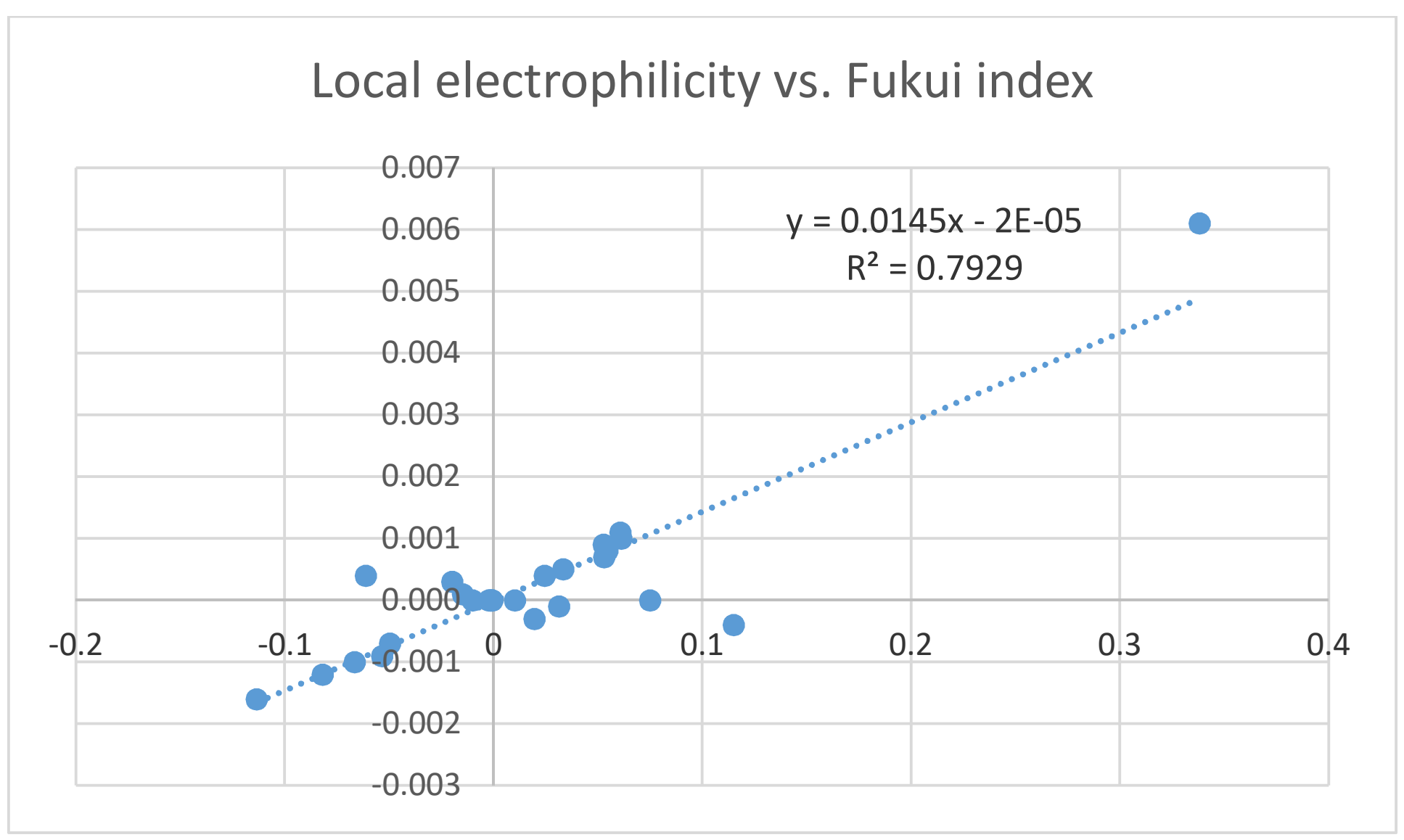

Figure S52. Linear regression of local electrophilicities against local Fukui indices for the CMOs of the $CH_3COOCH_3$ molecule. Four outliers corresponding to CMOs 20, 21, 26 and 28 have been removed.

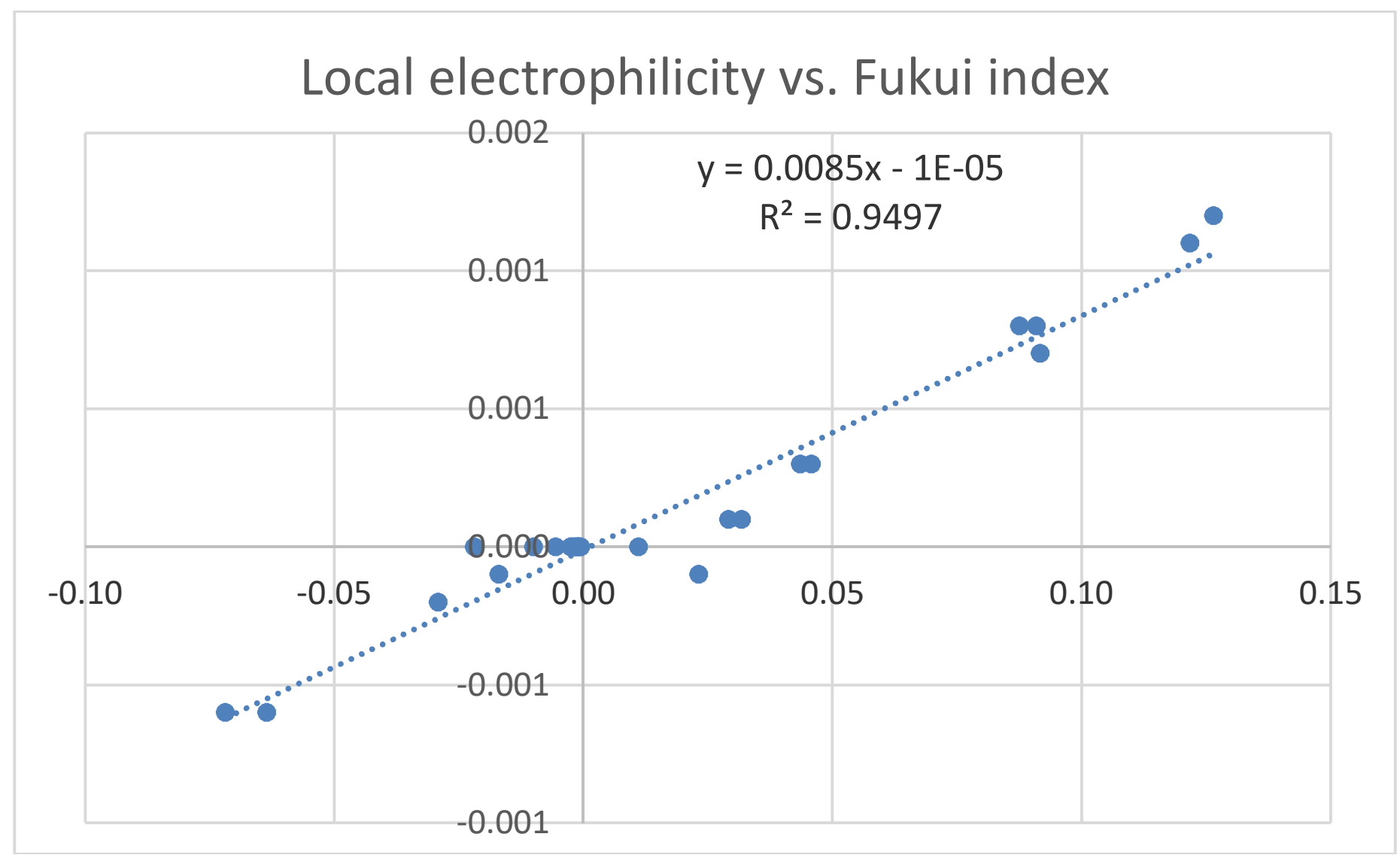

Figure S53. Linear regression of local electrophilicities against local Fukui indices for the CMOs of the $CH_2CHOCH_3$ molecule. Two outliers corresponding to CMOs 16 and 17 have been removed.

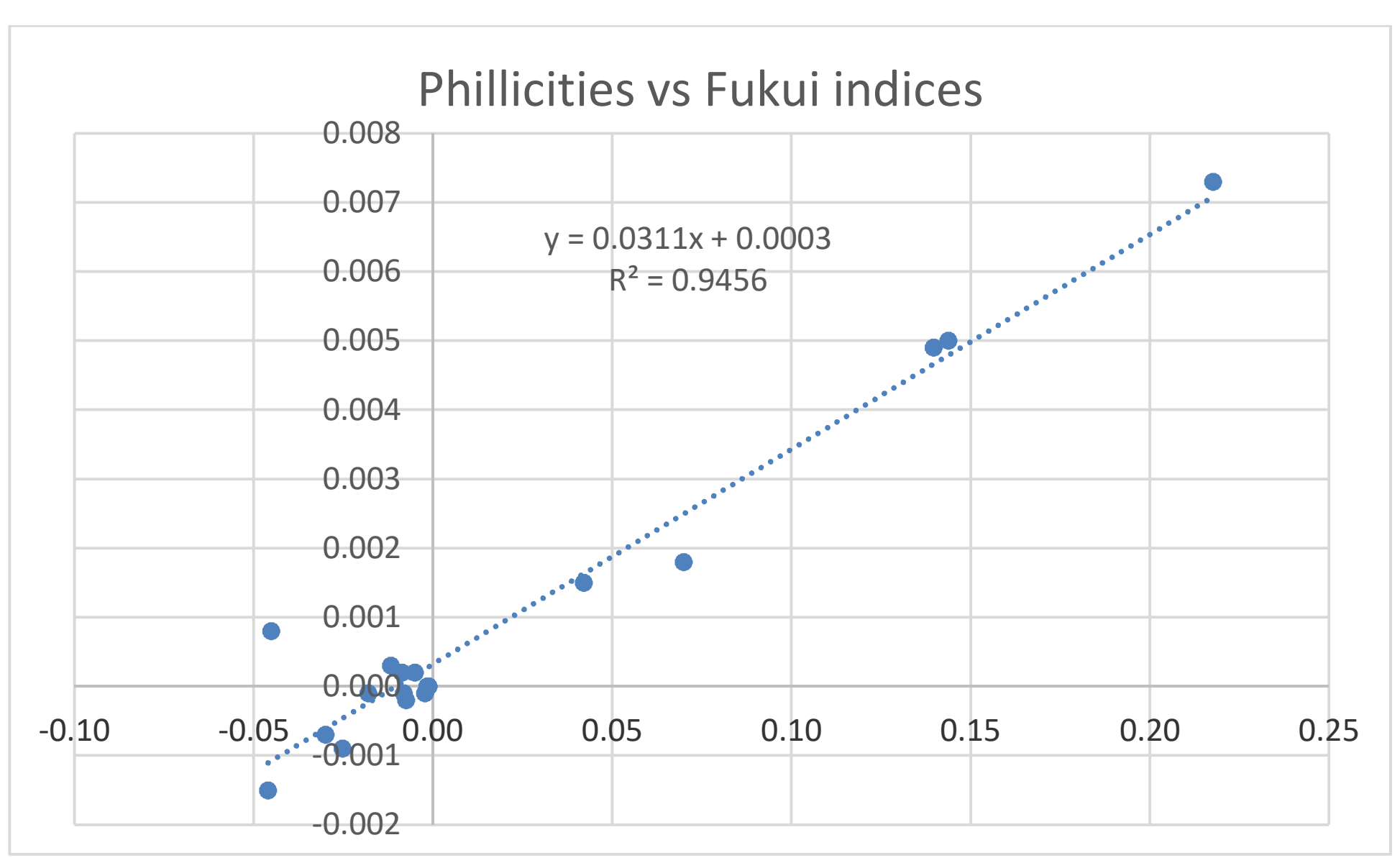

Figure S54. Linear regression of local electrophilicities against local Fukui indices for the CMOs of the $CH_2CHCHO$ molecule. Two outliers corresponding to CMOs 15 and 16 have been removed.

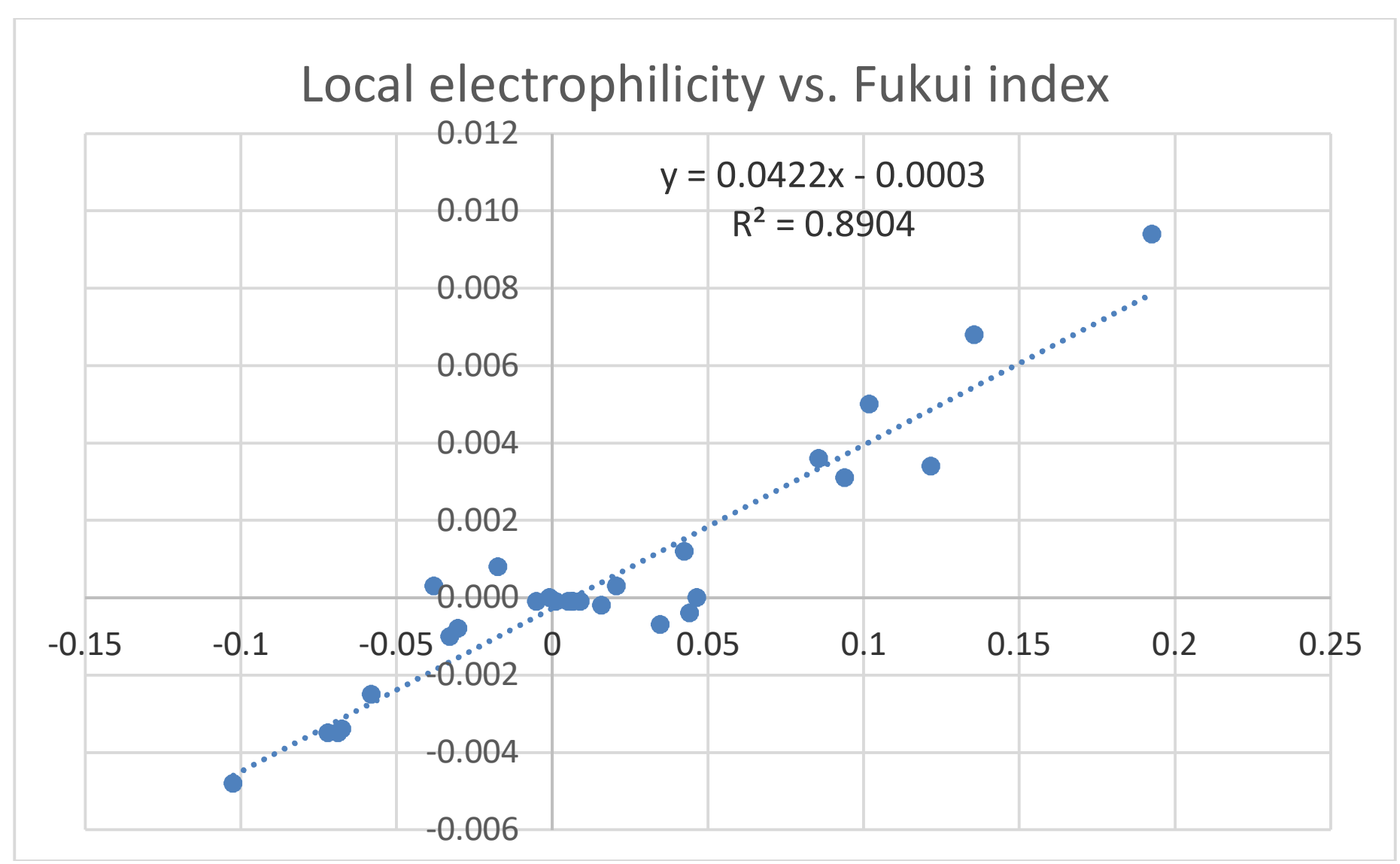

Figure S55. Linear regression of local electrophilicities against local Fukui indices for the CMOs of the $CH_2CHNO_2$ molecule. Two outliers corresponding to CMOs 19 and 20 have been removed.

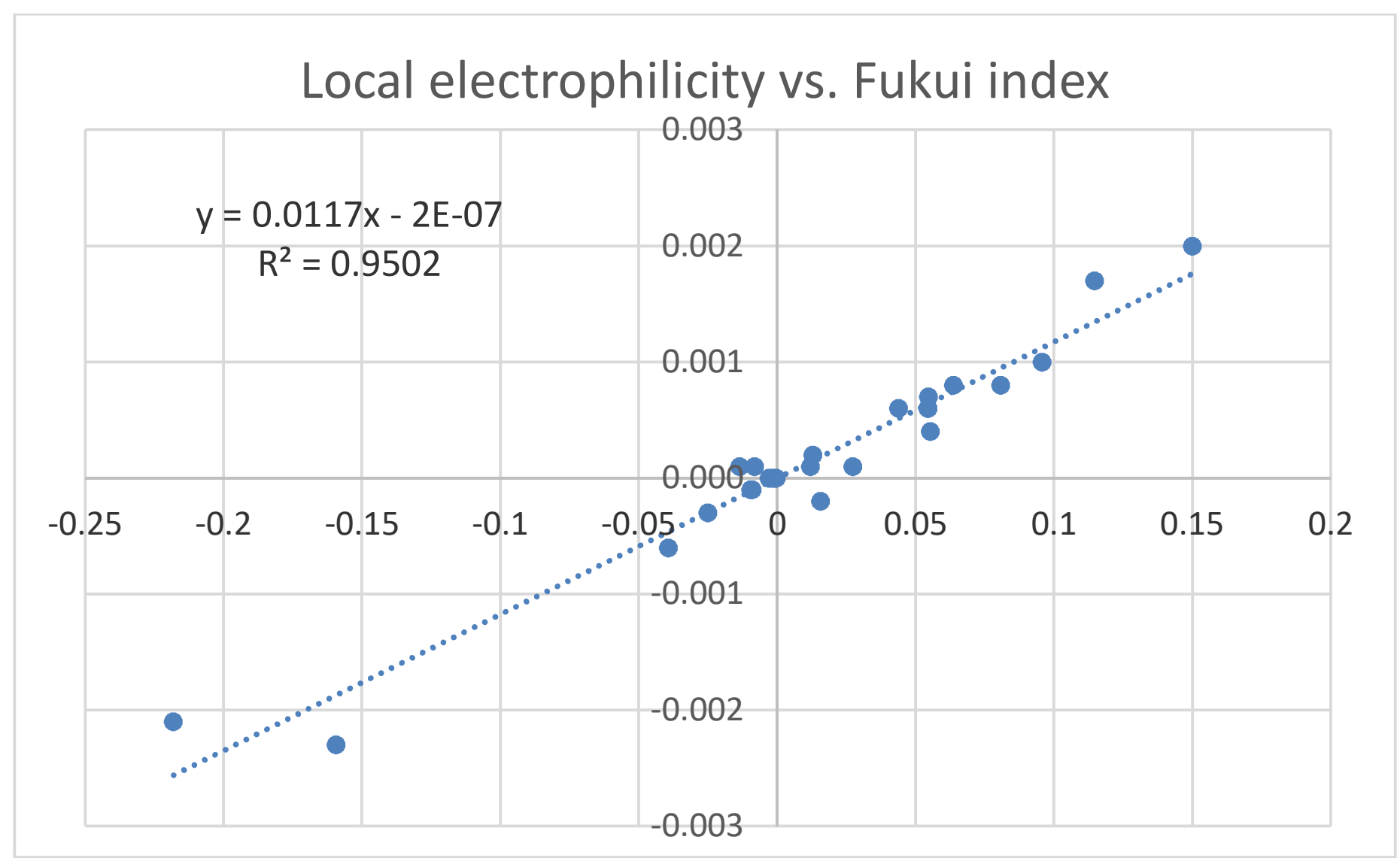

Figure S56. Linear regression of local electrophilicities against local Fukui indices for the CMOs of the $CH_3CHSH$ molecule. Two outliers corresponding to CMOs 17 and 18 have been removed.

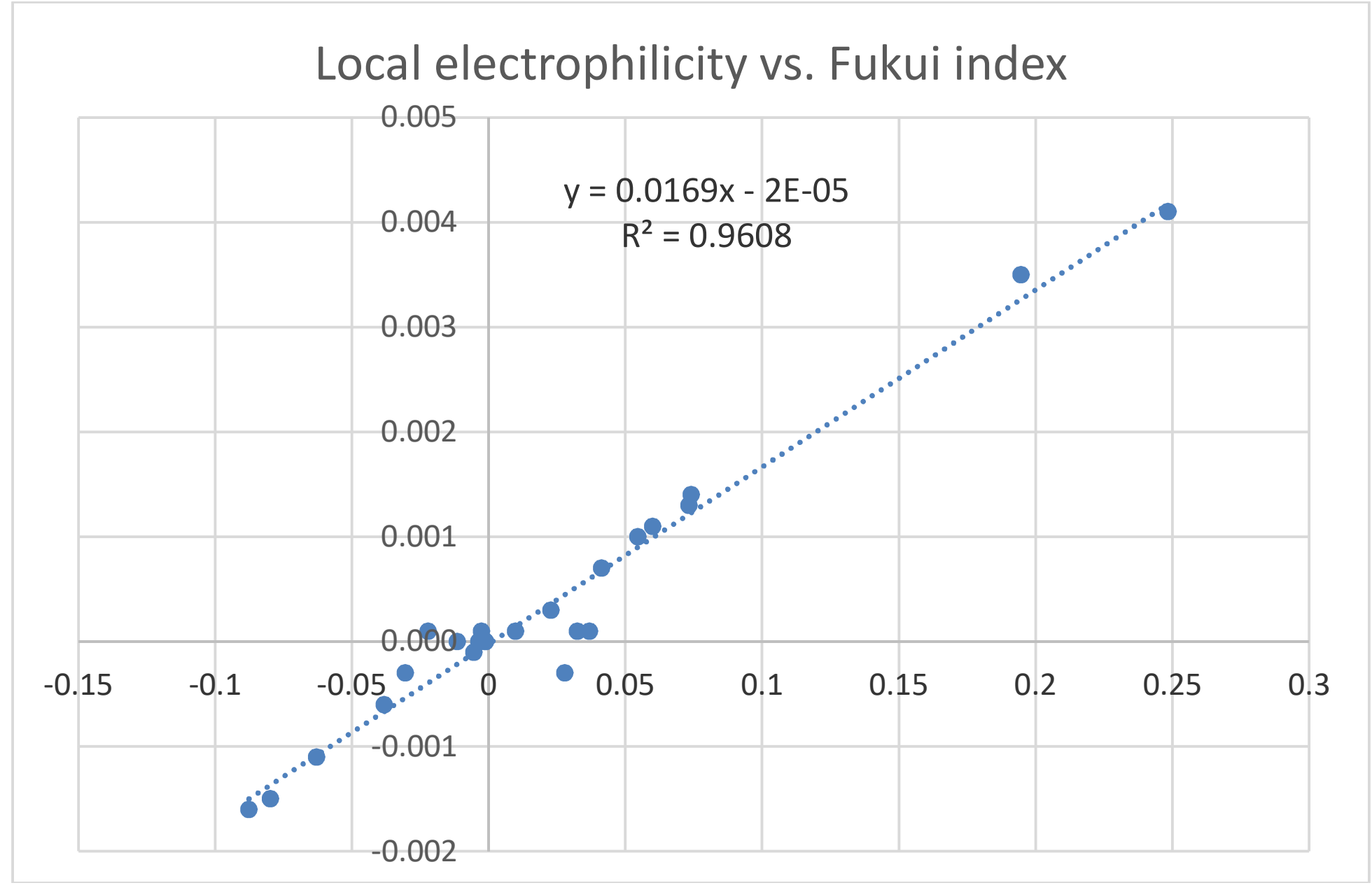

Figure S57. Linear regression of local electrophilicities against local Fukui indices for the CMOs of the $CH_3COOH$ molecule. Two outliers corresponding to NBOs 16 and 17 have been removed.

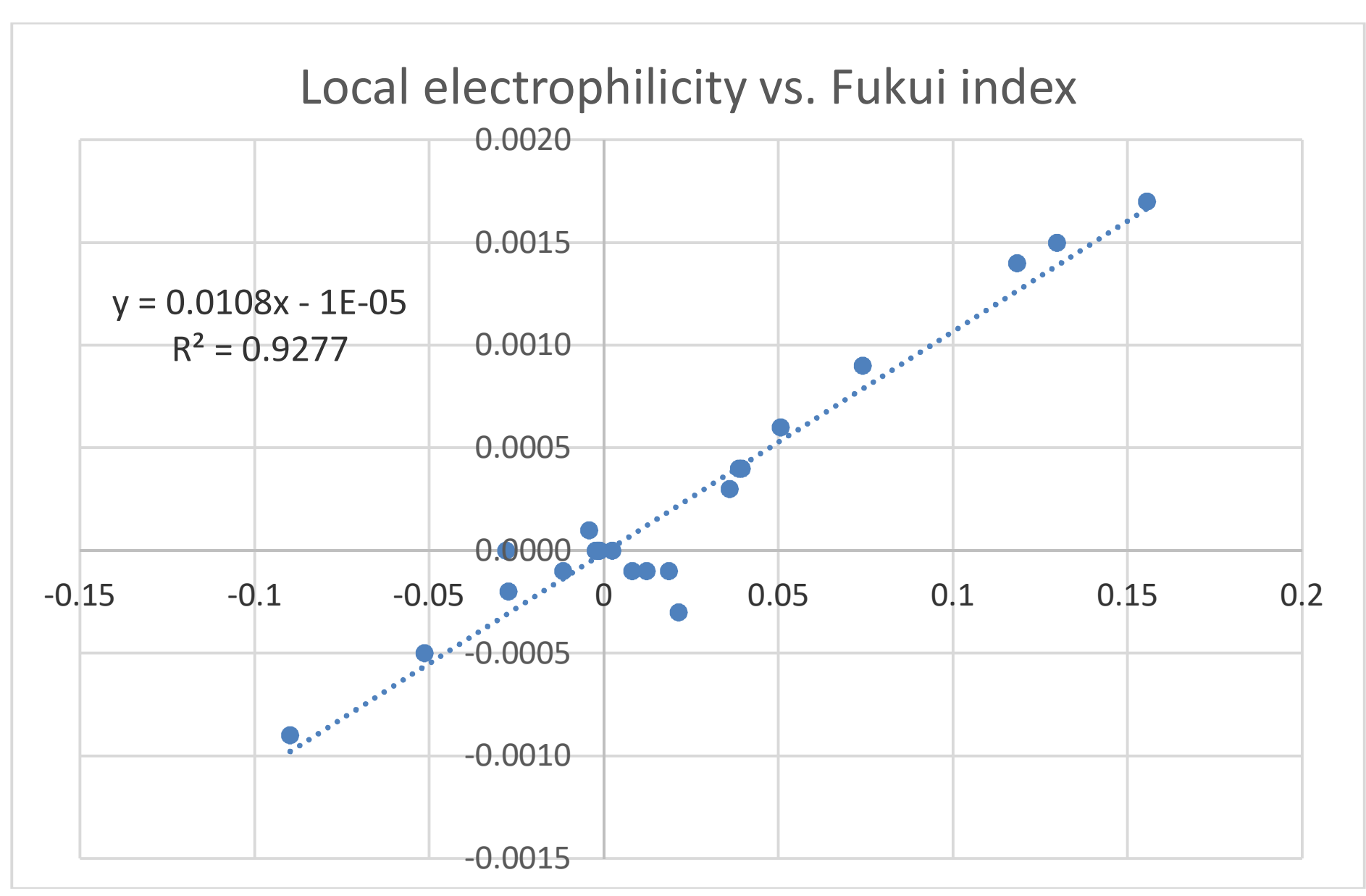

Figure S58. Linear regression of local electrophilicities against local Fukui indices for the CMOs of the $CH_3CONH_2$ molecule. Three outliers corresponding to NBOs 11, 16 and 17 have been removed.

## 8. Local $\eta_i^{-1}$ values versus the local Fukui indices:

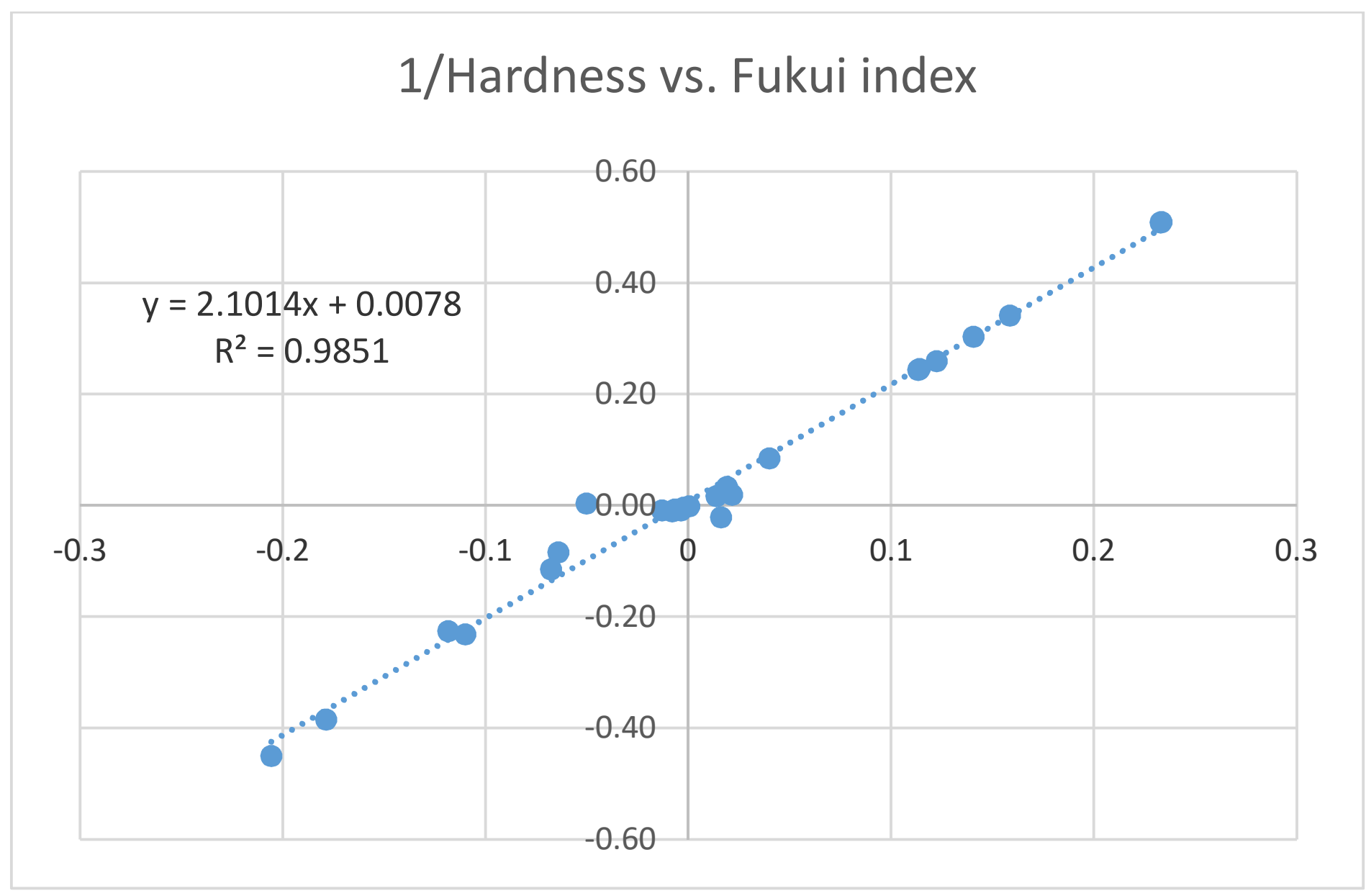

Figure S59. Lineal regression for local $\eta_i^{-1}$ values versus the local Fukui indices for the CMOs of the $CH_2CHCl$ molecule. Two outliers corresponding to NBOs 16 and 17 have been removed.

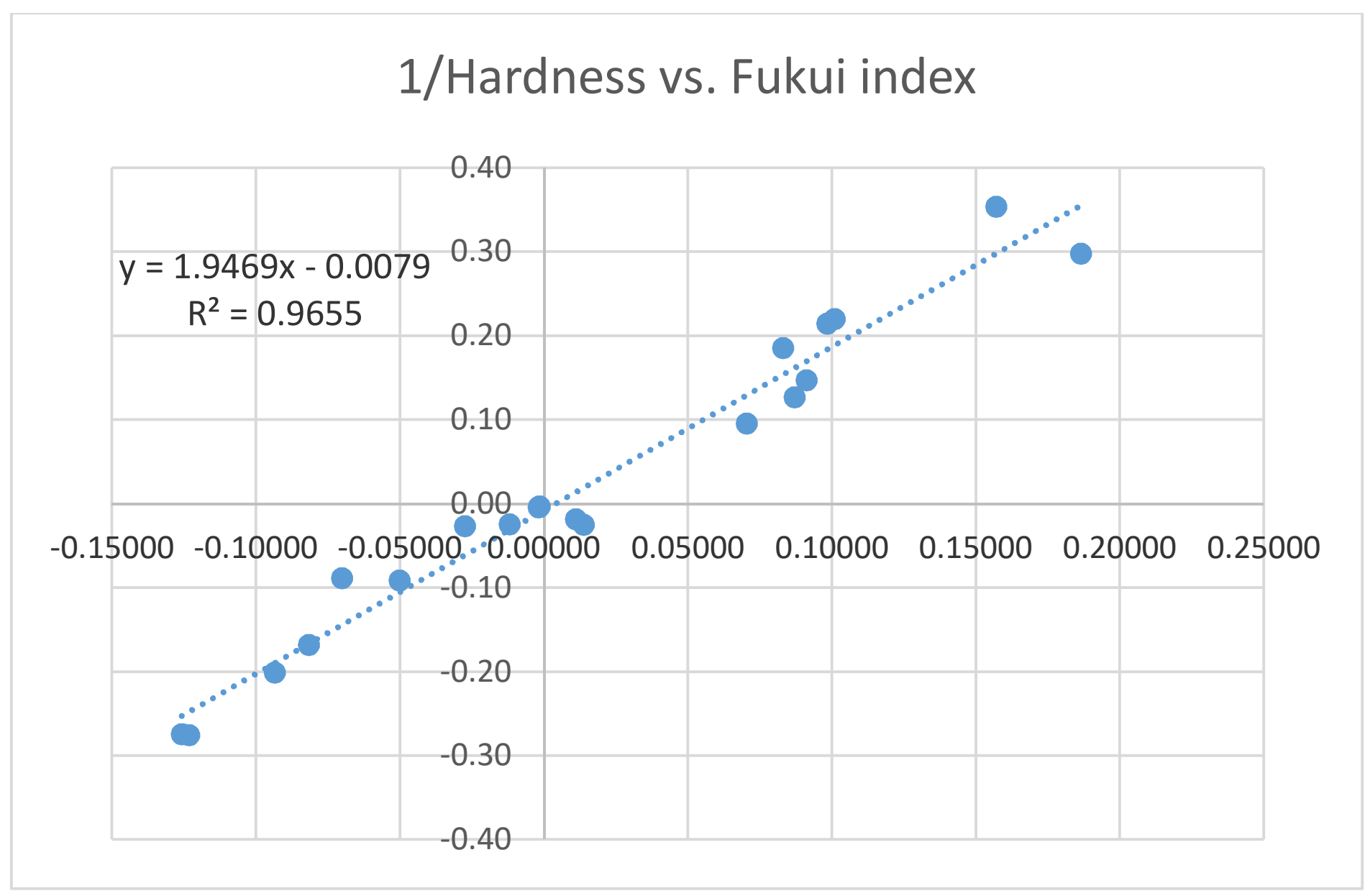

Figure S60. Lineal regression for local $\eta_i^{-1}$ values versus the local Fukui indices for the CMOs of the $CH_2CHNH_2$ molecule. Two outliers corresponding to NBOs 12 and 13 have been removed.

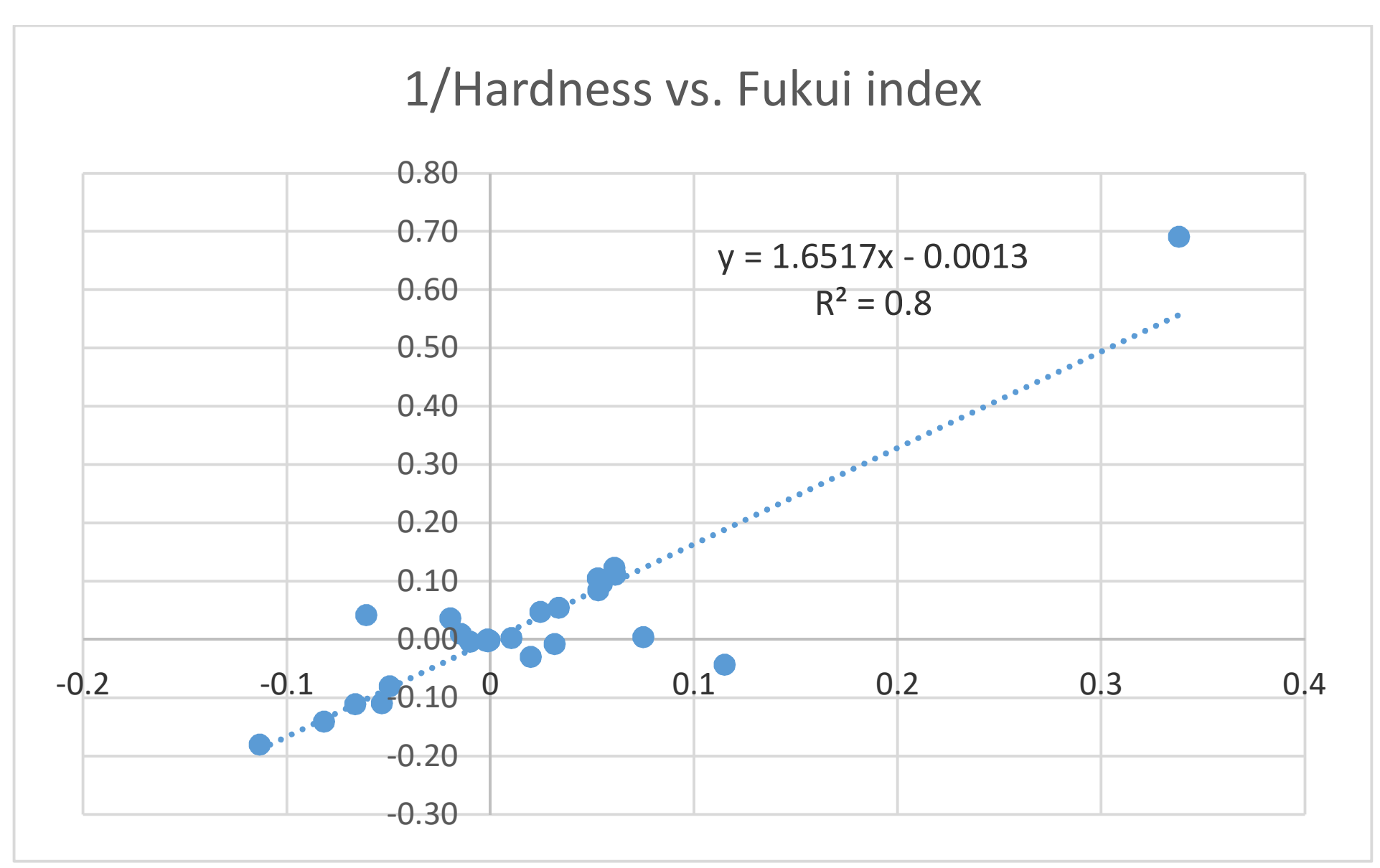

Figure S61. Lineal regression for local $\eta_i^{-1}$ values versus the local Fukui indices for the CMOs of the $CH_3COOCH_3$molecule. Four outliers corresponding to CMOs 20, 21, 26 and 28 have been removed.

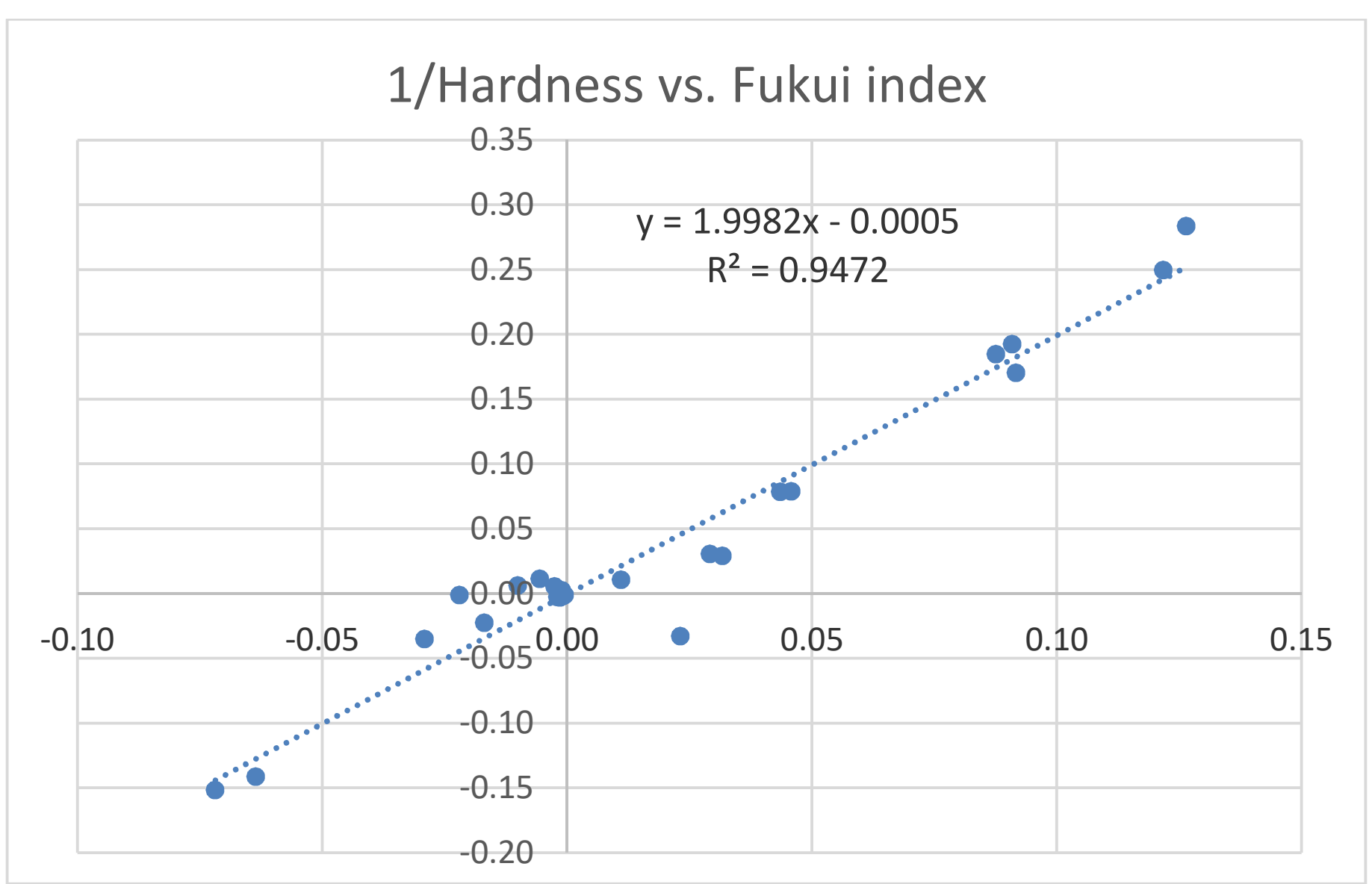

Figure S62. Lineal regression for local $\eta_i^{-1}$ values versus the local Fukui indices for the CMOs of the $CH_2CHOCH_3$ molecule. Two outliers corresponding to CMOs 16 and 17 have been removed.

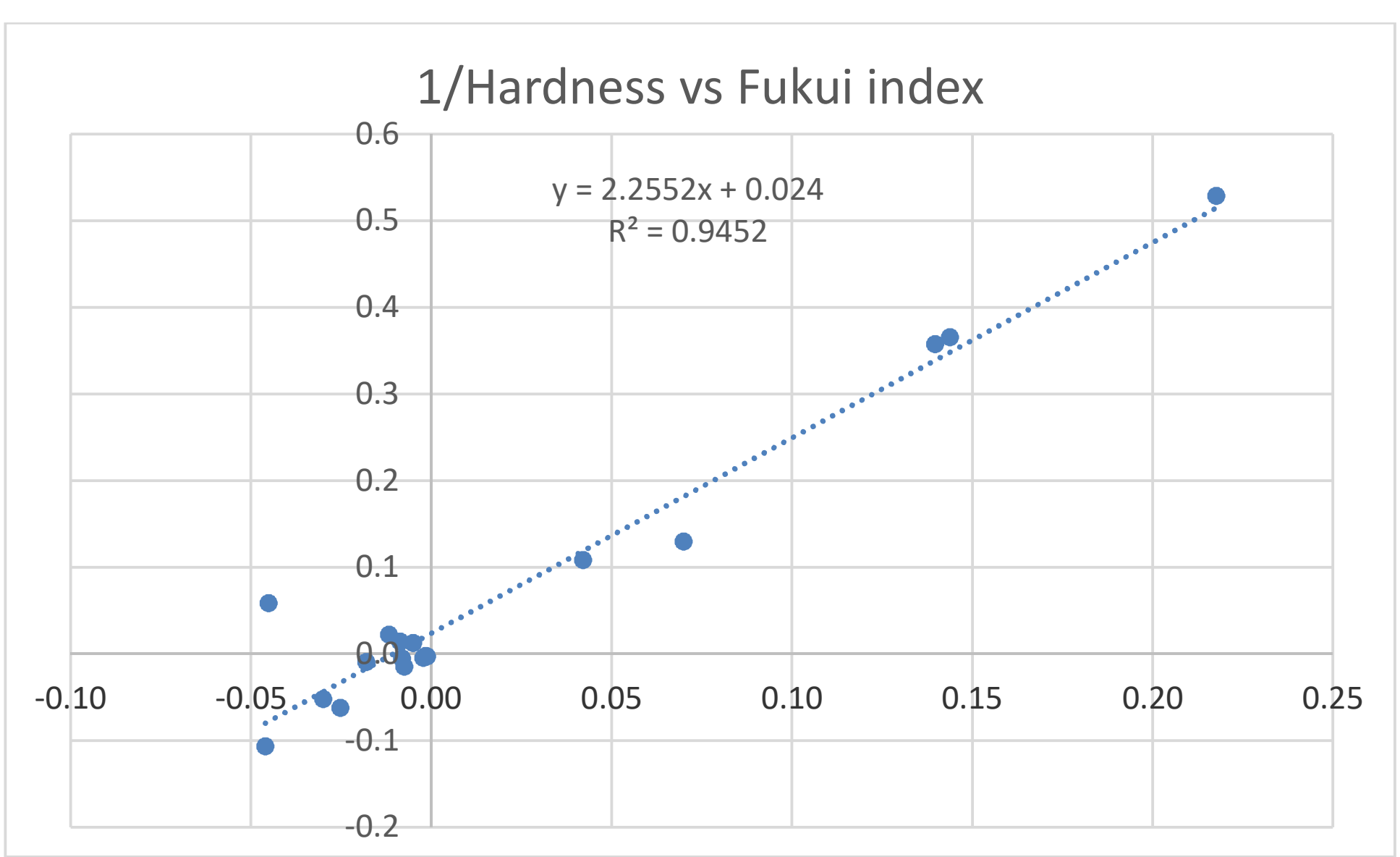

Figure S63. Lineal regression for local $\eta_i^{-1}$ values versus the local Fukui indices for the CMOs of the $CH_2CHCHO$ molecule. Two outliers corresponding to CMOs 15 and 16 have been removed.

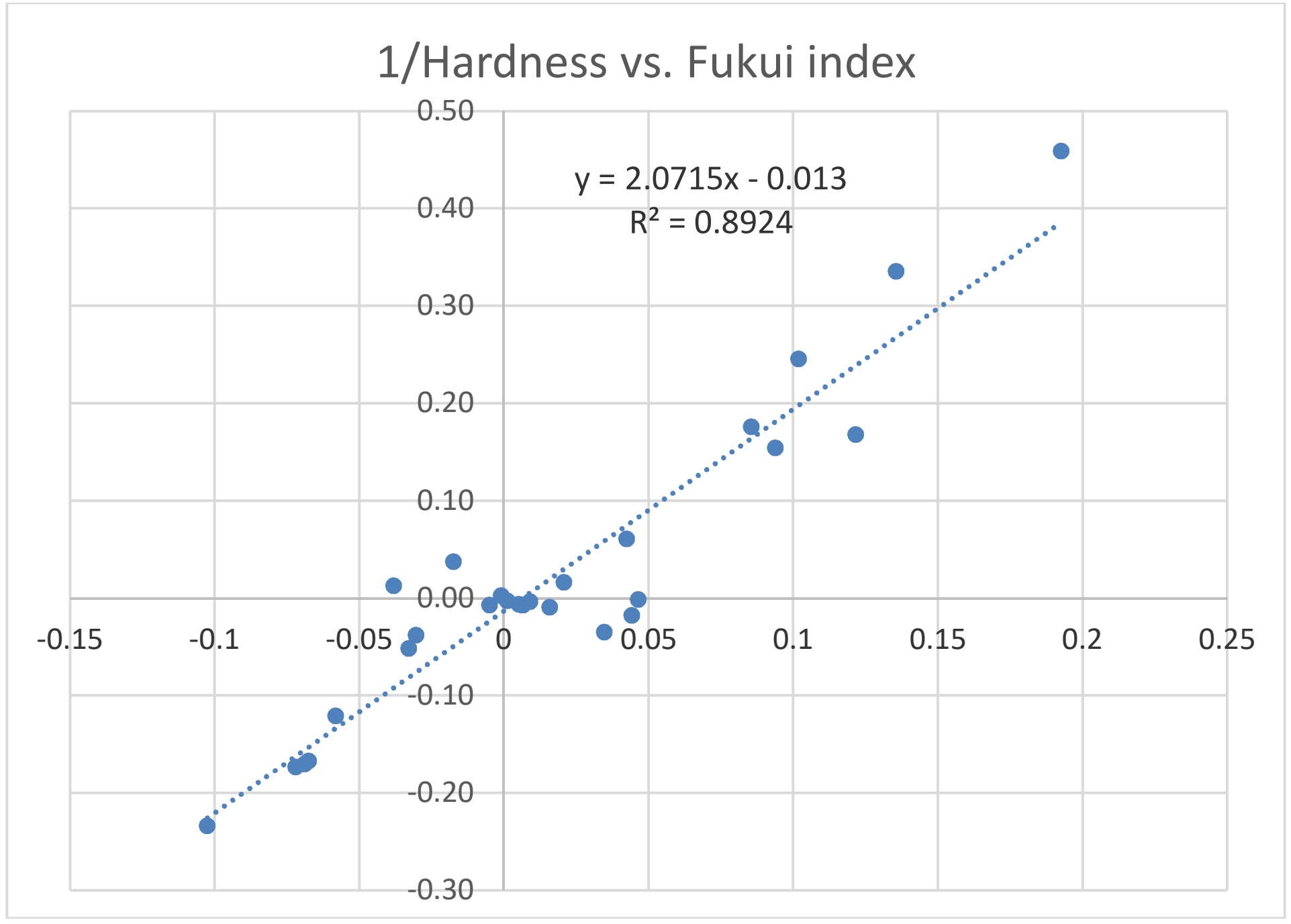

Figure S64. Lineal regression for local $\eta_i^{-1}$ values versus the local Fukui indices for the CMOs of the $CH_2CHNO_2$ molecule. Two outliers corresponding to CMOs 19 and 20 have been removed.

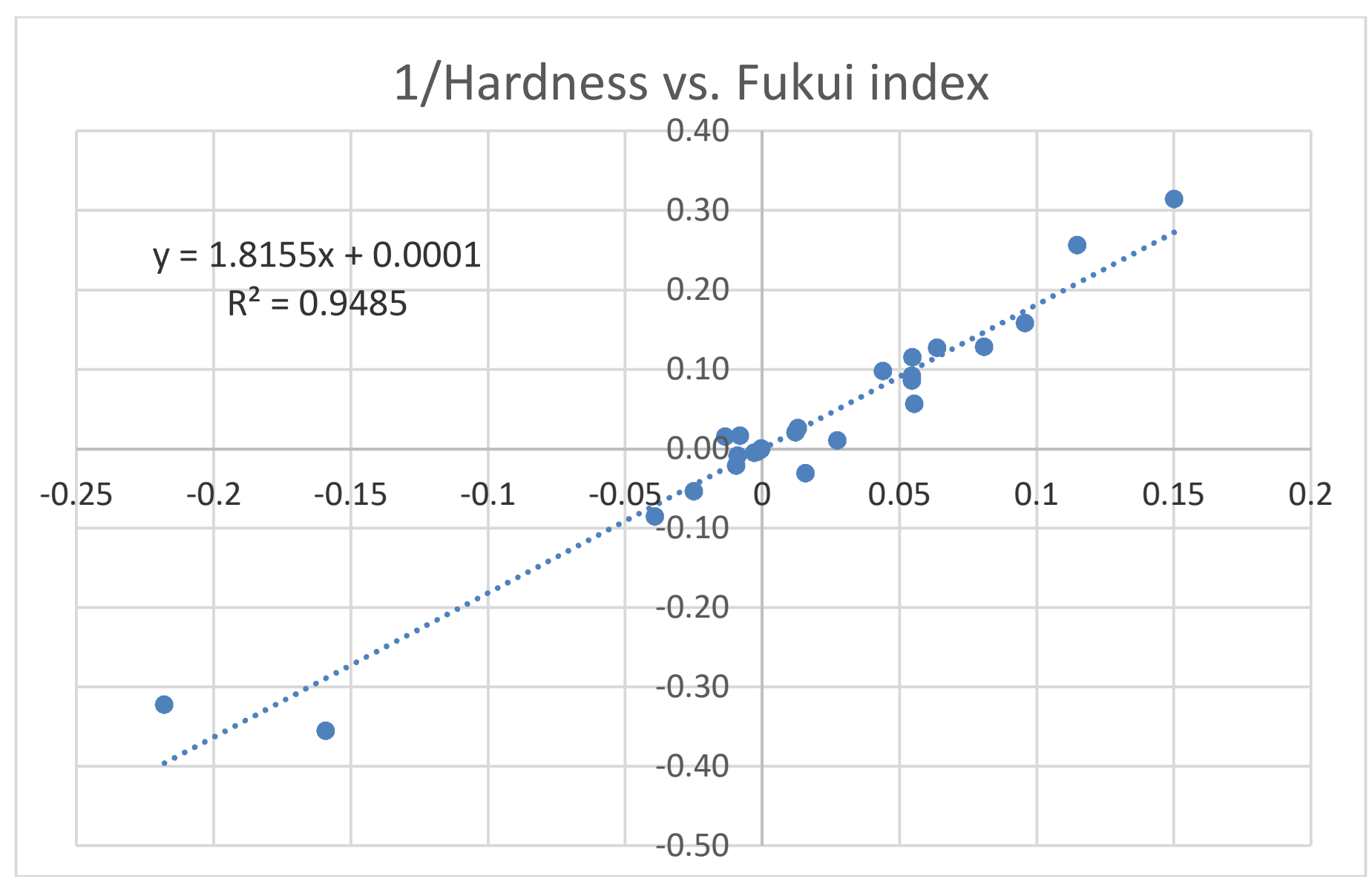

Figure S65. Lineal regression for local $\eta_i^{-1}$ values versus the local Fukui indices for the CMOs of the $CH_3CHSH$ molecule. Two outliers corresponding to CMOs 17 and 18 have been removed.

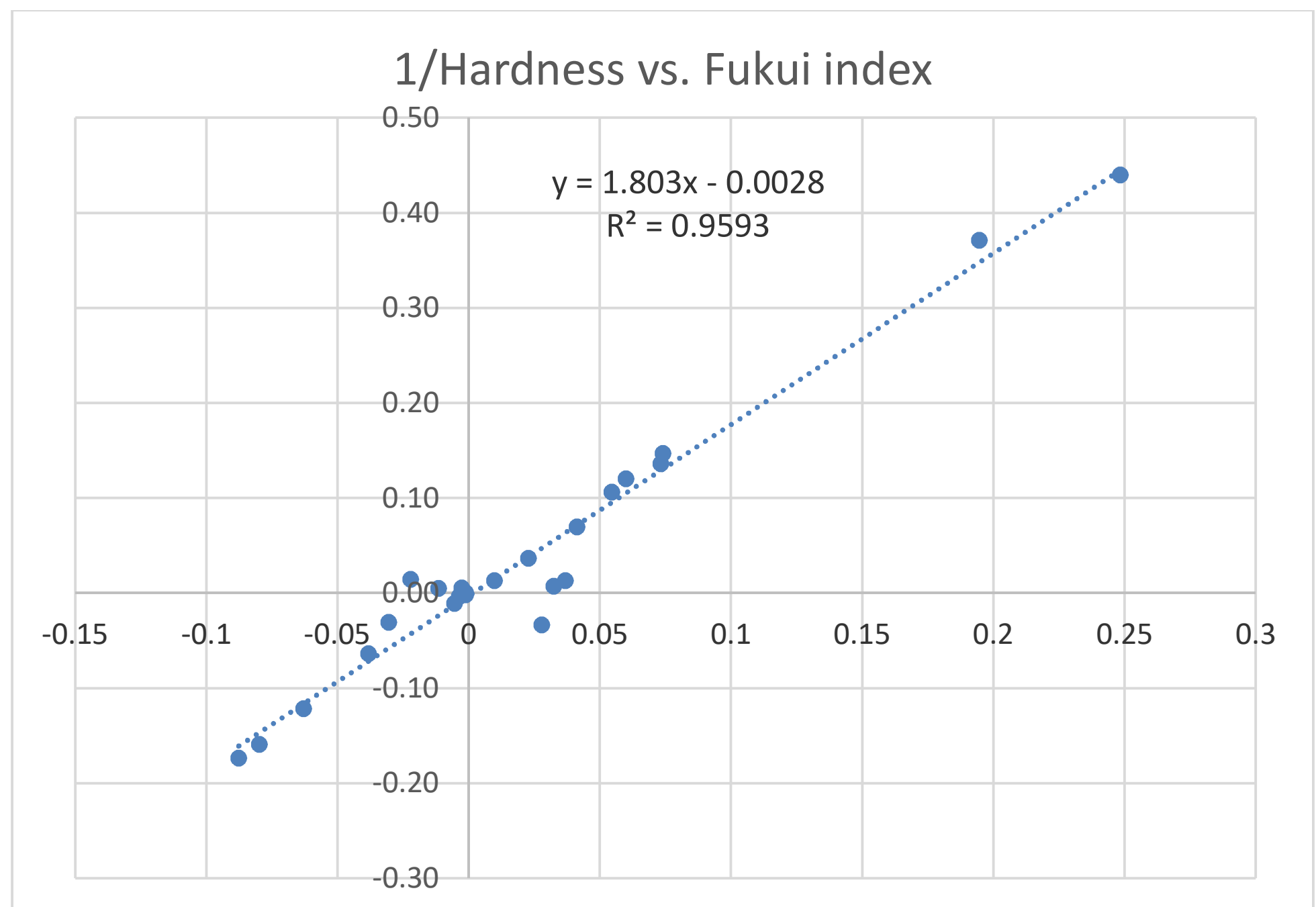

Figure S66. Lineal regression for local $\eta_i^{-1}$ values versus the local Fukui indices for the CMOs of the $CH_3COOH$ molecule. Two outliers corresponding to NBOs 16 and 17 have been removed.

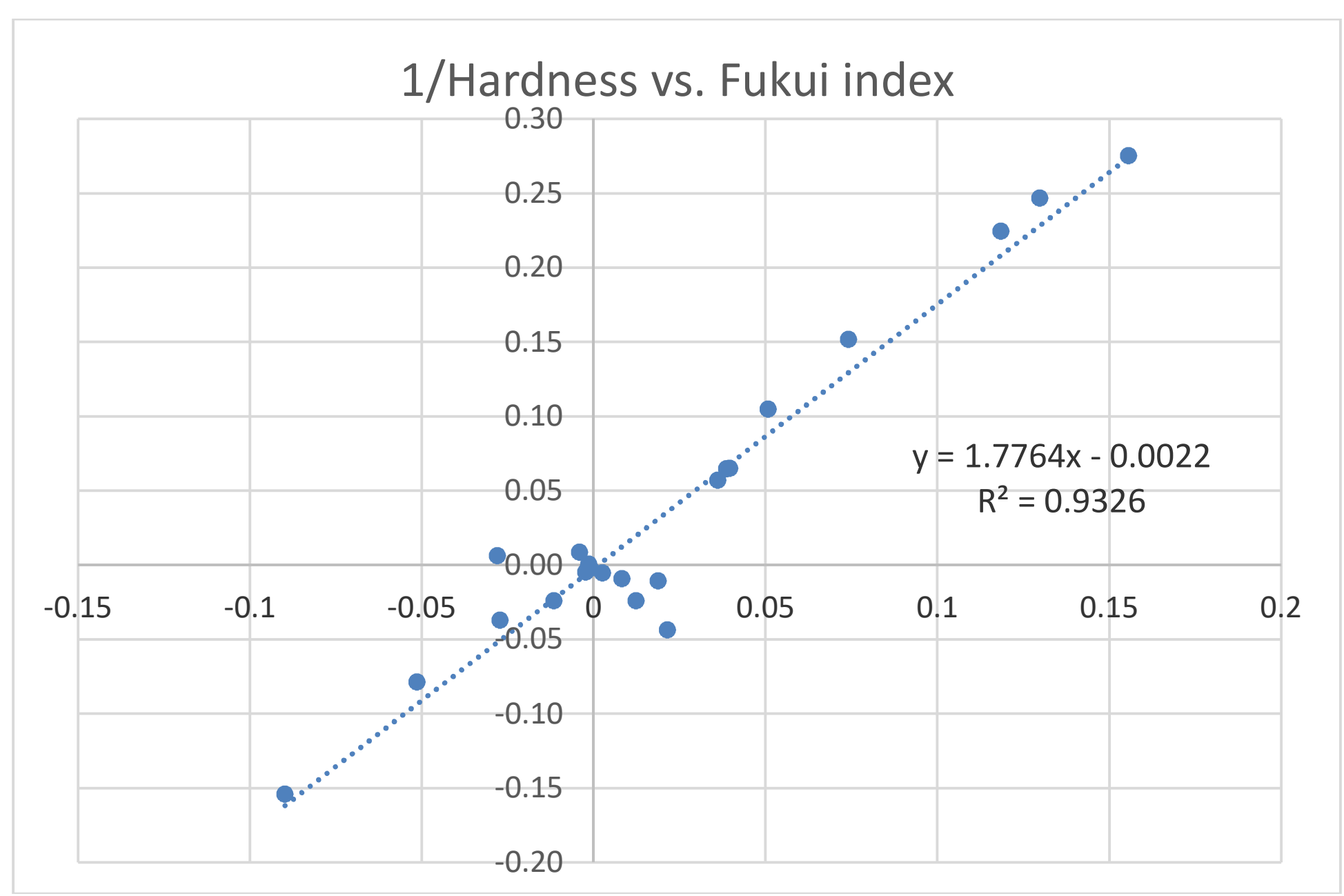

Figure S67. Lineal regression for local $\eta_i^{-1}$ values versus the local Fukui indices for the CMOs of the $CH_3CONH_2$ molecule Three outliers corresponding to NBOs 11, 16 and 17 have been removed.

## 9. Local $\eta_i^{-1}$ values versus the maximum local charge variations:

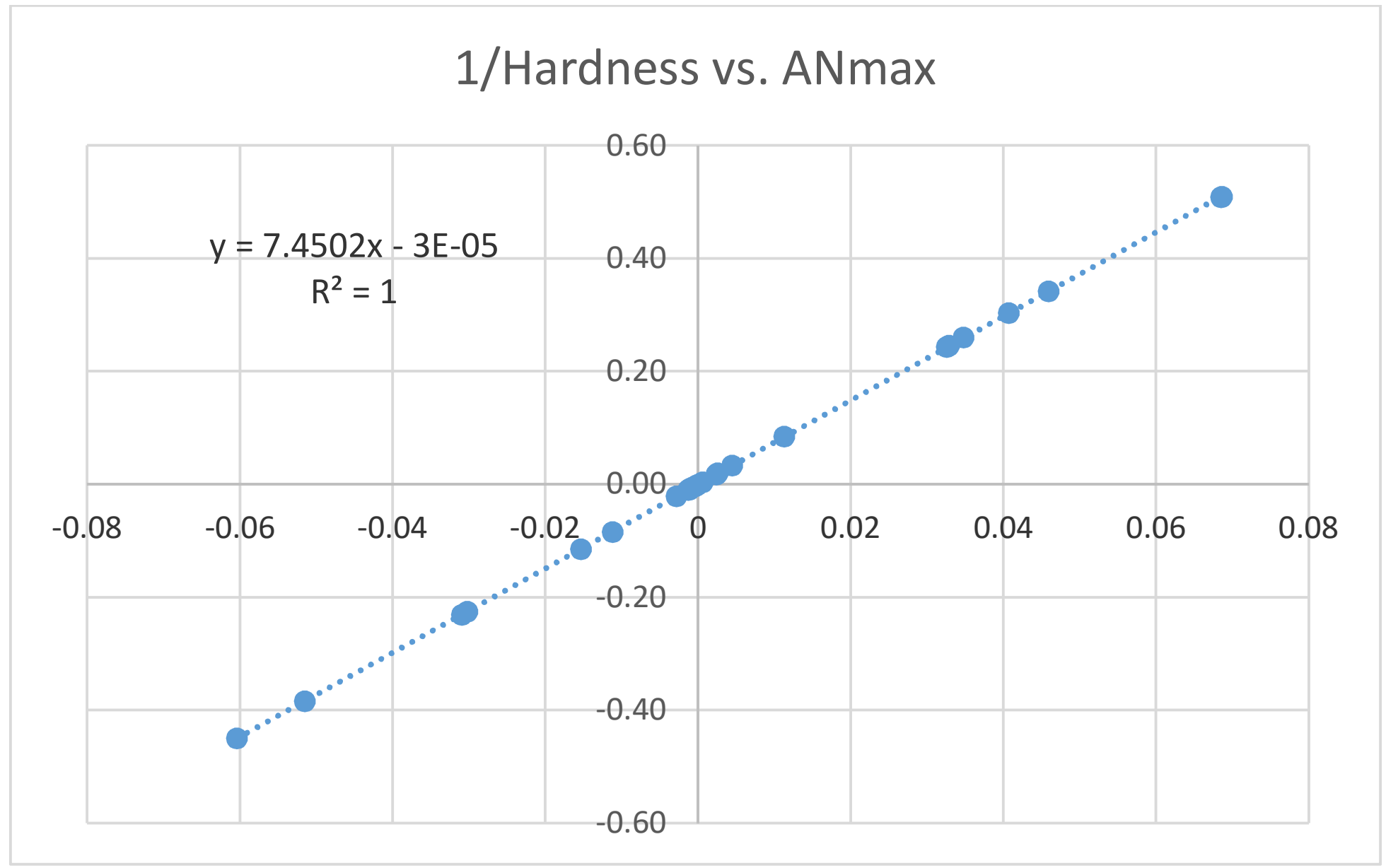

Figure S68. Lineal regression for local values $\eta_i^{-1}$ versus the maximum local charge variations for the CMOs of the $CH_2CHCl$ molecule. Two outliers corresponding to NBOs 16 and 17 have been removed.

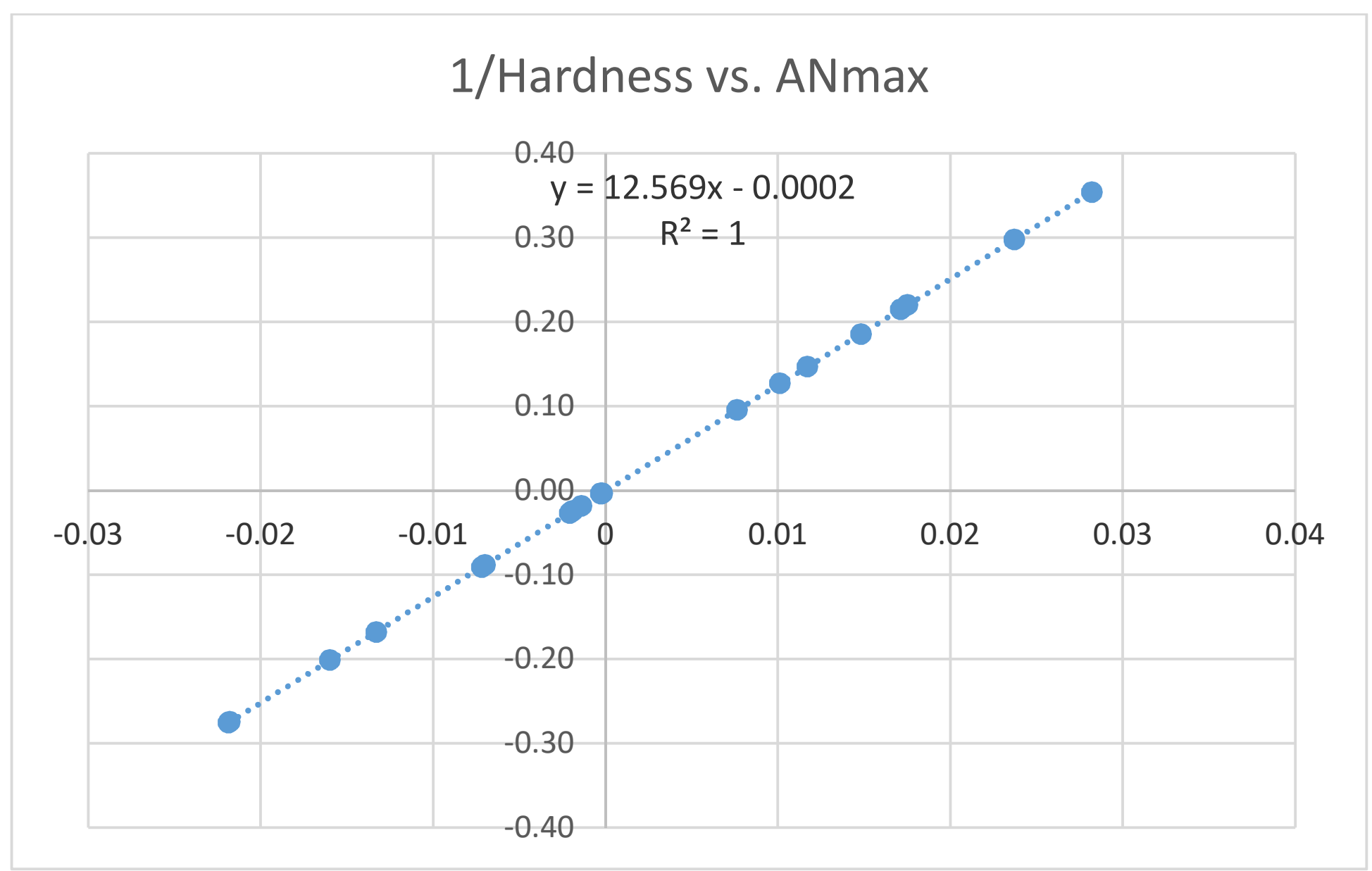

Figure S69. Lineal regression for local values $\eta_i^{-1}$ versus the maximum local charge variations for the CMOs of the $CH_2CHNH_2$ molecule. Two outliers corresponding to NBOs 12 and 13 have been removed.

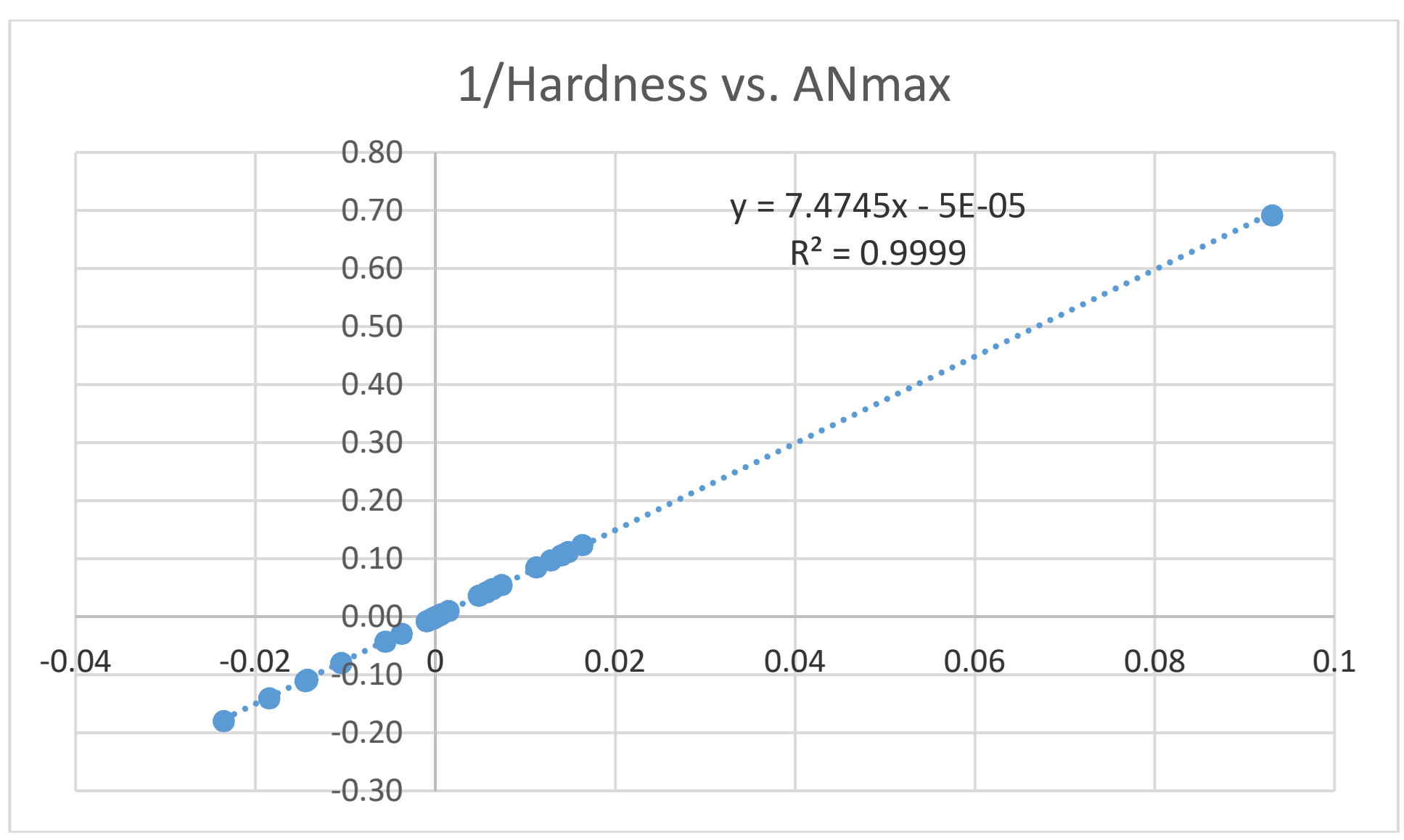

Figure S70. Lineal regression for local values $\eta_i^{-1}$ versus the maximum local charge variations for the CMOs of the $CH_3COOCH_3$ molecule. Four outliers corresponding to CMOs 20, 21, 26 and 28 have been removed.

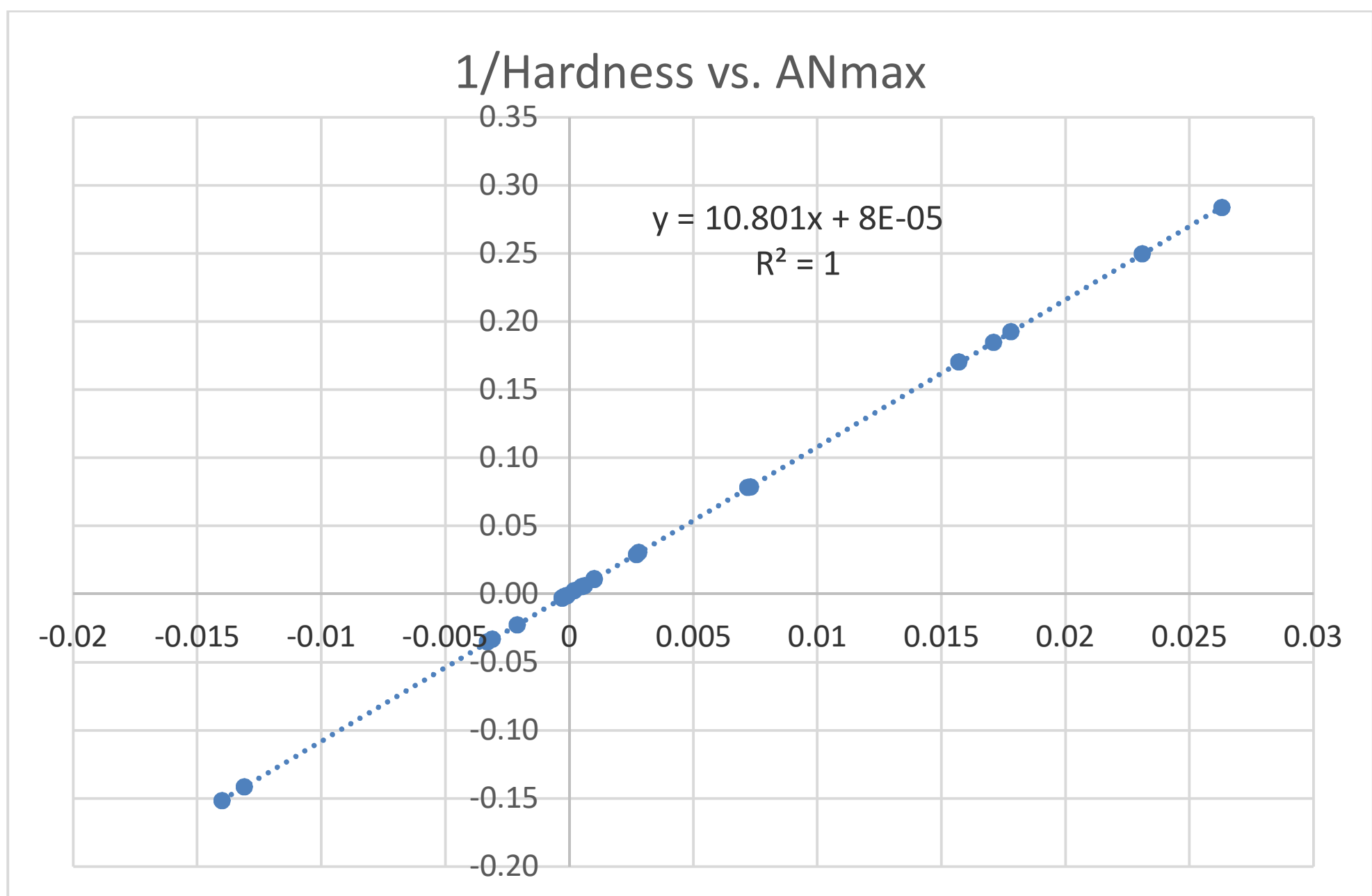

Figure S71. Lineal regression for local values $\eta_i^{-1}$ versus the maximum local charge variations for the CMOs of the $CH_2CHOCH_3$ molecule. An outlier corresponding to NBO 17 has been removed. Two outliers corresponding to CMOs 16 and 17 have been removed.

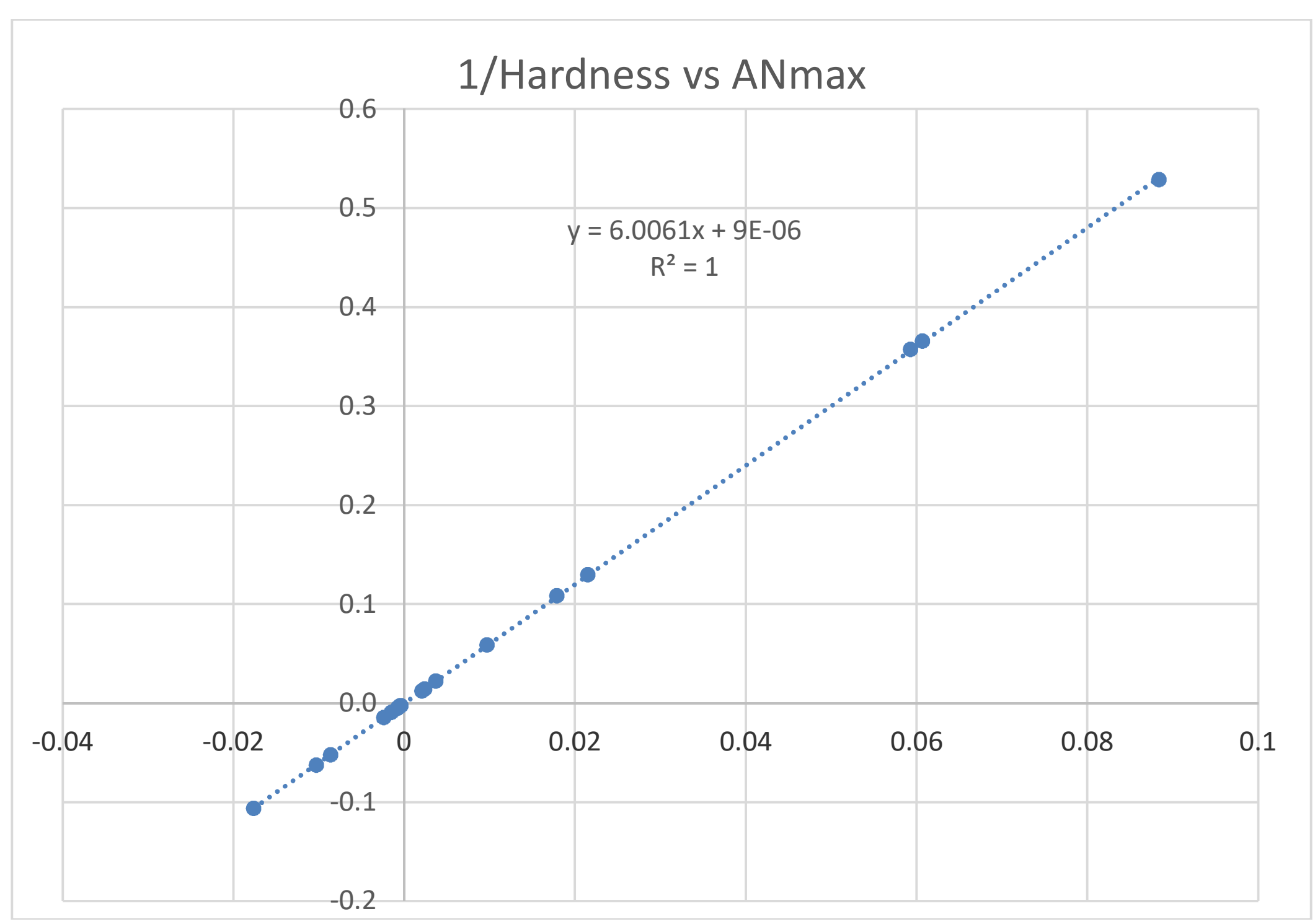

Figure S72. Lineal regression for local values $\eta_i^{-1}$ versus the maximum local charge variations for the CMOs of the $CH_2CHCHO$ molecule. Two outliers corresponding to CMOs 15 and 16 have been removed.

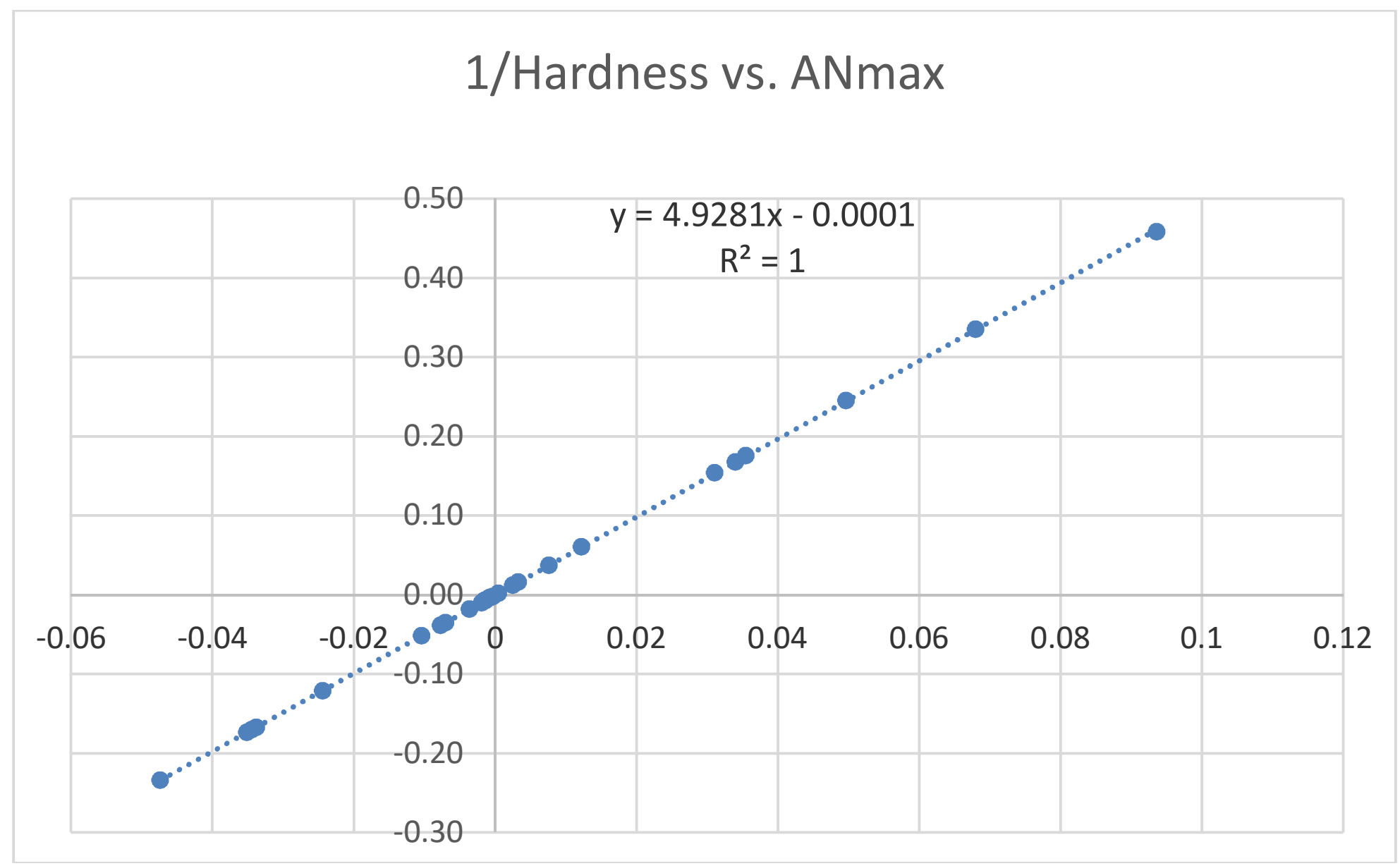

Figure S73. Lineal regression for local values $\eta_i^{-1}$ versus the maximum local charge variations for the CMOs of the $CH_2CHNO_2$ molecule. Two outliers corresponding to CMOs 19 and 20 have been removed.

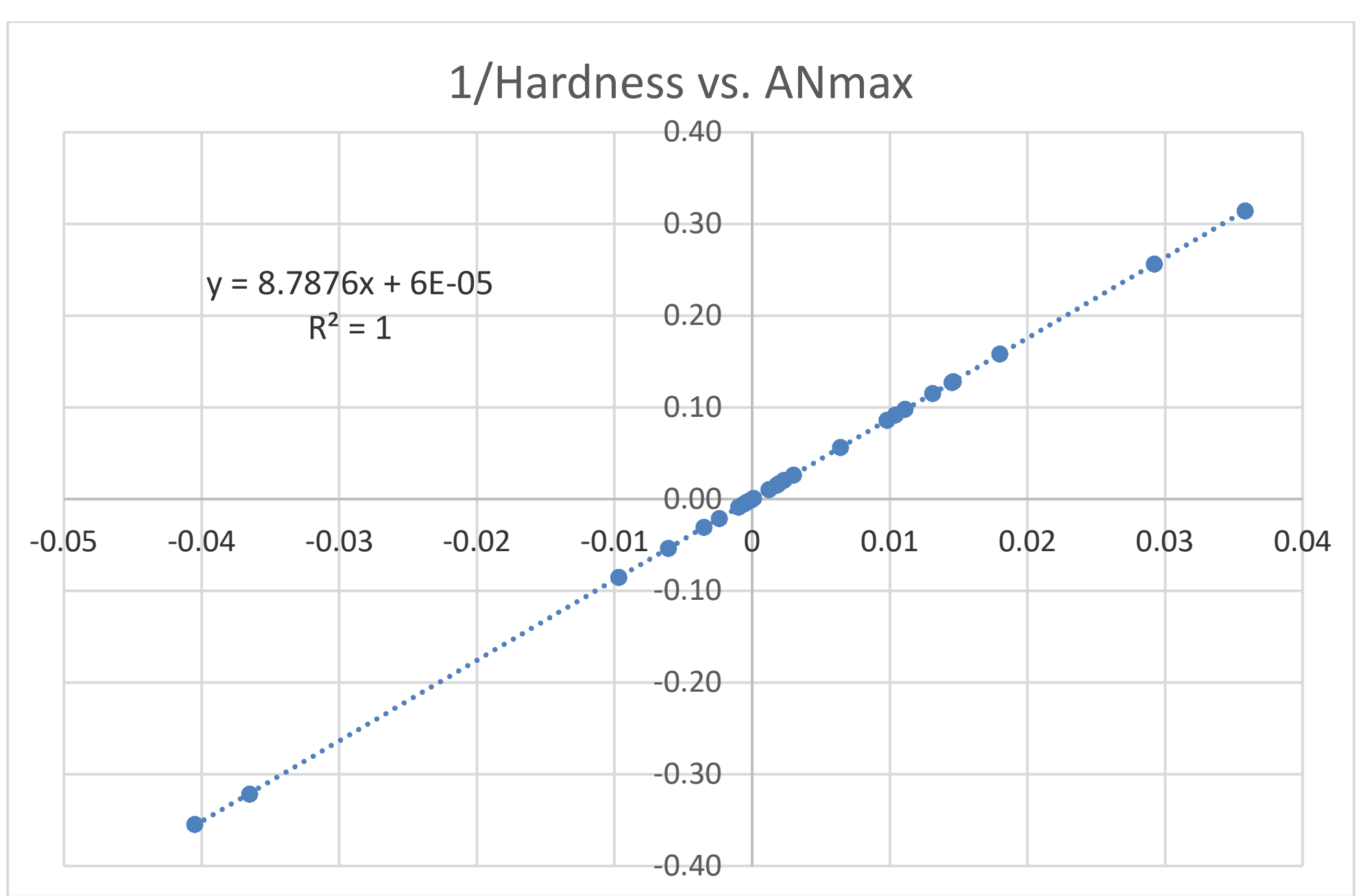

Figure S74. Lineal regression for local values $\eta_i^{-1}$ versus the maximum local charge variations for the CMOs of the $CH_3CHSH$ molecule. Two outliers corresponding to CMOs 17 and 18 have been removed.

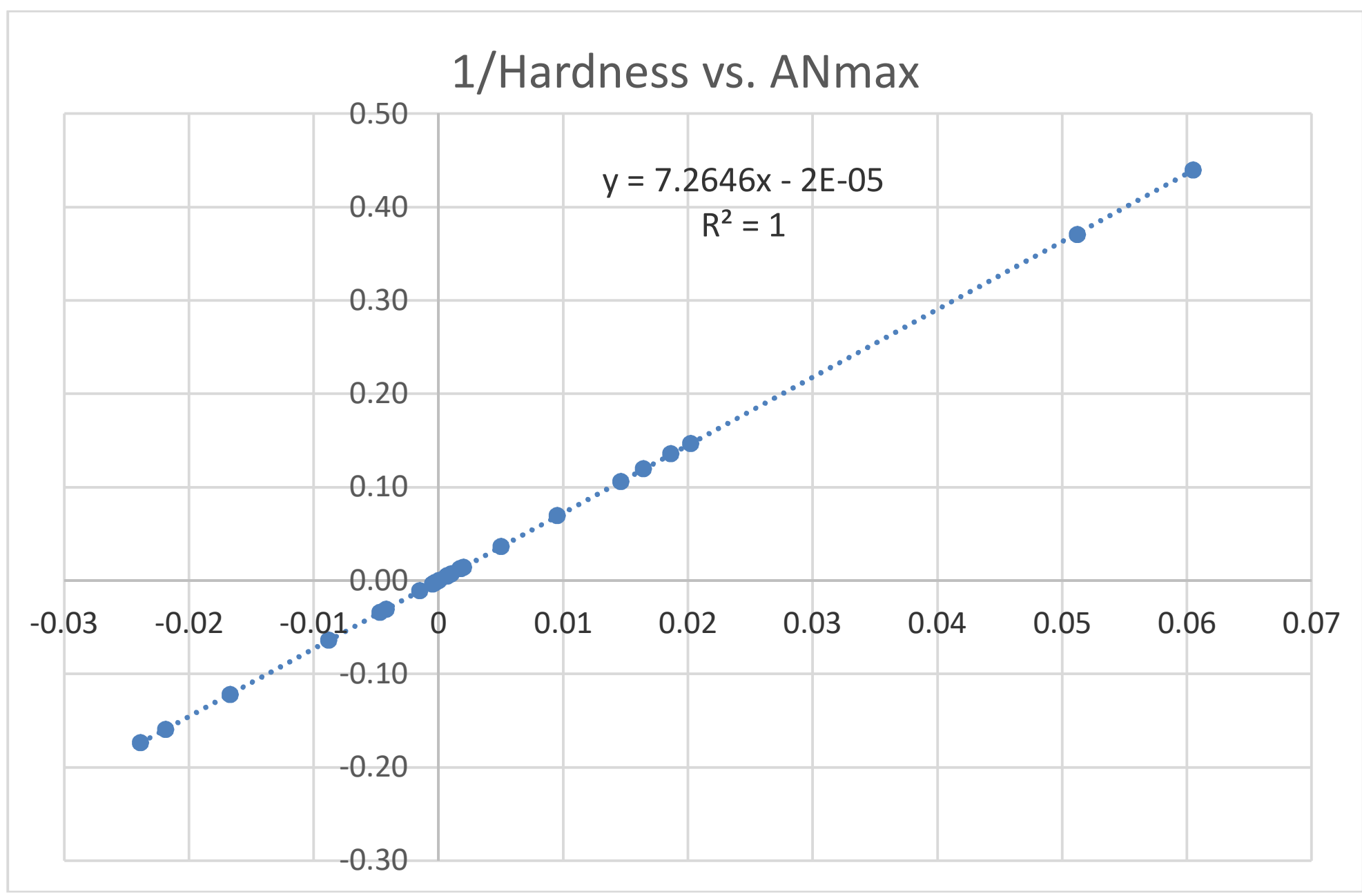

Figure S75. Lineal regression for local values $\eta_i^{-1}$ versus the maximum local charge variations for the CMOs of the $CH_3COOH$ molecule. Two outliers corresponding to NBOs 16 and 17 have been removed.

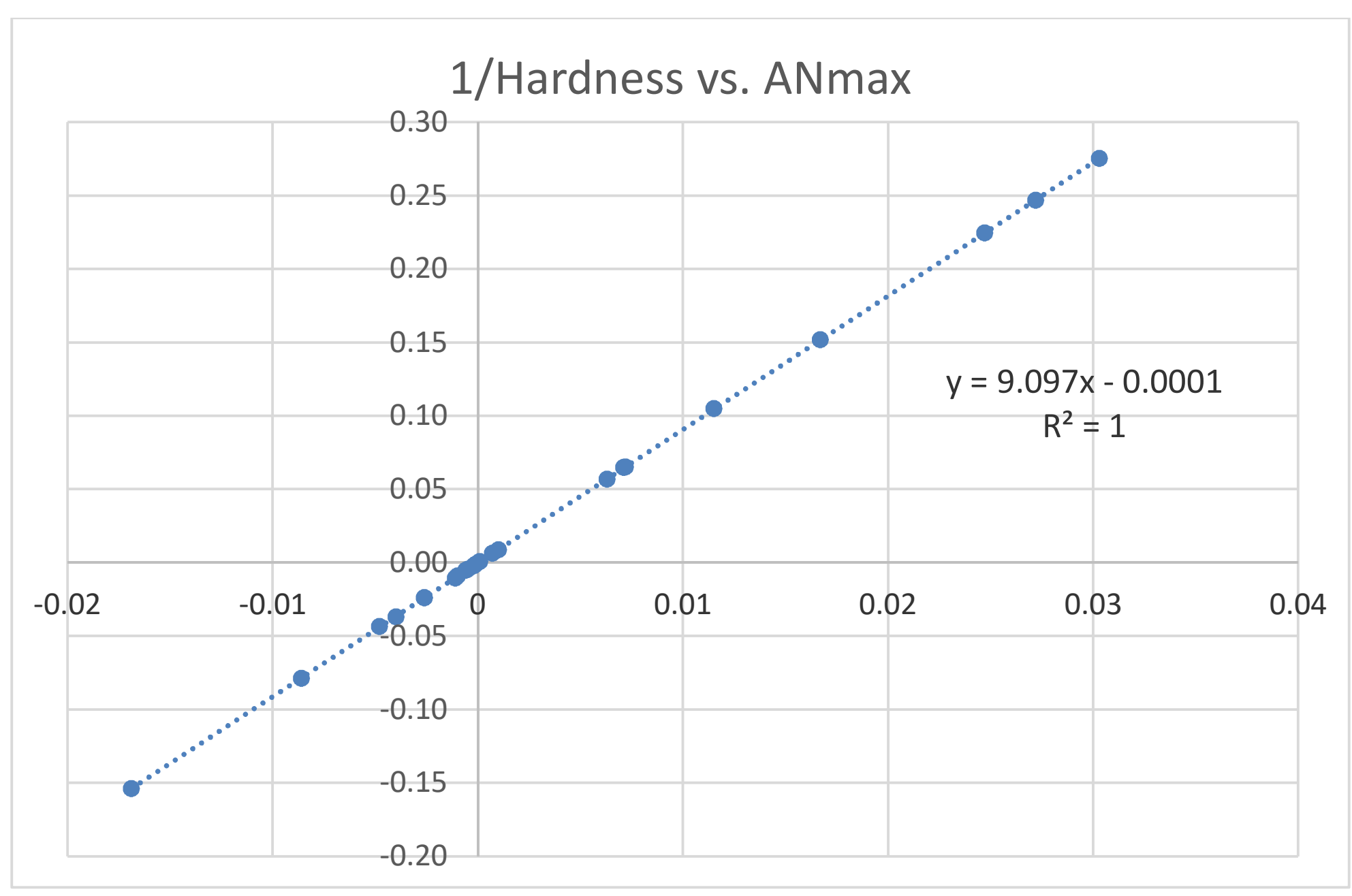

Figure S76. Lineal regression for local values $\eta_i^{-1}$ versus the maximum local charge variations for the CMOs of the $CH_3CONH_2$ molecule. Three outliers corresponding to NBOs 11, 16 and 17 have been removed.

## 10.-Tables:

| | $C_1$ | $C_2$ | $H_3$ | $H_4$ | $H_5$ | $C_6$ | $H_7$ | $H_8$ | $H_9$ |
|---|---|---|---|---|---|---|---|---|---|
| $f_i^-$ | 0.281 | 0.224 | 0.084 | 0.081 | 0.074 | 0.063 | 0.076 | 0.040 | 0.076 |
| $f_i^+$ | 0.263 | 0.223 | 0.097 | 0.094 | 0.086 | 0.057 | 0.071 | 0.038 | 0.071 |
| $f_i$ | 0.272 | 0.224 | 0.090 | 0.088 | 0.080 | 0.060 | 0.074 | 0.039 | 0.074 |
| $f_i^{(2)}$ | -0.018 | -0.002 | 0.012 | 0.013 | 0.012 | -0.006 | -0.005 | -0.002 | -0.005 |
| $\Delta N_i^{\max}$ | 0.065 | 0.054 | 0.022 | 0.021 | 0.020 | 0.014 | 0.018 | 0.009 | 0.018 |
| $\eta_i$ | 1.753 | 2.127 | 5.316 | 5.491 | 5.997 | 7.995 | 6.463 | 12.166 | 6.464 |
| $\gamma_i$ | -0.481 | -0.498 | -0.541 | -0.544 | -0.544 | -0.424 | -0.446 | -0.455 | -0.446 |
| $\Delta N_i^{\max\,\prime}$ | 0.066 | 0.054 | 0.022 | 0.021 | 0.019 | 0.014 | 0.018 | 0.009 | 0.018 |
| $\omega_i$ | 0.0037 | 0.0031 | 0.0012 | 0.0012 | 0.0011 | 0.0008 | 0.0010 | 0.0005 | 0.0010 |
| Volume (iso.:0.002) | 81.22 | 70.86 | 35.62 | 36.34 | 36.07 | 64.34 | 36.37 | 36.68 | 36.33 |
| Volume (iso.:0.001) | 97.05 | 81.54 | 47.56 | 48.13 | 48.42 | 72.99 | 48.55 | 48.27 | 48.54 |
| Volume (iso.:0.00004) | 118.38 | 94.92 | 66.83 | 66.56 | 67.15 | 83.44 | 67.52 | 65.94 | 67.58 |

Table S1. Reactivity descriptors calculated for $CH_2CHCH_3$ using the 9-element partition set, all the values are a.u.. The global values obtained were (in a.u.) $\mu$: -0.1139, $\eta$: 0.4756, $\Delta N^{\max}$ : 0.2396, and $\omega$: 0.0136.